\documentclass[11pt, openright, twoside]{report}
\usepackage{subfiles}
\usepackage[nohyperlinks]{acronym}
\usepackage{graphicx}
\usepackage[colorlinks = true, linkcolor = black, urlcolor = black, citecolor = blue]{hyperref}
\usepackage[utf8]{inputenc}
\usepackage[T1]{fontenc} 
\usepackage{geometry}
\usepackage{chngcntr}
\counterwithin{table}{section}
\counterwithin{figure}{section}
\usepackage{afterpage}
\usepackage[export]{adjustbox}
\usepackage{datetime}
\usepackage{lipsum}
\usepackage{setspace}
\usepackage{booktabs}
\usepackage{multicol}
\usepackage{multirow}
\usepackage{fontspec}
\usepackage{unicode-math}
\usepackage{tocbibind}
\usepackage{nomencl}
\usepackage{etoolbox}
\usepackage{graphicx}
\usepackage{dcolumn}
\usepackage{bm}
\usepackage{braket}
\usepackage{physics}
\usepackage{amsmath}
\usepackage{array}
\usepackage{float}
\usepackage{multirow}
\usepackage{tabularx}
\usepackage{longtable}
\usepackage{booktabs}
\usepackage[square, authoryear]{natbib}
\usepackage{tocloft}
\makenomenclature

\newdateformat{monthyeardate}{%
  \monthname[\THEMONTH], \THEYEAR}
\newcommand\myemptypage{
    \null
    \thispagestyle{empty}
    \newpage
    }
\newgeometry{
	paper=a4paper, 
        total={210mm,297mm },
	inner=3.23cm, 
	outer=1.5cm, 
	bindingoffset=.5cm, 
	top=2.5cm, 
	bottom=2.5cm, 
}
\usepackage{setspace}
\usepackage[explicit]{titlesec}  

\titleformat{\chapter}[display]
{\fontsize{27}{27}\sffamily\bfseries\filleft}{\hfil \thechapter}{}{{#1}}  

\titleformat{\section}{\Large\bfseries\sffamily}{\thesection}{0.5em}{\MakeUppercase{#1}}
\titleformat{\subsection}{\large\bfseries\sffamily}{\thesubsection}{0.5em}{\MakeUppercase{#1}}
\titleformat{\subsubsection}{\sffamily}{\thesubsubsection}{0.5em}{\MakeUppercase{#1}}

\newcommand{\ttitle}{Discrete and Continuous Wigner Functions in Open Quantum Systems: Non-Markovian and Thermodynamic Effects}
 
\newcommand{\degree}{Doctor of Philosophy (Ph.D.)}
\newcommand{\authorname}{Jai Lalita}

\newcommand{\deptname}{Department of Physics}

\begin{document}

\setmainfont[ Path = font/,
 BoldFont={CandaraBI}, 
 ItalicFont={Candarai},
 BoldItalicFont={CandaraB}
 ]{Candarai}
 
\setstretch{1}
\begin{titlepage}
\begin{center}
\begin{flushright}
\Huge{\textbf{{\ttitle}}}
\end{flushright}
\vfill
\begin{flushright}
\Large{\textit{A thesis submitted by}}\\
\huge{\textbf{\authorname}}
\end{flushright}
\vfill
\begin{flushright}
\Large{\textit{in partial fulfillment of the requirements for the award of the degree of}}\\
\huge{\textbf{\degree}}
\end{flushright}
\vfill
\includegraphics[width=0.40\textwidth,right]{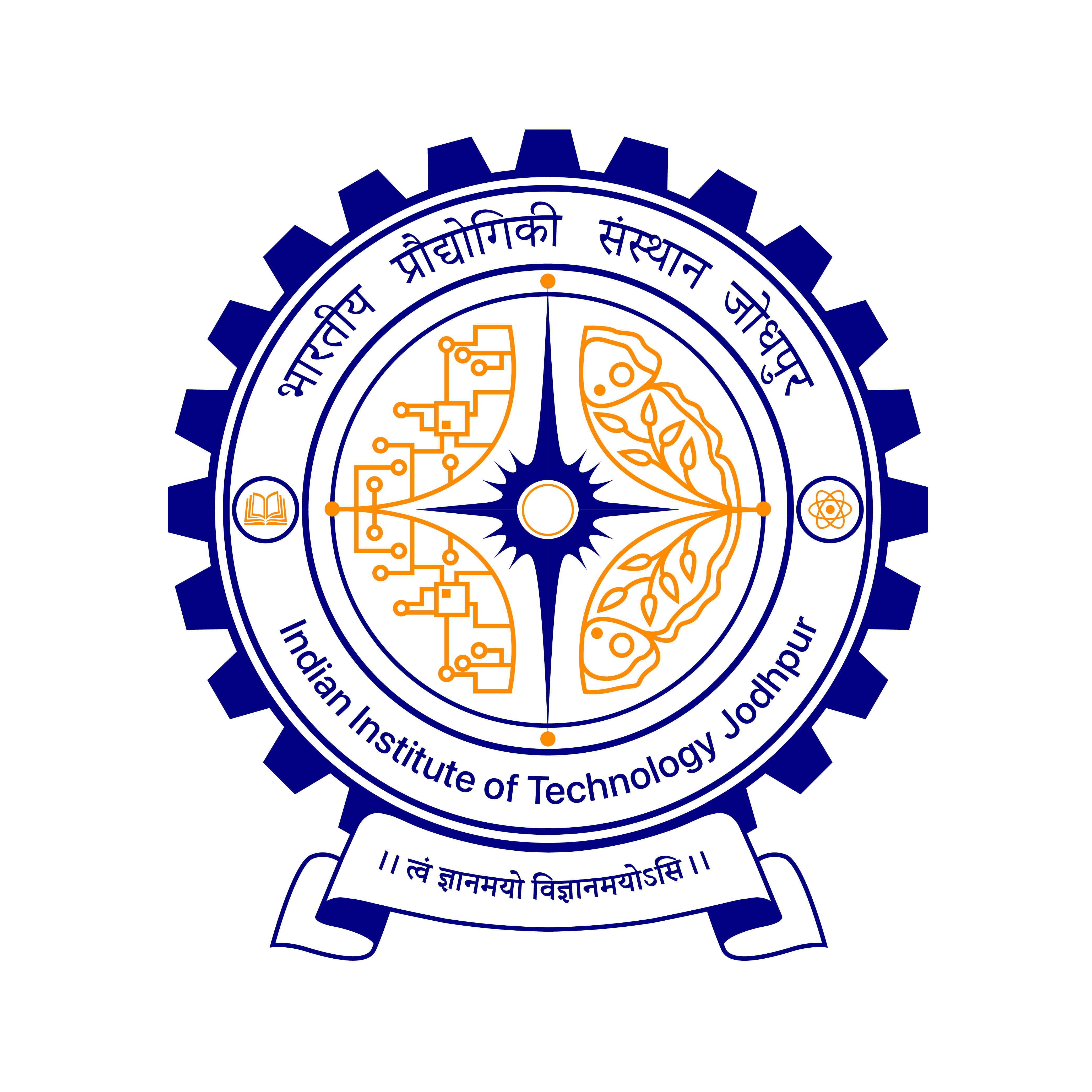}
\begin{flushright}
\Large{\textbf{Indian Institute of Technology Jodhpur}}\\
\Large{\textbf{\deptname}}\\ 
\Large{\emph{\monthyeardate\today}}
\end{flushright}

 
\end{center}
\end{titlepage}

\myemptypage
\pagenumbering{roman}
\setcounter{page}{3}
\setmainfont[ Path = font/,
 BoldFont={CandaraBI}, 
 ItalicFont={booki},
 BoldItalicFont={bookbi}
 ]{bookr}

\begin{flushright}
\huge{\textbf{Abstract}}
\end{flushright}
\addcontentsline{toc}{chapter}{Abstract}


The central aim of the thesis is to examine how non-classical resources in finite-dimensional quantum systems can be identified, characterized, and protected for practical use in the presence of realistic noise. Using the discrete Wigner functions (DWFs) framework, we introduce negative quantum states and examine how their Wigner negativity, mana, coherence, and teleportation fidelity evolve under unital and non-unital channels, with particular attention to non-Markovian random-telegraph and amplitude-damping dynamics. We also analyze protection strategies based on weak measurement and quantum measurement reversal, showing that these methods can enhance quantum correlations, reduce fidelity deviation, and improve teleportation performance for two-qubit negative states in memory-bearing environments. Moreover, we demonstrate that certain negative states, derived from phase-space point operators, exhibit greater resilience than Bell states in measures of entanglement under non-Markovian noise. Further, this thesis focuses on developing and implementing quantum circuits for generating these states on superconducting hardware, and realizing them for the first time on IBM’s \emph{ibm\_brisbane} device. Their preparation is verified using quantum state tomography, demonstrating high fidelity under realistic noise conditions. We propose a teleportation scheme that leverages one of the two-qubit negative quantum states as a resource. Moreover, these two-qubit negative quantum states are also found to perform better than the Bell states for maximal CHSH violation and Fisher information in noisy conditions. We believe that these negative quantum states have the potential to be used in place of the traditional Bell states in scenarios where non-Markovian errors are prevalent. 

The thesis further investigates a two-qubit quantum system in contact with an environment modeled by a microscopic collision model with added ancilla-ancilla collisions in the non-Markovian regime. Two schemes of the two-qubit collision model with carried-forward correlations are introduced. In one scheme, a single stream of ancillae interacts with only one of the qubits of the two-qubit system; in the other, both the qubits interact with two independent sequences of ancillae, which could be at the same or different temperatures. The system's non-Markovian evolution is examined using the trace distance measure, and the non-classicality of the system is studied using the Wigner function, non-classical volume, and concurrence. Also, interesting steady-state behavior is observed when both the independent ancillae are kept at the same temperature.

Finally, the interplay of non-classical volume, von Neumann entropy, entropy production, and ergotropy is explored in various open quantum systems. Two categories of open quantum system models are utilized to study this interplay: spin-spin and spin-boson interaction models. The spin-spin interaction models include the quantum collision and central spin models. On the other hand, the spin-boson interaction models consist of the non-Markovian amplitude damping channel, the Markovian generalized amplitude damping channel, and the Jaynes-Cummings model. Across these various open quantum systems, universal interrelations emerge, where the non-classical volume shows contrasting evolution with entropy, and entropy production contrasts with ergotropy. The initial state of the reservoir in these open quantum systems is shown to have an impact on these interrelations. These findings establish an interesting link between quantum information and the thermodynamics of open quantum systems. Overall, by bridging phase-space methods, quantum information tasks, experimental implementation, and thermodynamic considerations, the work advances both the foundational understanding and practical deployment of non-classical resources beyond conventional Bell-state paradigms.

\clearpage
\myemptypage
\newpage
\thispagestyle{empty}
\chapter*{\centerline{}  }

\vspace*{1in}

\textit{\Large This thesis is dedicated to the loving memory of my grandmother, Smt. Bhanmati Devi, and to my parents, Sh. Satish Kumar and Smt. Anita Devi, in gratitude for their unwavering support and encouragement, and to the reader.}\\\\

\myemptypage 
\begin{flushright}
\tableofcontents
\clearpage
\myemptypage
\listoffigures
\clearpage
\myemptypage
\listoftables
\clearpage
\myemptypage
\renewcommand{\acsfont}[1]{\textsc{#1}}

\setlength{\columnsep}{4pc}
\setstretch{0.5}
\begin{flushright}
\Huge{\textbf{List of Abbreviations}}
\end{flushright}
\addcontentsline{toc}{chapter}{List of Abbreviations}

\begin{acronym}[TDMA] 
\setstretch{0.2}
\acro{GKSL}{Gorini-Kossakowski-Sudarshan-Lindblad}
\acro{OQS}{Open Quantum Systems}
\acro{CP}{Completely Positive}
\acro{CPTP}{Completely Positive Trace Preserving}
\acro{QDs}{Quasi-probability Distributions}
\acro{CV}{Continuous Variable}
\acro{DWFs}{Discrete Wigner Functions}
\acro{MUBs}{Mutually Unbiased Bases}
\acro{EPR}{Einstein-Podolsky-Rosen}
\acro{NS}{Negative Quantum State}
\acro{AD}{Amplitude Damping}
\acro{NMAD}{non-Markovian Amplitude Damping}
\acro{GAD}{Generalized Amplitude Damping}
\acro{JCM}{Jaynes-Cummings Model}
\acro{RTN}{Random Telegraph noise}
\acro{WM}{Weak Measurement}
\acro{QMR}{Quantum Measurement Reversal}
\acro{QT}{Quantum Teleportation}
\acro{UQT}{Universal Quantum Teleportation}
\acro{CHSH}{Clauser-Horne-Shimony-Holt}
\acro{QFI}{Quantum Fisher Information}
\acro{BLP}{Breuer-Laine-Pillo}


\end{acronym}
  
\myemptypage
\end{flushright}

\clearpage
\pagenumbering{arabic}
\setcounter{page}{1} 


\newpage
\setcounter{chapter}{0} 

\titleformat{\chapter}[display]
{\fontsize{27}{27}\bfseries\filleft}{ \thechapter}{0pt}{{#1}}  

\thispagestyle{empty}

\chapter{Introduction}\label{chap1:introduction}

Quantum technologies rely critically on the preservation of non-classical phenomena and quantum correlations~\citep{bennett1993teleporting, bouwmeester1997experimental, adhikari2012operational, masanes2011secure, nielsen2010quantum, giovannetti2011advances, thapliyal2017quantum}, yet any physical system unavoidably interacts with an environment that degrades these features~\citep{nielsen2010quantum, breuer2002theory, Banerjee2018, weiss2012quantum, czerwinski2022dynamics}. Understanding how memory effects, environmental structure, and system–bath correlations influence the dynamics of quantum states is therefore essential for advancing quantum communication, computation, and quantum thermodynamics~\citep{Ekert1991Quantumcryptography, Scarani2009The_security, Quantumerrorcorrection2013, Pezz2018Quantummetrology, Braun2018Quantum_enhanced, arisoy2019thermalization, Landi2021Irreversibleentropy}. The fundamental basis for examining how an environment influences a quantum system is provided by the theory of open quantum systems, where one examines the evolution of a system’s reduced density matrix after tracing over environmental variables~\citep{breuer2002theory, Banerjee2018}. 

Historically, the study of quantum dynamics began with closed systems governed by unitary evolution. However, the recognition that no quantum system is truly isolated led to the development of the theory of open quantum systems.
The formal description typically follows either master equations or completely positive trace-preserving (CPTP) maps. Master equations such as the Gorini–Kossakowski–Sudarshan–Lindblad (GKSL) form encode Markovian dynamics, where the environment has no memory and loses any information it extracts from the system. However, in many physical settings, such as solid-state architectures, structured reservoirs, biological complexes, and engineered quantum devices, environmental correlations persist~\citep{nielsen2010quantum, breuer2002theory, Banerjee2018, weiss2012quantum}. In such cases, the system’s past influences its future, leading to non-Markovian behavior characterized by information backflow, memory kernels, and deviations from the CP-divisibility principle. Understanding the distinctions between these dynamical regimes is crucial for predicting when entanglement or coherence is likely to persist, revive, or irreversibly disappear~\citep{de2017dynamics, rivas2014quantum, li2018concepts, vacchini2012classical, Breuer2016Colloquium, daffer2004depolarizing, kumar2018non, utagi2020ping, Naikoo2019Facets, utagi2020temporal, Thomas2018thermodynamics, paulson2021hierarchy, Campbell2018Systemenvironment, Rodrigues2019thermodynamics}. To capture this rich behavior, the thesis employs several canonical noise models. Amplitude damping channels describe energy relaxation and spontaneous emission, and both Markovian and non-Markovian variants are used to highlight the impact of environmental memory on the quantum correlations~\citep{Bellomo2007NMAD, Naikoo2019Facets}. Random telegraph noise accounts for stochastic phase fluctuations with a tunable memory parameter, allowing transitions between noise regimes~\citep{daffer2004depolarizing, kumar2018non}. Additional channels, such as depolarizing and generalized amplitude damping, provide benchmarks for hardware-level noise and finite-temperature environments~\citep{nielsen2010quantum, Breuer2007, Srikanth2008Squeezed, omkar2013dissipative}. Complementing these models, collision models serve as microscopic descriptions in which sequential interactions between the system and environmental ancillae naturally generate tunable memory effects when ancilla–ancilla correlations are introduced~\citep{ThermalizingQuantumMachines_2002, ziman2005description, Rybár_2012, McCloskey2014Non-Markovianity, Ciccarello2013Collision-model, Kretschmer2016Collisionmodel, Saha_2024_quantum, Li2024Witnessing}. 

Further, to advance quantum communication and computation, tackling the degradation of quantum correlations is crucial. For this, quantum measurements, $\textit{i.e.}$, weak measurement (WM), and quantum measurement reversal (QMR) have been used. The WM extracts information from the system without causing it to collapse into an eigenstate, unlike the conventional von Neumann measurement \citep{aharonov1988result, oreshkov2005weak, korotkov2006undoing, katz2008reversal, kim2009reversing, korotkov2010decoherence, kim2012protecting, dressel2014colloquium, lahiri2021exploring, sabale2023towards}. A proper QMR can reconstruct the state with a certain probability, also called the success probability \citep{pramanik2013improving, he2020enhancing}. Therefore, it has been shown that the WM and QMR can enhance and protect the quantum correlations of the qubit and qutrit quantum systems from the effects of noise, in particular, non-Markovian noise \citep{korotkov2010decoherence, kim2012protecting, xiao2013protecting, sun2017recovering}. Further, these quantum measurements can also protect and elevate universal quantum teleportation (UQT) requirements for two-qubit quantum systems \citep{horodecki1996teleportation, badziag2000local, bang2018fidelity, ghosal2020optimal}. Additionally, using photonic and superconducting quantum systems, WM and QMR can be implemented experimentally \citep{monroe2021weak, katz2008reversal, kim2009reversing}.

Along with them, the notion of phase space is essential in investigating classical systems' dynamics. The uncertainty principle, however, limits its straightforward application to the quantum scenario. Nevertheless, it is still conceivable to create quasi-probability distributions (QDs) for quantum mechanical systems analogous to their classical counterparts. Wigner created the first QD, now known as the Wigner function ($W(q, p)$) \citep{Wigner1932Quantum, hillery1984distribution}. It is not only real-valued and normalized but also gives the correct value of the probability density for the quadrature $a{Q} + b{P}$ (here ${Q}$ and ${P}$ are canonical positions and momentum operators) when integrated along the phase-space line $aq + bp$. However, unlike probability densities, the Wigner function can assume negative values for some quantum states; thus making it a quasi-probability distribution function. Classical light states, like coherent states, have positive Wigner functions \citep{hudson1974wigner}, whereas quantum light states, like photon-added/subtracted coherent and entangled states, do not \citep{zavatta2004quantum, malpani2020impact}. The appearance of negative regions in the Wigner function is a hallmark of genuinely quantum behavior~\citep{kenfack2004negativity}. Moreover, through homodyne measurements, Wigner functions can be experimentally recreated, and the visual representation of the recreated state effectively highlights quantum interference processes~\citep{leonhardt1997measuring}.

The original Wigner function only applies to the continuous situation, whereas density operators can also express discrete degrees of freedom like spin. Therefore, given the significance of Wigner functions for continuous variable (CV) systems~\citep{agarwal1981relation, agarwal1998state, Thapliyal2015Quasiprobability, Thapliyal2016tomograms}, much emphasis has been paid to creating their finite-dimensional analogues as we generally deal
with finite-dimensional Hilbert space systems in quantum information and processing. For example, for a system of $n$ qubits, the dimension of the Hilbert space of states is $d = 2^n$. For such systems, various discrete analogues of the Wigner function have been proposed~\citep{cohen1986joint, wootters1987wigner, galetti1988extended, leonhardt1996discrete, wootters2004picturing, gibbons2004discrete, chaturvedi2005wigner}. The Wigner function formulation applied to arbitrary spin systems of prime dimensions was developed in~\citep{wootters1987wigner}. It is defined on an explicitly geometrical phase space over a finite mathematical field. The integers $(0,..., d-1)$, with addition and multiplication mod $d$, make up the finite mathematical field (where $d$ is the dimension of the system's Hilbert space). Later, this formulation has been reformulated for the power of the prime dimension Hilbert space~\citep{wootters2004picturing, gibbons2004discrete}. A tomographical scheme was proposed to infer the quantum states of finite—dimensional systems from experiments by developing a new discrete Wigner formalism~\citep{leonhardt1996discrete}. An algebraic approach was also provided to find the Wigner distributions for finite odd-dimensional quantum systems~\citep{chaturvedi2005wigner}. Withal, discrete Wigner functions (DWFs) have been used to investigate a variety of exciting problems connected with quantum computation, such as magic state distillation~\citep{howard2014contextuality, veitch2014resource, schmid2022uniqueness}, separability~\citep{pittenger2005wigner}, quantum state tomography~\citep{wootters2004picturing, paz2004quantum}, teleportation~\citep{koniorczyk2001wigner, paz2002discrete}, decoherence~\citep{lopez2003phase}, and error correction~\citep{paz2005qubits}. 

Here we focus on the class of DWFs defined in~\citep{wootters2004picturing, gibbons2004discrete} for power-of-prime dimensions. This class defines DWFs by associating lines in discrete phase space to projectors belonging to a fixed set of mutually unbiased bases (MUBs)~\citep{wootters1989MUB, Lawrence2002MUB, bandyopadhyay2002MUB, pittenger2004mutually}. The advantage of this formulation is that, here, DWFs transparently correspond to the expectation values of the phase-space point operators, detailed in chapter~\ref{chap2:Preliminaries}. The method of extremizing DWFs was contemplated by finding states corresponding to the eigenstates of the minimum and maximum eigenvalues of the phase space point operator~\citep{casaccino2008extrema}. Thereafter, this idea was extended for all odd prime dimensions to find the maximally negative quantum states~\citep{van2011noise}, $\textit{i.e.}$, states corresponding to a minimal eigenvalue of the phase space point operator's eigenstate. It has been shown that these states are maximally robust against depolarizing noise. In the present thesis, we introduce negative quantum states generated from the spectra of phase-space point operators for the Hilbert space of dimension $d = 2, 3, 2^2$~\citep{lalita2023harnessing}, i.e., for qubit, qutrit, and two-qubit quantum systems. Their construction follows naturally from the algebraic properties of phase-space point operators defining the discrete Wigner representation~\citep{gibbons2004discrete}. The eigenstates associated with negative eigenvalues inherit intrinsic non-classical features, and the state corresponding to the most negative eigenvalue is identified as the maximally negative, or “first negative,” quantum state ($\ket{NS_1}$ state)~\citep{lalita2023harnessing}. With the motivation to understand the impact of noise on the DWFs under the action of (non)-Markovian, non-unital (depicted by the amplitude damping noise) as well as unital (illustrated by the random telegraph noise), we calculate the DWFs for the qubit, qutrit, and two-qubit quantum systems. We examine how the DWFs vary with time under the impact of both unital and non-unital noisy channels, in the (non)-Markovian regimes. In particular, we study the variation of DWFs corresponding to the maximally negative quantum state, $\textit{i.e.}$, the $\ket{NS_{1}}$ state of the single-qubit, single-qutrit, and two-qubit systems with time under the (non)-Markovian amplitude damping and random telegraph noise. Further, we investigate how the discrete Wigner negativity $|N_{G}(\pmb{\rho})|$ of a maximally negative quantum state of a qubit, qutrit, and two-qubit varies with time under the impact of both (non)-Markovian unital and non-unital noisy channels. A concept connected with the states having negative discrete Wigner functions is the mana~\citep{veitch2014resource}, which has been used in the literature to compute magic associated with non-stabilizer states. Magic states are found to be ideal resources for quantum computational speedup and fault-tolerant quantum computation. Here, we compute the mana of the qutrit's first and second negative quantum states~\citep{jain2020qutritmagic}. We also study the variation of mana with time under the aforementioned (non)-Markovian channels. Moreover, as we know, quantum coherence is a fundamental prerequisite for all quantum correlations, including entanglement, and it is a crucial physical resource in quantum computation and information processing~\citep{baumgratz2014quantifying, xi2015quantum, streltsov2017colloquium, hu2018quantum, zhao20191,paulson2022quantum}. Also, entanglement is a premium quantum correlation with numerous operational applications. We will study the dynamics of quantum coherence and entanglement, \textit{i.e.}, concurrence~\citep{Wootters1998Entanglement}, utilizing two-qubits' first, second, and third negative quantum states using DWFs and compare them with the corresponding dynamical evolution of the Bell state. Further, the average fidelity is commonly used to gauge a channel's performance~\citep{horodecki1996teleportation, ghosal2021characterizing}. We will use DWFs to compare the average fidelity of the two-qubit system's first, second, and third negative quantum states with the Bell states when subjected to the above-specified (non)-Markovian noise channels. 

To a greater extent, in order to find the suitable two-qubit quantum states, apart from the Bell states, for universal quantum teleportation (UQT), introduced below, we study the impact of weak measurement (WM) and quantum measurement reversal (QMR)~\citep{korotkov2006undoing, katz2008reversal, kim2009reversing, korotkov2010decoherence, kim2012protecting, sabale2023towards} on the quantum correlations \citep{Wootters1998Entanglement, horodecki2009quantum, ollivier2001quantum, henderson2001classical, bennett1999quantum, schrodinger1935discussion, schrodinger1936probability, brunner2014bell, fan2022quantum, costa2016quantification} and UQT requirements \citep{horodecki1996teleportation, badziag2000local, bang2018fidelity, ghosal2020optimal} of the negative quantum states of two-qubit systems proposed in~\citep{lalita2023harnessing}. For quantum teleportation~\citep{gisin2007quantum, jin2010experimental, zeilinger2018quantum, jozsa1993teleporting, bennett1993teleporting, boschi1998experimental}, the average fidelity is the standard figure of merit~\citep{horodecki1996teleportation, badziag2000local}. It is the average overlap or closeness between Bob's input state and Charlie's received output state, calculated by averaging across all potential input states. In reality, all quantum systems are open and constantly interact with their immediate environment, which causes the degradation of non-local correlations~\citep{Wootters1998Entanglement, horodecki2009quantum, ollivier2001quantum, henderson2001classical, bennett1999quantum, schrodinger1935discussion, schrodinger1936probability, brunner2014bell, fan2022quantum, costa2016quantification, Chakrabarty2010study, luo2008quantum, ramkarthik2020quantum} and, hence, average fidelity. Moreover, that introduces variance in average fidelity values for various possible input states. So, to further explore the effect of noise on quantum teleportation, we employ fidelity deviation in addition to average fidelity~\citep{bang2018fidelity}. The highest achievable average fidelity value across all feasible local unitary operations in the standard teleportation protocol is the maximal fidelity~\citep{horodecki1996teleportation}. The goal is to minimize fidelity deviation while maintaining the highest feasible average fidelity~\citep{ghosal2020optimal}. So, for quantum teleportation, a suitable state is any two-qubit entangled state where the maximal fidelity is strictly greater than the classical constraint~\citep{horodecki1996teleportation, horodecki1999general}. On the contrary, if and only if a state exhibits zero fidelity deviation, it is considered useful for universal quantum teleportation (UQT)~\citep{ghosal2021characterizing}. A central driving element for realizing universal quantum teleportation was its prospective application in the technological advancement of communication~\citep{gisin2007quantum, jin2010experimental, zeilinger2018quantum, jozsa1993teleporting, bennett1993teleporting, boschi1998experimental}. We compare the variations of quantum correlations, maximal fidelity, fidelity deviation, and success probability of two-qubit negative quantum states with that of the maximally correlated state, i.e., the Bell $\ket{\phi^{+}}$ state, in the presence and absence of WM and QMR. The influence of both non-Markovian non-unital (specified by non-Markovian amplitude damping (AD)) and unital  (specified by non-Markovian random telegraph noise (RTN)) channels is taken into consideration.

As the quantum states that exhibit negativity of the discrete Wigner function have been identified to be robust to a wide variety of noise in quantum systems captured by non-Markovian errors~\citep{lalita2023harnessing, Lalita_2024ProtectingQC}. These two-qubit negative quantum states are also recognized as optimal candidates for universal quantum teleportation using weak measurements~\citep{Lalita_2024ProtectingQC} within non-Markovian noise environments. Consequently, the physical realization of these two-qubit negative quantum states is imperative, as they hold the potential to augment quantum information protocols by providing resilient resources. In this thesis, we present methods to generate these states using operations native to superconducting hardware, specifically single-qubit gates and the $CZ$ gate. Through quantum state tomography, we demonstrate that these states can be prepared with high fidelity on IBM’s quantum hardware under realistic noise conditions. We employ various approaches to assess the noise resilience of the negative quantum states and benchmark them against the standard Bell states. These approaches include fidelity variation analysis, phase sensitivity under $SU(2)$ rotations using quantum Fisher information (QFI)~\citep{Bollinger1996Optimalfrequency,peters1999measurement}, violations of the Bell-CHSH inequality~\citep{HORODECKI1995340}, and quantum information measures such as concurrence and teleportation fidelity. Our results highlight the role of negative quantum states as robust resources for entanglement in various quantum computing and communication applications. Furthermore, we propose a teleportation circuit that utilizes one of the two-qubit negative quantum states as its entanglement resource.

Furthermore, beyond conventional phenomenological noise models, collision models offer a fully microscopic and operational description of open quantum system dynamics. Additionally, the mathematical framework for describing open quantum systems can become cumbersome, both in deriving and solving the equations that regulate the system's dynamics. An alternate method that is easy to understand, adaptable, and simple for tracking the time evolution of the open system is the collision model approach~\citep{CM_1963, Ciccarello2022Quantumcollisionmodel}. Collision models have been extensively used in recent years to tackle various problems like decoherence, dissipation, and non-Markovian behavior in a controlled and tractable way~\citep{ziman2005description, Rybár_2012, Ciccarello2013Collision-model, McCloskey2014Non-Markovianity, Kretschmer2016Collisionmodel, CM_nm_2017, Campbell2018Systemenvironment, Rodrigues2019thermodynamics, Landi2021Irreversibleentropy, csenyacsa2022entropy}. They have also found applications in quantum thermodynamics for modeling thermalization and heat exchange processes~\citep{Strasberg2017Quantum, Pezzutto2016Implications, Campbell2021Collision, DeChiara_2018Reconciliation}. In this thesis, we examine such a two-qubit collision model by simulating a sequence of ``collisions" between the two-qubit system and environmental ancillae to describe the system-environment interaction in a controllable manner. Depending on how the environmental degrees of freedom interact, such a ``collision" model can simulate both Markovian and non-Markovian dynamics~\citep{ThermalizingQuantumMachines_2002, ziman2005description, Rybár_2012, McCloskey2014Non-Markovianity, Ciccarello2013Collision-model, Kretschmer2016Collisionmodel, Saha_2024_quantum, Li2024Witnessing}. We also introduce two different schemes to investigate the effects of manipulating the interactions between environmental ancillae and the two-qubit system on the non-Markovianity and non-classicality of the two-qubit system. For that, we study trace distance, Wigner function, non-classical volume, and dynamics of entanglement generation of the two-qubit system using both schemes. The trace distance is diagnostic for non-Markovianity witnessed by information flow, while the Wigner function and non-classical volume reveal phase-space signatures of quantum behavior and deviations from classicality~\citep{Wigner1932Quantum, zavatta2004quantum, Thapliyal2015Quasiprobability}. Moreover, entanglement generation with the number of collisions allows the identification of regimes where quantum correlations are enhanced or degraded by the environment~\citep{Naikoo2019Facets, Tiwari2023QuantumCorrelations}. We also investigate the steady-state behavior under varying physical conditions for potential thermodynamic applications~\citep{Hanggi_talkner_review, tiwari2024strong, Thomas2018thermodynamics, Ashutosh2023thermodynamics}.

Moreover, the two-qubit collision model provides a more comprehensive framework than its single-qubit counterpart for simulating open quantum dynamics, as it captures both inter-qubit correlations and system-environment interactions. Unlike single-qubit models, which are restricted to individual decoherence processes, the two-qubit setting enables the study of entanglement dynamics, collective behavior, and environment-induced correlations between system qubits, which are key features in quantum communication and computation protocols. These advantages make the two-qubit collision model particularly suitable for exploring realistic quantum systems, where the coherence and interaction of multi-qubit systems with the environment are central.

Further, a comprehensive characterization of open-system dynamics aims to bridge the informational and thermodynamic viewpoints. Four quantities, non-classical volume, von Neumann entropy, entropy production, and ergotropy, jointly provide complementary insights into this connection. The non-classical volume quantifies coherence and superposition~\citep{Anatole2004Negativity}; the von Neumann entropy characterizes statistical uncertainty and information loss~\citep{nielsen2010quantum}; entropy production captures the degree of irreversibility~\citep{Esposito2010Threefaces}; and ergotropy represents the extractable work through cyclic unitary operations~\citep{AEAllahverdyan_2004Maximalwork}. Together, these quantities elucidate how quantum resources transform into thermodynamic quantities. Recent investigations in quantum information, quantum thermodynamics, and quantum batteries have further highlighted the intricate connections between them ~\citep{Manfredi2000Entropy, Perarnau2015Extractable, Francica2020QuantumCoherence, Medina2025Anomalous, pathania2025quantum}. Building upon these insights, this thesis systematically explores the interplay among non-classicality, von Neumann entropy, entropy production, and ergotropy across both spin–spin and spin–boson interaction models. By comparing the collision and central spin models, the non-Markovian amplitude damping and Markovian generalized amplitude damping channels, and the Jaynes–Cummings model, this work aims to uncover universal correspondences linking non-classicality, irreversibility, and extractable work. Although the inverse behavior of non-classical volume and von Neumann entropy might appear intuitively expected, there is no a priori guarantee that this trend persists across non-Markovian memory effects, thermal vs. non-thermal reservoirs, or strong-coupling regimes. We aim to demonstrate its robustness across five fundamentally different models and establish it as a structural feature of open-system evolution. Further, the comparison between ergotropy (a state function) and entropy production (a process-dependent quantity) is not intended as a quantitative relation. Rather, our motivation is to test whether a qualitative thermodynamic correspondence exists, namely, whether states that experience larger irreversible information loss during evolution tend to exhibit systematically lower extractable work, irrespective of the microscopic mechanism generating irreversibility. Such an investigation contributes toward a deeper understanding of how environmental structure, memory effects, and system–bath coupling jointly determine open quantum systems' informational and thermodynamic evolution.

\subsubsection{Organization of thesis}

This thesis is organized to progress systematically from foundational concepts to advanced applications, experimental realizations, and thermodynamic interpretations of non-classicality in open quantum systems.\\

Chapter~\ref{chap2:Preliminaries} serves as a comprehensive preliminary chapter, establishing the theoretical and conceptual framework required for the rest of the thesis. It introduces both continuous and discrete Wigner functions, detailing their construction, properties, and relevance as phase-space representations of quantum states. The chapter further discusses the theory of open quantum systems, including the distinction between Markovian and non-Markovian dynamics, standard noisy quantum channels, and microscopic descriptions based on collision models. Since these tools and concepts are employed rigorously in all subsequent chapters, this chapter provides the necessary mathematical and physical background to ensure clarity and coherence throughout the thesis.\\

Chapter~\ref{chap3:Harnessing} is devoted to the investigation presented in Harnessing Quantumness of States Using Discrete Wigner Functions under (Non)-Markovian Quantum Channels~\citep{lalita2023harnessing}. The chapter introduces, for the first time, negative quantum states in finite-dimensional Hilbert spaces with dimensions d = 2, 3, and 4. It systematically analyzes the evolution of discrete Wigner negativity, mana, and the DWFs for qubits, qutrits, and two-qubit systems subjected to non-unital noise, such as amplitude damping, and unital noise, including random telegraph noise. Furthermore, the chapter examines how quantum coherence, entanglement (quantified by concurrence), and fidelity evolve for the first, second, and third negative quantum states of two-qubit systems in comparison with Bell states, thereby elucidating the impact of environmental memory on non-classical phase-space features.\\

Chapter~\ref{chap4:Protecting} focuses on Protecting Quantum Correlations of Negative Quantum States Using Weak Measurement under Non-Markovian Noise~\citep{Lalita_2024ProtectingQC}. Here, weak measurement and measurement-reversal protocols are introduced as practical tools for mitigating decoherence. The chapter analyzes how these techniques enhance the robustness of entanglement and other quantum correlations in two-qubit negative quantum states, particularly in non-Markovian environments, and evaluates their impact on teleportation performance and fidelity measures.\\

Chapter~\ref{chap5:Physical_realization} presents the work titled Noise-Resilient Negative Quantum States \citep{lalita2025realizingnegativequantumstates}. This chapter bridges the gap between theory and experiment by demonstrating the preparation, characterization, and benchmarking of two-qubit negative quantum states on superconducting quantum hardware. The robustness of these states is systematically compared with conventional Bell states using metrics such as concurrence, CHSH inequality violation, quantum Fisher information, and teleportation fidelity under realistic noise models. The results establish two-qubit negative quantum states as experimentally viable and resilient to noise resources for quantum information tasks.\\

Chapter~\ref{chap6:two_qubit_collision_model} examines A Two-Qubit Quantum Collision Model: Non-Markovianity and Non-Classicality~\citep{lalita2025non_classicality}. In this chapter, a microscopic collision-model framework is employed to investigate how structured environmental interactions and ancilla–ancilla correlations lead to non-Markovian dynamics. The chapter introduces the two-qubit collision model and two different interaction schemes for the first time. The analysis reveals how information backflow influences entanglement generation, non-classical volume, and steady-state behavior, thereby providing a deeper understanding of memory effects at the microscopic level.\\

Chapter~\ref{chap7:Interrelation} is devoted to Interrelation of Non-Classicality, Entropy, Irreversibility and Work Extraction in Open Quantum Systems~\citep{lalita2025interrelation}. This chapter extends the discussion beyond information-theoretic quantities to thermodynamic aspects, exploring how non-classicality correlates with entropy production, irreversibility, and ergotropy. By comparing the collision and central spin models, the non-Markovian amplitude damping and Markovian generalized amplitude damping channels, and the Jaynes–Cummings model, this work aims to uncover universal correspondences linking non-classicality, irreversibility, and extractable work.\\

Finally, Chapter~\ref{chap8:conclusion} summarizes the main findings of the thesis, consolidates the insights gained across different dynamical models and physical settings, and discusses their implications for future quantum technologies. The chapter also outlines potential directions for further research.

\subsubsection{SUMMARY of contributions}
The following contributions represent the central original results of this thesis. While the individual mathematical tools employed, discrete Wigner functions, non-Markovian channels, weak measurement protocols, collision models, and standard quantum correlations quantifiers, are established in the literature, their integration and the specific results derived below are new contributions.
\begin{itemize}
    \item \textit{First introduction of negative quantum states for $d = 2, 3$, and $4$}: Chapter~\ref{chap3:Harnessing} (Figs.~\ref{negativityNMRTN}–\ref{fidelityNMAD}) introduces, for the first time, negative quantum states in finite-dimensional Hilbert spaces of dimensions $d = 2$ (qubit), $d = 3$ (qutrit), and $d = 4$ (two-qubit) as eigenstates of phase-space point operators~\citep{lalita2023harnessing}.
    
    \item \textit{Comparative resource analysis under non-Markovian channels}: Chapter~\ref{chap3:Harnessing} demonstrates that certain negative quantum states retain coherence, entanglement, and teleportation fidelity for longer durations than Bell states under the non-Markovian amplitude damping channel (Figs.~\ref{coherence_NMRTN}–\ref{fidelityNMAD}). The advantage is metric and regime-specific: it holds under non-Markovian AD for coherence and entanglement, while Bell states dominate under dephasing channels.
    
    \item \textit{WM and QMR protection of negative quantum states}: Chapter~\ref{chap4:Protecting} (Figs.~\ref{concur_NMAD}–\ref{P_success}, TABLES~\ref{table1}–\ref{table2}) provides a first systematic study of how weak measurement (WM) and quantum measurement reversal (QMR) protocols protect quantum correlations and enhance universal quantum teleportation performance for negative quantum states in non-Markovian environments, including explicit trade-off analysis with success probability~\citep{Lalita_2024ProtectingQC}.
    
    \item \textit{First experimental realization on IBM superconducting hardware}: Chapter~\ref{chap5:Physical_realization} (Figs.~\ref{NS1_NS2_NS3__NS3_prime_circuit}–\ref{city_plot_NS2_miti}, TABLE~\ref{New_table}) demonstrates the first preparation of two-qubit negative quantum states on IBM's $ibm_brisbane$ quantum processor using native-gate circuits, verified by quantum state tomography with and without error mitigation. For the first time, we present two-qubit negative quantum states as better noise-resilient resource states than Bell states for teleportation and quantum metrology (Figs. \ref{DP_NS1_NS2_NS3}-\ref{Teleportation_NS3_double_prime}) in the presence of non-Markovian AD noise~\citep{lalita2025realizingnegativequantumstates}.

    \item \textit{First two-qubit collision model with two interaction schemes}: Chapter~\ref{chap6:two_qubit_collision_model} (Figs.~\ref{approach_1}–\ref{ss_fidelity_all_cases}) introduces, for the first time, a two-qubit quantum collision model with two distinct interaction schemes (Scheme A and Scheme B), linking ancilla-ancilla partial-swap interactions to non-Markovian dynamics, phase-space non-classicality, and entanglement generation~\citep{lalita2025non_classicality}.

    \item \textit{Cross-model thermodynamic correspondence}: Chapter~\ref{chap7:Interrelation} (Figs.~\ref{Single_qubit_collision_model}–\ref{Jaynes_cummings_model}) establishes a qualitative but structurally robust correspondence among non-classical volume, von Neumann entropy, entropy production, and ergotropy across five distinct open-system models (collision model, central spin model, non-Markovian AD channel, generalized AD channel, and Jaynes-Cummings model)~\citep{lalita2025interrelation}.
\end{itemize}



\newpage
\setcounter{chapter}{1} 

\titleformat{\chapter}[display]
{\sffamily\fontsize{27}{27}\bfseries\filleft}{\thechapter}{0pt}{{#1}}  
  
\thispagestyle{empty}

\chapter{Preliminaries}\label{chap2:Preliminaries}
The purpose of this chapter is to establish the mathematical and conceptual foundations required for the analysis carried out in the subsequent chapters of this thesis. The central themes of this work are phase-space representations of finite-dimensional quantum systems and the dynamics of such systems under realistic environmental noise. Since these topics draw from distinct but complementary areas of quantum theory, it is essential to present a unified and self-contained set of preliminaries.

The chapter is organized into two main parts. The first part develops the phase-space formulation of quantum mechanics, beginning with a concise review of quasi-probability distributions and the continuous Wigner function, followed by a detailed construction of discrete Wigner functions for finite-dimensional Hilbert spaces. Particular emphasis is placed on the finite field-based discrete phase-space formalism, which provides a natural framework for identifying and quantifying non-classicality through Wigner negativity. The second part of the chapter focuses on the dynamical description of quantum systems. After reviewing unitary evolution and the density-operator formalism, we introduce reduced dynamics, completely positive dynamical maps, quantum channels, and the distinction between Markovian and non-Markovian processes. In addition to the standard master-equation approach, a microscopic description of open-system dynamics based on collisional (repeated-interaction) models is presented. This framework plays a central role in the later analysis, as it provides a transparent physical interpretation of memory effects and offers a natural setting for studying the evolution of phase-space structures under environmental interactions. Standard results from quantum mechanics and open-system theory are reviewed only to the extent necessary for later developments. More specialized constructions, such as discrete phase-space point operators, finite-field labeling schemes, non-Markovian noise models, and collisional dynamics, are presented in detail, as they form the technical backbone of the thesis.

\section*{Notation Summary}
The following tables summarize the principal symbols used throughout this thesis.

\subsection{Unique Symbols}

\renewcommand{\arraystretch}{1.2}

\begin{longtable}{p{0.18\textwidth} p{0.52\textwidth} p{0.22\textwidth}}
\toprule
\textbf{Symbol} & \textbf{Definition} & \textbf{First Used In} \\
\midrule
\endfirsthead

\toprule
\textbf{Symbol} & \textbf{Definition} & \textbf{First Used In} \\
\midrule
\endhead

$\rho$ & Density operator of the system & Sec. 2.1 \\
$|\psi\rangle$ & Pure quantum state vector & Sec. 2.1 \\
$\mathcal{H}$ & Hilbert space of the system & Sec. 2.1 \\

$W(q,p)$ & Continuous Wigner function & Sec. 2.1.2 \\
$W_{\alpha}$ & Discrete Wigner function & Sec. 2.1.4 \\
$A_\alpha$ & Phase-space point operator & Sec. 2.1.4 \\
$\mathrm{Tr}(\cdot)$ & Trace operation & Sec. 2.1 \\

$N_G(\rho)$ & Wigner negativity & Sec. 3.2.1 \\
$M(\rho)$ & Mana (measure of magic) & Sec. 3.3 \\
$\delta$ & Non-classical volume & Sec. 6.4.1 \\

$|NS_i\rangle$ & Negative quantum states & Sec. 3.2.2 \\

$\Phi_t$ & Dynamical map at time $t$ & Sec. 2.6 \\
$\mathcal{E}$ & Quantum channel (CPTP map) & Sec. 2.4 \\
$M_\mu$ & Kraus operators & Sec. 2.5 \\
$\mathcal{L}$ & Lindblad generator & Sec. 2.7.1 \\

$H_S$ & System Hamiltonian & Sec. 2.3 \\
$H_{SE}$ & Interaction Hamiltonian & Sec. 2.3 \\

$\gamma$ & Decay rate & Sec. 3.4 \\
$g, b$ & System-environment coupling strength & Sec. 3.4 \\

$p$ & Weak measurement strength & Sec. 4.2 \\
$q$ & Measurement reversal strength & Sec. 4.2 \\
$P_succ$ & Success probability & Sec. 4.2 \\

$\Theta$ & Intra-ancilla interaction strength & Sec. 6.2 \\
$\Delta t$ & Collision time interval & Sec. 6.2 \\

$C_{l_!}(\rho)$ & $l_1$ norm od coherence & Sec. 3.6 \\
$C(\rho)$ & Concurrence & Sec. 3.6 \\
$d(A : B)$ & Quantum discord & Sec. 4.3 \\
$S_n(\rho)$ & Quantum steering measure & Sec. 4.3 \\

$S(\rho)$ & Von Neumann entropy & Sec. 7.2 \\
$\Sigma$ & Entropy production & Sec. 7.2 \\
$\mathcal{W}(\rho)$ & Ergotropy & Sec. 7.2 \\

$\beta$ & Inverse temperature ($\beta = 1/k_B T$) & Sec. 6.6 \\
$T$ & Temperature & Sec. 6.6\\
$\mathcal{N}$ & Non-Markovianity measure & Sec. 6.3\\

\bottomrule
\end{longtable}

\subsection{Context-Dependent Symbols}

\renewcommand{\arraystretch}{1.2}
\begin{longtable}{p{0.15\textwidth} p{0.55\textwidth} p{0.22\textwidth}}
\toprule
\textbf{Symbol} & \textbf{Different Meanings} & \textbf{Where Used} \\
\midrule
\endfirsthead

\toprule
\textbf{Symbol} & \textbf{Different Meanings} & \textbf{Where Used} \\
\midrule
\endhead

$p$ & (i) Weak measurement strength; (ii) Probability parameter & Sec. 4.2, Sec. 5.3.2 \\

$q$ & (i) Phase-space coordinate; (ii) Measurement reversal strength & Sec. 2.1.2, Sec. 4.2 \\

\bottomrule
\end{longtable}

\section{Phase-Space Formulation of Quantum Mechanics}
\subsection{Quasi-Probability Distribution Functions}
The description of quantum states through quasi-probability distributions provides an alternative viewpoint to the conventional Hilbert-space formulation. Among the various representations of this kind, the Wigner function stands out because it resembles a classical phase-space distribution while still encoding the intrinsically quantum features of a state, including interference and non-classical correlations~\citep{Wigner1932Quantum, hillery1984distribution}. Although initially developed for continuous-variable systems, the underlying logic of the Wigner formalism can be extended to finite-dimensional quantum systems. This chapter reviews the essential ingredients of the continuous Wigner function and then develops the discrete counterpart that will be used throughout this thesis. The aim is to establish a coherent mathematical foundation before introducing its applications in later chapters. The presentation draws on standard results from phase-space quantum mechanics and on the finite-field-based scheme proposed in~\citep{wootters2004picturing, gibbons2004discrete}.
\subsection{Introduction to Wigner Functions}
Classical mechanics describes the state of a particle through a probability density $w_{\mathrm{cl}}(q,p)$ on phase space, where the expectation value of an observable $A_{\mathrm{cl}}(q,p)$ is obtained via
\begin{equation}
    \langle A_{\mathrm{cl}} \rangle = \int dq \int dp \; A_{\mathrm{cl}}(q,p)\, w_{\mathrm{cl}}(q,p).
\end{equation}
Quantum mechanics, however, prohibits assigning a genuine probability distribution to position and momentum simultaneously because these variables correspond to non-commuting operators. Despite this, one can still construct a real-valued function on phase space, the Wigner function, that reproduces the correct marginal distributions and expectation values, while allowing for negative regions indicative of inherently quantum interference~\citep{kenfack2004negativity}.

To make such a representation operational, one needs a systematic way of connecting Hilbert-space operators with phase-space functions. This link is provided by the Weyl transform~\citep{case2008wigner, Schleich2001}, which enables the computation of quantum expectation values through integrals over phase-space quasi-distributions. Once this bridge is established, the Wigner function can be viewed as the phase-space representation of the density operator.
\subsubsection{Average of an Operator and the Weyl Transform}
The Weyl correspondence assigns to any operator $\hat{A}$ a function $\tilde{A}(q,p)$ on phase space. This object is defined via the integral transform
\begin{equation}
    \tilde{A}(q,p) = 
    \int d\xi \, e^{-ip\xi/\hbar}
    \left\langle q + \frac{\xi}{2} \right| \hat{A} \left| q - \frac{\xi}{2} \right\rangle.
\end{equation}
In analogy with the classical setting, the expectation value of $\hat{A}$ in a state $\rho$ may then be expressed as
\begin{equation}
    \langle \hat{A} \rangle = 
    \int dq \int dp \; \tilde{A}(q,p)\, W_\rho(q,p),
\end{equation}
where $W_\rho(q,p)$ denotes the Wigner function associated with the state $\rho$.
The introduction of the Weyl transform thus ensures that quantum averages can be evaluated in terms of phase-space functions. This mapping also provides a means of reconstructing operators from their phase-space symbols, ensuring that the formalism respects Hermiticity and linearity. Together, these properties make the Weyl correspondence an essential tool for understanding the relationship between operators and quasi-probability distributions.
\subsubsection{Wigner Functions}
The Wigner function corresponding to a pure state $|\psi\rangle$ with wavefunction $\psi(q)$ is given by
\begin{equation}
    W(q,p) = \frac{1}{h} 
    \int d\xi \, 
    e^{-ip\xi/\hbar}
    \, \psi\!\left(q + \frac{\xi}{2}\right)
    \psi^{*}\!\left(q - \frac{\xi}{2}\right).
\end{equation}
Its definition for mixed states follows by linearity, using a density operator $\rho$ written in its spectral decomposition,
\begin{equation}
    W_\rho(q,p) = \sum_i p_i W_{\psi_i}(q,p),
\end{equation}
where $\{p_i,|\psi_i\rangle\}$ are the eigenvalues and eigenvectors of $\rho$.
The Wigner function is real, properly normalized,
\begin{equation}
    \int dq \int dp \; W(q,p) = 1,
\end{equation}
and reproduces correct marginal distributions, for example,
\begin{equation}
    \int dp \; W(q,p) = |\psi(q)|^2.
\end{equation}
It also satisfies the overlap identity
\begin{equation}
    \mathrm{Tr}(\rho_1 \rho_2) 
    = 2\pi\hbar \int dq \int dp \, W_1(q,p)W_2(q,p),
\end{equation}
which allows one to express state overlaps directly in phase space.
A striking feature of $W(q,p)$ is that it may assume negative values. These regions are often regarded as signatures of non-classical behavior and arise from interference among different components of the wave function. Nevertheless, expectation values and marginal probabilities computed from $W(q,p)$ always remain physically meaningful. All these properties motivate the use of the Wigner function as a diagnostic tool for quantum behavior and serve as guiding principles for constructing discrete analogues.
\subsection{Discrete Wigner Functions for Finite-Dimensional Systems}
Many physical systems of interest, including spin systems, qudits, and multiqubit registers, are described by finite-dimensional Hilbert spaces. Extending phase-space ideas to such systems requires replacing the continuous phase space by a finite geometry while preserving key structural properties such as completeness, covariance, and well-defined marginals. The discrete Wigner functions (DWFs) provide a phase-space representation for finite-dimensional quantum systems analogous to the continuous Wigner function discussed earlier. In this formalism, a real number is associated with each point in a finite phase space, capturing information about the quantum state. The entire set of values forms a complete representation of the density operator. This section provides a brief review of the main approaches to constructing DWFs, as well as several contexts in which they have proven useful.
\subsubsection{Survey of Discrete Wigner Function Constructions}
Multiple strategies have been proposed for defining discrete Wigner functions, each tailored to particular types of systems or specific mathematical constraints~\citep{cohen1986joint, wootters1987wigner, galetti1988extended, leonhardt1996discrete, wootters2004picturing, gibbons2004discrete, chaturvedi2005wigner}. Early attempts relied on generalizations of the Weyl correspondence using finite-dimensional analogues of displacement operators~\citep{schwinger1960special}. Subsequent contributions introduced constructions based on operator bases with properties mirroring those required in the continuous setting, such as hermiticity, orthogonality, and trace completeness. 

A significant step forward was made in~\citep{wootters2004picturing}, where a discrete phase space for systems of prime dimension was formulated using the structure of finite fields. In his approach, phase-space points are organized into lines and striations, with each line corresponding to the projector associated with a vector of a mutually unbiased basis (MUB). Later, this finite-field viewpoint to dimensions that are powers of primes is generalized~\citep{gibbons2004discrete}. This construction clarified the geometric underpinnings of discrete phase space through the affine geometry of $\mathbb{F}_{p^{m}}\times\mathbb{F}_{p^{m}}$. Other notable methods include~\citep{cohen1986joint, wootters1987wigner, galetti1988extended, leonhardt1996discrete}. While technically distinct, all these approaches share a common goal: to represent quantum states in a form that resembles classical probability distributions while retaining quantum properties, such as interference and negativity. The remainder of the chapter focuses on the specific construction proposed~\citep{gibbons2004discrete}, which will be used throughout this thesis.
\subsection{Gibbons et al. Construction Over a Finite Field}
The discrete Wigner function formalism, developed in~\citep{gibbons2004discrete}, provides a unified geometric approach for all finite-dimensional systems whose Hilbert space dimension is a power of a prime. The key mathematical object in this construction is a finite field (also known as a Galois field), which is used to label points in phase space and to define the associated lines, striations, and mutually unbiased bases.
\subsubsection{Galois fields}
Any set $\mathbb{F}$ in conjunction with two laws of composition ($+$, $\times$) is said to be a field if it has the following properties,
\\1. The set $\mathbb{F}$ is an abelian group $\mathbb{F}^+$, with the addition operation “$+$,” and its identity element is 0.
\\2. The set $\mathbb{F} - \{0\}$ is an abelian
group $\mathbb{F}^\times$ = $\mathbb{F}$ - \{0\}, with the multiplication operation “$\times$”,  and its identity element is 1.
\\3. Addition distributes over multiplication: $a.(b + c) = a.b + a.c$  for all $a, b, c \in \mathbb{F}$.
\\The order of the Galois field is given by the number of elements in the field. Such a field exists if and only if the order of the field is a prime or a power of a prime. These finite fields are denoted by $\mathbb{F}_N$; here, $N$ is the order of the field. Thus, the Galois field $\mathbb{F}_{N}$ contains exactly $N=p^{m}$ elements, where $p$ is a prime number and $m$ is a positive integer. Further, the extension of the finite field to the power of a prime dimension requires the use of an irreducible polynomial, which is discussed below.
\subsubsection{Irreduciable polynomial}
The choice of an irreducible polynomial is essential for constructing $\mathbb{F}_{p^{m}}$. 
For any given prime field $\mathbb{F}_p$ and positive integer $m$, let $a_0 +a_1 x +· · ·+ a_n x^m = 0$, where $a_i \in \mathbb{F}_p$, be an irreducible polynomial of order $m$ if no solution exists within $\mathbb{F}_p$.
\\For any general prime power $N = p^m$ , the finite field $\mathbb{F}_N$ is generated by finding the irreducible polynomial of order $m$. Let's say $\omega$ is one of the solutions of these irreducible polynomials which is not an element of $\mathbb{F}_p$ then, the other powers of $\omega$ are also solutions of this polynomial and the set $\{0, 1, \omega, \omega^2, · · ·, \omega^{N - 2}\}$  satisfies all the properties of the field. In this way, the elements of $\mathbb{F}_N$ are given by $\mathbb{F}_N = \{0, 1, \omega, \omega^2 , · · · , \omega^{N - 2}\} $ discussed in \citep{gibbons2004discrete}. The finite field $\mathbb{F}_{p^m}$ may also be considered as a vector space of dimension $m$. These finite fields form the algebraic backbone of the discrete phase space used in the subsequent sections. For illustration, consider the case $N=4=2^{2}$, where a suitable irreducible polynomial is
\begin{equation}
    \pi(\omega)=\omega^{2} + \omega + 1.
\end{equation}
The field $\mathbb{F}_{4}$ can then be represented using the elements
\[
\{0,\,1,\,\omega,\,\omega+1\},
\]
with addition performed modulo 2 on the coefficients of $\omega$ and multiplication determined by the relation $\omega^{2}=\omega+1$.
Its addition and multiplication tables are given by:
\begin{table}[H]
\centering
\begin{tabular}{c|cccc}
$+$ & $0$ & $1$ & $\omega$ & $\omega+1$ \\
\hline
$0$ & $0$ & $1$ & $\omega$ & $\omega+1$ \\
$1$ & $1$ & $0$ & $\omega+1$ & $\omega$ \\
$\omega$ & $\omega$ & $\omega+1$ & $0$ & $1$ \\
$\omega+1$ & $\omega+1$ & $\omega$ & $1$ & $0$
\end{tabular}
\caption{Addition table for $\mathbb{F}_{4}$.}
\end{table}  

\begin{table}[H]
\centering
\begin{tabular}{c|cccc}
$\times$ & $0$ & $1$ & $\omega$ & $\omega+1$ \\
\hline
$0$ & $0$ & $0$ & $0$ & $0$ \\
$1$ & $0$ & $1$ & $\omega$ & $\omega+1$ \\
$\omega$ & $0$ & $\omega$ & $\omega+1$ & $1$ \\
$\omega+1$ & $0$ & $\omega+1$ & $1$ & $\omega$
\end{tabular}
\caption{Multiplication table for $\mathbb{F}_{4}$.}
\end{table}
\subsubsection{Trace of Field Elements}
The Galois-field trace is a map from $\mathbb{F}_{p^{m}}$ to the prime field $\mathbb{Z}_{p}$, defined by
\begin{equation}
    \mathrm{Tr}(x)=x + x^{p} + x^{p^{2}} + \cdots + x^{p^{m-1}}.
\end{equation}
This function is linear over $\mathbb{F}_{p}$. It plays a central role in defining the additive characters of the field, which in turn are used to construct mutually unbiased bases and the associated phase-space point operators. For the specific case of $\mathbb{F}_{4}$, one obtains
\begin{align}
    \mathrm{Tr}(0) &= 0, \\
    \mathrm{Tr}(1) &= 0, \\
    \mathrm{Tr}(\omega) &= 1, \\
    \mathrm{Tr}(\omega+1) &= 1.
\end{align}
These values follow from the relation $\omega^{2}=\omega+1$ along with the characteristic-2 property that $1+1=0$. The trace operation will reappear prominently in the construction of displacement operators and in establishing the correspondence between lines in phase space and projectors in Hilbert space. Next, we discuss the concept of a mutually unbiased basis.

\subsubsection{Mutually Unbiased Bases}
Mutually unbiased bases (MUBs)~\citep{wootters1989MUB, Lawrence2002MUB, bandyopadhyay2002MUB, pittenger2004mutually} play a central role in the discrete Wigner function formalism for systems whose Hilbert-space dimension is a prime power. Let's consider two distinct orthonormal bases, $B_{1}$ and $B_{2}$, such that
\begin{equation}
B_{1} = \{\ket{\beta_{1,1}},\ket{\beta_{1,2}},....,\ket{\beta_{1,N}}\}, |\bra{\beta_{1,i}\ket{\beta_{1,j}}}^2 = \delta_{i,j}, 
\end{equation}
\begin{equation}
B_{2} = \{\ket{\beta_{2,1}},\ket{\beta_{2,2}},....,\ket{\beta_{2,N}}\}, |\bra{\beta_{2,i}\ket{\beta_{2,j}}}^2 = \delta_{i,j}, 
\end{equation}
These are mutually unbiased if,
\begin{equation}
|\bra{\beta_{i,j}\ket{\beta_{i',j'}}}^2 = \frac{1}{N}(1 - \delta_{i,i'}) + \delta_{i,i'}\delta_{j,j'}.
\end{equation}
A complete set of MUBs in dimension $N$ consists of exactly $(N+1)$ pairwise unbiased bases. Such sets exist whenever $N$ is a prime power, and their construction typically relies on the algebraic structure of finite fields. We now discuss the discrete phase space and discrete Wigner functions in detail for a prime-dimensional Hilbert space using the formalism provided in~\citep{gibbons2004discrete}.

\subsection{Discrete Phase Space and Wigner Functions for Power of a Prime Dimensions}\label{DWF_def}
A $N\times N$ real array is the discrete equivalent of phase space in an $N$-dimensional Hilbert space. Suppose we are describing a quantum state whose dimension $N$ in Hilbert space is a power of a prime ($N = p^m$). In these circumstances, label the position and momentum coordinates of the $N\times N$ grid with finite Galois field GF($p^m$) elements~\citep{lidl1994introduction}. This is because if we do so, we can endow the phase-space grid with the same geometric properties as the ordinary plane. For instance, we can define the finite $N\times N$ grid linear equations of the form $aq + bp = c$ and lines as its solutions (where all the components and actions in this equation are contained in GF ($p^m$) ). In contrast to the continuous phase space, this discrete arrangement has no geometrical lines of points. Instead, a line is a collection of $N$ distinct points in phase space. Our discrete phase space may then be divided into various parallel line collections. Each of these collections is referred to as a striation \citep{wootters2004picturing}. A method for creating $(N+1)$ striations of a $N\times N$ phase-space array~ \citep{gibbons2004discrete}, produces striations with the following three properties:\\
$(i)$ For a given pair of points in the discrete phase space, there is exactly one line containing both points.\\
$(ii)$ Two non-parallel lines intersect exactly at one point, $\textit{i.e.}$, they share only one common point.\\
$(iii)$ For any phase-space point $\alpha(q, p)$ which is not contained in the line $\lambda$, there is exactly one line parallel to $\lambda$ containing the phase-point $\alpha(q, p)$.\\
Further in this section, we will see that these striations are crucial for formulating the discrete Wigner function $W_{\alpha}$. If the dimension of the space of states is a power of a prime integer ($N = p^m$), then it is known that there exists a complete set of $(N + 1)$ MUBs. Note that this is exactly the number of striations one can find with properties (i)–(iii) discussed above.

\subsubsection{\label{DWF} Discrete Wigner functions (DWFs)} 
Now, we have a collection of $(N + 1)$ mutually unbiased bases $(B_1, B_2,..., B_{N+1})$ and a set of $(N + 1)$ striations $(S_1, S_2,..., S_{N+1})$ having $N$ parallel lines of the $N\times N$ phase space. We must select two one-to-one mappings to build a DWF, $\textit{i.e.}$, each basis set $B_i$ is associated with each striation $S_i$, and each basis vector $\ket{\beta_{i, j}}$ is associated with a line $\lambda_{i,j}$ (the $j^{th}$ line of the $i^{th}$ striation).  Each line $\lambda_{i, j}$ of $i^{th}$ striation is associated to a projector ${P}_{i,j} = \ketbra{\beta_{i,j}}{\beta_{i,j}}$, defines a quantum net. This may be accomplished in various ways, each producing a different quantum net and, thus, a different definition of the discrete Wigner function $W_{\alpha}$, as detailed in \citep{gibbons2004discrete}. For a $N = p^{m}$ dimensional Hilbert space, there are passible $d^{d+1}$ quantum nets. Thus, $d^{d+1}$ possible definitions of the DWFs correspond to the same density matrix. Given that the above-discussed linkages exist and having fixed a quantum net, the DWFs are uniquely defined as 
\begin{equation}
p_{i,j} \equiv \Tr[\ketbra{\beta_{i,j}}{\beta_{i,j}}{\rho}] = \sum_{\alpha \in \lambda_{i,j}} W_{\alpha}, 
\end{equation}
In other words, we want the probability of projecting onto the basis vector corresponding to each line to be equal to the sum of the discrete Wigner function elements corresponding to that line. Further, it can be demonstrated that the resultant discrete Wigner function at any phase-space point $\alpha(q, p)$ is \citep{gibbons2004discrete},
\begin{equation}
    \begin{aligned}
      W_{\alpha} = \frac{1}{N} \Tr[ {A}_{\alpha} {\rho} ],
    \end{aligned}
    \label{DWFformula}
\end{equation}
where
\begin{equation}
    \begin{aligned}
      {A}_{\alpha} = \sum_{\alpha \in \lambda_{i,j}} {P}_{i,j} - \textbf{I}.
    \end{aligned}\label{A_formula}
\end{equation}
The operators ${A}_{\alpha}$, known as phase-space point operators, are Hermitian operators. Therefore, all their eigenvalues are real. They also form a complete basis for the space of operators, which are orthogonal in the Schmidt inner product (i.e.,  $\Tr[{A}_{\alpha}{A}_{\beta}] = \delta_{\alpha,\beta}/N$ ) and $\Tr({A}_{\alpha}) = 1$. The sum of phase space point operators ${A}_{\alpha}$ along any line $\lambda$ is equal to the projectors associated with it, $\textit{i.e.}$, ${P}(\lambda) = \sum_{\lambda \ni \alpha} {A}_{\alpha}$. Moreover, any density operator can also be written as, 
\begin{equation}
    \begin{aligned}
      {\rho} = \sum_{\alpha} W_{\alpha}{A}_{\alpha},
    \end{aligned}
    \label{rho-decomposition-in-A}
\end{equation}
where $W_{\alpha}$, i.e., the discrete Wigner functions, are the expansion coefficients. It can be shown that the discrete Wigner function $W_{\alpha}$ shares many characteristics with the continuous Wigner function $W(q, p)$ \citep{gibbons2004discrete}, such as it is real (but can also be negative), normalized, and gets its values from measurements onto MUB using Eq. (\ref{DWFformula}). In this case, the MUB projectors take on the role that the quadratures $a{Q} + b{P}$ play in $W(q, p)$, providing a highly symmetric collection of observables whose measurement results fully define the state (quantum tomography). In Refs. \citep{gibbons2004discrete, paz2005qubits, pittenger2004mutually, galvao2005discrete, cormick2006classicality}, more features of $W_{\alpha}$ are examined. All together, these properties ensure that the discrete Wigner function retains both the operational interpretability and the geometric structure expected of a quasi-probability representation. They also make the formalism particularly effective for analyzing correlations, contextuality, stabilizer structure, and other forms of non-classicality in finite-dimensional quantum systems. Further, negative values of $W_{\alpha}$ signal non-classical features and will be central to the analysis in later chapters.
\subsection{Labeling Scheme for Multiqubit Discrete Wigner Functions}
For multiqubit systems, the dimension is $N=2^{m}$, and the phase space is constructed from the finite field $\mathbb{F}_{2^{m}}$. Each phase-space coordinate is represented as an element of this field, and the labeling of points requires a mapping between the binary vector representation of $\mathbb{F}_{2}^{m}$ and the algebraic representation in $\mathbb{F}_{2^{m}}$. A standard approach is to fix a primitive element $\omega\in\mathbb{F}_{2^{m}}$ generated by an irreducible polynomial and express each field element as a polynomial in $\omega$ with binary coefficients. For instance, in $\mathbb{F}_{4}$ one writes:
\[
0,\; 1,\; \omega,\; \omega+1.
\]
This labeling allows tensor-product bases in Hilbert space to align with the structure of the discrete phase space. The correspondence becomes essential when constructing phase-space point operators for multiqubit systems, as the lines and striations of the phase space encode the MUBs associated with entangled as well as separable bases. Notably, the labeling scheme ensures that the combinatorial properties of the phase space respect the underlying tensor-product structure of the Hilbert space, which is crucial for applications involving entanglement and quantum correlations.
\section{Advantages and Limitations of the Gibbons et al.\ Construction}
The discrete Wigner function formalism introduced in~\citep{gibbons2004discrete} offers a unified geometric and algebraic framework for representing finite-dimensional quantum states. Its most compelling advantage lies in the use of finite-field arithmetic to impose a consistent phase-space structure across all powers of a prime dimension. This structure ensures that each striation of the discrete phase space is associated with a mutually unbiased basis, thereby providing a natural bridge between phase-space geometry and Hilbert-space measurements.

A key strength of the construction is that the phase-space point operators $\{A_{\alpha}\}$ form a complete, orthogonal operator basis. As a result, every density operator can be represented uniquely in terms of its discrete Wigner function values:
\[
\rho = \sum_{\alpha} W_{\alpha} A_{\alpha}.
\]
Moreover, the line-sum property of these operators ensures that marginals of the Wigner function correspond directly to measurement probabilities in the MUBs. This preserves the operational interpretation of the representation, making the discrete Wigner formalism well-suited for analyzing quantum correlations, contextuality, and non-classicality in finite-dimensional systems~\citep{bennett1993teleporting, bouwmeester1997experimental, adhikari2012operational, masanes2011secure, nielsen2010quantum, giovannetti2011advances, thapliyal2017quantum}.

Another advantage is the geometric transparency of the formalism. The affine-plane structure of the phase space, the organization into striations, and the correspondence with MUBs together provide an intuitive visualization of quantum states. This geometric insight has proved particularly valuable in studies of stabilizer states, entanglement, and quantum error correction, where regular patterns in the phase space often reflect structural features of the underlying quantum state. Despite these strengths, the construction provided in~\citep{gibbons2004discrete} is not unique. Different assignments of MUB projectors to striations result in distinct sets of phase-space point operators, and thus in different discrete Wigner functions for the same quantum state. These alternatives, sometimes referred to as quantum nets, are physically equivalent in the sense that they yield the same measurement probabilities; however, they may differ in the distribution of Wigner function values. Such differences can influence visual interpretations or computational properties associated with quasi-probability negativity. Another limitation stems from the fact that the construction relies on the existence of a finite field of order $p^{m}$. Consequently, the method applies only when the Hilbert-space dimension is a power of a prime. Systems of composite dimension that are not prime powers do not admit this type of discrete phase space with all the desired properties, and their discrete Wigner function constructions typically require different or less uniform mathematical machinery. Even so, within its range of applicability, this framework provides a powerful and mathematically elegant means of representing finite-dimensional quantum states in phase space. Its combination of algebraic structure and geometric clarity makes it a valuable tool across many areas of quantum information theory.
\section{Dynamics of Closed and Open Quantum Systems}
The state of a closed physical system is described by a state vector $|\psi\rangle$ which is an element of some Hilbert space $\mathcal{H}$. The norm of $|\psi\rangle$ is defined as $||\psi||=\sqrt{\langle\psi|\psi\rangle}$~\citep{shankar2012principles}. Let $q$ be the set of various parameters apart from time $t$ on which the state vector depends; one shows this dependence as $|\psi(q,t)\rangle$. The time evolution is given by the Schrödinger equation
\begin{equation}
    i\hbar\frac{\partial}{\partial t}|\psi(q,t)\rangle=H|\psi(q,t)\rangle
    \label{schorodinger_eq}
\end{equation}
Here, $H$ is the generator of time evolution known as the Hamiltonian of the system and is generally a Hermitian operator. The solution of Eq.~\eqref{schorodinger_eq} can be represented in terms of a unitary operator $U(t,t_{0})$ such that $|\psi(q,t)\rangle=U(t,t_{0})|\psi(q,t_{0})\rangle$. If the Hamiltonian is time-independent, the system is said to be closed and isolated, and the unitary operator is $U(t-t_{0})=\exp[-iH(t-t_{0})]$. If the system is under the influence of external driving, the Hamiltonian is time-dependent, $H(t)$, and the time evolution involves the time-ordered operator $\mathcal{T}$ as $U(t,t_{0})=\mathcal{T}\exp[-i\int_{t_{0}}^{t}dsH(s)]$, which means that the operators at earlier times are at the left of the operators at later times. For an ensemble of pure states $\{p_{i},|\psi_{i}(q,t)\rangle\}$, one resorts to the density matrix description. In the density matrix formulation, the dynamics is expressed through the Liouville--von Neumann equation
\begin{equation}
\frac{d\rho(t)}{dt} = -\frac{i}{\hbar}[H(t),\rho(t)] \equiv \mathcal{L}(t)\rho(t),
\end{equation}
where $\mathcal{L}(t)$ denotes the Liouvillian superoperator. Its formal solution is
\begin{equation}
\rho(t) = \mathcal{T}\exp\left[\int_{t_0}^{t} \mathcal{L}(s),ds\right]\rho(t_0).
\end{equation}

Moreover, realistic quantum systems are inevitably influenced by their surroundings and therefore cannot be treated as perfectly isolated. Such systems are referred to as \emph{open quantum systems}. The dynamics of an open system arise from its interaction with external degrees of freedom, collectively referred to as the environment or reservoir. While the system alone exhibits non-unitary and generally irreversible behavior, the composite system consisting of the system and its environment is assumed to evolve according to unitary quantum mechanics~\citep{Breuer2007, Banerjee2018}.

Let the Hilbert spaces associated with the system and environment be denoted by $\mathcal{H}_S$ and $\mathcal{H}_E$, respectively. The total system resides in the tensor product space $\mathcal{H}_S \otimes \mathcal{H}_E$. The total Hamiltonian governs the joint evolution
\begin{equation}
H(t) = H_S \otimes I_E + I_S \otimes H_E + H_{SE},
\end{equation}
where $H_S$ and $H_E$ describe the free dynamics of the system and environment, respectively, and $H_{SE}$ accounts for their mutual interaction.

If the joint density operator at an initial time $t_0$ is $\rho_{SE}(t_0)$, the evolution of the composite system at a later time $t$ is given by
\begin{equation}
\rho_{SE}(t) = U(t,t_0)\,\rho_{SE}(t_0)\,U^{\dagger}(t,t_0),
\end{equation}
where $U(t,t_0)$ is the unitary time-evolution operator generated by $H(t)$. The state of the system alone is obtained by tracing over the environmental degrees of freedom,
\begin{equation}
\rho_S(t) = \mathrm{Tr}_E\big[\rho_{SE}(t)\big].
\end{equation}
The partial trace operation is defined such that expectation values of all system observables $M_S$ satisfy
\begin{equation}
\mathrm{Tr}_{SE}\big[(M_S \otimes I_E)\rho_{SE}\big] = \mathrm{Tr}_S\big[M_S \rho_S\big],
\end{equation}
ensuring a consistent description of system measurements. Differentiating the reduced state with respect to time yields the formal equation of motion
\begin{equation}
\frac{d}{dt}\rho_S(t) = -i\,\mathrm{Tr}_E\big[H(t),\rho_{SE}(t)\big].
\end{equation}
Unlike the closed-system case, this equation does not generally close in terms of $\rho_S(t)$ alone, reflecting the fact that system–environment correlations play a nontrivial role in the evolution.

A crucial distinction between closed and open quantum dynamics lies in their reversibility. The unitary evolution of the composite system is time-reversal symmetric, satisfying $U^{-1}(t,t_0) = U(t_0,t)$. However, after tracing out the environment, this reversibility is typically lost. The reduced dynamics of the system become effectively irreversible due to the flow of information into the environment and the buildup of correlations that are inaccessible when only system observables are monitored. The open-system framework thus provides a natural explanation for phenomena such as decoherence, dissipation, and relaxation. These effects are not fundamental violations of quantum mechanics but rather emergent consequences of restricting attention to a subsystem of a larger, closed quantum universe. This perspective forms the basis for the dynamical descriptions developed in later sections, including dynamical maps, quantum channels, and master equation approaches.
\section{Dynamical Maps}
Consider the composite state $\rho(0)=\rho_{S}(0)\otimes\rho_{E}$, assuming the system and environment are initially uncorrelated. The evolution of the system state from time $t=0$ to $t$ is described by a dynamical map $\mathcal{E}_{(t,0)}$ as follows
\begin{equation}
    \rho_{S}(t)=\mathcal{E}_{(t,0)}[\rho(0)]= \mathrm{Tr}_E[U(t)\rho_s(0)\otimes\rho_E U^\dagger(t)].
\end{equation}
This map must satisfy the following properties:
\begin{enumerate}
    \item \textbf{Complete Positivity (CP):} The map $\mathcal{E}$ is positive, and the combined operation $\mathcal{E}\otimes \mathbb{I}_{n}$ is also positive for all dimensions $n$.
    \item \textbf{Trace Preserving (TP):} $\mathrm{Tr}[\mathcal{E}_{(t,0)}[\rho]]= \mathrm{Tr}[\rho]$ for all $\rho \in \mathcal{H}$.
\end{enumerate}
According to Kraus's theorem, any CP map can be written in the form~\citep{kraus1983states}
\begin{equation}
    \rho_{S}(t)=\sum_{\mu}K_{\mu}(t,t_{0})\rho_{S}(t_{0})K_{\mu}^{\dagger}(t,t_{0}),
\end{equation}
where the operators $K_{\mu}$ satisfy $\sum_{\mu}K_{\mu}^{\dagger}K_{\mu}=\mathbb{I}$. It implies that $\Tr[\rho_s(t)] = 1$ for any input state $\rho_s(t_0)$.
\section{Quantum Channel}
A \emph{quantum channel} in the Schrödinger picture is a completely positive and trace-preserving map $\Phi:\mathcal{T}(\mathcal{H}_{A})\rightarrow\mathcal{T}(\mathcal{H}_{B})$. Here $\mathcal{T}(\mathcal{H_A})$ and $\mathcal{T}(\mathcal{H_B})$ denotes the set of operators in the Hilbert space $\mathcal{H_A}$ and $\mathcal{H_B}$, respectively. The operator sum representation of a channel is given as
\begin{equation}
    \Phi[\rho]=\sum_{\mu}M_{\mu}\rho M_{\mu}^{\dagger}
\end{equation}
where $M_{\mu}$ are Kraus operators obeying $\sum_{\mu}M_{\mu}^{\dagger}M_{\mu}= \mathbb{I}$ and $\rho$ need not to be a pure state. A quantum channel is characterized by the following properties,
\begin{itemize}
    \item \textbf{Linearity:} $\Phi[\alpha\rho_{1}+\beta\rho_{2}]=\alpha\Phi(\rho_{1})+\beta\Phi(\rho_{2})$.
    \item \textbf{Hermiticity Preserving:} $\rho=\rho^{\dagger}\Rightarrow\Phi[\rho]=\Phi[\rho]^{\dagger}$.
    \item \textbf{Positivity Preserving:} $\rho\ge0\Rightarrow\Phi[\rho]\ge0$.
    \item \textbf{Trace Preserving:} $Tr(\Phi[\rho])=Tr(\rho)$.
    \item \textbf{Complete Positivity:} $I_{k\times k}\otimes\Phi[\rho]\ge0$,~~ for all $k$.
\end{itemize}
\section{Markovian and Non-Markovian Processes}
A stochastic process is described by an infinite hierarchy of joint probabilities. If the probability of a random variable taking value $x_{n}$ at time $t_{n}$ is conditioned only to the values $x_{n-1}$ at time $t_{n-1}$, the process is called a Markov process:
\begin{equation}
    p_{n}(x_{n},t_{n}|x_{n-1},t_{n-1};...;x_{0},t_{0})=p_{2}(x_{n},t_{n}|x_{n-1},t_{n-1})
\end{equation}
For stationary processes, shifting the origin of time does not change the probabilities. This leads to the Chapman-Kolmogorov equation:
\begin{equation}
    p(x_{k},t|x_{j})=\sum_{l}p(x_{k},t-t^{\prime})p(x_{l},t^{\prime}|x_{j})
\end{equation}
Considering small time differences leads to the Master equation involving transition rates $w(x_k|x_j)$.

In the context of quantum systems, a simple quintessential example of a Markov process is one where the family of maps has a quantum dynamical semigroup (QDS) structure~\citep{alicki2007quantum}
\begin{equation}
    \Phi_{(t_{2}+t_{1},0)}=\Phi_{(t_{2},0)}\Phi_{(t_{1},0)}, \quad t_{1},t_{2}\ge0
\end{equation}
\section{Theory of Markovian and Non-Markovian Open Quantum Systems}
The equation of motion for the state of the system $\rho_{S}(t)$ is generally given by an equation involving a memory kernel $\mathcal{K}_{t,t^{\prime}}$ as follows
\begin{equation}
    \frac{d}{dt}\rho_{S}(t)=-i[H_{S},\rho_{S}(t)]+\int_{t_{0}}^{t}\mathcal{K}_{t,t^{\prime}}[\rho_{S}(t^{\prime})]dt^{\prime}
\end{equation}
\subsection{Markovian Dynamics}
Markovian dynamics involves two approximations~\citep{Breuer2007, Banerjee2018, Hall2014Canonical}:
\begin{enumerate}
    \item \textbf{Born-Markov approximation:} Neglecting memory effects by approximating $\mathcal{K}_{t,t^{\prime}}[\rho_{S}(t^{\prime})]=\mathcal{K}\delta(t-t^{\prime})[\rho_{S}(t^{\prime})]$.
    \item \textbf{Rotating wave approximation:} Ignoring fast-rotating terms in the memory kernel.
\end{enumerate}
This leads to the Lindblad master equation with generator $\mathcal{L}$~\citep{Breuer2009Measure} given as
\begin{equation}
    \mathcal{L}\rho_{S}=-i[H,\rho_{S}]+\sum_{i}\gamma_{i}\left(A_{i}\rho_{S}A_{i}^{\dagger}-\frac{1}{2}\{A_{i}^{\dagger}A_{i},\rho_{S}\}\right)
\end{equation}
where $\gamma_{i}\ge0$ are decay rates. Any generator of this form guarantees physically consistent and feasible solutions (CP maps).

\subsection{Non-Markovian Dynamics}
While the Lindblad master equation serves as the universally accepted prototype for Markovian dynamics, characterized by memoryless evolution where future states are independent of the past, the characterization of non-Markovian dynamics is more complex. There is a divided opinion within the scientific community regarding a precise definition, leading to the perspective that non-Markovian maps may exhibit memory effects through various distinct mechanisms. Consequently, this regime is often described quantitatively through diverse measures of non-Markovianity.

The fundamental distinction lies in the semi-group property defined by $\Phi_{(t_{2}+t_{1},0)}=\Phi_{(t_{2},0)}\Phi_{(t_{1},0)}$. Rigorous studies have established a one-to-one correspondence between completely positive trace-preserving (CPTP) dynamical maps possessing this semi-group property and master equations in the Lindblad form~\citep{Lindblad1976Onthegenerators, gorini1976completely}. Therefore, any deviation from this semi-group property is proposed as the principal characteristic of non-Markovian dynamics~\citep{Wolf2008Assissing}. Several quantitative measures have been constructed based on this deviation, such as calculating the minimum amount of isotropic noise required to render the open system's dynamics Markovian, or quantifying the departure from self-similarity in the evolution. Physically, non-Markovian effects become prominent when the timescales of the system-environment interaction are smaller than those of the environment~\citep{daffer2004depolarizing}. Under these conditions, the system couples to specific frequencies of the environment rather than a flat, white noise spectrum. Consequently, the generator of the dynamics, $\mathcal{L}$, acquires explicit time dependence, manifesting in time-dependent Hamiltonians $H_{S}(t)$, jump operators $A_{i}(t)$, and decay rates $\gamma_{i}(t)$~\citep{Sudarshan1961Stochastic}.

The nature of these time-dependent rates is critical in classifying the dynamics. A non-trivial time dependence of $\gamma_{i}(t)$ inherently breaks the quantum dynamical semigroup (QDS) structure. However, a distinction must be made regarding the sign of these rates:
\begin{itemize}
    \item If $\gamma_{i}(t) \ge 0$ for all $t$, the map remains CP-divisible. This scenario represents a time-dependent Markov process rather than true non-Markovian dynamics.
    \item If $\gamma_{i}(t)$ temporarily takes negative values, the CP-divisibility is violated, signifying a non-Markovian process often associated with information backflow from the environment to the system.
\end{itemize}
Mathematically, this results in dynamics described by a two-parameter family of CPTP maps with a time-inhomogeneous composition law $\Phi_{t,t_{0}}=\Phi_{t,t^{\prime}}\Phi_{t^{\prime},t_{0}}$ for ordered times $t>t^{\prime}>t_{0}$. The evolution obeys the master equation $d\Phi_{t,t_{0}}/dt=\mathcal{L}(t)\Phi_{t,t_{0}}$. Crucially, while the full map $\Phi_{t,t_{0}}$ depends on the interval $t-t_{0}$, it violates the strict semigroup property because the intermediate map does not depend solely on the time difference, i.e., $\Phi_{t^{\prime},t_{0}} \ne \Phi(t^{\prime}-t_{0})$. Finally, it is essential to acknowledge the absence of a unified mathematical guarantee for non-Markovian equations. While the Lindblad form guarantees a family of physically consistent CP maps regardless of derivation, no such elegant mathematical framework currently exists for general non-Markovian processes.
\subsection{Noise models}\label{ch2_noise_models}
This thesis focuses on several paradigmatic noise models, including depolarizing noise, amplitude damping, and random telegraph noise with memory effects. These channels are chosen because they admit clear physical interpretations and allow controlled interpolation between Markovian and non-Markovian regimes.
\subsubsection{Deploarizing noise}
Depolarizing noise is a prevalent type of quantum noise in quantum computing. For single qubit, the depolarizing noise channel, $\mathcal{E}(\rho)$, with error probability $p$ can be written as~\citep{nielsen2010quantum},
\begin{equation}
    \mathcal{E(\rho)} = (1 - p)\rho + \frac{p}{3}( X \rho X + Y \rho Y + Z \rho Z) = \frac{p}{2}I + (1 - p)\rho,
\end{equation}
where $X$, $Y$, and $Z$ are Pauli operators and $\rho$ is the density matrix of a qubit. This channel depolarizes the qubit with probability $p$ and leaves the qubit intact with probability $(1 - p)$. 
The Kraus operators of this channel are
\begin{equation}
    K_{0} = \sqrt{1 - p}I, K_{1} = \sqrt{\frac{p}{3}}X, K_{2} = \sqrt{\frac{p}{3}}Y, K_{3} = \sqrt{\frac{p}{3}}Z.
\end{equation}
\subsubsection{Non-Markovian amplitude damping noise}\label{preli_AD}
Amplitude-damping (AD) noise has been employed to address various phenomena. Attenuation, energy dissipation, spontaneous photon emission, and idle errors in quantum computing are a few examples of these phenomena in two-level systems~\citep{Breuer2007, nielsen2010quantum, Breuer2016Colloquium}. A $d$-dimensional generalization of this was introduced in \citep{dutta2016entanglement}. The Kraus operators of the (non)-Markovian AD channel (non-unital) for a single qubit system are given as \citep{Bellomo2007NMAD},
\begin{equation}
    \mathbf{K_0} = \begin{pmatrix}
     1 & 0\\
     0 & \sqrt{1 - \lambda(t)}
    \end{pmatrix},
    \mathbf{K_1} = \begin{pmatrix}
    0 & \sqrt{\lambda(t)}\\
    0 & 0
\end{pmatrix},
\label{NMAD_Kraus_operators}
\end{equation}
where, $\lambda(t) = 1 - e^{-gt}\left(\frac{g}{l} \sinh{\frac{lt}{2}} + \cosh{\frac{lt}{2}}\right)^2$, and $l = \sqrt{g(g - 2\gamma)}$. The coupling strength $\gamma$ is related to the qubit relaxation time ($\tau_s = \frac{1}{\gamma}$), and $g$ is the line width that depends on the reservoir correlation time ($\tau_r = \frac{1}{g}$). The system exhibits Markovian and non-Markovian evolution of a state if $2\gamma << g$ and $2\gamma >> g$, respectively \citep{Naikoo2019Facets}. The dynamical map for a single-qubit system is:

\begin{equation}
    \begin{aligned}
      \mathcal{E}^{NMAD}({\rho}) = \mathbf{K}_0{\rho}\mathbf{K}_0^{\dag} + \mathbf{K}_1{\rho}\mathbf{K}_1^{\dag}.  
      \end{aligned}\label{ADCfinalrho}
\end{equation}

The Kraus operators that define the (non)-Markovian AD channel for qutrits are \citep{ghosal2021characterizing, utagi2020ping}
\begin{eqnarray}
\nonumber
    \mathbf{K}_{0} &=& \begin{pmatrix}
     1 & 0 & 0\\
     0 & \sqrt{1 - \lambda(t)} & 0\\
     0 & 0 & \sqrt{1 - \lambda(t)}
    \end{pmatrix},\\
    \mathbf{K}_{1} &=& \begin{pmatrix}
    0 & \sqrt{\lambda(t)} & 0\\
    0 & 0 & 0\\
    0 & 0 & 0
    \end{pmatrix},\nonumber\\
    \mathbf{K}_{2} &=& \begin{pmatrix}
    0 & 0 & \sqrt{\lambda(t)}\\
    0 & 0 & 0\\
    0 & 0 & 0
\end{pmatrix}.
\end{eqnarray}
The dynamical map form for a qutrit is
\begin{equation}
    \begin{aligned}
      \mathcal{E}^{NMAD}({\rho}) = \mathbf{K}_0{\rho}\mathbf{K}^{\dag}_0 + \mathbf{K}_1{\rho}\mathbf{K}^{\dag}_1 + \mathbf{K}_2{\rho}\mathbf{K}^{\dag}_2.    
      \end{aligned}\label{NMADqutritfinalrho}
\end{equation}
The expression for $\lambda(t)$ and the regimes of (non)-Markovian behaviour remain unchanged, as discussed above for the single qubit case. Further, the dynamical map for the local interaction of two-qubit systems with (non)-Markovian AD channel acts as
\begin{equation}
   \begin{aligned}
      \mathcal{E}^{NMAD}({\rho}) = \sum_{i = 0}^{1}\sum_{j = 0}^{1} (\mathbf{K}_i \otimes \mathbf{K}_j) {\rho} (\mathbf{K}_i \otimes \mathbf{K}_j)^{\dag}.
      \end{aligned}\label{2qubitNMADfinalrho}
\end{equation}
The Kraus operators, $\textbf{K}_{0}$ and $\textbf{K}_{1}$, are as defined for the qubit's (non)-Markovian AD.
\subsubsection{Random telegraph noise}\label{preli_RTN}
When a system is exposed to a bi-fluctuating classical noise that generates random telegraph noise (RTN) with pure dephasing \citep{daffer2004depolarizing, kumar2018non}, this channel characterizes the system's dynamics. We try to understand how the single-qubit's $NS_{1}$ state DWFs evolve in the presence of (non)-Markovian random telegraph noise (RTN). The dynamical map for a single-qubit system under the action of (non)-Markovian RTN channel is:
\begin{equation}
    \begin{aligned}
      \mathcal{E}^{RTN}({\rho}) = \mathbf{R}_0{\rho}\mathbf{R}^{\dag}_0 + \mathbf{R}_1{\rho}\mathbf{R}^{\dag}_1,   
      \end{aligned}\label{RTNfinalrho}
\end{equation}
where the two Kraus operators are given as
\begin{eqnarray}
      \mathbf{R}_0 = \sqrt{\frac{1 + \Lambda(t)}{2}}\textbf{I}_{2},
      \mathbf{R}_1 = \sqrt{\frac{1 - \Lambda(t)}{2}}\pmb{\sigma_z}.
      \label{NMRTN_Kraus_operators}
\end{eqnarray}
Here, $\Lambda(t)$ is the memory kernel
\begin{equation}
      \Lambda(t) = e^{-\gamma^{RTN} t}\left[ \cos\left(\zeta \;\gamma^{RTN} t\right) + \frac{\sin\left(\zeta \;\gamma^{RTN} t\right)}{\zeta}\right],
\end{equation}
where $\zeta = \sqrt{\left(\frac{2b}{\gamma^{RTN}}\right)^2 - 1}$, and $b$, $\gamma^{RTN}$ quantifies the system–environment coupling strength and fluctuation rate, respectively. The dynamics is Markovian if $(4 b \tau)^2 < 1$ and non-Markovian if $(4 b \tau)^2 > 1$ (here, $\tau = \frac{1}{2\gamma^{RTN}}$ as discussed in~\citep{kumar2018non}).
For implementing the (non)-Markovian RTN channel for a qutrit, the Kraus operators are given by~\citep{daffer2004depolarizing, kumar2018non},
\begin{eqnarray}
      \mathbf{R}_0 = \sqrt{\frac{1 + \Lambda(t)}{2}}\textbf{I}_{3},
      \mathbf{R}_1 = \sqrt{\frac{1 - \Lambda(t)}{2}}\pmb{S_z},\nonumber\\
      \mathbf{R}_2 = \sqrt{\frac{1 - \Lambda(t)}{2}}\pmb{S},
\end{eqnarray}
where,
\begin{equation}
\pmb{S} = \frac{1}{2} \left(\pmb{S_x S_x} + \pmb{S_y S_y} - \pmb{S_z S_z}\right),
\end{equation}
and, 
\begin{eqnarray}
    \pmb{S_x} = \frac{1}{\sqrt{2}}\begin{pmatrix}
                               0 & 1 & 0\\
                               1 & 0 & 1\\
                               0 & 1 & 0
                               \end{pmatrix},
    \pmb{S_y} = \frac{1}{\sqrt{2}}\begin{pmatrix}
                               0 & -\iota & 0\\
                               \iota & 0 & -\iota\\
                               0 & \iota & 0
                               \end{pmatrix},\nonumber\\
    \pmb{S_z} =                 \begin{pmatrix}
                               1 & 0 & 0\\
                               0 & 0 & 0\\
                               0 & 0 & -1
                               \end{pmatrix},
     \textbf{I}_{3} =              \begin{pmatrix}
                               1 & 0 & 0\\
                               0 & 1 & 0\\
                               0 & 0 & 1
                               \end{pmatrix}.                         
\end{eqnarray}
The prior description of the qubit's memory kernel $\Lambda(t)$ and criteria of (non)-Markovianity still holds. Further, the dynamical map for a single-qutrit system is 
\begin{equation}
    \begin{aligned}
      \mathcal{E}^{RTN}({\rho}) = \mathbf{R}_0{\rho}\mathbf{R}^{\dag}_0 + \mathbf{R}_1{\rho}\mathbf{R}^{\dag}_1 + \mathbf{R}_2{\rho}\mathbf{R}^{\dag}_2.    
      \end{aligned}\label{RTNqutritfinalrho}
\end{equation}
Moreover, the dynamical map of the local interaction of two qubits with (non)-Markovian RTN channel is 
\begin{equation}
   \begin{aligned}
      \mathcal{E}^{RTN}({\rho}) = \sum_{i = 0}^{1}\sum_{j = 0}^{1} (\mathbf{R}_i \otimes \mathbf{R}_j) {\rho} (\mathbf{R}_i \otimes \mathbf{R}_j)^{\dag}.
      \end{aligned}\label{2qubitRTNfinalrho}
\end{equation}
Here, $\mathbf{R}_i$ and $\mathbf{R}_j$ are as discussed for the qubit case above. 

Next, we discuss the collision models. The collision models play a central role in this thesis for several reasons. First, they provide a physically grounded alternative to phenomenological master equations, allowing reduced dynamics to be derived explicitly from microscopic interactions between the system and its environment. This strengthens the conceptual foundations of the open-system analysis employed throughout this work. Second, collisional models offer a natural and intuitive framework for studying non-Markovian effects. Since non-Markovianity is a key theme in later chapters, particularly in connection with information backflow and memory-assisted dynamics, the collisional approach provides essential insight into the physical mechanisms underlying such behavior.
\section{Collisional Models}
Collisional models, also known as repeated-interaction models, provide a microscopic and operationally transparent framework for describing the dynamics of open quantum systems~\citep{CM_1963, Ciccarello2022Quantumcollisionmodel}. Unlike phenomenological master-equation approaches, collisional models explicitly construct the reduced dynamics of a system from a sequence of elementary interactions with environmental degrees of freedom. This framework provides a clear physical interpretation of both Markovian and non-Markovian dynamics, establishing a direct link between unitary system-environment interactions and effective dynamical maps~\citep{ThermalizingQuantumMachines_2002, ziman2005description, Rybár_2012, McCloskey2014Non-Markovianity, Ciccarello2013Collision-model, Kretschmer2016Collisionmodel, Saha_2024_quantum, Li2024Witnessing}.

In a collisional model, the environment is represented as a large collection of identical ancillary subsystems, referred to as ancillas, which are initially uncorrelated with the system and with each other. The system interacts sequentially with these ancillas for short, fixed durations. After each interaction, the ancilla is discarded, and the system proceeds to interact with the next environmental unit. The total Hamiltonian governing a single collision can be written as
\begin{equation}
H_{\mathrm{tot}} = H_S + H_A + H_{SA},
\end{equation}
where $H_S$ and $H_A$ denote the free Hamiltonians of the system and the ancilla, respectively, and $H_{SA}$ describes their interaction. The joint unitary evolution associated with a single collision of duration $\tau$ is given by
\begin{equation}
U = \exp\left(-\frac{i}{\hbar} H_{\mathrm{tot}} \, \tau \right).
\end{equation}
Assuming that each ancilla is prepared in the same initial state $\rho_A$ and that no correlations exist between the system and future ancillas, the reduced state of the system after a single collision is obtained by tracing out the ancilla degrees of freedom,
\begin{equation}
\rho_S' = \mathrm{Tr}_A \left[ U (\rho_S \otimes \rho_A) U^\dagger \right].
\end{equation}
This transformation defines a completely positive and trace-preserving (CPTP) map acting on the system state. Repeated application of this map generates the discrete-time evolution of the open quantum system.
\subsection{Markovian Limit and Connection to Master Equations}
When ancillas do not interact with one another and each system--ancilla interaction is statistically identical, the resulting dynamics is memoryless. In this regime, the state of the system after each collision depends only on its immediate past, and the evolution is Markovian. The corresponding dynamical map satisfies a semigroup property and admits a continuous-time limit. By taking the interaction time $\tau$ to be small and appropriately rescaling the system--ancilla coupling strength, the discrete-time dynamics generated by the collisional model converges to a continuous-time master equation of Lindblad form. This procedure provides a microscopic derivation of Markovian quantum dynamical semigroups, clarifying the physical origin of complete positivity and trace preservation in standard master-equation approaches.
\subsection{Non-Markovian Dynamics and Environmental Memory}
Non-Markovian dynamics naturally arise in collisional models when the assumptions of independent and memoryless ancillas are relaxed. Memory effects can be introduced, for example, by allowing interactions between environmental ancillas or by permitting a given ancilla to interact with the system more than once. In such scenarios, correlations established during earlier collisions can influence subsequent system dynamics~\citep{ziman2005description, Rybár_2012, Ciccarello2013Collision-model, McCloskey2014Non-Markovianity, Kretschmer2016Collisionmodel, CM_nm_2017, Campbell2018Systemenvironment, Rodrigues2019thermodynamics, Landi2021Irreversibleentropy, csenyacsa2022entropy}. As a consequence, the reduced evolution of the system can no longer be described by a completely positive divisible family of dynamical maps. Time-dependent decay rates may temporarily assume negative values, indicating a reversal of information flow from the system back to the environment. Within the collisional framework, this information backflow admits a clear microscopic interpretation in terms of environmental correlations and feedback mechanisms. An important advantage of collisional models is that the degree of non-Markovianity can be systematically controlled by tuning interaction strengths, collision times, or the structure of ancilla--ancilla couplings. This tunability makes collisional models particularly suitable for exploring different non-Markovian regimes within a unified theoretical setting.


\newpage
\setcounter{chapter}{2} 

\titleformat{\chapter}[display]
{\sffamily\fontsize{27}{27}\bfseries\filleft}{\thechapter}{0pt}{{#1}}  
  
\thispagestyle{empty}

\chapter{Harnessing quantumness of states using discrete Wigner functions under (non)-Markovian quantum channels}\label{chap3:Harnessing}

\section{Introduction}
In this work, we focus on the negative quantum states of a qubit, qutrit, and two-qubit systems to examine how their DWFs and discrete Wigner negativity $|N_{G}(\pmb{\rho})|$ change under the impact of a variety of noisy channels, both unital and non-unital, in the (non)-Markovian regimes. With the motivation to understand the impact of noise on the DWFs under the action of (non)-Markovian, unital (illustrated by the random telegraph noise) as well as non-unital (depicted by the amplitude damping noise), we calculate the DWFs for the qubit, qutrit, and two-qubit systems. In particular, we study the variation of DWFs corresponding to the maximally negative quantum state, $\textit{i.e.}$, the first negative quantum ($NS_{1}$) state of the qubit, qutrit, and two-qubit systems under the same (non)-Markovian noisy channels. Negative quantum states are discussed in greater detail in Sec.~\ref{ch3_DWF_negativity}. A concept connected with the states having negative discrete Wigner functions is the mana~\citep{veitch2014resource}, which has been used in the literature to compute magic associated with non-stabilizer states. Magic states are found to be ideal resources for quantum computational speedup and fault-tolerant quantum computation. Here, we compute the mana of the qutrit's first and second negative quantum states \citep{jain2020qutritmagic}. We also study the variation of mana under the aforementioned (non)-Markovian channels. We also examine discrete Wigner negativity for the power of prime dimension systems (for $d = 2, 3, 2^2$) under the same (non)-Markovian channels. Quantum coherence is a fundamental prerequisite for all quantum correlations, including entanglement, and it is a crucial physical resource in quantum computation and information processing \citep{baumgratz2014quantifying, xi2015quantum, streltsov2017colloquium, hu2018quantum, zhao20191,paulson2022quantum}. Additionally, entanglement is a premium quantum correlation with numerous operational applications. We will study the dynamics of quantum coherence and entanglement, \textit{i.e.}, concurrence \citep{Wootters1998Entanglement}, utilizing two-qubits' first, second, and third negative quantum states using DWFs and compare them with the corresponding dynamical evolution of Bell states. Average fidelity is commonly used to gauge a channel's performance \citep{horodecki1996teleportation}. The notion of fidelity is a qualitative metric for differentiating between two quantum states \citep{ghosal2021characterizing}. We will use DWFs to compare the average fidelity of the two-qubit system's first, second, and third negative quantum states with the Bell states when subjected to the above noisy channels.  

The chapter is organized as follows. In Sec .~\ref {ch3_DWF_negativity}, we discuss the discrete Wigner negativity, followed by an introduction to negative quantum states for the Hilbert space of $d = 2, 3, 4$. We define the discrete counterpart of non-classical volume in Sec.~\ref{ch3_mana}. Sections~\ref{ch3_neg} and~\ref{ch3_DWF_noise} study the variation of discrete Wigner negativity $|N_G(\pmb{\rho})|$ and mana and the behaviour of DWFs of quantum systems (particularly qubit, qutrit, and two-qubit) under (non)-Markovian unital (random telegraph noise) and non-unital (amplitude damping) channels, respectively. In Secs.~\ref{ch3_coh_con} and~\ref{ch3_fid}, we study the variation of quantum coherence, concurrence, and fidelity of the first, second, and third negative quantum state of the two-qubit and Bell state using DWFs under the same (non)-Markovian channels, followed by the conclusion in Sec.~\ref{ch3_conclusion}. \textit{The contents of this chapter are based on~\citep{lalita2023harnessing}. \copyright Wiley-VCH GmbH. Adapted and reproduced with permission.}
 
\section{From Discrete Wigner Negativity to Negative Quantum States}\label{ch3_DWF_negativity}
In this section, we discuss the discrete Wigner negativity and introduce negative quantum states using the discrete Wigner formalism elaborated in chapter~\ref{chap2:Preliminaries}. Moreover, we also discuss the discrete counterpart of non-classical volume, i.e., mana. Furthermore, we investigate how discrete Wigner negativity and mana vary in the presence of non-Markovian amplitude-damped (AD) and random telegraph noise (RTN).
\subsection{Discrete Wigner Negativity}
The discrete Wigner negativity $|N_{G}({\rho})|$ of a state ${\rho}$, using DWFs, is defined in the following way \citep{van2011noise}, 
 \begin{equation}
 |N_{G}({\rho})| = 
 \begin{cases}
    \left| \min_{\alpha \in Z^{D+1}_D} { (\Tr[{A}_{\alpha}{\rho}])}\right|, &   \Tr[{A}_{\alpha}{\rho}] < 0, \\
      0, &   \Tr[{A}_{\alpha}{\rho}] \geq 0.  
\end{cases}\label{negativity}
\end{equation}
Furthermore, an expression for the robustness of $D$-prime dimensional states toward depolarizing noise having error probability $a$, using the discrete Wigner negativity of states $|N_G({\rho})|$ was developed in \citep{van2011noise},
\begin{equation}
    \begin{aligned}
      a^{*}({\rho}) = 1 - \frac{1}{D^2 |N_{G}({\rho})| + 1},  
      \end{aligned}\label{robust}
\end{equation}
where,
$a^{*}({\rho})$ is $min(a)$ such that
\begin{equation}
    \begin{aligned}
      (1 - a){\rho} + a\frac{I}{D} = \sum_{i} c_{i}\ketbra{S_i}. 
      \end{aligned}
\end{equation}
Here $0 \leq c_i \leq 1$,  $ \sum_{i}c_{i} = 1$ and $\ket{S_i}$ are D-dimensional stabilizer states.
Using Eq. (\ref{robust}), it was shown that for all $D$-prime dimensions, the eigenstates corresponding to the most negative eigenvalue of the phase space point operators are maximally robust to depolarizing noise.\\ 
We will take this up to introduce the negative quantum states of the power of prime-dimensional systems ($d = p^n$), especially qubit, qutrit, and two-qubit systems, using the formalism detailed in~\ref{DWF_def}. Further, discrete Wigner negativity certifies that a state is a non-stabilizer state; it cannot be prepared from a stabilizer state using stabilizer operations alone. However, Wigner negativity does not, by itself, certify entanglement (a product state can have negative DWF), nor does it certify contextuality in the Spekkens-Kochen-Specker sense for all dimensions~\citep{Kochen_Specker_contextuality2022Budroni}. Conversely, a state with zero Wigner negativity (positive DWF) admits an efficient classical simulation via phase-space sampling. However, a state with high Wigner negativity is not automatically useful for every quantum information task; operational advantage depends on the specific task and noise model. The conclusions in this thesis are therefore stated in terms of specific metrics (coherence, concurrence, teleportation fidelity, CHSH violation) under specific noise channels. They should not be interpreted as universal statements about all non-classical resources.

\subsection{Negative Quantum States}\label{ch3_NQS}
Negative quantum states are defined as the normalized eigenvectors associated with the negative eigenvalues of the phase space point operator $A_\alpha$ at a given phase space coordinate $\alpha(q, p)$. The operator $A_\alpha$ plays a pivotal role in discrete phase space formulations, as it directly determines the discrete Wigner function (DWF) via the relation $W_\alpha = \frac{1}{d} \text{Tr}(\rho A_\alpha)$, where $d$ is the system's Hilbert space dimension~\citep{van2011noise, casaccino2008extrema, lalita2023harnessing} elaborated in~\ref{DWF_def}. Particularly, the eigenvectors corresponding to the negative eigenvalues of $A_\alpha$ are known to minimize the DWF, thereby revealing signatures of quantum non-classicality~\citep{van2011noise, lalita2023harnessing}. Using the framework developed in~\citep{wootters2004picturing, gibbons2004discrete}, the negative quantum states are obtained for $d = 2, 3, 2^2$~\citep{lalita2023harnessing, Lalita_2024ProtectingQC}. Moreover, the eigenvector corresponding to the most negative eigenvalue of the phase space point operator $A_\alpha$ is referred to as the first negative quantum state and is denoted by $|NS_1\rangle$. Similarly, the second and third negative quantum states, denoted by $|NS_2\rangle$ and $|NS_3\rangle$, respectively, correspond to the normalized eigenvectors associated with the second and third most negative eigenvalues of $A_\alpha$. This pattern extends to further negative eigenvalues, yielding additional negative quantum states. 

For the single-qubit, there are possible $2^{2+1}$ $A_{\alpha}$'s having spectrum ($\frac{1+\sqrt{3}}{2}$, $\frac{1-\sqrt{3}}{2}$). Out of all possible $\ket{NS_1}$ states for the single-qubit, we shall use the following form of the $\ket{NS_1}$ state, which in the Bloch vector form is given by
\begin{equation}
    \rho_{\ket{NS_1}} = \tfrac{1}{2}( I_2 + a_1\sigma_x + a_2\sigma_y + a_3\sigma_z),
\end{equation}
where the Bloch vector components $a_1$, $a_2$, $a_3$ take the following approximate values $a_1 = 0.50$, $a_2 = 0.56$, $a_3 = -0.66$. Also, for the  single-qutrit case, we have, 
\begin{equation}
    \rho_{\ket{NS_1}} = \frac{1}{3}({I}_{3} + \sqrt{3} \textbf{n} . {{\lambda}}),
\end{equation}
where, $\textbf{n} \in \textbf{R}^8$ have the following approximate values, $n_1 = 0$, $n_2 = 0$, $n_3 = -0.5$, $n_4 = 0$, $n_5 = 0$, $n_6 = 0.4$, $n_7 = 0.7$, $n_8 = -0.3$, ${I}_{3}$ is an identity operator and ${\lambda}$'s are eight Gell-Mann matrices to describe a generalization of the Bloch ball representation of a qubit to the case of a qutrit~\citep{goyal2016geometry}. 

In the case of two-qubit quantum systems, the number of possible $A_{\alpha}$'s and probable DWFs is $4^{4+1}$. Among all, 320 $A_{\alpha}$'s exhibit the spectrum ($-0.5000$, $-0.5000$, $0.1339$, $1.866$), another 320 have the spectrum ($-0.8661$, $-0.5000$, $0.8661$, $1.5000$), and the remaining 384 have the spectrum ($-0.8968$, $-0.1420$, $0.2787$, $1.7601$) as described in~\citep{casaccino2008extrema}. By exploring various combinations of MUB vectors and striations, we identify three distinct $A_{\alpha}$'s, each having one of the above possible spectra. One of the possible $A_{\alpha}$ at the phase space point $\alpha(1, 1)$ with spectrum ($-0.8968$, $-0.1420$, $0.2787$, $1.7601$) is given by
\begin{equation}
   \begin{aligned}
    A_{(1, 1)} = \left(
\begin{array}{cccc}
 0 & -\frac{1}{2}-\frac{i}{2} & \frac{1}{2}-\frac{i}{2} & -\frac{1}{2} \\
 -\frac{1}{2}+\frac{i}{2} & 0 & \frac{i}{2} & 0 \\
 \frac{1}{2}+\frac{i}{2} & -\frac{i}{2} & 1 & 0 \\
 -\frac{1}{2} & 0 & 0 & 0 \\
\end{array}
\right).
      \end{aligned}
      \label{A_NS1}
\end{equation}
The two-qubit $\ket{NS_1}$ state is the normalized eigenvector corresponding to the above phase space point operator $A_{(1, 1)}$'s most negative eigenvalue $-0.8968$. Similarly, the $\ket{NS_2}$, $\ket{NS_3}$, and $\ket{NS_3^{\prime}}$ states can be derived from the other phase space point operators' eigenvalue spectrum. The spectrum ($-0.5000$, $-0.5000$, $0.1339$, $1.866$) possess eigenvalue $-0.5$ with multiplicity $2$. Thus, the eigenvalue $-0.5$ has multiple linearly independent standard eigenvectors. The approximate explicit expressions of the two-qubit negative quantum states are
\begin{eqnarray}
    \begin{aligned}
        \ket{NS_1} &= \left(~a~~b~~c~~d~\right)^T;
        \ket{NS_2} = \left(~p~~q~~r~~s~\right)^T; \nonumber \\
        \ket{NS_3} &= \left(~-l~~m~~n~~l~\right)^T,
        \ket{NS_3^{\prime}} = \left(~-x~~y~~z~~x~\right)^T, \nonumber \\
        \text{and}~~\ket{NS_3^{\prime\prime}} &= \left(~0~~i k~~k~~0~\right)^T.                            
    \end{aligned}
    \label{negative_quantum_states}
\end{eqnarray}
Here, $a = -0.743$, $b = -0.357(1 - i)$, $c = 0.102(1 + i)$, $d = -0.414$, $p = 0.788$, $q = -0.288(1 - i)$, $r = -0.288(1 + i)$, $s = -0.211$, $l = 0.0508$, $m = 0.631 - 0.228i$, $n = -0.279 - 0.682i$, $x = 0.575$, $y = -0.346 + 0.310i$, $z = -0.265 - 0.229i$, and $k = \frac{1}{\sqrt{2}}$. These expressions represent numerically derived approximations of the two-qubit negative quantum states associated with distinct spectral profiles of the phase space point operators.

\subsection{Resource-Theory Context of Negative Quantum States}

\section{Discrete counterpart of non-classical volume}\label{ch3_mana}
The mana ${M}({\rho})$ of a state ${\rho}$ provides information about its applicability in magic state distillation protocols \citep{veitch2014resource}. It is defined as
\begin{equation}
    \begin{aligned}
     {M}({\rho}) \equiv \log \left[ \sum_{\alpha} |W_{\alpha}| \right] = \log(2 Sn({\rho}) + 1),  
      \end{aligned}\label{mana}
\end{equation}
where, $Sn({\rho})$ is sum negativity given by,
\begin{equation}
    \begin{aligned}
     Sn({\rho}) \equiv \sum_{\alpha:W_{\alpha}<0}|W_{\alpha}| \equiv \frac{1}{2} \left(\sum_{\alpha}| W_{\alpha}| - 1\right).  
      \end{aligned}
\end{equation}
A physical interpretation of mana is provided by $Sn({\rho})$. It is the absolute value of the sum of negative entries in the DWFs of a state ${\rho}$. These negative entries are a hindrance to classical computation, and hence they motivate the development of quantum computation. Further, it is the discrete counterpart of the nonclassical volume \citep{kenfack2004negativity,Thapliyal2015Quasiprobability,teklu2015nonlinearity}, defined using the Wigner function in the CV regime.

Discrete Wigner negativity $|N_G({\rho})|$ (for $d = 2, 3, 4$) and mana (for $d = 3$) are studied next under (non)-Markovian noisy channels using Eq. (\ref{negativity}) and Eq. (\ref{mana}), respectively.

\section{\label{ch3_neg}Discrete Wigner negativity and mana variation under different noisy channels}
\begin{figure}[!htpb]
    \centering
    \includegraphics[height=55mm,width=0.75\columnwidth]{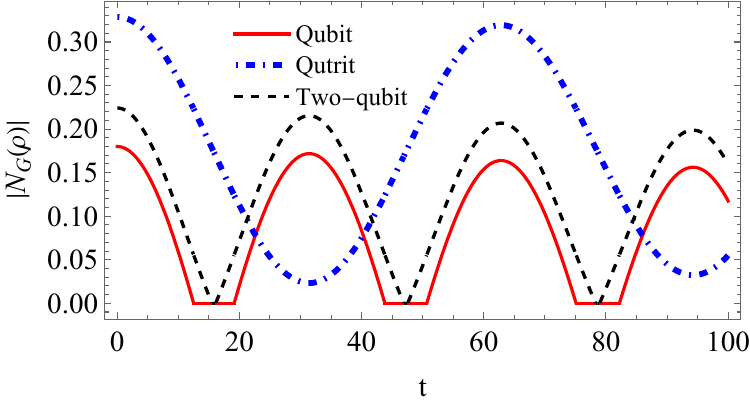}
    \caption{Variation of discrete Wigner negativity for a qubit, qutrit, and two-qubit systems, non-Markovian RTN with time. For $\gamma = 0.001$ and $b = 0.05$.}
    \label{negativityNMRTN}
\end{figure}

\begin{figure}[!htpb]
    \centering
    \includegraphics[height=55mm,width=0.75\columnwidth]{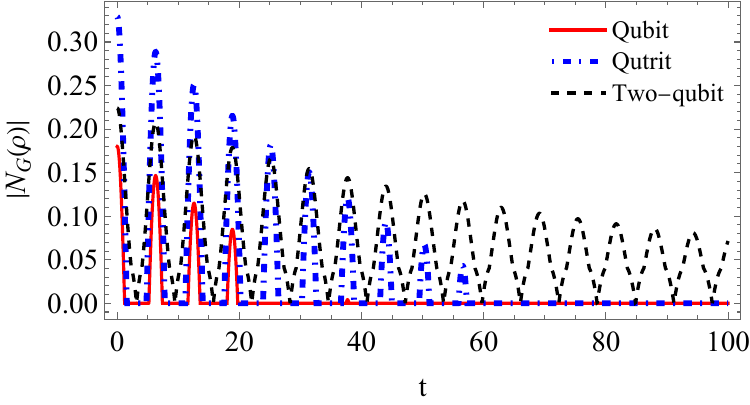}
    \caption{Variation of discrete Wigner negativity for a qubit, qutrit, and two-qubit systems under non-Markovian AD noise with time. For $\gamma = 50$, $g = 0.01$.}
    \label{negativityNMAD}
\end{figure}

Figures~\ref{negativityNMRTN}~ and~\ref{negativityNMAD} show how $|N_G({\rho})|$ changes when subjected to non-Markovian AD and RTN noise. Discrete Wigner negativity is highest for a qutrit compared to a qubit and a two-qubit. Under the influence of (non)-Markovian AD noise, it falls rapidly compared to the two qubits and sustains for a longer duration than the single qubit, which is depicted by Fig.~\ref{negativityNMAD}. Under non-Markovian RTN, all 
The cases {\it, i.e.,} qubit, qutrit, two-qubit, show expected oscillatory behavior, with the peaks and dips of qubit and two-qubit in synchronization, with an alternate pattern with the qutrit, see Fig.~\ref{negativityNMRTN}. 

\begin{figure}[!htpb]
    \centering
    \includegraphics[height=75mm,width=0.85\columnwidth]{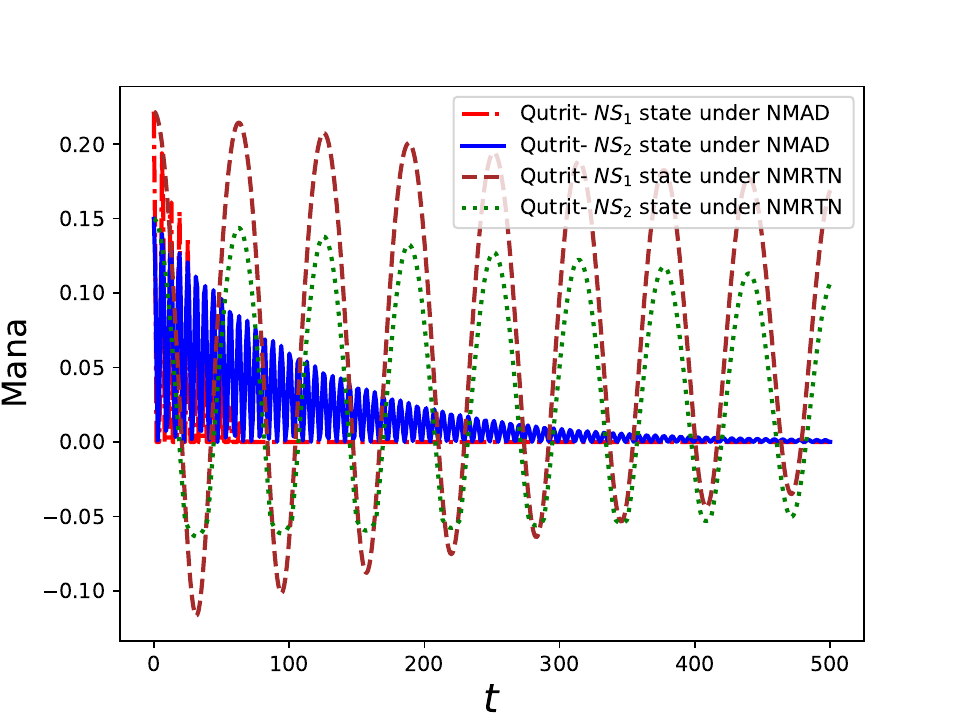}
    \caption{Variation of mana of a qutrit's $\ket{NS_1}$ and $\ket{NS_2}$ state under non-Markovian AD ($\gamma = 50$, $g = 0.01$) and non-Markovian RTN ($\gamma = 0.001$, $b = 0.05$) noise with time.}
    \label{mana-qutrit-NMAD}
\end{figure}

Figure (\ref{mana-qutrit-NMAD}) displays how mana varies for a qutrit's $\ket{NS_1}$ and $\ket{NS_2}$ state when subjected to non-Markovian AD and RTN noise. As we can see from Fig.~\ref{mana-qutrit-NMAD}, initially, the $\ket{NS_1}$ state has a higher value of mana than the $\ket{NS_2}$ state. However, it dies off very quickly in comparison to the $\ket{NS_2}$ state. Hence, the $\ket{NS_2}$ state persists longer and has a finite mana value. Under the non-Markovian RTN, mana for both the negative quantum states of the qutrit show expected oscillatory behaviour, which is persistent for much longer than the non-Markovian AD. Interestingly, the action of the phase S-gate \citep{li2023optimal} produces the conjugate of both the negative quantum states of the qutrit.

\section{\label{ch3_DWF_noise}DWFs of maximally negative quantum states under noisy channels}  
In this section, we calculate the DWFs for single-qubit, single-qutrit, and two-qubit systems, using the formalism given in chapter~\ref{chap2:Preliminaries}, Sec.~\ref{DWF_def}. We then identify the DWFs for the first negative quantum states of single-qubit, single-qutrit, and two-qubit quantum states. Further, we examine their fluctuations under various (non)-Markovian channels.
\subsection{Single-qubit} 
\begin{figure}[!htpb]
    \centering
    \includegraphics[width = 0.4\textwidth, height = 65mm]{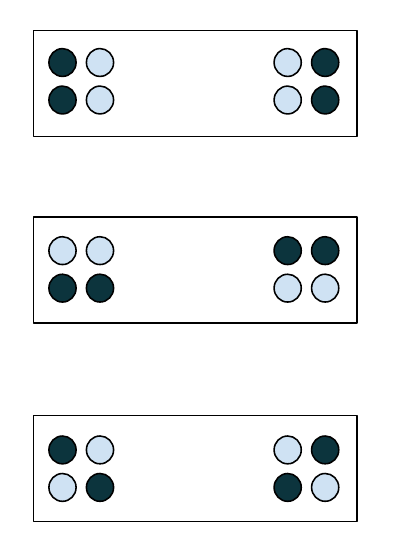}
    \caption{Lines and striations of the $2 \times 2$ phase space.}
    \label{striation1}
\end{figure}

$\label{single-qubit}$
The discrete phase space for a single-qubit system is defined on a $2 \times 2$ real array. The points in this discrete phase space are labeled by elements of the Galois field $\mathbb{F}_2 = \{0, 1\}$. The eigenstates of Pauli operators, $\sigma_x$, $\sigma_y$, and $\sigma_z$ can conveniently be chosen as MUBs for single-qubit systems \citep{galvao2005discrete}, as seen from TABLE~\ref{single_qubit_MUBs}. This $2 \times 2$ phase space has three striations, each having the necessary properties $(i), (ii)$, and $(iii)$ listed in Sec.~\ref {DWF_def} of chapter~\ref{chap2:Preliminaries} and displayed in Fig.~\ref{striation1}. Using the Bloch vector representation for a single-qubit system, {\it i.e.}, $\rho = \frac{1}{2}({I}_2 + \textbf{a} . {\sigma})$, (here $\textbf{a} \in \textbf{R}^3$ and ${\sigma}$'s are Pauli spin matrices) and Eq.~(\ref{DWFformula}) of Sec.~\ref{DWF_def}, chapter~\ref{chap2:Preliminaries}, we find the expressions for single-qubit DWFs for a given association of MUB's as follows 
\begin{eqnarray}
\nonumber
W_{1, 1} = \frac{1}{4} (1 - a_2 + a_3),
 W_{1, 2} = \frac{1}{4} (1 + a_2 - a_3),\\ 
 W_{2, 1} = \frac{1}{4} (1 + a_2 + a_3),
W_{2, 2} = \frac{1}{4} (1 - a_2 - a_3), 
\end{eqnarray}
where $a_1$, $a_2$, and $a_3$ are the components of the single-qubit Bloch vector $\textbf{a}$.
\begin{table}
\centering
\begin{tabular}{ | m{2cm}| m{5cm} | }
  \hline
  \textbf{Striation} & \textbf{MUBs associated with striation}\\ 
  \hline
  1 & 
  $\begin{pmatrix}
  0\\
  1
\end{pmatrix}$,
$\begin{pmatrix}
  1\\
  0
\end{pmatrix}$\\
\hline
  2 &
  $\frac{1}{\sqrt{2}}\begin{pmatrix}
  1\\
  1
\end{pmatrix}$,
$\frac{1}{\sqrt{2}}\begin{pmatrix}
  1\\
  -1
\end{pmatrix}$\\
\hline
  3 & 
  $\frac{1}{\sqrt{2}}\begin{pmatrix}
  1\\ 
  \iota
\end{pmatrix}$,       
$\frac{1}{\sqrt{2}}\begin{pmatrix}
  1\\ 
  -\iota
\end{pmatrix}$\\
\hline
\end{tabular}
\caption{\label{single_qubit_MUBs} The MUBs associated with lines of the $2 \times 2$ discrete phase space of single-qubit systems.}
\end{table}
\subsubsection{\label{qubitNMAD} Amplitude Damping Noise}
\begin{figure}[!htpb]
    \centering
    \includegraphics[height=85mm,width=0.75\columnwidth]{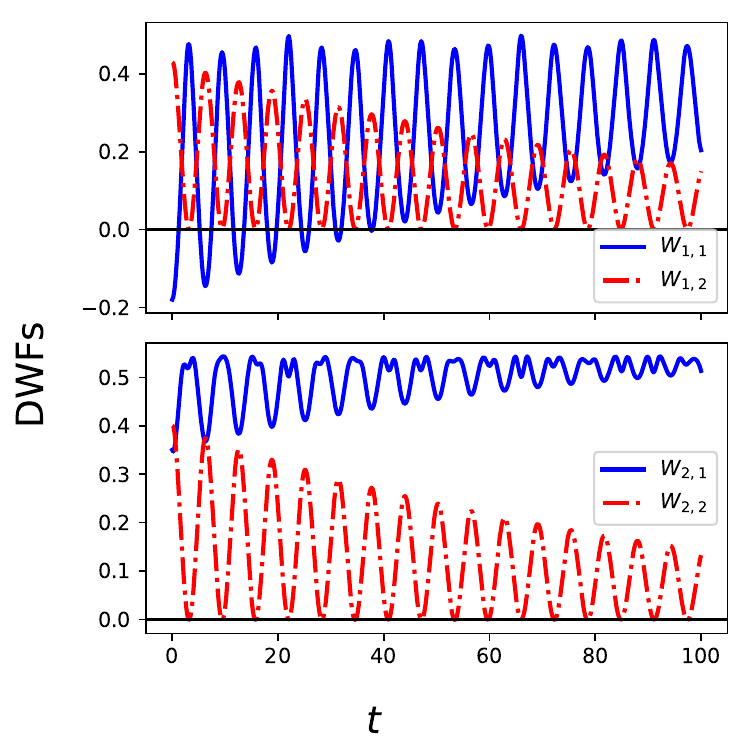}
    \caption{Variation of DWFs corresponding to the qubit's $\ket{NS_1}$ state (for $a_1 = 0.50$, $a_2 = 0.56$, and $a_3 = -0.66$), under non-Markovian AD noise (for $\gamma = 50$, $g = 0.01$) with time.}
    \label{qubitDWFNMAD}
\end{figure}

\begin{figure}[!htpb]
    \centering
    \includegraphics[height=65mm,width=0.75\columnwidth]{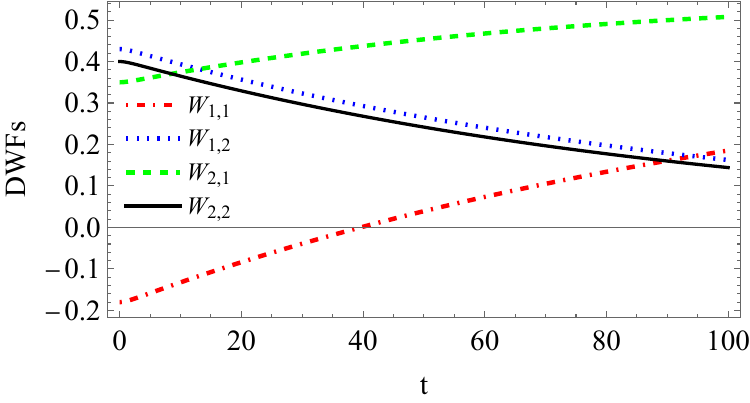}
    \caption{Variation of DWFs corresponding to the qubit's $\ket{NS_1}$ state (for $a_1 = 0.50$, $a_2 = 0.56$, and $a_3 = -0.66$), under Markovian AD noise (for $\gamma = 0.01$, $g = 1$) with time.}
    \label{qubitDWFMAD}
\end{figure}

The evolution of a single-qubit system DWFs under the (non)-Markovian AD noise using the Bloch vector representation, given above in Sec. \ref{single-qubit}, with  Eq. (\ref{ADCfinalrho}) and Eq. (\ref{DWFformula}) detailed in Sec.~\ref{DWF_def} for a particular association of MUBs, can be seen to be 
\begin{align}
    W_{1, 1} &= \frac{-(-1 + a_3)e^{-gt}(\gamma + (-g + \gamma)\cosh(l t) - l \sinh{(l t)}}{4(g - 2\gamma)}
    + \frac{1}{4}\left( 2 - (a_1 + a_2)T(g, \gamma, t)\right),\nonumber \\
    W_{1, 2} &= -\frac{1}{8}(-1 + a_3)e^{-gt}(1 + \cosh{(l t)}) \nonumber \\
    &- \frac{e^{-gt}\left( (-1 + a_3)g \sinh{(l t/2)}^2 + (-1 + a_3) l \sinh{(l t)} + (a_1 - a_2) e^{gt} (g - 2\gamma) T(g, \gamma, t) \right)}{4(g-2\gamma)},\nonumber \\
    W_{2, 1} &= \frac{-(-1 + a_3)e^{-gt}(\gamma + (-g + \gamma)\cosh(l t) - l \sinh{(l t)}}{4(g - 2\gamma)}
    + \frac{1}{4}\left( 2 + (a_1 + a_2)T(g, \gamma, t)\right),\nonumber \\
    W_{2, 2} &= -\frac{1}{8}(-1 + a_3)e^{-gt}(1 + \cosh{(l t)}) \nonumber \\
    &- \frac{e^{-gt}\left( (-1 + a_3)g \sinh{(l t/2)}^2 + (-1 + a_3) l \sinh{(l t)} - (a_1 - a_2) e^{gt} (g - 2\gamma) T(g, \gamma, t)\right)}{4(g-2\gamma)},
\end{align}
where, $T(g, \gamma, t) = \frac{\sqrt{e^{-g t}(-\gamma + (g - \gamma)\cosh{l t} + l\sinh{l t})}}{g - 2\gamma}$ and $a_1$, $a_2$, and $a_3$ are the componenets of the single-qubit Bloch vector $\textbf{a}$.
Plots of the DWFs for the $\ket{NS_1}$ state of the single-qubit are shown in Fig.~\ref{qubitDWFNMAD} and Fig.~\ref{qubitDWFMAD} for non-Markovian and Markovian cases, respectively. The kinks at the peaks in the variation of $W_{2, 1}$ in Fig.~\ref{qubitDWFNMAD} are due to the normalization property of the DWFs.
\subsubsection{\label{RTNqubit}Random Telegraph Noise} 
\begin{figure}[!htpb]
    \centering
    \includegraphics[height=65mm,width=0.75\columnwidth]{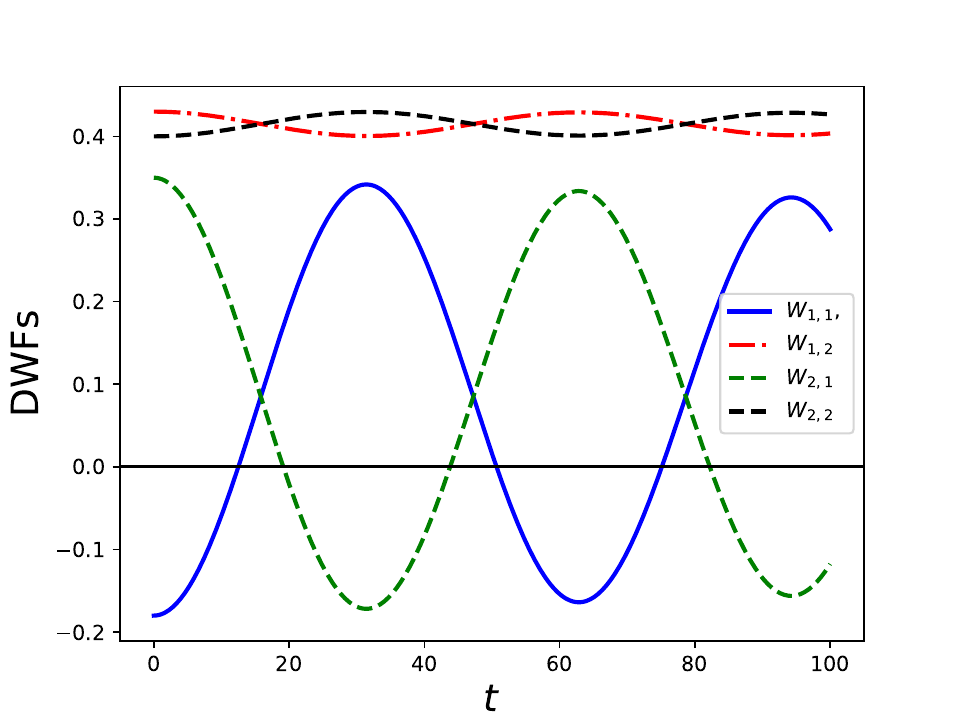}
    \caption{Variation of DWFs corresponding to the qubit's $\ket{NS_1}$ state (when $a_1 = 0.50$, $a_2 = 0.56$ and $a_3 = -0.66$), under non-Markovian RTN (for $\gamma = 0.001$, $b = 0.05$) with time.}
    \label{qubitNMRTN}
\end{figure}

\begin{figure}[!htpb]
    \centering
    \includegraphics[height=65mm,width=0.75\columnwidth]{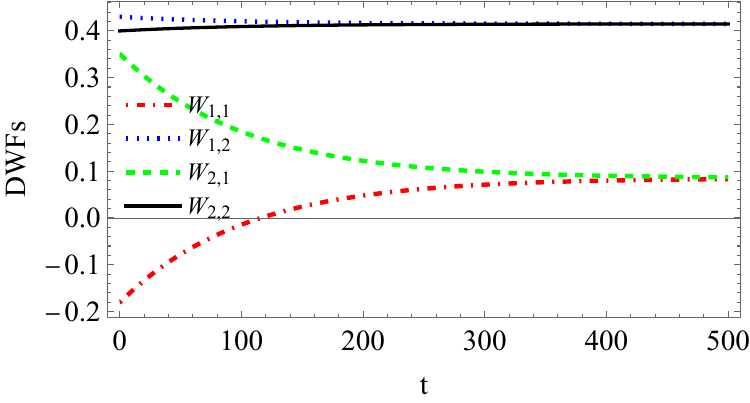}
    \caption{Variation of DWFs corresponding to the qubit's $\ket{NS_1}$ state (when $a_1 = 0.50$, $a_2 = 0.56$ and $a_3 = -0.66$), under Markovian RTN (for $\gamma = 1$ and, $b = 0.07$) with time.}
    \label{qubitMRTN}
\end{figure}

To determine the DWFs of a single-qubit quantum system under the action of (non)-Markovian RTN channel, we employ the Bloch vector representation of single-qubit systems given in Sec.~\ref{single-qubit} with Eq.~(\ref{RTNfinalrho}) and Eq. (\ref{DWFformula}) elaborated earlier in chapter~\ref{chap2:Preliminaries}. For a particular association of MUBs, the final expressions of DWFs for a single-qubit are given as  
\begin{eqnarray}
     W_{1, 1} &=& \frac{1}{4}\left(1 + a_3 - (a_1 + a_2)e^{-\gamma t}\cos\left({\zeta\gamma t}\right)\right. - \left.\frac{(a_1 + a_2)e^{-\gamma t}\sin\left({\zeta\gamma t}\right)}{\zeta}\right),\nonumber\\
      W_{1, 2} &=& \frac{1}{4}\left(1 - a_3 + (-a_1 + a_2)e^{-\gamma t}\cos\left({\zeta\gamma t}\right) \right. + \left.\frac{(-a_1 + a_2)e^{-\gamma t}sin\left({\zeta\gamma t}\right)}{\zeta}\right),\nonumber\\
      W_{2, 1} &=& \frac{1}{4}\left(1 + a_3 + (a_1 + a_2)e^{-\gamma t}\cos\left({\zeta\gamma t}\right) \right. + \left.\frac{(a_1 + a_2)e^{-\gamma t}\sin\left({\zeta\gamma t}\right)}{\zeta}\right),\nonumber\\
     W_{2, 2} &=& \frac{1}{4}\left(1 - a_3 + (a_1 - a_2)e^{-\gamma t}\cos\left({\zeta\gamma t}\right) \right. + \left.\frac{(a_1 - a_2)e^{-\gamma t}\sin\left({\zeta\gamma t}\right)}{\zeta}\right),
\end{eqnarray}
where $a_1$, $a_2$, and $a_3$ are the componenets of the single-qubit Bloch vector $\textbf{a}$.
Figs.~\ref{qubitNMRTN}, \ref{qubitMRTN}, depict the variation of the single-qubit $\ket{NS_{1}}$ state (for $a_1$ = 0.50, $a_2$ = 0.56, and $a_3$ = -0.66) DWFs, in the presence of non-Markovian and Markovian RTN noise, respectively. The characteristic oscillatory features in the non-Markovian regime are clearly visible. 

\subsection{\label{qutrit}Single-qutrit} 
\begin{figure}[!htpb]
    \centering
    \includegraphics[width = 0.4\textwidth, height = 65mm]{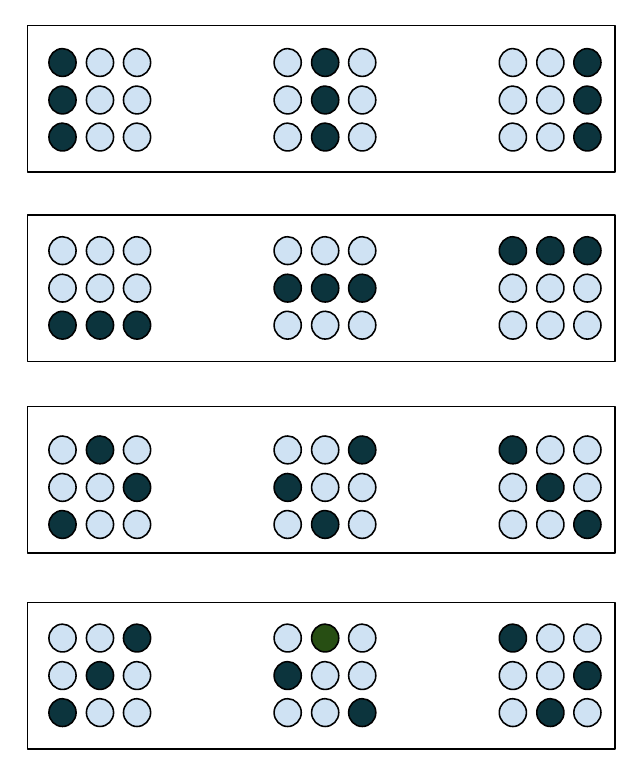}
    \caption{Lines and striations of the $3 \times 3$ phase space.}
    \label{striation2}
\end{figure}

A $3 \times 3$ real array defines the discrete phase space for single-qutrit systems. Elements of the Galois field $\mathbb{F}_3 = \{0, 1, \omega\}$ are used to label the points in this discrete phase space. This phase space has four possible striations, each of which possesses the predefined characteristics $(i)$, $(ii)$, and $(iii)$, as elaborated in Sec.~\ref{DWF_def} and shown in Fig.~\ref{striation2}. A set of four possible MUBs given in TABLE~\ref{qutrit_MUB} is required for a one-to-one mapping with striations to calculate DWFs of single-qutrit systems~\citep{brierley2009all}. 
The Bloch vector representation for a single-qutrit $\it, i.e.,$ $\rho = \frac{1}{3}({I}_{3} + \sqrt{3} \textbf{n} . {\lambda})$, (here, $\textbf{n} \in \textbf{R}^8$, ${I}_{3}$ is an identity operator and ${\lambda}$'s are eight Gell-Mann matrices to describe a generalization of the Bloch ball representation of qubit to the case of qutrit given in~\citep{goyal2016geometry} 
). Using the Bloch vector representation of a single-qutrit together with Eq.~(\ref{DWFformula}), we determine the DWFs of a single-qutrit for a particular association of MUBs given in TABLE~\ref{qutrit_MUB} as follows,
\begin{eqnarray}
      W_{1, 1} &=& \frac{1}{9} (1 + \sqrt{3} n_3 -\sqrt{3} n_6 - 3n_7 + n_8),\nonumber\\
      W_{1, 2} &=& \frac{1}{9} (1 - \sqrt{3} n_1 -\sqrt{3} n_3 - 3n_2 + n_8),\nonumber\\
     W_{1, 3} &=& \frac{1}{9} (1 - \sqrt{3} n_4 + 3n_5 - 2n_8),\nonumber\\
     W_{2, 1} &=& \frac{1}{9} (1 + \sqrt{3} n_3 -\sqrt{3} n_6 + 3n_7 + n_8),\nonumber\\
     W_{2, 2} &=& \frac{1}{9} (1 - \sqrt{3} n_1 -\sqrt{3} n_3 + 3n_2 + n_8),\nonumber\\
    W_{2, 3} &=& \frac{1}{9} (1 - \sqrt{3} n_4 - 3n_5 - 2n_8),\nonumber\\
     W_{3, 1} &=& \frac{1}{9} (1 + \sqrt{3} n_3 + 2\sqrt{3} n_6 + n_8),\nonumber\\
      W_{3, 2} &=& \frac{1}{9} (1 + 2\sqrt{3} n_1 - \sqrt{3} n_3 + n_8),\nonumber\\
     W_{3, 3} &=& \frac{1}{9} (1 + 2\sqrt{3} n_4 - 2n_8).
\end{eqnarray}
\begin{table}
\begin{center}
\begin{tabular}{ | m{2cm}| m{5cm} | }

  \hline
  \textbf{Striation} & \textbf{MUBs associated with striation}\\ 
  \hline
  1 &  
  $\begin{pmatrix}
  1\\ 
  0\\
  0
\end{pmatrix}$,
$\begin{pmatrix}
  0\\ 
  1\\
  0
\end{pmatrix}$,
$\begin{pmatrix}
  0\\ 
  0\\
  1
\end{pmatrix}$\\
\hline
  2 &  
  $\frac{1}{\sqrt{3}}\begin{pmatrix}
  1\\ 
  1\\
  1
\end{pmatrix}$,
$\frac{1}{\sqrt{3}}\begin{pmatrix}
  1\\ 
  \omega\\
  \omega^2
\end{pmatrix}$,
$\frac{1}{\sqrt{3}}\begin{pmatrix}
  1\\ 
  \omega^2\\
  \omega
\end{pmatrix}$\\
\hline
  3 &  
  $\frac{1}{\sqrt{3}}\begin{pmatrix}
  1\\ 
  \omega^2\\
  \omega^2
\end{pmatrix}$,
$\frac{1}{\sqrt{3}}\begin{pmatrix}
  1\\ 
  1\\
  \omega
\end{pmatrix}$,
$\frac{1}{\sqrt{3}}\begin{pmatrix}
  1\\ 
  \omega\\
  1
\end{pmatrix}$\\
\hline
  4 &  
  $\frac{1}{\sqrt{3}}\begin{pmatrix}
  1\\ 
  \omega\\
  \omega
\end{pmatrix}$,
$\frac{1}{\sqrt{3}}\begin{pmatrix}
  1\\ 
  \omega^2\\
  1
\end{pmatrix}$,
$\frac{1}{\sqrt{3}}\begin{pmatrix}
  1\\ 
  1\\
  \omega^2
\end{pmatrix}$\\
\hline
\end{tabular}
\end{center}
\caption{\label{qutrit_MUB} The MUBs associated with lines of the $3 \times 3$ discrete phase space of single-qutrit systems. Here $\omega = e^{2\pi\iota/3}$ is a cube root of unity.}
\end{table}

\subsubsection{Amplitude Damping Noise}
\begin{figure}[!htpb]
    \centering
    \includegraphics[height=65mm,width=0.75\columnwidth]{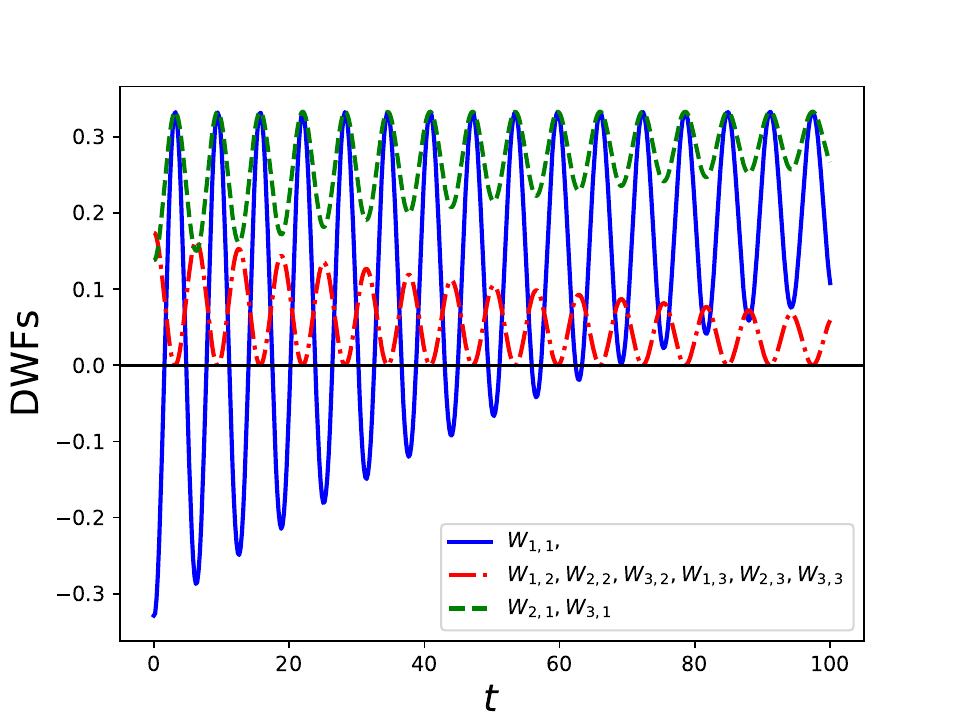}
    \caption{Variation of DWFs corresponding to the the qutrit's $\ket{NS_1}$ state (for $n_1 = 0$, $n_2 = 0$, $n_3 = -0.5$, $n_4 = 0$, $n_5 = 0$, $n_6 = 0.4$, $n_7 = 0.7$, $n_8 = -0.3$), under non-Markovian AD (for $\gamma = 50$, $g = 0.01$) with time.}
    \label{qutritDWFNMAD}
\end{figure}

To study the behavior of single-qutrit DWFs under the (non)-Markovian AD channel, we use its dynamical map Eqs.~(\ref{NMADqutritfinalrho}) and (\ref{DWFformula}) (elaborated earlier in chapter~\ref{chap2:Preliminaries}) for a particular association of MUBs given in TABLE \ref{table2}. For the qutrit's $\ket{NS_1}$ state, Fig.~\ref{qutritDWFNMAD} displays the DWFs variation for the non-Markovian AD case. Compared to the single-qubit's $\ket{NS_1}$ state DWFs, single-qutrit's $\ket{NS_1}$ state DWFs have a higher negative value and remain negative for a longer time, as depicted by Fig.~\ref{qutritDWFNMAD}.
\subsubsection{Random Telegraph noise}
\begin{figure}[!htpb]
    \centering
    \includegraphics[height=65mm,width=0.75\columnwidth]{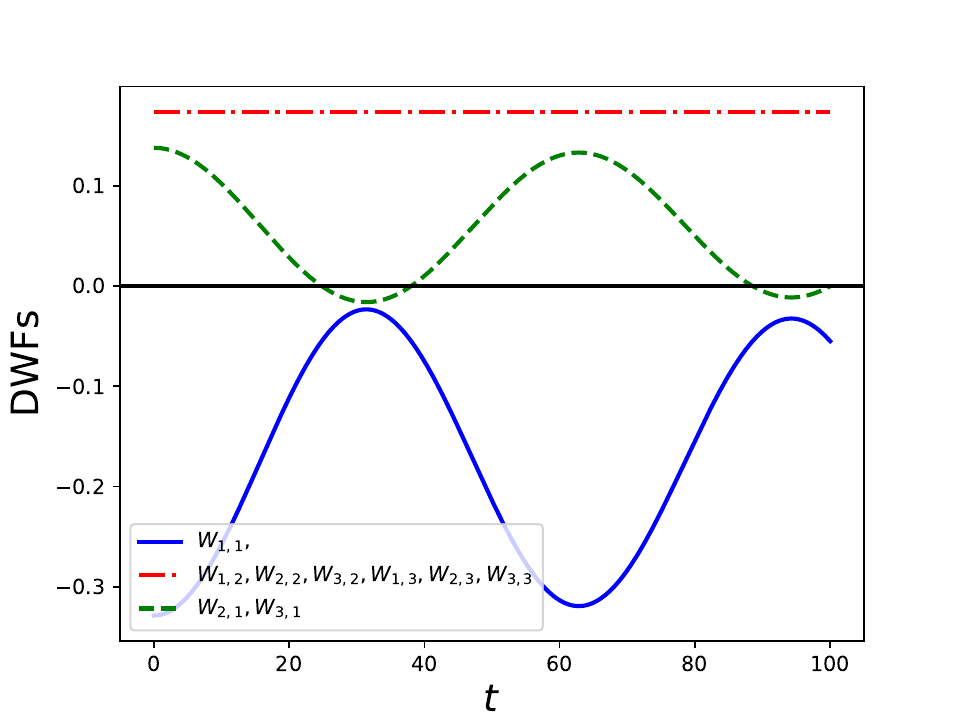}
    \caption{Variation of DWFs corresponding to the qutrit's $\ket{NS_1}$ state (when $n_1 = 0$, $n_2 = 0$, $n_3 = -0.5$, $n_4 = 0$, $n_5 = 0$, $n_6 = 0.4$, $n_7 = 0.7$, $n_8 = -0.3$), under (non)-Markovian RTN (for $\gamma = 0.001$, $b = 0.05$) with time.}
    \label{qutritDWFNMRTN}
\end{figure}

The DWFs of the single-qutrit system under (non)-Markovian RTN channel are determined by first calculating its dynamical form $\mathcal{E}^{RTN}({\rho})$ (discussed in Sec.~\ref{preli_RTN}) using the Bloch vector representation of qutrit described in Sec.~\ref{qutrit} and then using Eq.~(\ref{DWFformula}) for a particular association of MUBs described in TABLE~\ref{qutrit_MUB}. Fig.~\ref{qutritDWFNMRTN} shows the variation of DWFs of the qutrit's $\ket{NS_1}$ state with time for the non-Markovian RTN channel. In the Markovian RTN regime, non-oscillatory behaviour is seen in contrast to the non-Markovian RTN case.
\subsection{\label{two-qubit}Two-qubit} 
\begin{figure}[!htpb]
    \centering
    \includegraphics[width = 0.4\textwidth, height = 65mm]{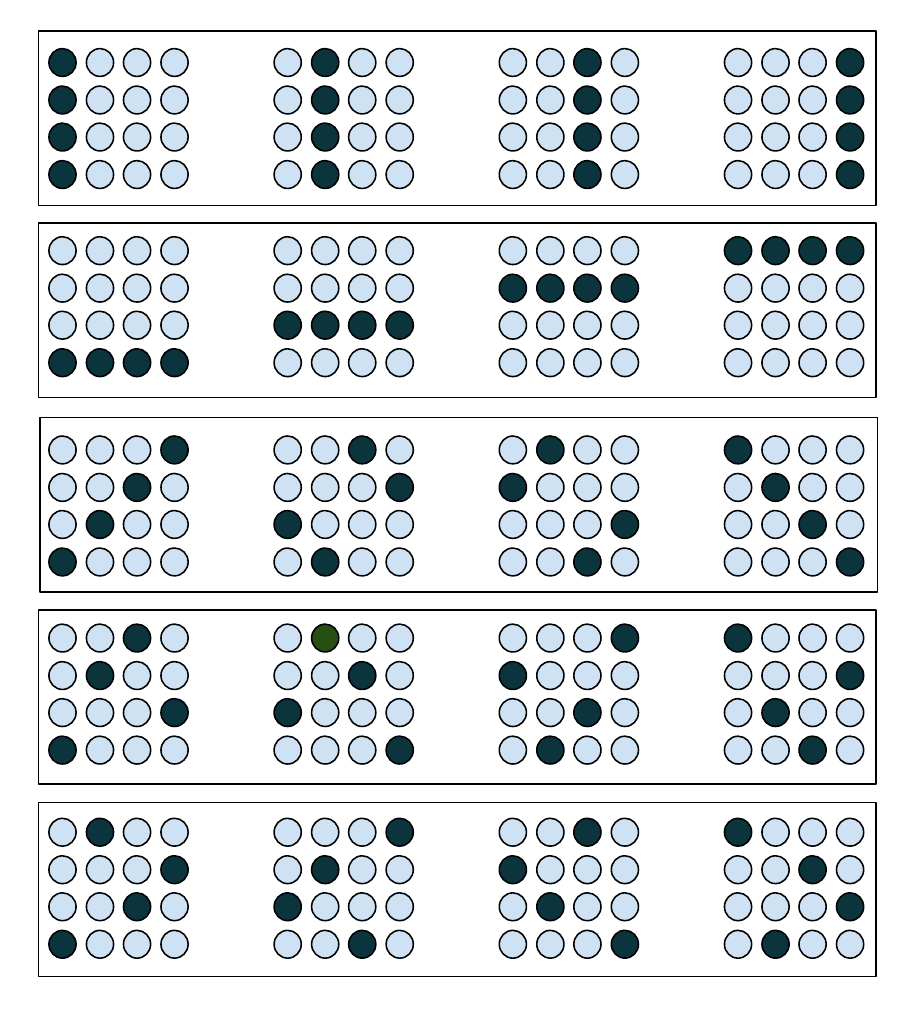}
    \caption{Lines and striations of the $4 \times 4$ phase space.}
    \label{striation3}
\end{figure}

The discrete phase space for two-qubit systems is defined on a $4 \times 4$ array. The Galois field, $\mathbb{F}_4 = \{0, 1, \omega, \omega^2\}$, is used to label the points in this discrete phase space. There are five possible sets of parallel lines (striations), each of which satisfies the ($i$), ($ii$), and ($iii$) properties listed in Sec.~\ref {DWF_def} and depicted by Fig.~\ref{striation3}. A set of five MUBs is needed for a one-to-one mapping with striations~\citep{durt2010mutually}, as provided in TABLE~\ref{two_qubit_MUB}. A two-qubit system is represented as: 
\small{
\begin{equation}
{\rho} = \frac{1}{4}({I_{2}} \otimes {I_{2}} + \sum_{i = 1} ^{3} a_{i} {\sigma_{i}} \otimes {I_{2}} + \sum_{i = 1} ^{3} s_{i} {I_{2}} \otimes {\sigma_{i}}  + \sum_{i,j = 1} ^{3} t_{ij} ({\sigma_{i}} \otimes {\sigma_{j}})),
\label{2-qubit-rho}
\end{equation}
where ${I_{2}}$ denotes identity operator, ${\sigma_{i}}$'s denote the standard Pauli matrices, $a_{i}$'s and $s_{i}$'s are components of vectors in $\textbf{R}^{3}$. The coefficients $t_{ij} = \Tr({\rho} ({\sigma_{i}} \otimes {\sigma_{j}}))$ combine to give a real matrix $\textbf{T}$, known as the correlation matrix. DWFs of two-qubit systems using Eq.~(\ref{DWFformula}) for a particular association of MUBs given in the TABLE~\ref{two_qubit_MUB} and Eq.~(\ref{2-qubit-rho}) are given below,
\begin{align}
W_{1,1} &= \frac{1}{16} \left( 1 - a_1 - a_2 + a_3 - s_1 + s_2 + s_3 + t_{11} - t_{12}\right. - \left. t_{13} + t_{21} - t_{22} - t_{23} - t_{31} + t_{32} + t_{33}\right),\nonumber\\
W_{1,2} &= \frac{1}{16} \left( 1 - a_1 - a_2 + a_3 - s_1 - s_2 - s_3 + t_{11} + t_{12}\right. + \left. t_{13} + t_{21} + t_{22} + t_{23} - t_{31} - t_{32} - t_{33}\right),\nonumber\\
W_{1,3} &= \frac{1}{16} \left( 1 - a_1 + a_2 - a_3 - s_1 + s_2 + s_3 + t_{11} - t_{12}\right. - \left. t_{13} - t_{21} + t_{22} + t_{23} + t_{31} - t_{32} - t_{33}\right),\nonumber\\
W_{1,4} &= \frac{1}{16} \left( 1 - a_1 + a_2 - a_3 - s_1 - s_2 - s_3 + t_{11} + t_{12}\right. + \left. t_{13} - t_{21} - t_{22} - t_{23} + t_{31} + t_{32} + t_{33}\right),\nonumber
\end{align}
\begin{align}
W_{2,1} &= \frac{1}{16} \left( 1 - a_1 - a_2 + a_3 + s_1 - s_2 + s_3 - t_{11} + t_{12}\right. - \left. t_{13} - t_{21} + t_{22} - t_{23} + t_{31} - t_{32} + t_{33}\right),\nonumber\\
W_{2,2} &= \frac{1}{16} \left( 1 - a_1 - a_2 + a_3 + s_1 + s_2 - s_3 - t_{11} - t_{12}\right. + \left. t_{13} - t_{21} - t_{22} + t_{23} + t_{31} + t_{32} - t_{33}\right),\nonumber\\
W_{2,3} &= \frac{1}{16} \left( 1 - a_1 + a_2 - a_3 + s_1 - s_2 + s_3 - t_{11} + t_{12}\right. - \left. t_{13} + t_{21} - t_{22} + t_{23} - t_{31} + t_{32} - t_{33}\right),\nonumber\\
W_{2,4} &= \frac{1}{16} \left( 1 - a_1 + a_2 - a_3 + s_1 + s_2 - s_3 - t_{11} - t_{12}\right. + \left. t_{13} + t_{21} + t_{22} - t_{23} - t_{31} - t_{32} + t_{33}\right),\nonumber    
\end{align}
\begin{align}
W_{3,1} &= \frac{1}{16} \left( 1 + a_1 + a_2 + a_3 - s_1 + s_2 + s_3 - t_{11} + t_{12}\right. + \left. t_{13} - t_{21} + t_{22} + t_{23} - t_{31} + t_{32} + t_{33}\right),\nonumber\\
W_{3,2} &= \frac{1}{16} \left( 1 + a_1 + a_2 + a_3 - s_1 - s_2 - s_3 - t_{11} - t_{12}\right. - \left. t_{13} - t_{21} - t_{22} - t_{23} - t_{31} - t_{32} - t_{33}\right),\nonumber\\
W_{3,3} &= \frac{1}{16} \left( 1 + a_1 - a_2 - a_3 - s_1 + s_2 + s_3 - t_{11} + t_{12}\right. + \left. t_{13} + t_{21} - t_{22} - t_{23} + t_{31} - t_{32} - t_{33}\right),\nonumber\\
W_{3,4} &= \frac{1}{16} \left( 1 + a_1 - a_2 - a_3 - s_1 - s_2 - s_3 - t_{11} - t_{12}\right. - \left. t_{13} + t_{21} + t_{22} + t_{23} + t_{31} + t_{32} + t_{33}\right),\nonumber   
\end{align}
\begin{align}
    W_{4,1} &= \frac{1}{16} \left( 1 + a_1 + a_2 + a_3 + s_1 - s_2 + s_3 + t_{11} - t_{12}\right. + \left. t_{13} + t_{21} - t_{22} + t_{23} + t_{31} - t_{32} + t_{33}\right),\nonumber\\
W_{4,2} &= \frac{1}{16} \left( 1 + a_1 + a_2 + a_3 + s_1 + s_2 - s_3 + t_{11} + t_{12}\right. - \left. t_{13} + t_{21} + t_{22} - t_{23} + t_{31} + t_{32} - t_{33},\right)\nonumber\\
W_{4,3} &= \frac{1}{16} \left( 1 + a_1 - a_2 - a_3 + s_1 - s_2 + s_3 + t_{11} - t_{12}\right. + \left. t_{13} - t_{21} + t_{22} - t_{23} - t_{31} + t_{32} - t_{33}\right),\nonumber\\
W_{4,4} &= \frac{1}{16} \left( 1 + a_1 - a_2 - a_3 + s_1 + s_2 - s_3 + t_{11} + t_{12}\right. - \left. t_{13} - t_{21} - t_{22} + t_{23} - t_{31} - t_{32} + t_{33}\right).\nonumber
\end{align}
}
\begin{table}
\begin{center}
\begin{tabular}{ | m{2cm}| m{8cm} | }

  \hline
  \textbf{Striation} & \textbf{MUBs associated with striation}\\ 
  \hline
  1 &  
  $\begin{pmatrix}
  1\\ 
  0\\
  0\\
  0
\end{pmatrix}$,
$\begin{pmatrix}
  0\\ 
  1\\
  0\\
  0
\end{pmatrix}$,
$\begin{pmatrix}
  0\\ 
  0\\
  1\\
  0
\end{pmatrix}$,
$\begin{pmatrix}
  0\\ 
  0\\
  0\\
  1
\end{pmatrix}$\\
\hline
  2 &  
  $\frac{1}{2}\begin{pmatrix}
  1\\ 
  1\\
  1\\
  1
\end{pmatrix}$,
$\frac{1}{2}\begin{pmatrix}
  1\\ 
 -1\\
  1\\
  -1
\end{pmatrix}$,
$\frac{1}{2}\begin{pmatrix}
  1\\ 
  1\\
  -1\\
  -1
\end{pmatrix}$,
$\frac{1}{2}\begin{pmatrix}
  1\\ 
  -1\\
  -1\\
  1
\end{pmatrix}$\\
\hline
  3 &  
$\frac{1}{2}\begin{pmatrix}
  1\\ 
  -i\\
   i\\
   1
\end{pmatrix}$,
$\frac{1}{2}\begin{pmatrix}
  1\\ 
  i\\
   i\\
   -1
\end{pmatrix}$,
$\frac{1}{2}\begin{pmatrix}
  1\\ 
  -i\\
  -i\\
   -1
\end{pmatrix}$,
$\frac{1}{2}\begin{pmatrix}
  1\\ 
  i\\
  -i\\
   1
\end{pmatrix}$\\
\hline
  4 &  
$\frac{1}{2}\begin{pmatrix}
  1\\ 
  1\\
  i\\
   -i
\end{pmatrix}$,
$\frac{1}{2}\begin{pmatrix}
   1\\ 
  -1\\
   i\\
   i
\end{pmatrix}$,
$\frac{1}{2}\begin{pmatrix}
  1\\ 
  1\\
  -i\\
   i
\end{pmatrix}$,
$\frac{1}{2}\begin{pmatrix}
  1\\ 
  -1\\
  -i\\
   -i
\end{pmatrix}$\\
\hline
  5 &  
$\frac{1}{2}\begin{pmatrix}
  1\\ 
  -i\\
   1\\
  i
\end{pmatrix}$,
$\frac{1}{2}\begin{pmatrix}
  1\\ 
  i\\
   1\\
   -i
\end{pmatrix}$,
$\frac{1}{2}\begin{pmatrix}
  1\\ 
  -i\\
  -i\\
   -i\\
\end{pmatrix}$,
$\frac{1}{2}\begin{pmatrix}
  1\\ 
  i\\
  -1\\
   i
\end{pmatrix}$\\
\hline
\end{tabular}
\end{center}
\caption{\label{two_qubit_MUB} The MUBs associated with lines of the $4 \times 4$ discrete phase space of two-qubit systems.}
\end{table}

\subsubsection{Amplitude Damping Noise}
\begin{figure}[!htpb]
    \centering
    \includegraphics[height=85mm,width=0.75\columnwidth]{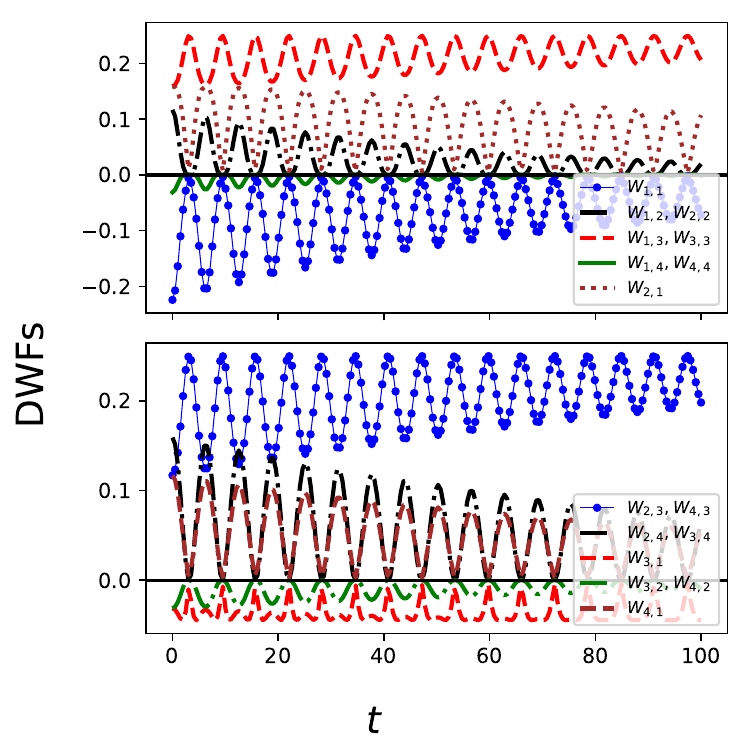}
    \caption{Variation of DWFs corresponding to the two-qubit's $\ket{NS_1}$ state ($a_1 = 0.14$, $a_2 = 0.14$, $a_3 = 0.61$, $s_1 = 0.44$, $s_2 = -0.44$, $s_3 = 0.14$, $t_{11} = 0.61$, $t_{12} = 0.14$, $t_{13} = -0.44$, $t_{21} = -0.14$, $t_{22} = -0.61$, $t_{23} = -0.44$, $t_{31} = 0.61$, $t_{32} = -0.61$ and, $t_{33} = 0.44$), under (non)-Markovian AD (for $\gamma = 50$, $g = 0.01$) with time.}
    \label{2qubitDWFNMAD}
\end{figure}

Using the dynamical form of the non-Markovian AD channel given in Eq.~(\ref{2qubitNMADfinalrho}) and Eq.~(\ref{DWFformula}) for a particular association of MUBs given in TABLE \ref{two_qubit_MUB}, we can study the variation of DWFs of a two-qubit system with time under the action of non-Markovian AD noise. Figure~\ref{2qubitDWFNMAD} illustrates the behaviour of the two-qubit $\ket{NS_1}$ state under the non-Markovian AD evolution with time. Compared to single-qubit's $\ket{NS_1}$ state DWFs, two-qubit's $\ket{NS_1}$ state DWFs have a higher negative value. It is less than the single-qutrit case but sustains negative values for much longer than the single qubit and qutrit, and is shown in Fig.~\ref{2qubitDWFNMAD}.   
\subsubsection{Random Telegraph Noise}
\begin{figure}[!htpb]
    \centering
    \includegraphics[height=85mm,width=0.75\columnwidth]{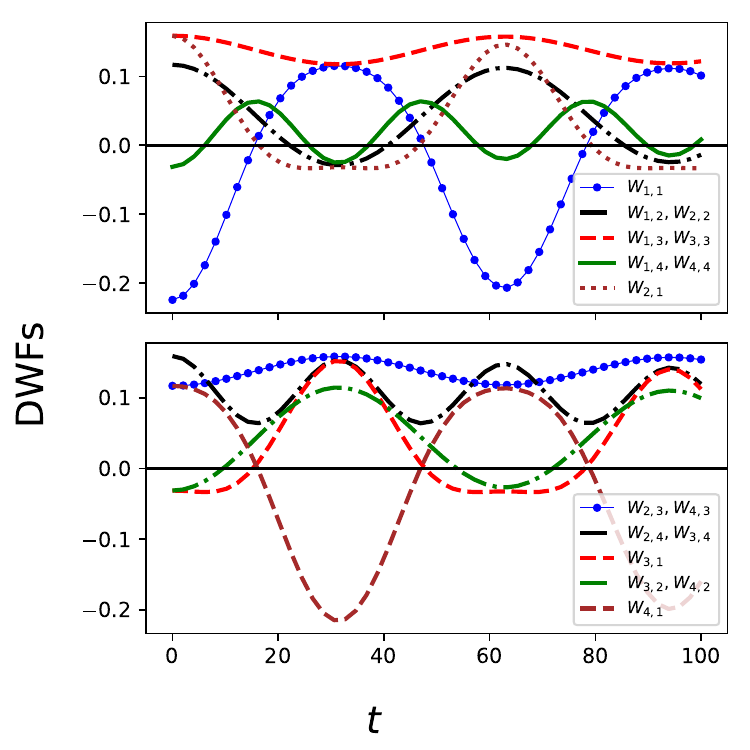}
    \caption{Variation of DWFs corresponding to the two-qubit's $\ket{NS_1}$ state ($a_1 = 0.14$, $a_2 = 0.14$, $a_3 = 0.61$, $s_1 = 0.44$, $s_2 = -0.44$, $s_3 = 0.14$, $t_{11} = 0.61$, $t_{12} = 0.14$, $t_{13} = -0.44$, $t_{21} = -0.14$, $t_{22} = -0.61$, $t_{23} = -0.44$, $t_{31} = 0.61$, $t_{32} = -0.61$ and, $t_{33} = 0.44$), under non-Markovian RTN (for $\gamma = 0.001$ and $b = 0.05$).}
    \label{2qubitDWFNMRTN}
\end{figure}

The evolution of DWFs of a two-qubit system for a particular association of MUBs given in TABLE~\ref{two_qubit_MUB} with the (non)-Markovian RTN channel can be constructed using Eq. (\ref{2qubitRTNfinalrho}) and Eq.(\ref{DWFformula}) discussed earlier in chapter~\ref{chap2:Preliminaries}. Figure~\ref{2qubitDWFNMRTN} depicts how the non-Markovian RTN scenario's two-qubit $\ket{NS_1}$ state changes over time.

\section{\label{ch3_coh_con}Quantum coherence and entanglement using DWFs under different noisy channels: two-qubit systems} 
\begin{figure}[!htpb]
    \centering
    \includegraphics[height=65mm,width=0.75\columnwidth]{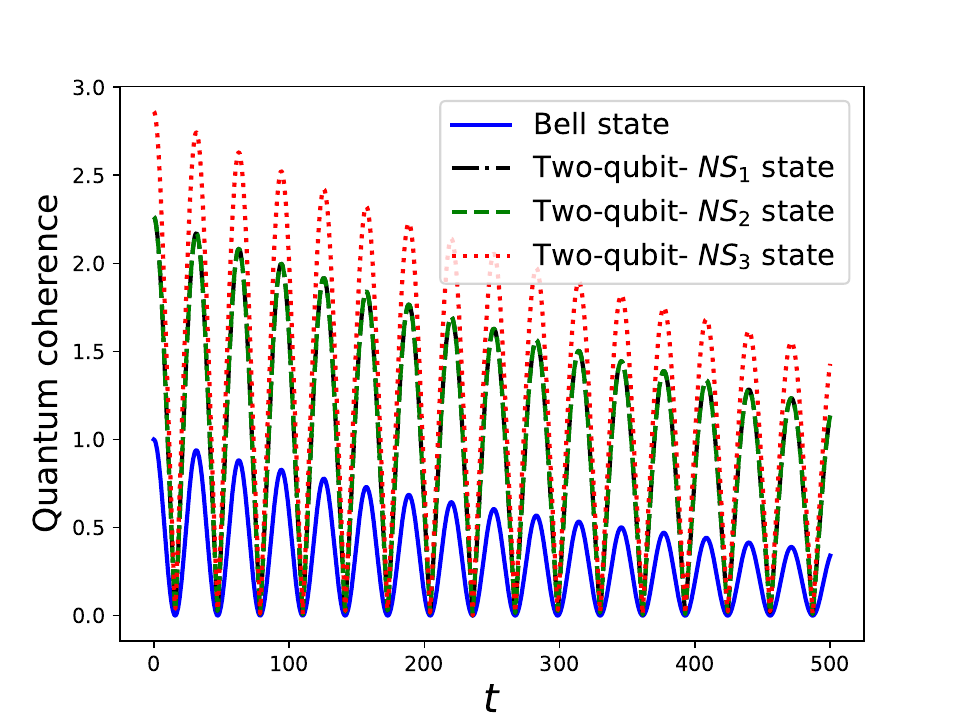}
    \caption{Variation of quantum coherence for the two-qubit's $\ket{NS_1}$, $\ket{NS_2}$, $\ket{NS_3}$ states, and Bell state under non-Markovian RTN noise with time. For $\gamma = 0.001$ and $b = 0.05$.}
    \label{coherence_NMRTN}
\end{figure}

\begin{figure}[!htpb]
    \centering
    \includegraphics[height=65mm,width=0.75\columnwidth]{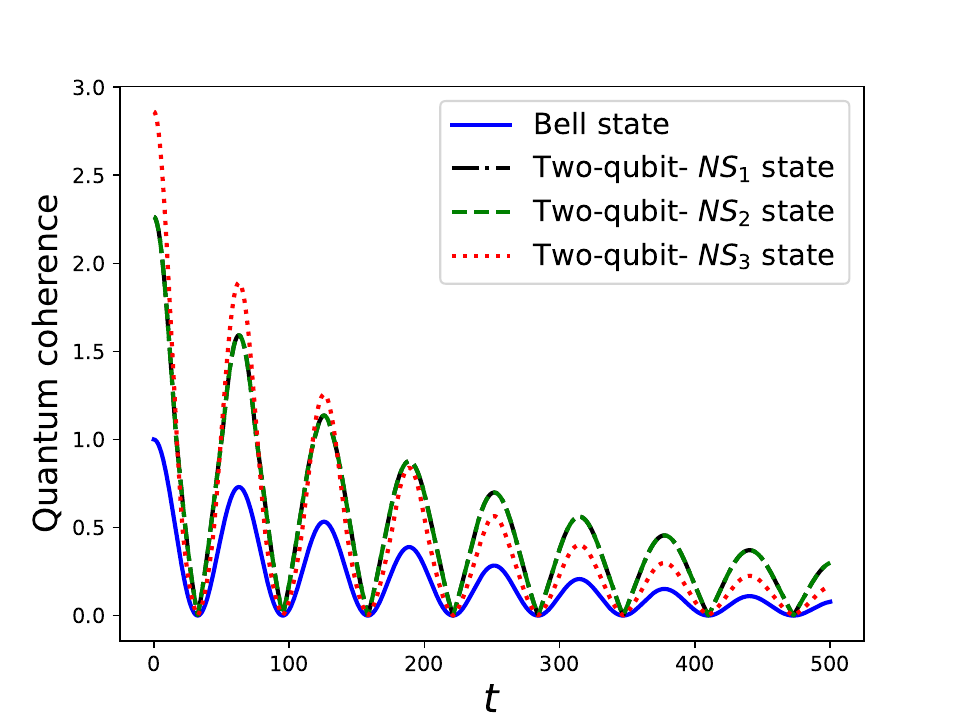}
    \caption{Variation of quantum coherence for the two-qubit's $\ket{NS_1}$, $\ket{NS_2}$, $\ket{NS_3}$ states, and Bell state under non-Markovian AD noise with time. For $\gamma = 1$, $g = 0.005$.}
    \label{coherence_NMAD}
\end{figure}

Several methods for determining the coherence of a quantum system are available in the literature \citep{baumgratz2014quantifying, girolami2014observable}. We particularly focus here on the $l_{1}$ norm of coherence, defined as the sum of the absolute values of all off-diagonal elements of ${\rho}$ as given below \citep{baumgratz2014quantifying}.
\begin{equation}
   \begin{aligned}
     C_{l_{1}}({\rho}) = \sum_{i \neq j} |{\rho}_{i,j}|.
      \end{aligned}
      \label{coherence_Eq.}
\end{equation}
The variation in quantum coherence for the two-qubit $\ket{NS_1}$, $\ket{NS_2}$, and $\ket{NS_3}$ states as well as Bell states under the operation of (non)-Markovian RTN and AD channels is next examined using Eq. (\ref{rho-decomposition-in-A}), $\textit{i.e.}$, the DWFs form of the states, and the above Eq. (\ref{coherence_Eq.}). From Figs. (\ref{coherence_NMRTN}) and (\ref{coherence_NMAD}), it is clear that the two-qubit negative quantum states, $\ket{NS_1}$, $\ket{NS_2}$, and $\ket{NS_3}'$ have quantum coherence greater than the Bell state under non-Markovian RTN and AD noise channels. Initially, the $NS_{3}$ state has maximum quantum coherence in comparison to the $NS_{1}$, $NS_{2}$, and Bell state, as can be seen from Figs. (\ref{coherence_NMRTN}) and (\ref{coherence_NMAD}). Moreover, all the states display anticipated decaying oscillatory behavior under the non-Markovian RTN and AD noise channels, as illustrated in Figs. (\ref{coherence_NMRTN}) and (\ref{coherence_NMAD}). Additionally, the quantum coherence of the $\ket{NS_1}$ and $\ket{NS_2}$ states under a non-Markovian AD noise channel is sustained for longer, as can be seen from Fig.~\ref{coherence_NMAD}.

\begin{figure}[!htpb]
    \centering
    \includegraphics[height=65mm,width=0.75\columnwidth]{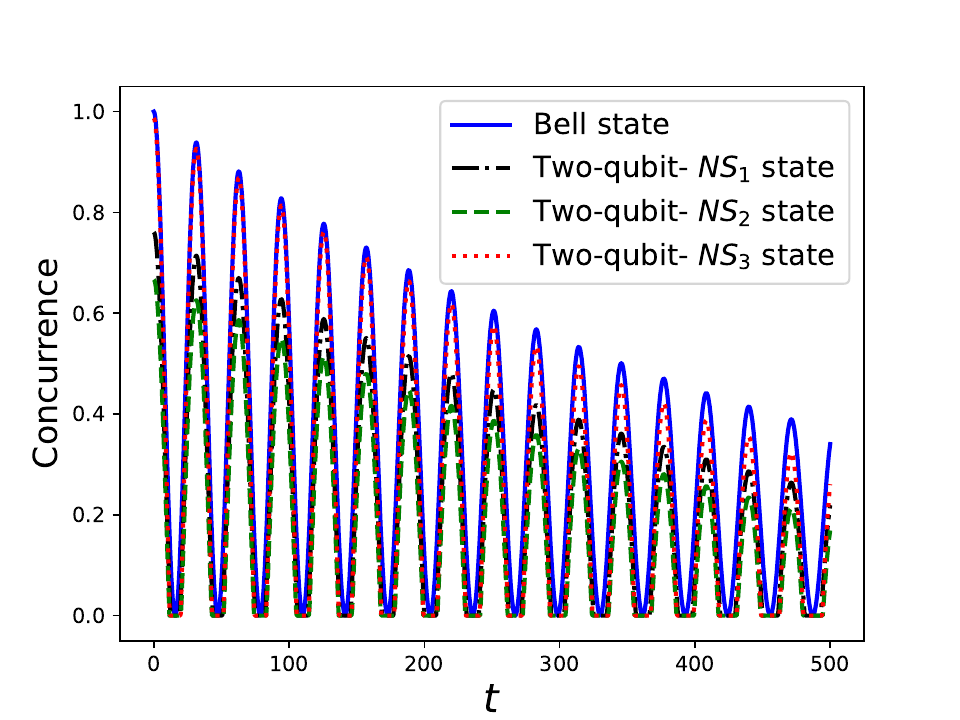}
    \caption{Concurrence variation for the two-qubit's $\ket{NS_1}$, $\ket{NS_2}$, $\ket{NS_3}$ states, and Bell state under non-Markovian RTN noise with time. For $\gamma = 0.001$ and $b = 0.05$.}
    \label{concurNMRTN}
\end{figure}

\begin{figure}[!htpb]
    \centering
    \includegraphics[height=65mm,width=0.75\columnwidth]{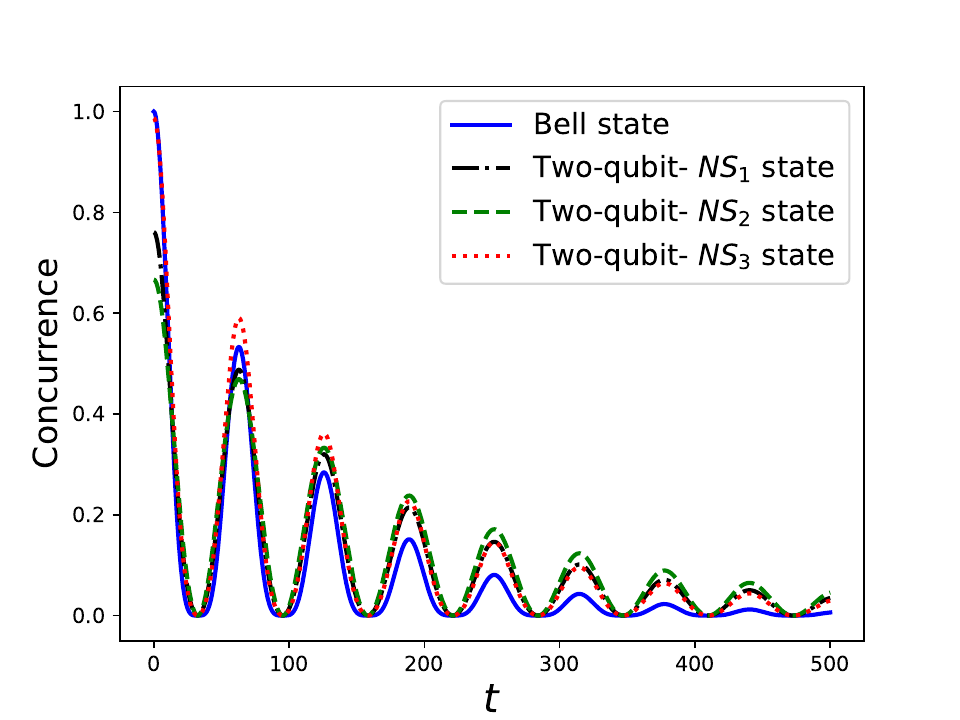}
    \caption{Concurrence variation for the two-qubit's $\ket{NS_1}$, $\ket{NS_2}$, $\ket{NS_3}$ states, and Bell state under non-Markovian AD noise with time. For $\gamma = 1$, $g = 0.005$.}
    \label{concurNMAD}
\end{figure}

We now examine the variation of entanglement under (non)-Markovian noisy channels, as it is one of the most crucial sources of quantum information. For a two-qubit system, concurrence is an entanglement metric \citep{Wootters1998Entanglement}, which is defined as
\begin{equation}
   \begin{aligned}
     C({\rho}_{AB}) = max \{0, \lambda_{1} - \lambda_{2} - \lambda_{3} - \lambda_{4}\},
      \end{aligned}
\end{equation}
Here $\lambda_{i}$'s are the eigenvalues of $\sqrt{\sqrt{{\rho}_{AB}} \tilde{{\rho}}_{AB} \sqrt{{\rho}_{AB}}}$ in the descending order and $\Tilde{{\rho}}_{AB} = (\sigma_{y} \otimes \sigma_{y}) {\rho}_{AB}^{*} (\sigma_{y} \otimes \sigma_{y})$, ${\rho}_{AB}^{*}$ is the complex conjugate of ${\rho}_{AB}$. We have $C({\rho}_{AB}) = 0$ for separable states and $0 < C({\rho}_{AB}) \leq 1$ for entangled states. Using the DWFs form of the states, Eq. (\ref{rho-decomposition-in-A}), we next analyze the variation in concurrence for the two-qubit's $\ket{NS_1}$, $\ket{NS_2}$, and $\ket{NS_3}$ states and Bell states under the operation of (non)-Markovian RTN and AD channels. We can see from Figs. (\ref{concurNMRTN}) and (\ref{concurNMAD}) that at $t = 0$, the $\ket{NS_1}$, $\ket{NS_2}$, and $\ket{NS_3}$ states have concurrence between zero and one, $\textit{i.e.}$, these states are entangled, an indicator of quantumness.  

\begin{figure}[!htpb]
    \centering
    \includegraphics[height=65mm,width=0.75\columnwidth]{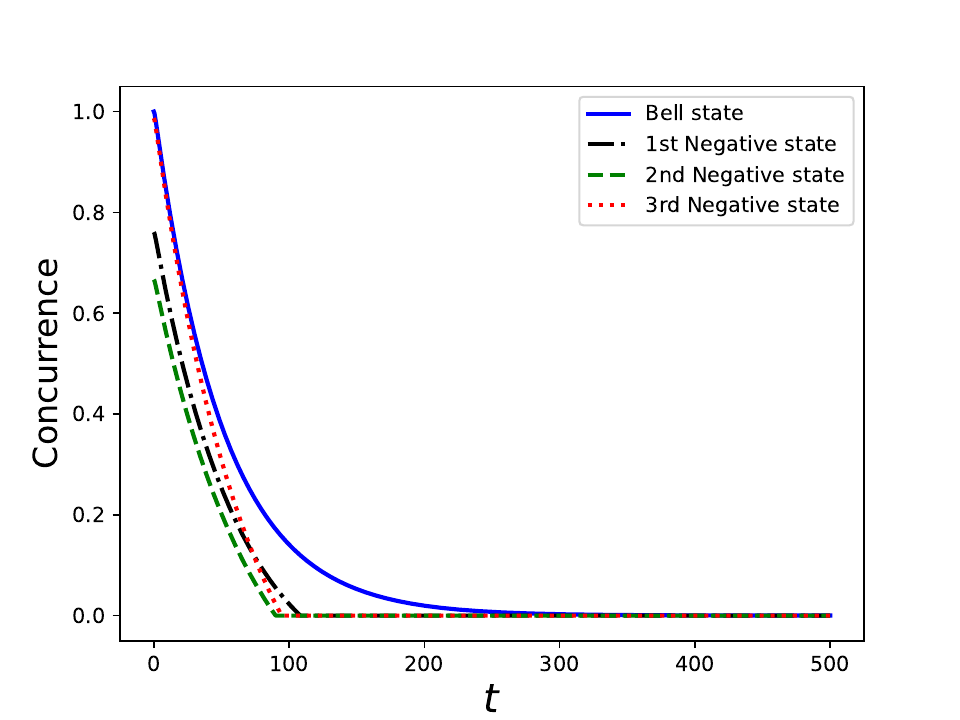}
    \caption{Concurrence variation for the two-qubit's $\ket{NS_1}$, $\ket{NS_2}$, $\ket{NS_3}$ states, and Bell state under Markovian RTN noise with time. For $\gamma = 1$ and $b = 0.07$.}
    \label{concurMRTN}
\end{figure}

\begin{figure}[!htpb]
    \centering
    \includegraphics[height=65mm,width=0.75\columnwidth]{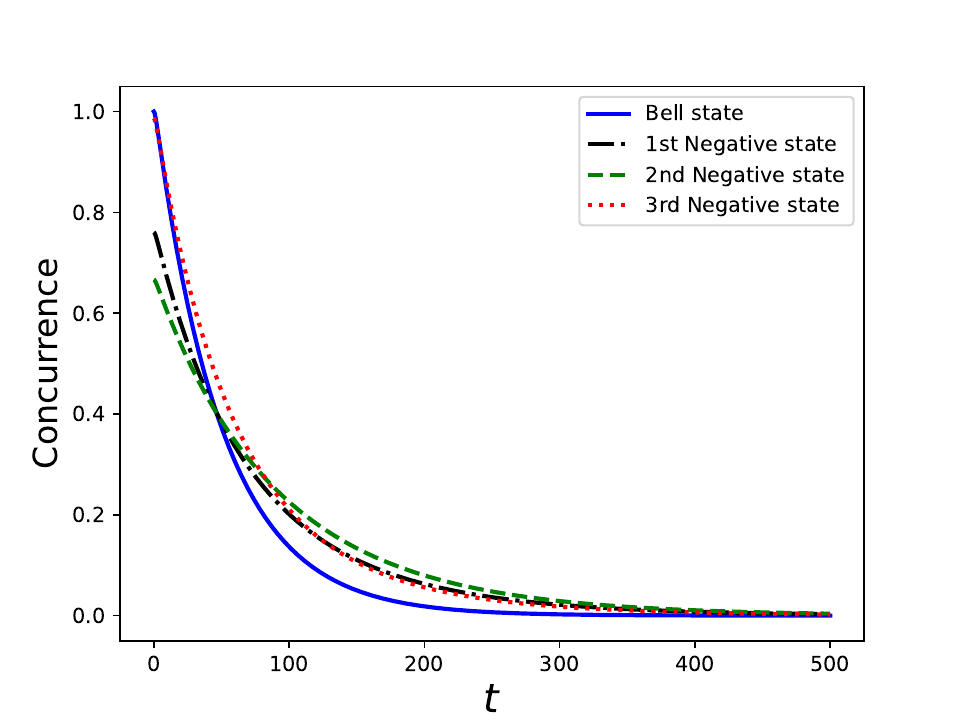}
    \caption{Concurrence variation for the two-qubit's $\ket{NS_1}$, $\ket{NS_2}$, $\ket{NS_3}$ states, and Bell state under Markovian AD noise with time. For $\gamma = 0.01$, $g = 1$.}
    \label{concurMAD}
\end{figure}

Figure~(\ref{concurMRTN}) indicates that, under Markovian RTN noise, the Bell state consistently exhibits higher concurrence than all considered negative states over both short and long time scales. In contrast, from Fig.~(\ref{concurMAD}) we can observe that under the Markovian AD channel, the negative quantum states, $\ket{NS_1}$, $\ket{NS_2}$, and $\ket{NS_3}$, surpass the Bell state over a substantial temporal interval, with $\ket{NS_3}$ maintaining a higher concurrence throughout the entire evolution. In the non-Markovian regime, Fig.~(\ref{concurNMRTN}) shows that both the Bell and negative states display synchronized, decaying oscillations in concurrence under RTN noise. Moreover, Fig.~(\ref{concurNMAD}) reveals that, although $\ket{NS_3}$ initially has concurrence comparable to the Bell state, it subsequently exceeds it under non-Markovian AD dynamics. Similarly, $\ket{NS_1}$ and $\ket{NS_2}$, despite starting with lower concurrence, eventually attain values greater than that of the Bell state at later times. 



\section{\label{ch3_fid}Teleportation fidelity variation using DWFs under different noisy channels}
Quantum teleportation uses two-qubit entangled states as a resource, and the teleportation fidelity \citep{horodecki1996teleportation} is determined as
\begin{equation}
   \begin{aligned}
     F({\rho}_{AB}) = \frac{1}{2}\left(1 + \frac{N_F({\rho}_{AB})}{3}\right),
      \end{aligned}\label{teleportationfidelity}
\end{equation}
where ${\rho}_{AB}$ is as in Sec. \ref{two-qubit}, and $N_{F}({\rho}_{AB}) = Tr\sqrt{T^{\dagger}T}$. The two-qubit state is advantageous for quantum teleportation iff $N_{F}({\rho}_{AB}) > 1$, that is, $F({\rho}_{AB}) > \frac{2}{3}$ (classical limit). Using the DWF form of states, Eq. (\ref{rho-decomposition-in-A}) discussed in chapter~\ref{chap2:Preliminaries}, we estimate the correlation matrix $T$ for the two-qubit's $\ket{NS_1}$, $\ket{NS_2}$, and $\ket{NS_3}$ states, and Bell states. The correlation matrix elements for the two qubits $NS_{1}$ state are 

\begin{align}
t_{11} =& 1 - 2\left(W_{1,1} + W_{1,2} + W_{1,3} + W_{1,4}\right.  + \left. W_{4,1} + W_{4,2} + W_{4,3} + W_{4,4}\right),\nonumber\\
t_{12} =& 1 - 2\left(W_{1,2} + W_{1,4} + W_{2,1} + W_{2,3}\right.  +\left. W_{3,1} + W_{3,3} + W_{4,2} + W_{4,4}\right),\nonumber\\
t_{13} =& 1 - 2\left(W_{1,2} + W_{1,4} + W_{2,2} + W_{2,4}\right.  +\left. W_{3,1} + W_{3,3} + W_{4,1} + W_{4,3}\right),\nonumber\\
t_{21} =& 1 - 2\left(W_{1,1} + W_{1,2} + W_{2,3} + W_{2,4}\right.  +\left. W_{3,3} + W_{3,4} + W_{4,1} + W_{4,2}\right),\nonumber\\
t_{22} =& 1 - 2\left(W_{1,2} + W_{1,3} + W_{2,1} + W_{2,4}\right.  +\left. W_{3,1} + W_{3,4} + W_{4,2} + W_{4,3}\right),\nonumber\\
t_{23} =& 1 - 2\left(W_{1,2} + W_{1,3} + W_{2,2} + W_{2,3}\right.  +\left. W_{3,1} + W_{3,4} + W_{4,1} + W_{4,4}\right),\nonumber\\
t_{31} =& 1 - 2\left(W_{1,1} + W_{1,2} + W_{2,3} + W_{2,4}\right.  +\left.  W_{3,1} + W_{3,2} + W_{4,3} + W_{4,4}\right),\nonumber\\
t_{32} =& 1 - 2\left(W_{1,2} + W_{1,3} + W_{2,1} + W_{2,4}\right.  +\left.  W_{3,2} + W_{3,3} + W_{4,1} + W_{4,4}\right),\nonumber\\
t_{33} =& 1 - 2\left(W_{1,1} + W_{1,4} + W_{2,1} + W_{2,4}\right.  +\left.  W_{3,1} + W_{3,4} + W_{4,1} + W_{4,4}\right).\nonumber\\
\label{T-matrix}
\end{align}
Teleportation fidelity is calculated using Eq. (\ref{teleportationfidelity}) and the above Eq.~(\ref{T-matrix}) correlation matrix elements to find the $N_{F}({\rho}_{AB})$. 

\begin{figure}[!htpb]
    \centering
    \includegraphics[height=65mm,width=0.75\columnwidth]{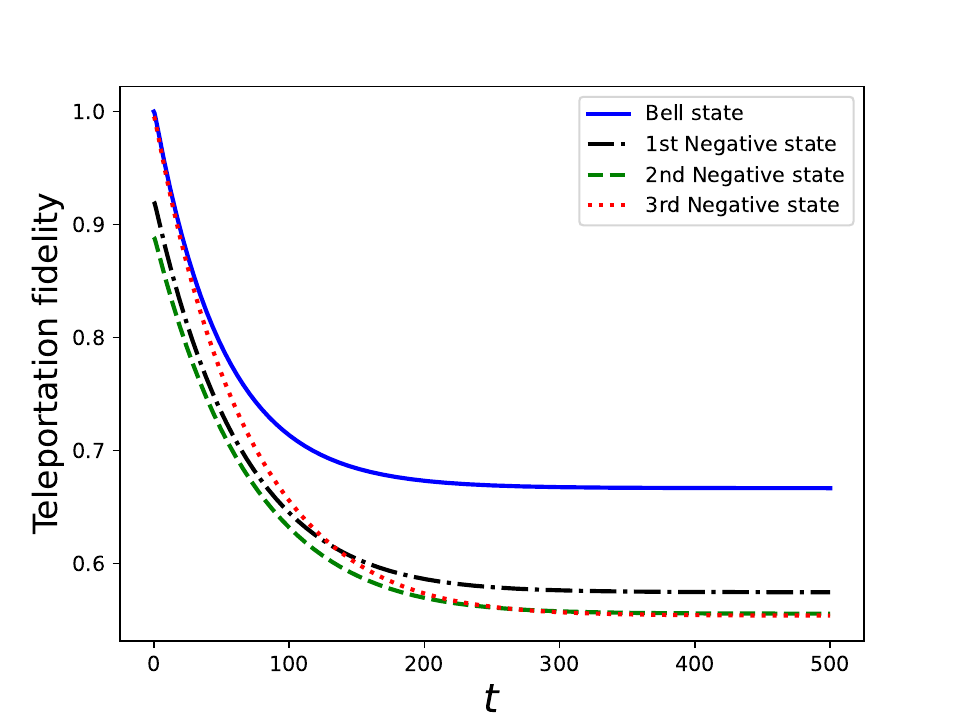}
    \caption{Variation of teleportation fidelity for the two-qubit's $\ket{NS_1}$, $\ket{NS_2}$, $\ket{NS_3}$ states, and Bell state under Markovian RTN noise with time. For $\gamma = 1$ and $b = 0.07$.}
    \label{mrtn_fid_final}
\end{figure}

\begin{figure}[!htpb]
    \centering
    \includegraphics[height=65mm,width=0.75\columnwidth]{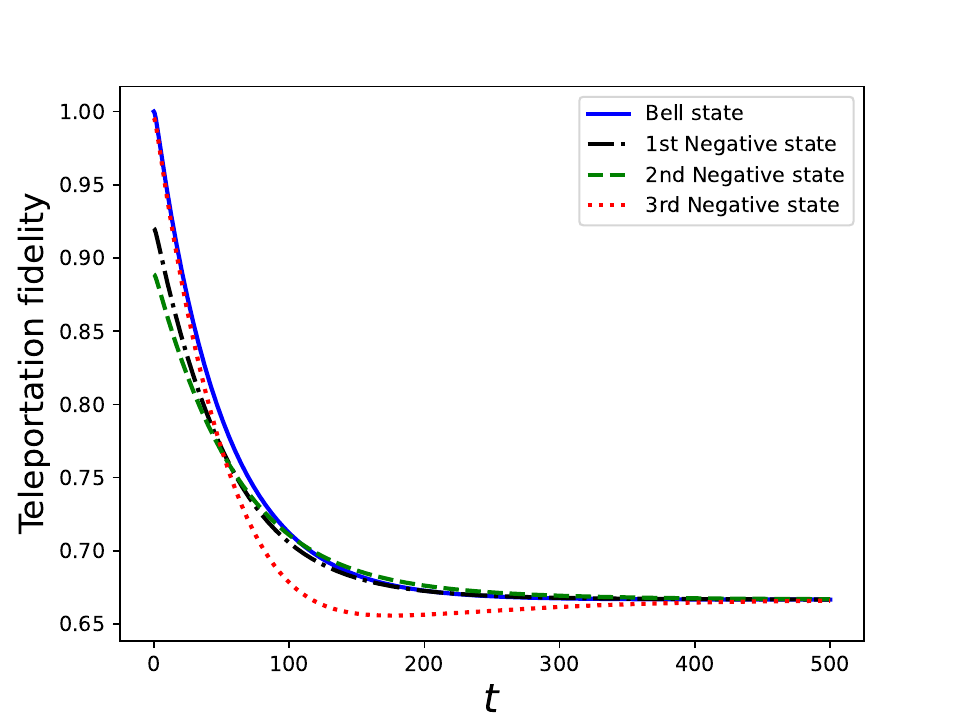}
    \caption{Variation of teleportation fidelity for the two-qubit's $\ket{NS_1}$, $\ket{NS_2}$, $\ket{NS_3}$ states, and Bell state under Markovian AD noise with time. For $\gamma = 0.01$, $g = 1$.}
    \label{mad_fid_final}
\end{figure}

\begin{figure}[!htpb]
    \centering
    \includegraphics[height=65mm,width=0.75\columnwidth]{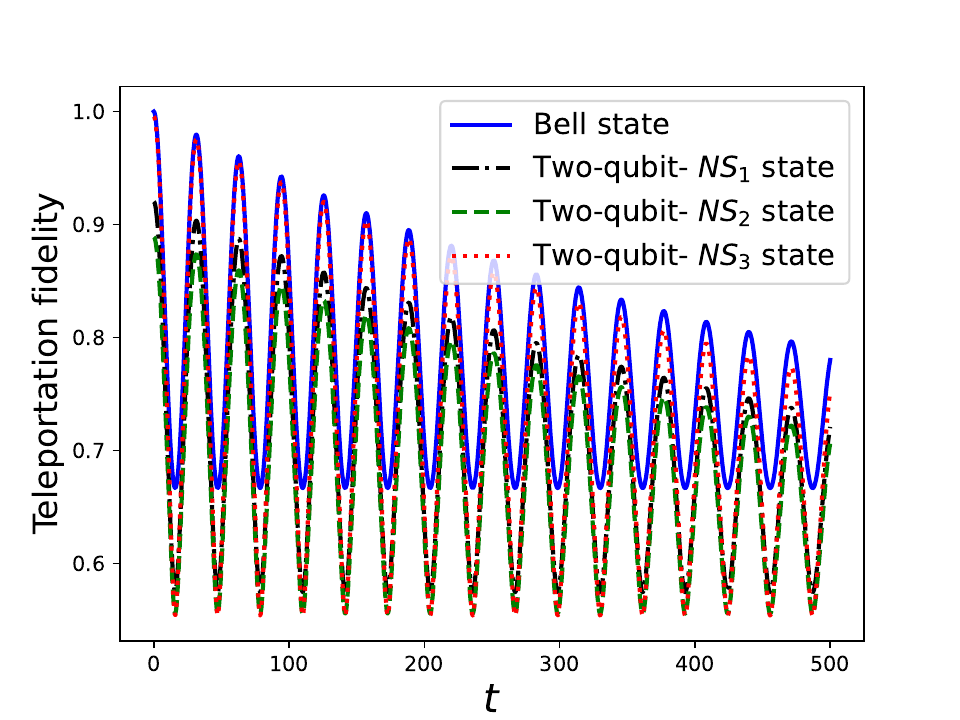}
    \caption{Variation of teleportation fidelity for the two-qubit's $\ket{NS_1}$, $\ket{NS_2}$, $\ket{NS_3}$ states, and Bell state under non-Markovian RTN noise with time. For $\gamma = 0.001$ and $b = 0.05$.}
    \label{fidelityNMRTN}
\end{figure}

\begin{figure}[!htpb]
    \centering
     \includegraphics[height=65mm,width=0.75\columnwidth]{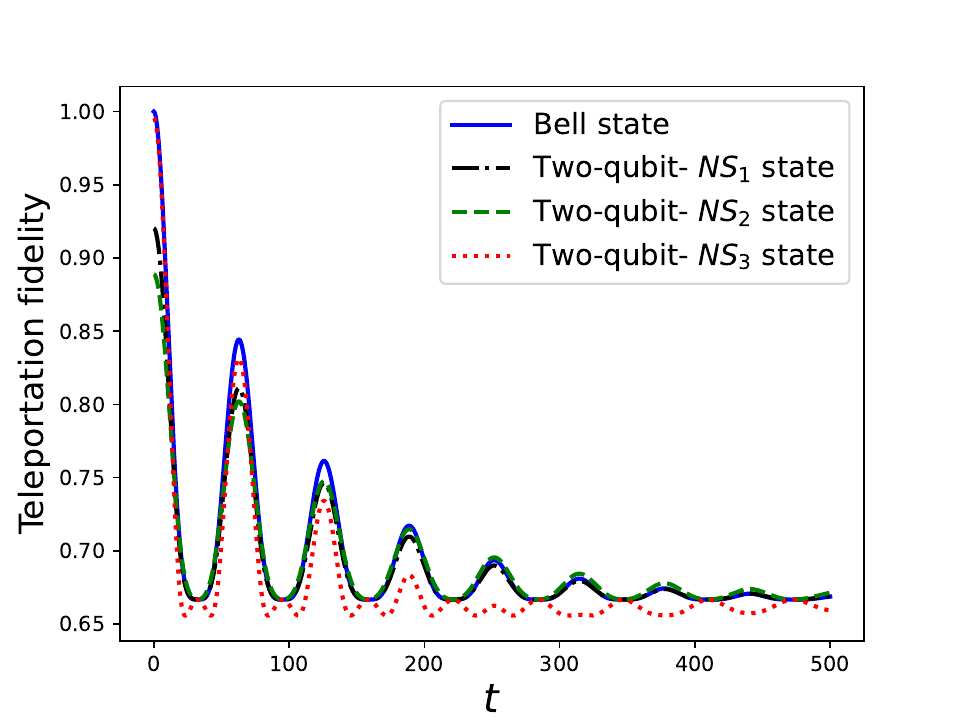}
    \caption{Variation of teleportation fidelity for the two-qubit's $\ket{NS_1}$, $\ket{NS_2}$, $\ket{NS_3}$ states, and Bell state under non-Markovian AD noise with time. For $\gamma = 1 , g = 0.005$.}
    \label{fidelityNMAD}
\end{figure}

Figure~(\ref{mrtn_fid_final}) shows that, under Markovian RTN noise, only the Bell state sustains teleportation fidelity above the classical threshold throughout the evolution, whereas the negative states remain useful for quantum teleportation only over a short time interval. In contrast, under Markovian AD noise (Fig.~(\ref{mad_fid_final})), all considered states achieve fidelities at or above the classical limit over time. Although $\ket{NS_1}$ and $\ket{NS_2}$ initially exhibit lower fidelity, they asymptotically approach values comparable to the Bell state. Further, Figs. (\ref{fidelityNMRTN}) and (\ref{mad_fid_final}) depict the variation of fidelity under non-Markovian RTN and non-Markovian AD noisy channels, respectively. The fidelity of the $\ket{NS_1}$, $\ket{NS_2}$, and $\ket{NS_3}$ states and the Bell state show decaying oscillations in synchronization, with the only difference that the $\ket{NS_1}$, $\ket{NS_2}$, and $\ket{NS_3}$ states are going below the upper bound of classical teleportation under the action of a non-Markovian RTN channel, as shown in Fig.~\ref{fidelityNMRTN}. From Fig.~\ref{fidelityNMAD}, it can be seen that under non-Markovian AD noise, the fidelity of the $\ket{NS_3}$ state is initially similar to the fidelity of the Bell state. In contrast, $\ket{NS_1}$ and $\ket{NS_2}$ states have lesser values. But at longer times, the fidelity of the $\ket{NS_1}$ and $\ket{NS_2}$ states is similar to the fidelity of the Bell state. In contrast, the $\ket{NS_3}$ state has lesser teleportation fidelity at long duration.


\section{\label{ch3_conclusion}Summary} 
The use of the discrete Wigner function to investigate the behavior of quantum states under different noisy channels is significant and could provide valuable insights into the robustness of quantum information in noisy environments. The behavior of DWFs of a qubit, qutrit, and two-qubit maximally negative quantum states was studied under different (non)-Markovian channels. Also studied was the variation of mana for qutrit's $\ket{NS_1}$ and $\ket{NS_2}$ states under (non)-Markovian evolution. Initially, the mana value of the qutrit's $\ket{NS_1}$ state is higher than that of the $\ket{NS_2}$. Yet, compared to the qutrit's $\ket{NS_2}$ state, the $\ket{NS_1}$ state quickly dissipates under non-Markovian AD. As a result, the qutrit's $\ket{NS_2}$ state endures for a longer time. Both negative quantum states of qutrit exhibit the anticipated oscillatory behavior under the non-Markovian RTN, which is persistent for a far more extended period than the non-Markovian AD. An interesting facet of this work was the behavior of the negative quantum states as compared to the Bell states for important quantum information aspects such as quantum coherence, entanglement, and teleportation fidelity. We investigated the quantum coherence, entanglement, and teleportation fidelity variation of the $\ket{NS_1}$, $\ket{NS_2}$, and $\ket{NS_3}$ states of the two-qubit system and compared this to the Bell state under different noisy channels using DWFs. Under the non-Markovian AD noise, the quantum coherence and entanglement of two-qubit $\ket{NS_1}$, $\ket{NS_2}$, and $\ket{NS_3}$ states are sustained for longer than the Bell state, but for teleportation fidelity, they behave like the Bell state. Moreover, the quantum coherence of the two-qubit $\ket{NS_1}$, $\ket{NS_2}$, and $\ket{NS_3}$ states persists for longer than the Bell state, but for the entanglement and teleportation fidelity, the Bell states dominate under noisy dephasing channels such as depolarising and non-Markovian RTN.

\subsection{Limitations and Scope}
The results presented in this chapter are obtained for specific parameter sets representative of the non-Markovian and Markovian regimes: for the amplitude damping channel, we use $\gamma = 50$, $g = 0.01$ (strong non-Markovian), $\gamma = 0.01$, $g = 1$ (Markovian), $\gamma = 1$, $g = 0.005$ (intermediate); for the RTN channel, $\gamma = 0.001$, $b = 0.05$ (non-Markovian) and $\gamma = 1$, $b = 0.07$ (Markovian). The observed advantages of negative quantum states over Bell states, particularly in terms of coherence and entanglement under non-Markovian AD, are confirmed for these parameter values but may not hold uniformly across all parameter regimes. A systematic parameter sweep to delineate the exact boundary between regimes where negative states do and do not outperform Bell states is beyond the scope of this chapter. However, it is an important direction for future work. Additionally, the results are restricted to dimensions $d = 2, 3,$ and $4$; extending them to higher dimensions would require generalizing the phase-space point-operator framework.


\newpage
\setcounter{chapter}{3} 

\titleformat{\chapter}[display]
{\sffamily\fontsize{27}{27}\bfseries\filleft}{\thechapter}{0pt}{{#1}}  
  
\thispagestyle{empty}

\chapter{Protecting quantum correlations of negative quantum states using weak measurement under non-Markovian noise}\label{chap4:Protecting}
\section{Introduction}
In this chapter, in order to find suitable two-qubit states, apart from the Bell states, for universal quantum teleportation, we study the impact of weak measurement (WM) and quantum measurement reversal (QMR)~\citep{korotkov2006undoing, katz2008reversal, kim2009reversing, korotkov2010decoherence, kim2012protecting, sabale2023towards} on the quantum correlations~\citep{Wootters1998Entanglement, horodecki2009quantum, ollivier2001quantum, henderson2001classical, bennett1999quantum, schrodinger1935discussion, schrodinger1936probability, brunner2014bell, fan2022quantum, costa2016quantification} and UQT requirements~\citep{horodecki1996teleportation, badziag2000local, bang2018fidelity, ghosal2020optimal} of the negative quantum states of two-qubit systems proposed in~\citep{lalita2023harnessing}.  The influence of both non-Markovian non-unital (specified by non-Markovian amplitude damping (AD)) and unital  (specified by non-Markovian random telegraph noise (RTN)) channels are taken into consideration. Precisely, negative quantum states are non-classical states corresponding to the normalized eigenvectors of the negative eigenvalues of the phase-space point operators of discrete Wigner functions, reviewed in Sec.~\ref{ch3_NQS} of chapter~\ref{chap3:Harnessing}~\citep{lalita2023harnessing}. We also compare the variations of quantum correlations, maximal fidelity, fidelity deviation, and success probability of two-qubit negative quantum states with that of the maximally correlated state, i.e., the Bell state, in the presence and absence of WM and QMR under the above-mentioned quantum channels.

The chapter outline is as follows. In Sec.~\ref {ch4_Model}, we briefly discuss a tentative physical model, which introduces the notions of weak measurement and quantum measurement reversal. Section~\ref{ch4_protectingQCs} presents a short review of quantum correlations, maximal fidelity, fidelity deviation, and universal quantum teleportation and studies their behavior for two-qubit negative quantum states and the Bell states under non-Markovian non-unital (amplitude damping) and unital (random telegraph noise) channels with(without) weak measurement and quantum measurement reversal. Our results are discussed in Sec.~\ref{ch4_result&discussion}, followed by the conclusions in Sec.~\ref{ch4_conclusion}. \textit{The content of this chapter is derived from\citep{Lalita_2024ProtectingQC}. \copyright IOP Publishing. Adapted and reproduced with permission.}

\section{\label{ch4_Model} Model}
\begin{figure}
    \centering
    \includegraphics[height=95mm,width=0.75\columnwidth]{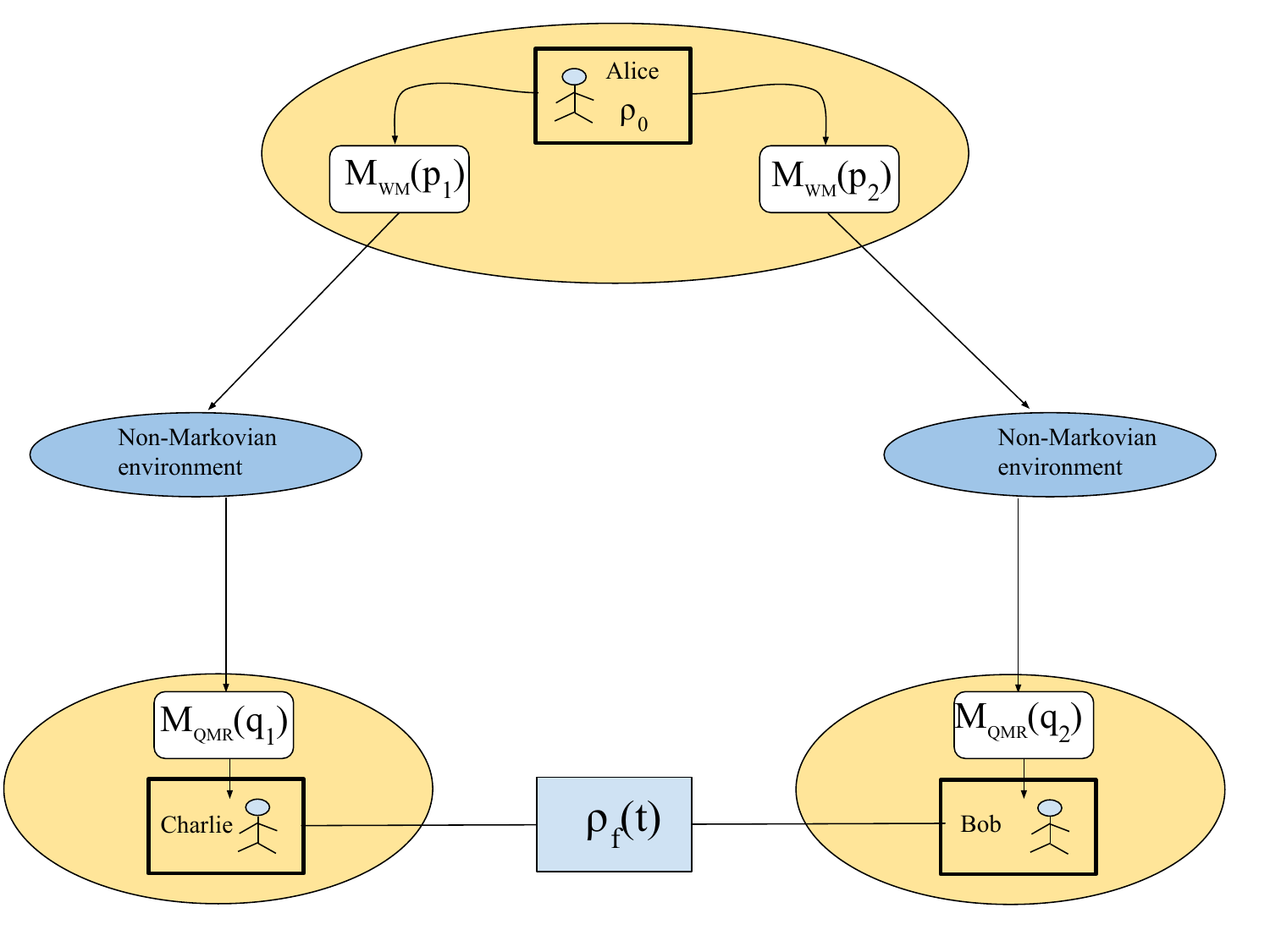}
    \caption{Schematic diagram for protecting quantum correlations of negative quantum states and Bell state using weak measurement (${M}_{\textit{WM}}$) and quantum measurement reversal (${M}_{\textit{QMR}}$).}
    \label{schematic_With_WM_QMR}
\end{figure}

This section presents the physical model for protecting quantum correlations and universal quantum teleportation protocol requirements of two-qubit quantum states using weak measurement and quantum measurement reversal in the non-Markovian environment. As shown in Fig.~\ref{schematic_With_WM_QMR}, Alice first performs weak measurement (${M}_{\textit{WM}}(p_1, p_2)$), as shown by Eq. (\ref{wm}), on the two-qubit negative quantum states elaborated in Sec.~\ref{ch3_NQS} of chapter~\ref{chap3:Harnessing} before distribution to Bob and Charlie via non-Markovian noisy quantum channels, discussed above~\ref{ch2_noise_models}. Bob and Charlie perform quantum measurement reversal (${M}_{\textit{QMR}}(q_1, q_2)$), given in Eq. (\ref{qmr}), on receiving the qubits. The resulting state ${\rho}_f(t)$, Eq. (\ref{eq:WM_QMR}), can be made maximally entangled by choosing appropriate ($p_1, p_2$) and ($q_1, q_2$). Further, the state ${\rho}_f(t)$ can also be used for QT between Bob and Charlie.

The non-unitary WM and QMR operations are given as
\begin{equation}
   \begin{aligned}
     {M}_{\textit{WM}}(p_1, p_2) = \begin{pmatrix}
                                1 & 0\\
                                0 & \sqrt{1-p_1}
                               \end{pmatrix} \otimes \begin{pmatrix}
                                1 & 0\\
                                0 & \sqrt{1-p_2}
                               \end{pmatrix},
      \end{aligned}
      \label{wm}
\end{equation}
\begin{equation}
   \begin{aligned}
     {M}_{\textit{QMR}}(q_1, q_2) = \begin{pmatrix}
                                \sqrt{1-q_1} & 0\\
                                0 & 1
                               \end{pmatrix} \otimes \begin{pmatrix}
                                \sqrt{1-q_2} & 0\\
                                0 & 1
                               \end{pmatrix}.
      \end{aligned}
      \label{qmr}
\end{equation}

Here, $(p_1, p_2)$ and $(q_1, q_2)$ are the WM and QMR strength parameters, respectively. In our work, we have considered that ($p_1 = p_2 = p$) and ($q_1 = q_2 = q$), $\textit{i.e.}$, the strength of WM and QMR parameters are equal for both the qubits. It is important to note that via WM, the state does not collapse towards $\ket{00}$ or $\ket{11}$, indicating that appropriate operations, such as QMR, can still recover the measured state. After the sequential WM, non-Markovian channel, and QMR, the final state is

\begin{equation}
 {\rho}_f(t) = \frac{{M}_{\textit{QMR}}\left( \sum_{i = 0}^{1}\sum_{j = 0}^{1} \mathbf{K}_{ij}[{M}_{\textit{WM}}{\rho}(0){M}_{\textit{WM}}^{\dag}] \mathbf{K}_{ij}^{\dag}\right) {M}_{\textit{QMR}}^{\dag}}{P^{succ}},
 \label{eq:WM_QMR}
\end{equation}

here $\mathbf{K}_{ij} = (\mathbf{K}_i\otimes \mathbf{K}_j)$ are the Kraus operators of the non-Markovian noise. Since the WM and QMR are probabilistic in nature, $P^{succ} = \Tr[{M}_{\textit{QMR}}\left( \sum_{i = 0}^{1}\sum_{j = 0}^{1} \mathbf{K}_{ij}[{M}_{\textit{WM}}{\rho}(0){M}_{\textit{WM}}^{\dag}] \mathbf{K}_{ij}^{\dag}\right) {M}_{\textit{QMR}}^{\dag}]$ is their success probability \citep{pramanik2013improving, he2020enhancing}. A lower success probability of the protocol is the price for higher quantum correlations, maximal fidelity, and lesser fidelity deviations.

\section{\label{ch4_protectingQCs}Protecting quantum correlations and universal quantum teleportation protocols under the non-Markovian AD and RTN channels using WM and QMR}
Here, we discuss the quantum correlations, maximal fidelity, fidelity deviation, and universal quantum teleportation protocols. We investigate the contribution of WM and QMR to the protection and enhancement of them. Further, we examine the variation of quantum concurrence, discord, steering, maximal fidelity, and fidelity deviation of the two-qubit negative quantum states, $\textit{i.e.}$, $\ket{NS_1}, \ket{NS_2}, \ket{NS_3'}$, and the Bell state under the non-Markovian AD and RTN channels, with(without) WM and QMR.

\subsection{Quantum correlations, maximal fidelity and fidelity deviation}
Quantum non-local correlations are one of the most remarkable and exclusive aspects of the quantum world, with no equivalence in the classical world. Below, we discuss the quantum concurrence, discord, and steering in brief. Furthermore, maximal fidelity, fidelity deviation, and UQT protocols are discussed.

\subsubsection{Concurrence}
Entanglement is one of the most important sources of quantum information. Entanglement is also a fundamental component of quantum correlation in compound quantum systems.  For a two-qubit system, concurrence is an entanglement measure \citep{Wootters1998Entanglement}, which is defined as 
\begin{equation}
   \begin{aligned}
     C({\rho}_{AB}) = \max \{0, \lambda_{1} - \lambda_{2} - \lambda_{3} - \lambda_{4}\},
      \end{aligned}
      \label{concur_eq.}
\end{equation}
Here $\lambda_{i}$'s are the eigenvalues of $\sqrt{\sqrt{{\rho}_{AB}} \tilde{{\rho}}_{AB} \sqrt{{\rho}_{AB}}}$, such that $\lambda_{1} \geq \lambda_{2} \geq \lambda_{3} \geq \lambda_{4}$, and $\Tilde{{\rho}}_{AB} = (\sigma_{y} \otimes \sigma_{y}) {\rho}_{AB}^{*} (\sigma_{y} \otimes \sigma_{y})$, where ${\rho}_{AB}^{*}$ is the complex conjugate of ${\rho}_{AB}$ and $\sigma_{y}$ is the Pauli bit-phase flip matrix.

\subsubsection{Discord}
There exist states that exhibit non-local behavior while remaining unentangled \citep{bennett1999quantum, luo2008quantum, ramkarthik2020quantum}. For quantifying such non-local correlations, Quantum Discord (QD) was developed  \citep{ollivier2001quantum, henderson2001classical}. It is a measure of a quantum system's overall non-local correlations. The QD for a bipartite composite quantum system ${\rho}_{AB}$, where $A$ and $B$ are the separate subsystems, is defined in the following manner,
\begin{equation}
    \begin{aligned}
        d(A:B) = t(A:B) - c(A|B),
    \end{aligned}
\end{equation}
where $t(A: B)$ and $c(A|B)$ are the total and classical correlations between the two subsystems, respectively. These correlations are defined as follows,

\begin{eqnarray}
    \nonumber
    t(A:B) &=& s(A) + s(B) - s(A, B),\\
    c(A|B) &=& s(A) - s(A|B).
    \label{corr_von-neumann_eqn}
\end{eqnarray}

Here $s(A) = -\Tr({\rho}_A ln {\rho}_A )$ and $s(B) = -\Tr({\rho}_B ln {\rho}_B )$ represent the von Neumann entropies of subsystem states ${\rho}_A$ and ${\rho}_B$ respectively. The $s(A, B)$ and $s(A|B)$ are the system's mutual and conditional quantum entropies, respectively, and are given by,
\begin{eqnarray}
    \nonumber
    s(A, B) &=& -\Tr({\rho}_{AB} ln {\rho}_{AB}),\\
    s(A|B) &=&  min_{\{{\pi}_i\}}\sum_{i=1}^{\mathbf{H}_B} p_i s({\rho}_{A|{\pi}_i}).
    \label{conditional_entropy}
\end{eqnarray}

Here, $\mathbf{H}_B$ is the Hilbert space dimension of subsystem B, and minimization is performed over all possible measurement operators ${\pi}_i$. The post-measurement state for subsystem $A$ when a measurement is performed on subsystem $B$ is ${\rho}_{A|{\pi}_i}$ and $p_i = \Tr({\pi}_i^{\dag} {\pi}_i {\rho}_{AB})$ is the probability of measurement operators ${\pi}_i$. It is possible to write the state ${\rho}_{A|{\pi}_i}$ explicitly as,
\begin{equation}
    \begin{aligned}
    {\rho}_{A|{\pi}_i} = \frac{1}{p_i} Tr({\pi}_i {\rho}_{AB}{\pi}_i)
    \end{aligned}
\end{equation}

It is essential to note that we can determine the QD by measuring either subsystem $A$ or $B$. In this case, we are measuring subsystem $B$ and restricting the measurement to one qubit because the depreciation in Eq. (\ref{conditional_entropy}) depends on $2^m$ parameters of the measurement operators $i$, where $m$ is the number of qubits in subsystem $B$. The general measurement parameters for $m = 1$ are 
\begin{eqnarray}
    \nonumber
    {\pi}_1 &=& I_A \otimes \ket{l}_{BB} \bra{l},\\
    {\pi}_2 &=& I_A \otimes \ket{m}_{BB} \bra{m},
\end{eqnarray}
where $\ket{l} = \cos{\frac{\theta}{2}}\ket{0} + e^{i\phi}\sin{\frac{\theta}{2}}\ket{1}$ and $\ket{m} = \sin{\frac{\theta}{2}}\ket{0} - e^{i\phi}\sin{\frac{\theta}{2}}\ket{1}$, and $0 \leq \theta \leq \pi$, $0 \leq \phi \leq 2\pi$. 
Using Eq. (\ref{corr_von-neumann_eqn}), the discord can be defined in terms of von Neumann, joint von Neumann, and conditional quantum entropies as

\begin{equation}
    \begin{aligned}
        d(A:B) = s(B) - s(A, B) + s(A|B).
    \end{aligned}
    \label{discord_von_neumann_entropy_eqn}
\end{equation}

\subsubsection{\label{steering}Steering}
The concept of steering was established in  \citep{schrodinger1935discussion, schrodinger1936probability}. If Alice and Bob share an entangled pair, Alice can remotely steer Bob's state by performing measurements exclusively on her half of the system. This type of quantum correlation is referred to as steering or EPR steering. It lies between the Bell non-locality \citep{brunner2014bell} and entanglement \citep{horodecki2009quantum}. It is also a resource of quantum teleportation \citep{fan2022quantum}. By considering how much a steering inequality is maximally violated, we can determine the degree of steerability of a particular quantum state \citep{costa2016quantification}. For two-qubit systems, the steering formula is given as,
\begin{equation}
   \begin{aligned}
     S_{n}({\rho}_{AB}) = \max \left\{0, \frac{\Omega_{n} - 1}{\sqrt{n} - 1}\right\}.
    \end{aligned}
    \label{steering_eq.}
\end{equation}
When $n = 2, 3$ per party measurements are involved, called two (three)-measurement steering, respectively, $\Omega_2 = \sqrt{c^2 - c_{min}^2}$, and $\Omega_3 = c$. Here $c = \sqrt{\textbf{c}^2}$, and $c_{min} \equiv min\{|c_i|\}$, $c_i$'s are the eigenvalues of correlation matrix $\textbf{T} = \{t_{ij}\}$, and $t_{ij} = Tr[{\rho}_{AB} ({\sigma_{i}} \otimes {\sigma_{j}})]$.

\subsubsection{\label{maxiaml fidelity and fidelity deviation sec.}Maximal fidelity and fidelity deviation}
A fundamental protocol for transmitting quantum information using shared entanglement and local operations and classical communication (LOCC) is quantum teleportation \citep{bennett1993teleporting}. The maximal average fidelity $(F_{{\rho}_{AB}})$ \citep{horodecki1996teleportation, badziag2000local}, and deviation in the fidelity $(\Delta_{{\rho}_{AB}})$ \citep{bang2018fidelity, ghosal2020optimal} are usually used to determine the quality of a teleportation protocol. 

The maximal average fidelity (or maximal fidelity) is the maximum of all average fidelities obtained by strategies using the standard protocol and local unitary operations. For two-qubit states with $det(\textbf{T}) < 0$ (here $\textbf{T}$ is the correlation matrix discussed in Sec. (\ref{steering}), the maximal average fidelity can be determined as \citep{horodecki1996teleportation, ghosal2020optimal}
\begin{equation}
   \begin{aligned}
     F_{{\rho}_{AB}} = \frac{1}{2} \left( 1 + \frac{1}{3} \sum_{i = 1}^{3}|e_{i}|\right),
    \end{aligned}
    \label{max_fid_eq.}
\end{equation}
here $e_{i}$'s are the eigenvalues of the correlation matrix $\textbf{T}$.

Fidelity deviation is described as the standard deviation of fidelity values across all possible input states. For a two-qubit state with $det(\textbf{T}) < 0$, fidelity deviation corresponding to the optimal protocol is \citep{bang2018fidelity, ghosal2020optimal} 
\begin{equation}
   \begin{aligned}
     \Delta_{{\rho}_{AB}} = \frac{1}{3\sqrt{10}} \sqrt{\sum_{i < j =1}^{3} (|e_{i}| - |e_{j}|)^2},
    \end{aligned}
    \label{FD_eq.}
\end{equation}
here also $e_{i}$'s are the eigenvalues of the correlation matrix $\textbf{T}$.

When $F_{{\rho}_{AB}} > \frac{2}{3}$, the particular two-qubit state ${\rho}_{AB}$ is beneficial for quantum teleportation (QT) \citep{horodecki1996teleportation, horodecki1999general}; here $\frac{2}{3}$ is the highest average fidelity possible with classical protocols. However, ${\rho}_{AB}$ is universal for quantum teleportation (UQT) iff $\Delta_{{\rho}_{AB}} = 0$ and $F_{{\rho}_{AB}} > \frac{2}{3}$. From these relations, some conditions are formulated in \citep{ghosal2020optimal} that can be checked for a two-qubit state ${\rho}_{AB}$ to verify its validity for universal quantum teleportation. These conditions are:
\\$(i)$ The states with the property $det(\textbf{T}) < 0$ form a subset of the states that are beneficial for QT; here, $\textbf{T}$ is the correlation matrix. 
\\$(ii)$ A two-qubit state ${\rho}_{AB}$ is useful for UQT iff, $|e_1| = |e_2| = |e_3| > 1/3$; here $|e_1|, |e_2|, |e_3|$ are the eigenvalues of the correlation matrix $\textbf{T}$.

All input states will be transported with the same fidelity if the two-qubit state satisfies the aforementioned universality condition. Just like the Bell state, all the considered two-qubit negative quantum states also have $det(\textbf{T}) < 0$. Subsequently, all the considered two-qubit negative quantum states approximately satisfy $|e_1| = |e_2| = |e_3| > 1/3$ except the $\ket{NS_2}$ state. Whereas, the $\ket{NS_2}$ state have $|e_1| = |e_2| \neq |e_3| > 1/3$.

We will now examine the performance of two-qubit negative quantum states for quantum correlations and UQT protocols with(without) WM and QMR under non-Markovian AD and RTN channels. The WM and QMR strength parameters $p$ and $q$ are optimized at $t = 0$ to obtain maximum concurrence for the two-qubit $\ket{NS_1}$, $\ket{NS_2}$, $\ket{NS_3'}$, and the Bell states. In the presence of non-Markovian AD and RTN channels, the optimal combinations for the $\ket{NS_1}$, $\ket{NS_2}$, $\ket{NS_3'}$, and the Bell states are ($p = 0.17$, $q = 0.54$), ($p = 0.05$, $q = 0.74$), ($p = 0.05$, $q = 0.05$), and ($p = 0.01$, $q = 0.01$), respectively. Furthermore, we use the same set of optimal parameters to study the quantum discord, steering, and UQT protocols. Also, these parameters are fixed during the evolution of the states under the non-Markovian AD and RTN channels.

\subsection{Concurrence under non-Markovian AD and RTN channels, with(without) WM and QMR}
We study the evolution of concurrence of two-qubit negative quantum states and the Bell state under the non-Markovian AD and RTN channels. Further, the effect of WM and QMR on the concurrence of two-qubit negative quantum states and the Bell state under the non-Markovian AD and RTN channel is analyzed.

\subsubsection{Under non-Markovian AD noise}
\begin{figure}
    \centering
    \includegraphics[height=65mm,width=0.95\columnwidth]{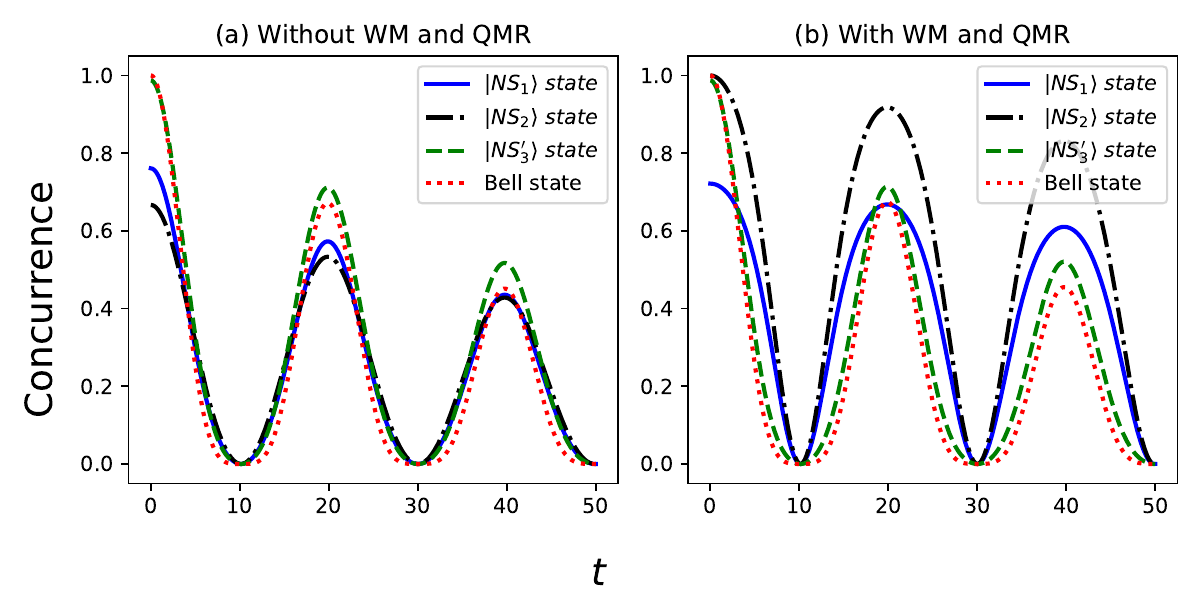}
    \caption{Variation of concurrence of $\ket{NS_1}$, $\ket{NS_2}$, $\ket{NS_3'}$, and Bell state under non-Markovian AD channel without WM and QMR in subplot (a), and with WM and QMR in subplot (b) with time. Here, for $\ket{NS_1}$ ($p = 0.17$, $q = 0.54$), for $\ket{NS_2}$ ($p = 0.05$, $q = 0.74$), for $\ket{NS_3'}$ ($p = 0.05$, $q = 0.05$), and for Bell state ($p = 0.01$, $q = 0.01$). The non-Markovian AD channel parameters are $g = 0.01$ and $\gamma = 5$.}
    \label{concur_NMAD}
\end{figure}

Figure (\ref{concur_NMAD}a) illustrates the behavior of the concurrence of the considered two-qubit negative quantum states and the Bell state without WM and QMR under the non-Markovian AD noise. It can be seen that the $\ket{NS_3'}$ state has entanglement equal to the Bell state, and both states have the highest entanglement in the initial period. But for a longer duration, the $\ket{NS_3'}$ state dominates the Bell state and the other two-qubit negative quantum states under the non-Markovian AD noise.

Upon optimizing WM and QMR strength parameters, the variation of concurrence is depicted in Fig.~\ref{concur_NMAD}(b). The WM and QMR can be seen to significantly improve and protect the concurrence of the $\ket{NS_1}$ and $\ket{NS_2}$ states. In fact, with WM and QMR, the $\ket{NS_2}$ state shows entanglement equal to the Bell state and the $\ket{NS_3'}$ state at $t = 0$. Moreover, over time, the $\ket{NS_2}$ state shows concurrence more than all the considered states. Additionally, the concurrence of the Bell state and $\ket{NS_3'}$ state remains unaltered with WM and QMR under non-Markovian AD evolution. It is also clear from Fig.~\ref{concur_NMAD}(b) that the two-qubit negative quantum states dominate the Bell state in terms of preserving their entanglement under non-Markovian AD noise for a longer duration.

\subsubsection{Under non-Markovian RTN channel}
\begin{figure}[!htpb]
    \centering
    \includegraphics[height=65mm,width=0.95\columnwidth]{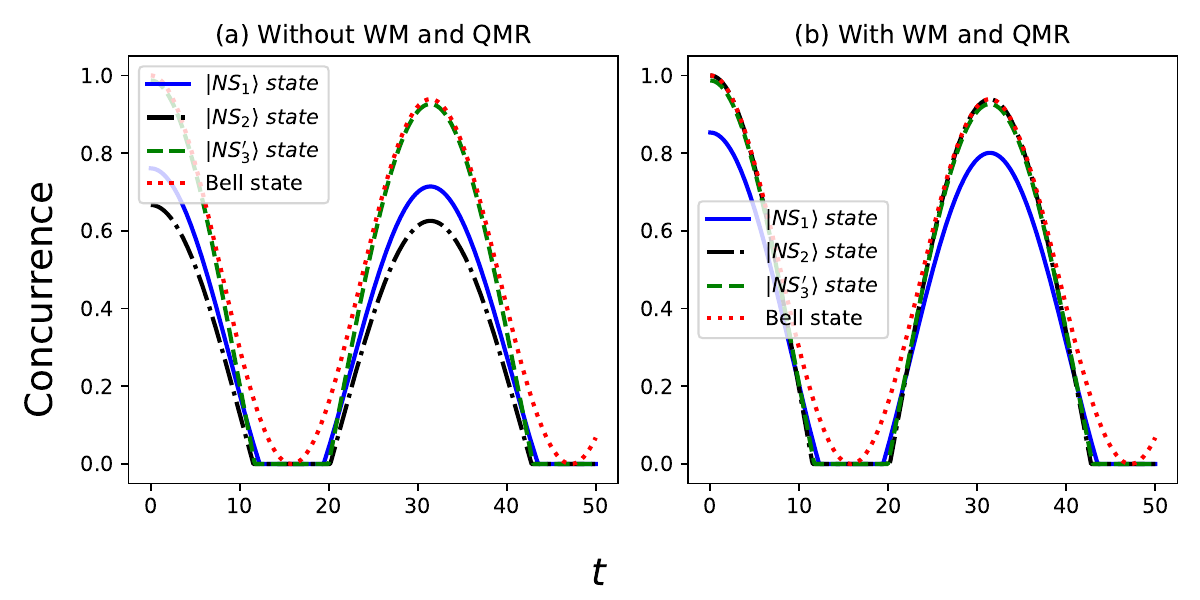}
    \caption{Variation of concurrence of $\ket{NS_1}$, $\ket{NS_2}$, $\ket{NS_3'}$, and Bell state under non-Markovian RTN channel without WM and QMR in subplot (a), and with WM and QMR in subplot (b) with time. Here, for $\ket{NS_1}$ ($p = 0.17$, $q = 0.54$), for $\ket{NS_2}$ ($p = 0.05$, $q = 0.74$), for $\ket{NS_3'}$ ($p = 0.05$, $q = 0.05$), and for Bell state ($p = 0.01$, $q = 0.01$). The non-Markovian RTN channel parameters are $b = 0.05$ and $\gamma = 0.001$.}
    \label{concur_NMRTN}
\end{figure}

The evolution of the concurrence in the absence of WM and QMR under the non-Markovian RTN channel is shown in Fig.~\ref{concur_NMRTN}(a). Under the non-Markovian RTN channel, the $\ket{NS_3'}$ state shows entanglement variations almost equal to the Bell state, and both have the highest entanglement over time. All the states show expectedly oscillatory, slowly decaying behavior under the non-Markovian RTN channel.

The concurrence variation in the presence of WM and QMR under the non-Markovian RTN channel can be seen in Fig.~\ref{concur_NMRTN}(b). The concurrence of the Bell state and $\ket{NS_3'}$ state remain unaltered with WM and QMR under non-Markovian RTN evolution. The concurrence of the $\ket{NS_1}$ and $\ket{NS_2}$ states have been improved. In fact, the $\ket{NS_2}$ state exhibits entanglement comparable to that of the Bell state and $\ket{NS_3'}$ state over time when WM and QMR are employed.

\subsection{Discord under non-Markovian AD and RTN channels, with(without) WM and QMR}
To study the dynamics of discord under non-Markovian AD and RTN channels, Eq. (\ref{discord_von_neumann_entropy_eqn}) and Eqs. \ref{NMAD_Kraus_operators}, (\ref{2qubitNMADfinalrho}, \ref{NMRTN_Kraus_operators}, \ref{2qubitRTNfinalrho}) are employed. Further, Eq. (\ref{eq:WM_QMR}) is utilized to understand the effect of WM and QMR on discord dynamics. 

\subsubsection{Under non-Markovian AD channel}
\begin{figure}[!htpb]
    \centering
    \includegraphics[height=65mm,width=0.95\columnwidth]{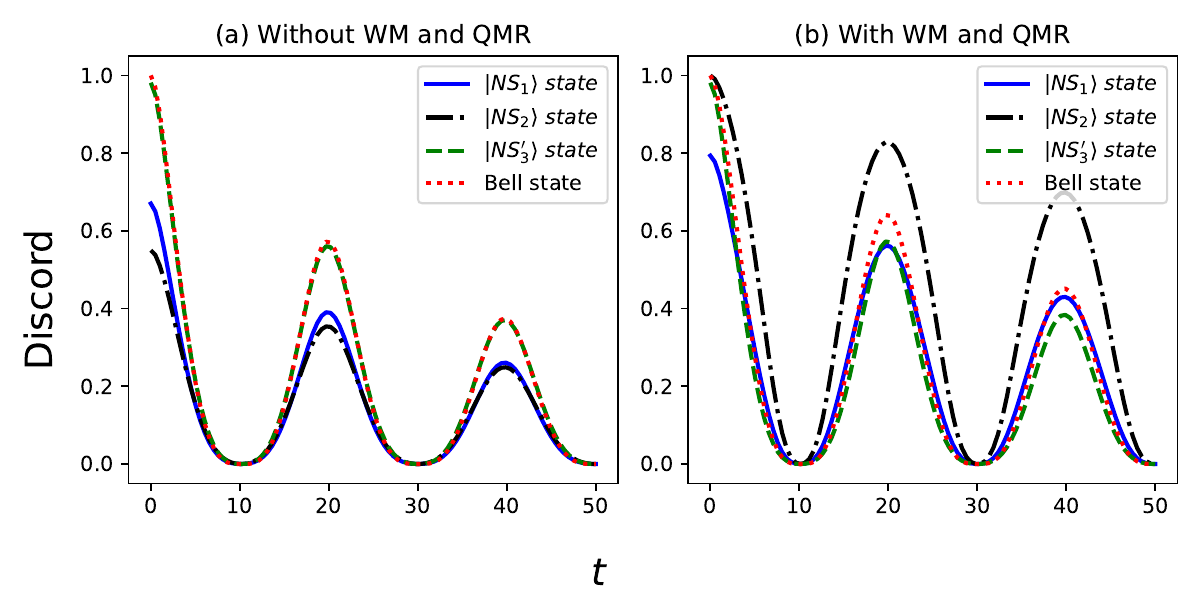}
    \caption{Variation of discord of $\ket{NS_1}$, $\ket{NS_2}$, $\ket{NS_3'}$, and Bell state under non-Markovian AD channel without WM and QMR in subplot (a), and with WM and QMR in subplot (b) with time. Here, for $\ket{NS_1}$ ($p = 0.17$, $q = 0.54$), for $\ket{NS_2}$ ($p = 0.05$, $q = 0.74$), for $\ket{NS_3'}$ ($p = 0.05$, $q = 0.05$), and for Bell state ($p = 0.01$, $q = 0.01$). The non-Markovian AD channel parameters are $g = 0.01$, and $\gamma = 5$.}
    \label{discord_NMAD}
\end{figure}

Figure~\ref{discord_NMAD}(a) shows the variation of discord of two-qubit $\ket{NS_1}$, $\ket{NS_2}$, $\ket{NS_3'}$, and the Bell state under non-Markovian AD noise without WM and QMR. The variation of discord of $\ket{NS_3'}$ state is similar to the Bell state under non-Markovian AD noise. Also, these states have the highest discord values among all the considered states.

The variation of discord under the non-Markovian AD noise with WM and QMR of the two-qubit $\ket{NS_1}$, $\ket{NS_2}$, $\ket{NS_3'}$, and the Bell state is depicted in Fig.~\ref{discord_NMAD}(b). The discord of the $\ket{NS_1}$ and $\ket{NS_2}$ states can be seen to be enhanced by the WM and QMR. The $\ket{NS_2}$ state exhibits discord at $t = 0$ comparable to the Bell state and the $\ket{NS_3'}$ state with WM and QMR. In addition, the $\ket{NS_2}$ state shows more discord over time than all other considered states. Furthermore, during non-Markovian AD evolution, the Bell state discord with WM and QMR remains unaffected. 

\subsubsection{Under non-Markovian RTN channel}
\begin{figure}[!htpb]
    \centering
    \includegraphics[height=65mm,width=0.95\columnwidth]{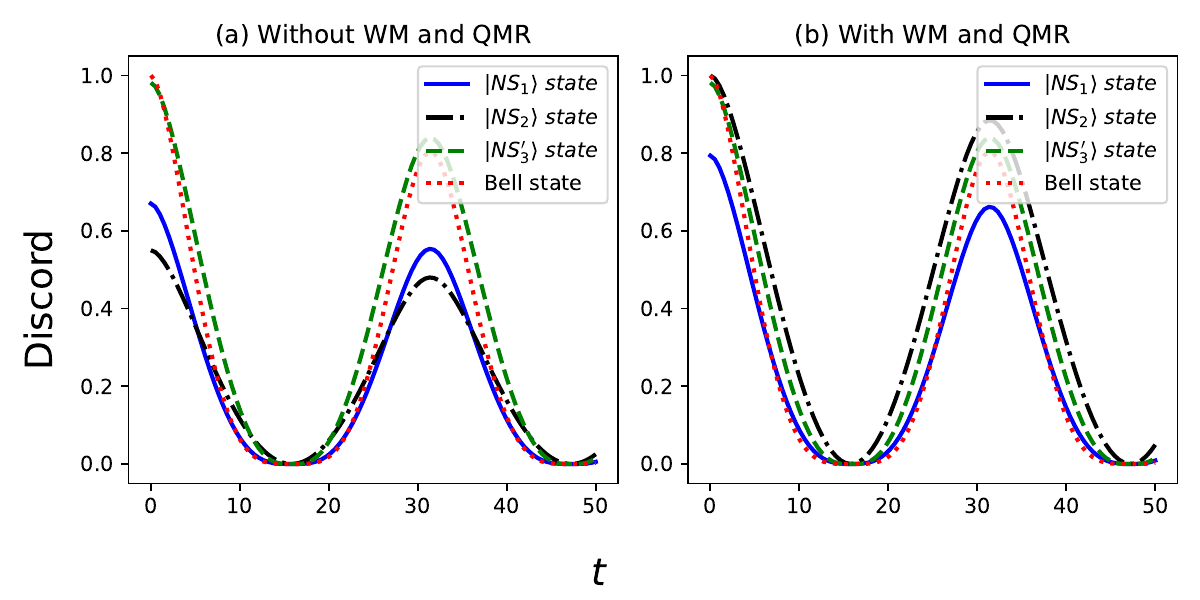}
    \caption{Variation of discord of $\ket{NS_1}$, $\ket{NS_2}$, $\ket{NS_3'}$, and Bell state under non-Markovian RTN channel without WM and QMR in subplot (a), and with WM and QMR in subplot (b) with time. Here, for $\ket{NS_1}$ ($p = 0.17$, $q = 0.54$), for $\ket{NS_2}$ ($p = 0.05$, $q = 0.74$), for $\ket{NS_3'}$ ($p = 0.05$, $q = 0.05$), and for Bell state ($p = 0.01$, $q = 0.01$). The non-Markovian RTN channel parameters are $b = 0.05$ and $\gamma = 0.001$.}
    \label{discord_NMRTN}
\end{figure}

Figure~\ref{discord_NMRTN}(a) depicts that without WM and QMR, at $t = 0$, the $\ket{NS_3'}$ state's discord is equal to the Bell state. It dominates the Bell state and other considered states over time under the non-Markovian RTN channel.

When the two-qubit $\ket{NS_1}$, $\ket{NS_2}$, $\ket{NS_3}$, and Bell $\ket{\phi^+}$ states are subjected to the non-Markovian RTN channel and WM and QMR, the variations in discord are represented by Fig.~\ref{discord_NMRTN}(b). At $t = 0$, the discord of $\ket{NS_1}$ and $\ket{NS_2}$ state is improved to a reasonable extent. In fact, with WM and QMR, at $t = 0$, the $\ket{NS_2}$ state shows discord equal to the Bell state and the $\ket{NS_3'}$ state. It also dominates all other considered states over time. At the same time, the Bell state and $\ket{NS_3'}$ state remain uninfluenced by WM and QMR.

\subsection{Steering under non-Markovian AD and RTN channels, with(without) WM and QMR}
To investigate the dynamics of two (three)-measurement steering of two-qubit $\ket{NS_1}$, $\ket{NS_2}$, $\ket{NS_3'}$, and the Bell states under non-Markovian AD and RTN channel, Eqs. (\ref{NMAD_Kraus_operators}, \ref{2qubitNMADfinalrho}, \ref{NMRTN_Kraus_operators}, \ref{2qubitRTNfinalrho}) and (\ref{steering_eq.}) for $n =2$, $3$ are utilized.  To comprehend how the WM and QMR affect the two (three)-measurement steering of the Bell state and the negative quantum states of the two-qubit system under non-Markovian AD and RTN noise, Eq. (\ref{eq:WM_QMR}) is utilized. 

\subsubsection{Under non-Markovian AD channel}
\begin{figure}[!htpb]
    \centering
    \includegraphics[height=100mm,width=0.95\columnwidth]{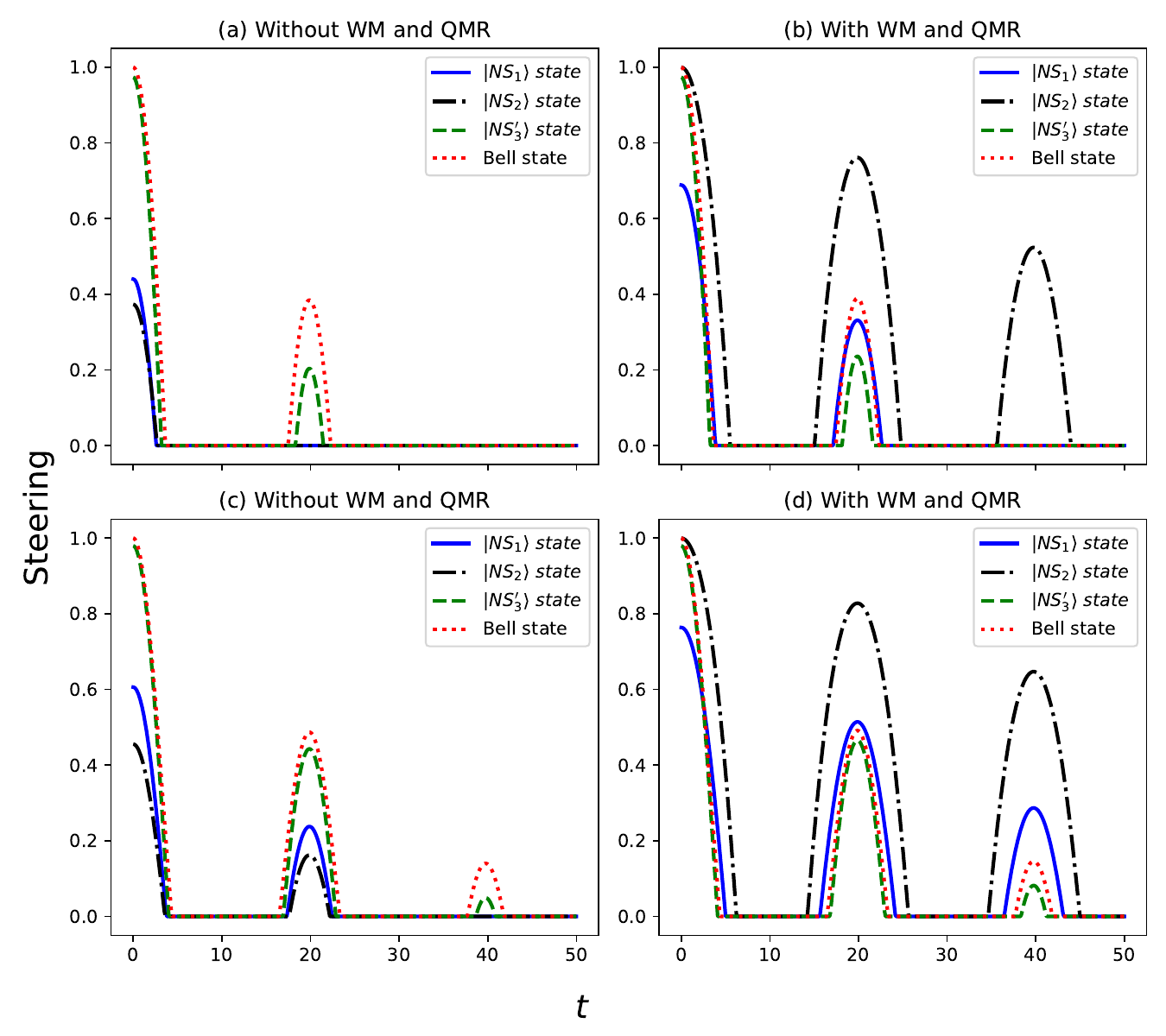}
    \caption{Variation of two (three)-measurement steering of $\ket{NS_1}$, $\ket{NS_2}$, $\ket{NS_3'}$, and Bell state under non-Markovian AD channel with time. Here, subplots (a) and (c) represent the two-measurement and three-measurement steering without WM and QMR, and subplots (b) and (d) represent the two-measurement and three-measurement steering with WM and QMR, respectively. Here, for $\ket{NS_1}$ ($p = 0.17$, $q = 0.54$), for $\ket{NS_2}$ ($p = 0.05$, $q = 0.74$), for $\ket{NS_3'}$ ($p = 0.05$, $q = 0.05$), and for Bell state ($p = 0.01$, $q = 0.01$). The non-Markovian AD channel parameters are $g = 0.01$, and $\gamma = 5$.}
    \label{steering_NMAD}
\end{figure}

At $t = 0$, the three-measurement steering of the $\ket{NS_1}$ and $\ket{NS_2}$ states is higher than their two-measurement steering, whereas their values are the same for the $\ket{NS_3'}$ state and the Bell state, as shown in Figs. (\ref{steering_NMAD}a) and (\ref{steering_NMAD}c). The $\ket{NS_3'}$ state and the Bell state show a maximum of two (three)-measurement steering for the initial period, whereas the Bell state dominates for a longer duration without WM and QMR. 

The two (three)-measurement steering of the $\ket{NS_1}$ and $\ket{NS_2}$ states can be seen to have improved significantly with WM and QMR, from Figs. (\ref{steering_NMAD}b) and (\ref{steering_NMAD}d). In fact, with WM and QMR, the $\ket{NS_2}$ state shows two (three)-measurement steering even higher than the Bell state over time. Furthermore, the two (three)-measurement steering of the Bell state remains unaffected. 

\subsubsection{Under non-Markovian RTN channel}
\begin{figure}[!htpb]
    \centering
    \includegraphics[height=100mm,width=0.95\columnwidth]{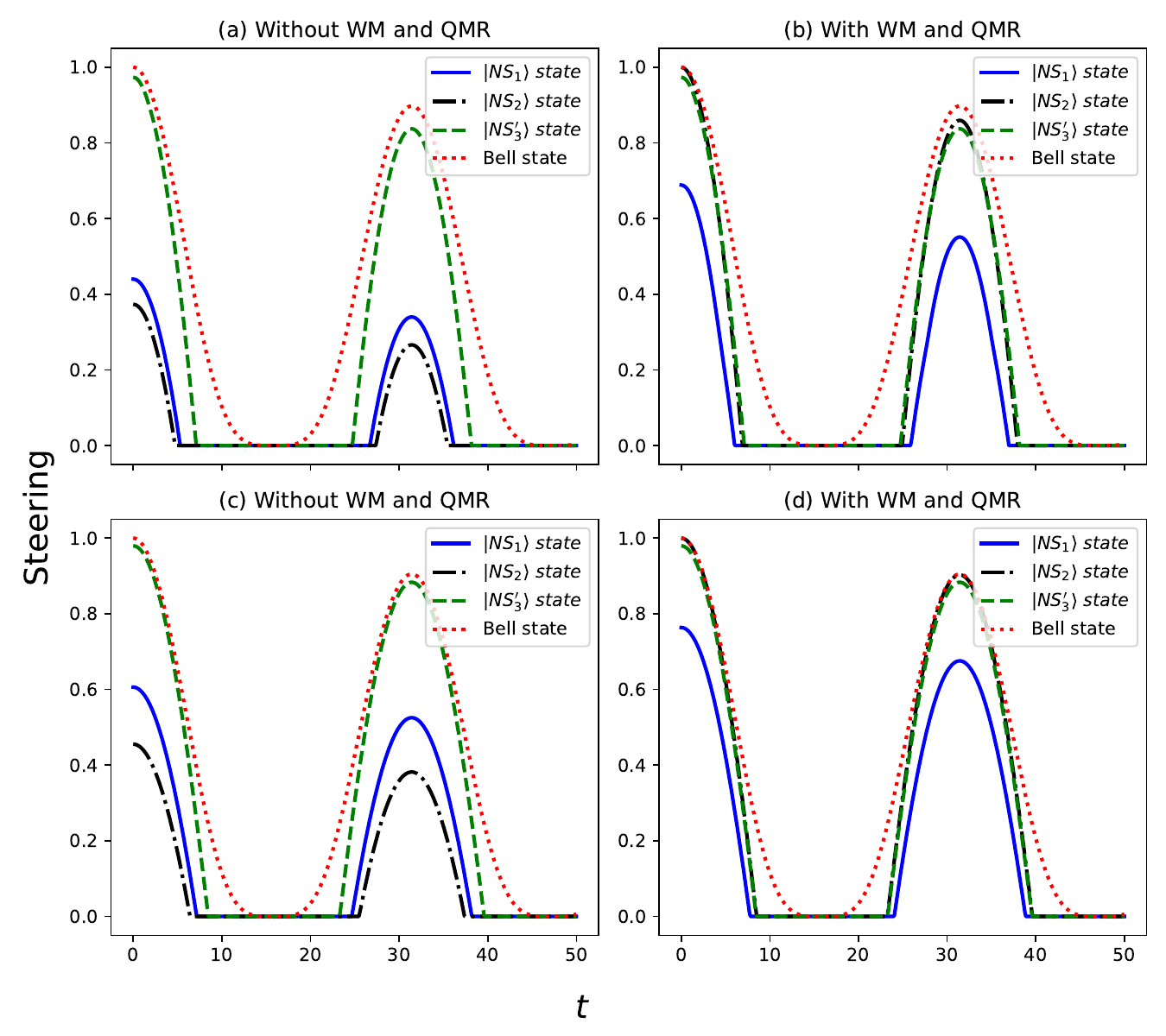}
    \caption{Variation of two (three)-measurement steering of $\ket{NS_1}$, $\ket{NS_2}$, $\ket{NS_3'}$, and Bell state under non-Markovian RTN channel with time. Here, subplots (a) and (b) represent the two-measurement and three-measurement steering without WM and QMR, and subplots (b) and (d) represent the two-measurement and three-measurement steering with WM and QMR, respectively. Here, for $\ket{NS_1}$ ($p = 0.17$, $q = 0.54$), for $\ket{NS_2}$ ($p = 0.05$, $q = 0.74$), for $\ket{NS_3'}$ ($p = 0.05$, $q = 0.05$), and for Bell state ($p = 0.01$, $q = 0.01$). The non-Markovian RTN channel parameters are $b = 0.05$ and $\gamma = 0.001$.}
    \label{steering_NMRTN}
\end{figure}

The decay of two (three)-measurement steering of the two-qubit negative quantum states and the Bell state under the non-Markovian RTN channel is less than the non-Markovian AD case. Like the non-Markovian AD case, the $\ket{NS_3'}$ state and the Bell state show a maximum of two (three)-measurement steering values with time under the non-Markovian RTN channel, which can be seen from Figs. (\ref{steering_NMRTN}a) and (\ref{steering_NMRTN}b) respectively. The three-measurement steering of $\ket{NS_1}$ and $\ket{NS_2}$ state at $t = 0$ is higher than their two-measurement steering.

The variations in two (three)-measurement steering of the $\ket{NS_1}$, $\ket{NS_2}$, $\ket{NS_3'}$, and the Bell state when subjected to a non-Markovian RTN channel with WM and QMR are shown in Figs. (\ref{steering_NMRTN}c) and (\ref{steering_NMRTN}d). From Figs. (\ref{steering_NMRTN}c) and (\ref{steering_NMRTN}d), it is evident that the two (three)-measurement steering of the $\ket{NS_1}$ and $\ket{NS_2}$ states has improved using WM and QMR. The $\ket{NS_2}$ state now exhibits two (three)-measurement steering equivalent to the $\ket{NS_3'}$ state with WM and QMR over time. Moreover, there is no change in the two (three)-measurement steering of the $\ket{NS_3'}$ and Bell states.

\subsection{Maximal fidelity under non-Markovian AD and RTN channels, with(without) WM and QMR}
We explore the dynamics of maximal fidelity of the two-qubit $\ket{NS_1}$, $\ket{NS_2}$, and $\ket{NS_3'}$, and the Bell state under non-Markovian AD and RTN channels. Further, Eq. (\ref{eq:WM_QMR}) is employed to understand the impact of the WM and QMR on the maximal fidelity in the presence of the non-Markovian AD and RTN channels.

\subsubsection{Under non-Markovian AD channel}
\begin{figure}[!htpb]
    \centering
    \includegraphics[height=65mm,width=0.95\columnwidth]{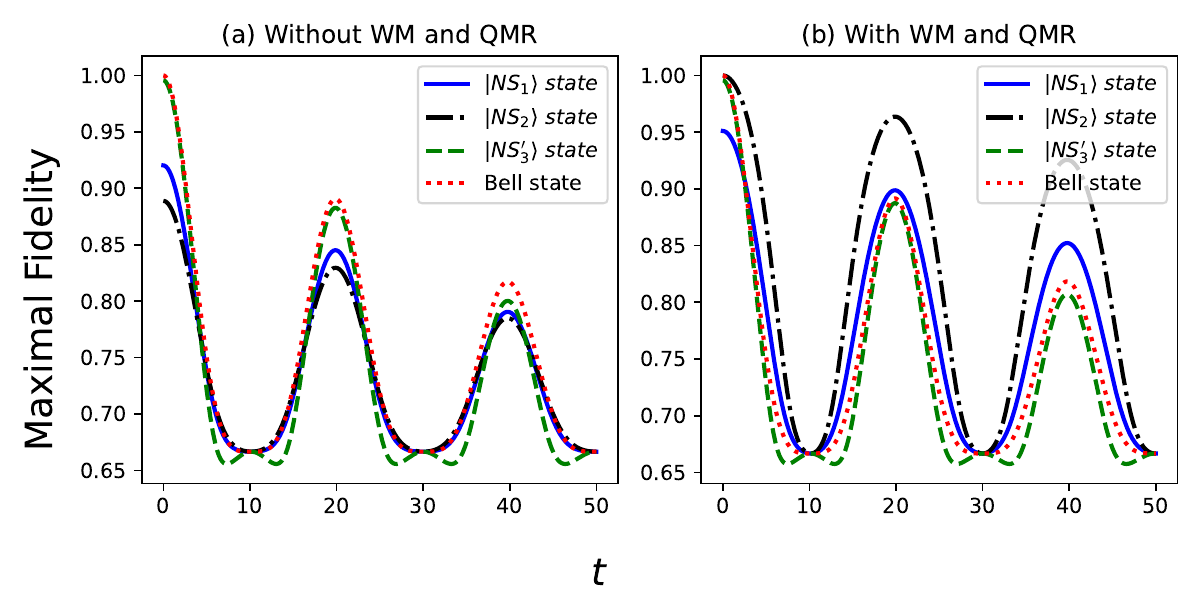}
    \caption{Variation of maximal average fidelity of $\ket{NS_1}$, $\ket{NS_2}$, $\ket{NS_3'}$, and Bell state under non-Markovian AD channel without WM and QMR in subplot (a), and with WM and QMR in subplot (b) with time. Here, for $\ket{NS_1}$ ($p = 0.17$, $q = 0.54$), for $\ket{NS_2}$ ($p = 0.05$, $q = 0.74$), for $\ket{NS_3'}$ ($p = 0.05$, $q = 0.05$), and for Bell state ($p = 0.01$, $q = 0.01$). The non-Markovian AD channel parameters are $g = 0.01$ and $\gamma = 5$.}
    \label{Maxiaml Fidelity_NMAD}
\end{figure}

From Fig.~\ref{Maxiaml Fidelity_NMAD}(a), it is clear that at $t = 0$, the Bell state and $\ket{NS_3'}$ state attain a maximal fidelity value of $1$. But the Bell state's maximal fidelity sustains longer with time. However, just like the Bell state, all the considered two-qubit negative quantum states have maximal fidelity always greater than $\frac{2}{3}$ under non-Markovian AD noise.

The WM and QMR successfully enhance and protect the maximal fidelity of the two-qubit $\ket{NS_1}$ and $\ket{NS_2}$ states. Indeed, the $\ket{NS_1}$ and $\ket{NS_2}$ states' fidelity leads the Bell state, as depicted in Fig.~\ref{Maxiaml Fidelity_NMAD}(b). On the other hand, the fidelity variations of the two-qubit $\ket{NS_3'}$, and the Bell state remain unaltered with WM and QMR.

\subsubsection{Under non-Markovian RTN channel}
\begin{figure}[!htpb]
    \centering
    \includegraphics[height=65mm,width=0.95\columnwidth]{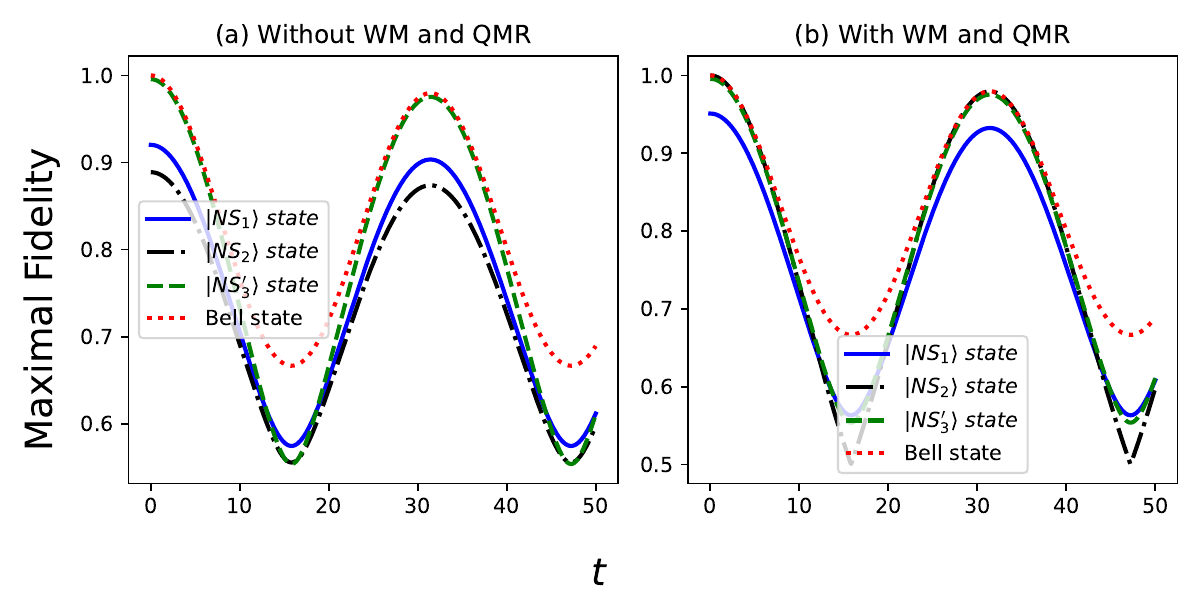}
    \caption{Variation of maximal average fidelity of $\ket{NS_1}$, $\ket{NS_2}$, $\ket{NS_3'}$, and Bell state under non-Markovian RTN channel without WM and QMR in subplot (a), and with WM and QMR in subplot (b) with time. Here, for $\ket{NS_1}$ ($p = 0.17$, $q = 0.54$), for $\ket{NS_2}$ ($p = 0.05$, $q = 0.74$), for $\ket{NS_3'}$ ($p = 0.05$, $q = 0.05$), and for Bell state ($p = 0.01$, $q = 0.01$). The non-Markovian RTN channel parameters are $b = 0.05$ and $\gamma = 0.001$.}
    \label{Maximal Fidelity_NMRTN}
\end{figure}

The two-qubit $\ket{NS_2}$ and Bell states show similar maximal fidelity behavior under the non-Markovian RTN channel for short evolution times. The decay in maximal fidelity over time of all the considered states is gradual compared to the non-Markovian AD noise, as shown in Fig.~\ref{Maximal Fidelity_NMRTN}(a).

As depicted in Fig.~\ref{Maximal Fidelity_NMRTN}(b), the maximal fidelity variations of the $\ket{NS_3'}$, and the Bell states, when subjected to a non-Markovian RTN channel, remain unaffected by WM and QMR. On the other hand, there is a significant improvement in the maximal fidelity of $\ket{NS_1}$ and $\ket{NS_2}$ states with WM and QMR. In fact, with WM and QMR, the $\ket{NS_2}$ state's maximal fidelity variations are comparable to the $\ket{NS_3'}$ state.  

\subsection{Fidelity deviation under non-Markovian AD and RTN channels, with(without) WM and QMR}
We analyze variation in the fidelity deviation of the Bell state and two-qubit negative quantum states in the presence of non-Markovian AD and RTN noise. Moreover, we study the impact of the WM and QMR on the fidelity deviation of the above-mentioned states.

\subsubsection{Under non-Markovian AD channel}
\begin{figure}[!htpb]
    \centering
    \includegraphics[height=65mm,width=0.95\columnwidth]{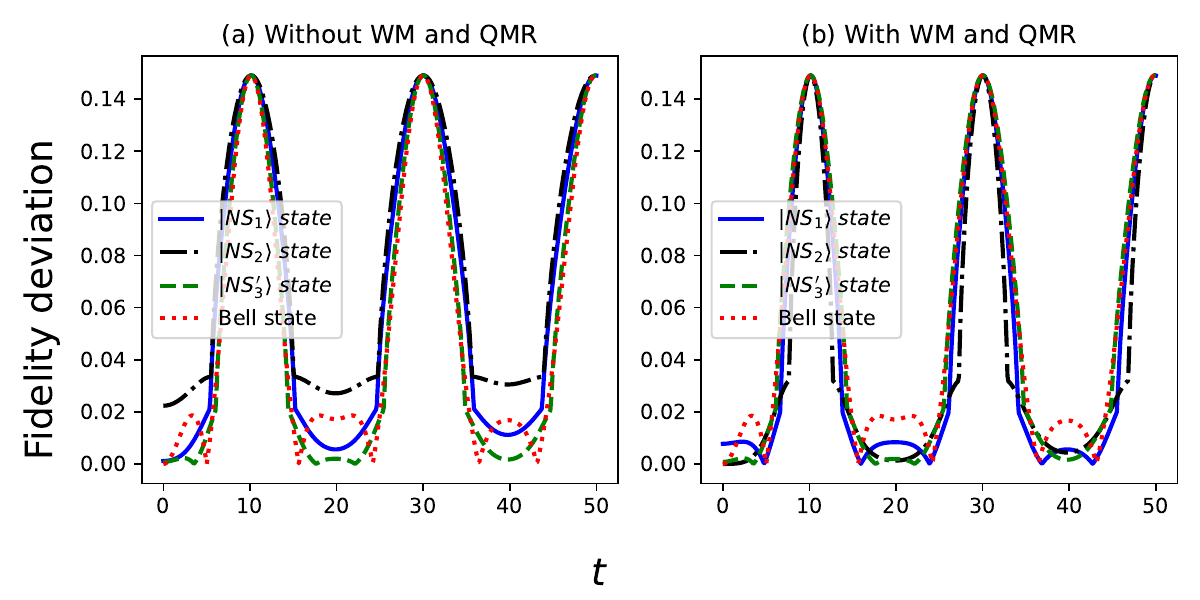}
    \caption{Variation of fidelity deviation of $\ket{NS_1}$, $\ket{NS_2}$, $\ket{NS_3'}$, and Bell state under non-Markovian AD channel without WM and QMR in subplot (a), and with WM and QMR in subplot (b) with time. Here, for $\ket{NS_1}$ ($p = 0.17$, $q = 0.54$), for $\ket{NS_2}$ ($p = 0.05$, $q = 0.74$), for $\ket{NS_3'}$ ($p = 0.05$, $q = 0.05$), and for Bell state ($p = 0.01$, $q = 0.01$). The non-Markovian AD channel parameters are $g = 0.01$ and $\gamma = 5$.}
    \label{FD_NMAD}
\end{figure}

Figure~\ref{FD_NMAD}(a) represents the variations in fidelity deviation of the states mentioned above under non-Markovian AD noise. We can observe from Fig.~\ref{FD_NMAD}(a) that at $t = 0$, all the considered states except the $\ket{NS_2}$ state show fidelity deviation approximately equal to zero. The reason for this is discussed in Sec. \ref{maxiaml fidelity and fidelity deviation sec.} that, except the $\ket{NS_2}$ state all the negative quantum states have $|e_1| = |e_2| = |e_3| > 1/3$. Moreover, all the two-qubit negative quantum states and the Bell state show oscillatory variations in fidelity deviation in synchronization, except for some kinks between the oscillations where the negative quantum states are approaching minimum deviation in fidelity. In these regions, the behavior of the Bell state is different from the two-qubit negative quantum states. If we compare the variation in fidelity deviation of all the states under the non-Markovian AD noise, the $\ket{NS_3'}$ state exhibits a smaller fidelity deviation than the other considered states. This shows that without WM and QMR, the $\ket{NS_3'}$ state is relatively better than the Bell state and other negative quantum states for UQT under non-Markovian AD noise.

The WM and QMR are seen to successfully reduce the deviation in the fidelity of all the considered two-qubit negative quantum states. The WM and QMR are able to squeeze the non-zero fidelity deviation area of two-qubit $\ket{NS_1}$ and $\ket{NS_2}$ states in contrast to the $\ket{NS_3'}$ and Bell states, as depicted in Fig.~\ref{FD_NMAD}(b). This makes them more suitable candidates for UQT under non-Markovian AD noise.

\subsubsection{Under non-Markovian RTN channel}
\begin{figure}[!htpb]
    \centering
    \includegraphics[height=65mm,width=0.95\columnwidth]{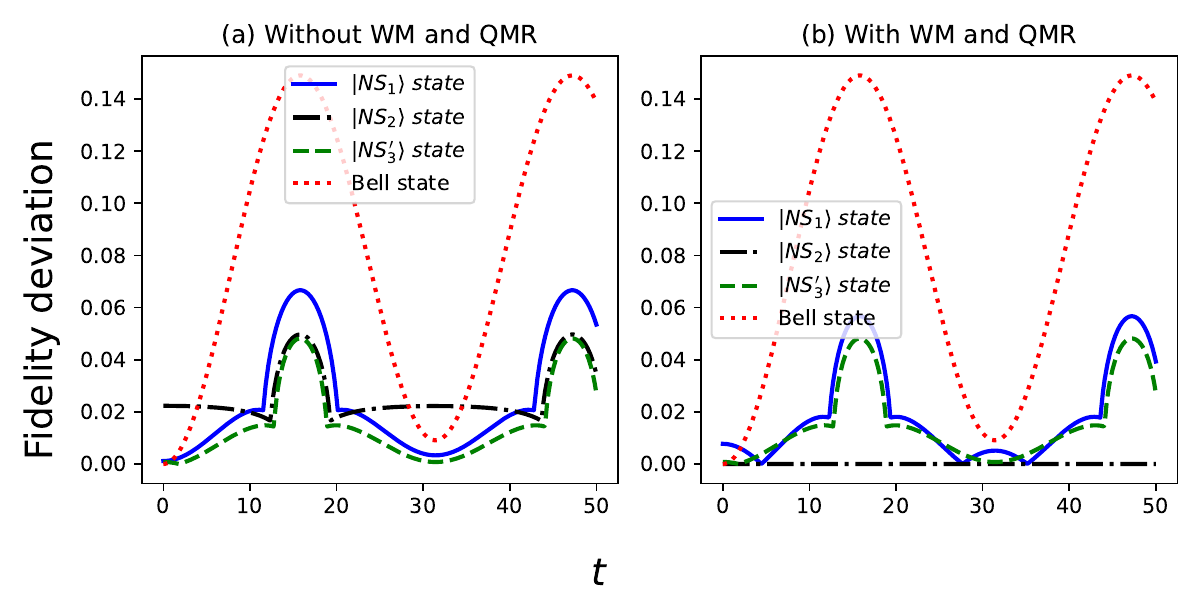}
    \caption{Variation of fidelity deviation of $\ket{NS_1}$, $\ket{NS_2}$, $\ket{NS_3'}$, and Bell state under non-Markovian RTN channel without WM and QMR in subplot (a), and with WM and QMR in subplot (b) with time. Here, for $\ket{NS_1}$ ($p = 0.17$, $q = 0.54$), for $\ket{NS_2}$ ($p = 0.05$, $q = 0.74$), for $\ket{NS_3'}$ ($p = 0.05$, $q = 0.05$), and for Bell state ($p = 0.01$, $q = 0.01$). The non-Markovian RTN channel parameters are $b = 0.05$ and $\gamma = 0.001$.}
    \label{FD_NMRTN}
\end{figure}

Under the non-Markovian RTN noise, all the two-qubit negative quantum states show less fidelity deviation than the Bell state, as shown in Fig.~\ref{FD_NMRTN}(a). Thus, compared to the Bell state, they are more suitable for UQT.

With the WM and QMR, the $\ket{NS_2}$ state shows zero deviation in fidelity under the non-Markovian RTN channel, making it an ideal state for UQT as demonstrated in Fig.~\ref{FD_NMRTN}(b). On the other hand, the WM and QMR have no impact on the fidelity deviation of the Bell state and $\ket{NS_3'}$ state under the non-Markovian RTN. Additionally, the deviation in the fidelity of $\ket{NS_1}$ state is reduced to some extent, as depicted by Fig.~\ref{FD_NMRTN}(b).

\begin{table}
\centering
\begin{tabular}{ | m{4.5cm}| m{4.5cm} | m{4.5cm} |}
  \hline 
  \textbf{Quantum correlations and UQT requirements} & \textbf{Without WM and QMR} &  \textbf{With WM and QMR}\\ 
  \hline
  Concurrence & \scriptsize$\ket{NS_3'}>BS>\ket{NS_1}>\ket{NS_2}$ & \scriptsize$\ket{NS_2}>\ket{NS_1}>\ket{NS_3'}>BS$ \\
  \hline
  Discord & \scriptsize$\ket{NS_3'}=BS>\ket{NS_1}>\ket{NS_2}$ & \scriptsize$\ket{NS_2}>\ket{NS_1} \approx BS>\ket{NS_3'}$\\
  \hline
  Two (three)-measurement steering & \scriptsize$BS>\ket{NS_3'}>\ket{NS_1}>\ket{NS_2}$ & \scriptsize$\ket{NS_2}>\ket{NS_1}>BS>\ket{NS_3'}$\\
  \hline
  Maximal Fidelity & \scriptsize$BS>\ket{NS_3'}>\ket{NS_1}>\ket{NS_2}$ & \scriptsize$\ket{NS_2}>\ket{NS_1}>BS>\ket{NS_3'}$\\
  \hline
  Fidelity deviation & \scriptsize$\ket{NS_3'}<BS<\ket{NS_1}<\ket{NS_2}$ & \scriptsize$\ket{NS_2} \approx \ket{NS_1}<\ket{NS_3'}<BS$\\
  \hline
\end{tabular}
\caption{\label{table1} Comparison of the quantum correlations, maximal fidelity, and fidelity deviation variations of two-qubit $\ket{NS_1}$, $\ket{NS_2}$, $\ket{NS_3'}$, and the Bell state (BS) under the non-Markovian AD channel ($t>0$).}
\end{table}

\begin{table}
\centering
\begin{tabular}{ | m{4.5cm}| m{4.5cm} | m{4.5cm} |}
  \hline 
  \textbf{Quantum correlations and UQT requirements} & \textbf{Without WM and QMR} &  \textbf{With WM and QMR}\\ 
  \hline
  Concurrence & \scriptsize$\ket{NS_3'} \approx BS>\ket{NS_1}>\ket{NS_2}$ & \scriptsize$\ket{NS_2} \approx \ket{NS_3'} \approx BS>\ket{NS_1}$ \\
  \hline
  Discord & \scriptsize$\ket{NS_3'}>BS>\ket{NS_1}>\ket{NS_2}$ & \scriptsize$\ket{NS_2}>\ket{NS_3'}>BS>\ket{NS_1}$\\
  \hline
  Two (three)-measurement steering & \scriptsize$BS>\ket{NS_3'}>\ket{NS_1}>\ket{NS_2}$ & \scriptsize$BS>\ket{NS_2} \approx NS_3>\ket{NS_1}$\\
  \hline
  Maximal Fidelity & \scriptsize$BS>\ket{NS_3'}>\ket{NS_1}>\ket{NS_2}$ & \scriptsize$BS>\ket{NS_2} \approx \ket{NS_3'}> \ket{NS_1}$\\
  \hline
  Fidelity deviation & \scriptsize$\ket{NS_3'}<\ket{NS_1}<\ket{NS_2}<BS$ & \scriptsize$\ket{NS_2}<\ket{NS_3'}<\ket{NS_1}<BS$\\
  \hline
\end{tabular}
\caption{\label{table2} Comparison of the quantum correlations, maximal fidelity, and fidelity deviation variations of two-qubit $\ket{NS_1}$, $\ket{NS_2}$, $\ket{NS_3'}$, and the Bell state (BS) under the non-Markovian RTN channel ($t>0$).}
\end{table}

\section{\label{ch4_result&discussion}Results and discussion}
\begin{figure}[!htpb]
    \centering
    \includegraphics[height=85mm,width=0.85\columnwidth]{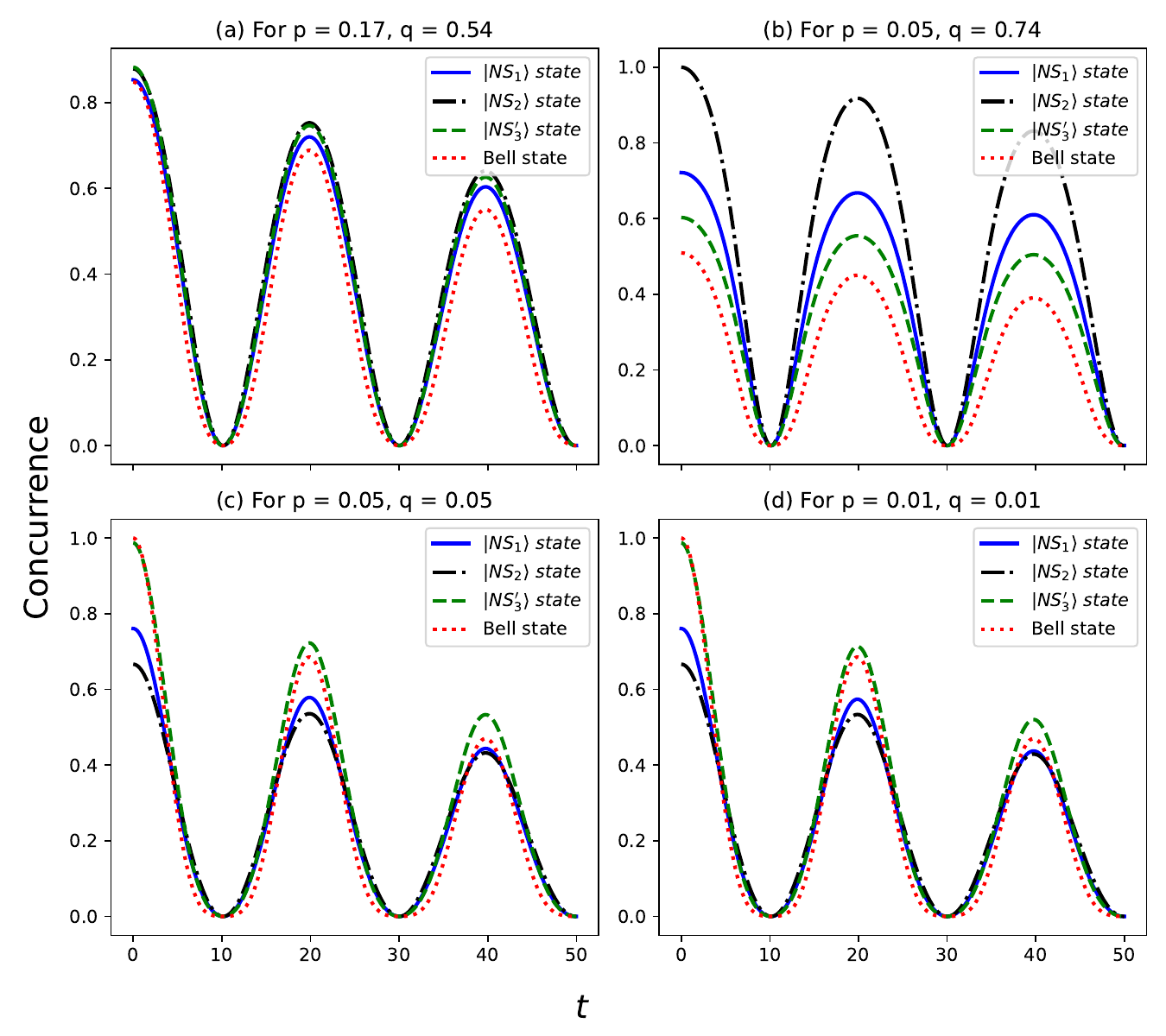}
    \caption{Variation of concurrence of $\ket{NS_1}$, $\ket{NS_2}$, $\ket{NS_3'}$, and Bell state under non-Markovian AD channel with WM and QMR in subplots (a), (b), (c), and (d). In subplot (a) $p = 0.17$, $q = 0.54$ for all states, in subplot (b) $p = 0.05$, $q = 0.74$ for all states, in subplot (c) $p = 0.05$, $q = 0.05$ for all states, and in subplot (d) $p = 0.01$, $q = 0.01$ for all states. The non-Markovian AD channel parameters are $g = 0.01$ and $\gamma = 5$.}
    \label{concur_NMAD_same_p_and_q}
\end{figure}

The WM and QMR strength parameters are optimized at time $t = 0$ for two-qubit $\ket{NS_1}$, $\ket{NS_2}$, $\ket{NS_3'}$, and the Bell states and optimal combinations of $(p, q)$ are obtained. If we keep the same values of $(p, q)$ for all the states, without picking up the optimal ones for every state, their behavior is depicted in Fig.~\ref{concur_NMAD_same_p_and_q}. It can be seen that the Bell state need not provide the best results. Moreover, by doing this, the main advantage of having WM and QMR would be lost. This is so because these parameters are in the control of the experimentalist and can be leveraged to get the optimal quantum correlations and UQT requirements for the states under consideration. This motivates the need to optimize the WM and QMR strength parameters at time $t = 0$ for two-qubit $\ket{NS_1}$, $\ket{NS_2}$, $\ket{NS_3'}$, and the Bell states. Under non-Markovian noisy quantum channels, the effect of WM and QMR on the quantum correlations and maximal fidelity of the two-qubit states is greater when the WM and QMR strength parameters are large, as discussed in \citep{sun2017recovering}, at the cost of success probability. This is corroborated by Fig.~\ref{P_success}. The WM and QMR also minimize the fidelity deviation of non-maximally entangled two-qubit states, consistent with  \citep{sabale2023towards}. We discuss these criteria below in conjunction with our results. The trade-off relation between success probability, discussed in Sec. (\ref{ch4_Model}), and WM and QMR strength parameters ($p, q$) can be observed in Fig.~\ref{P_success} under non-Markovian AD and RTN channels, respectively. 

Due to WM and QMR, the two-qubit $\ket{NS_2}$ state's (non-maximally entangled state at time $t = 0$) concurrence, discord, steering, maximal fidelity, and fidelity deviation are seen to improve significantly as we can observe that the strength parameters ($p$, $q$) are highest for this particular state with some non-zero finite success probability. The $\ket{NS_2}$ state's concurrence, discord, steering, and maximal fidelity were observed to be higher than all other considered states under non-Markovian AD noise. This pattern was followed by the behavior of discord under non-Markovian RTN noise. Also, there is a notable improvement in the concurrence, steering, and maximal fidelity of the $\ket{NS_2}$ state under the non-Markovian RTN channel, which is attributed to the WM and QMR. In fact, under the non-Markovian RTN channel, the $\ket{NS_2}$ state's concurrence, steering, and maximal fidelity were seen to be equivalent to the $\ket{NS_3'}$ state. The WM and QMR reduced the $\ket{NS_2}$ state's fidelity deviation under non-Markovian AD and RTN channels. In fact, under the non-Markovian RTN channel, its fidelity deviation is almost zero with time, making it an ideal candidate for UQT.

The WM and QMR positively impact the two-qubit $\ket{NS_1}$ state's quantum correlations and UQT requirements under non-Markovian AD and RTN channels. Moreover, with WM and QMR, this state maintains its concurrence, steering, maximal fidelity, and fidelity deviation to higher values than the $\ket{NS_3'}$ and Bell states under non-Markovian AD noise at the expense of low success probability. Whereas under the non-Markovian RTN channel, its fidelity deviation is less than the Bell state with(without) WM and QMR. Moreover, the fidelity deviation of all the considered two-qubit negative quantum states is less than the Bell state under the non-Markovian RTN channel, with(without) WM and QMR. 

The WM and QMR strength parameters are quite low for two-qubit $\ket{NS_3'}$ and Bell states with high success probability, which makes sense as the two-qubit $\ket{NS_3'}$ and Bell states are maximally entangled states at time $t = 0$. This explains why the quantum correlations, maximal fidelity, and fidelity deviation of the $\ket{NS_3'}$ and Bell states were not changed with WM and QMR. Additionally, the fidelity deviation of $\ket{NS_3'}$ state is lesser than the Bell state with(without) WM and QMR, making it a more suitable candidate for UQT.

The overall pattern hierarchy of quantum correlations, maximal fidelity, and fidelity deviation with(without) WM and QMR under non-Markovian AD and RTN channels is summarized in TABLE \ref{table1} and TABLE \ref{table2} respectively, for optimal combinations of WM and QMR strength parameters as follows $\ket{NS_1}$ ($p = 0.17$, $q = 0.54$), $\ket{NS_2}$ ($p = 0.05$, $q = 0.74$), $\ket{NS_3'}$ ($p = 0.05$, $q = 0.05$), and Bell state ($p = 0.01$, $q = 0.01$) under both non-Markovian AD and RTN channels.

\begin{figure}[!htpb]
    \centering
    \includegraphics[height=85mm,width=0.85\columnwidth]{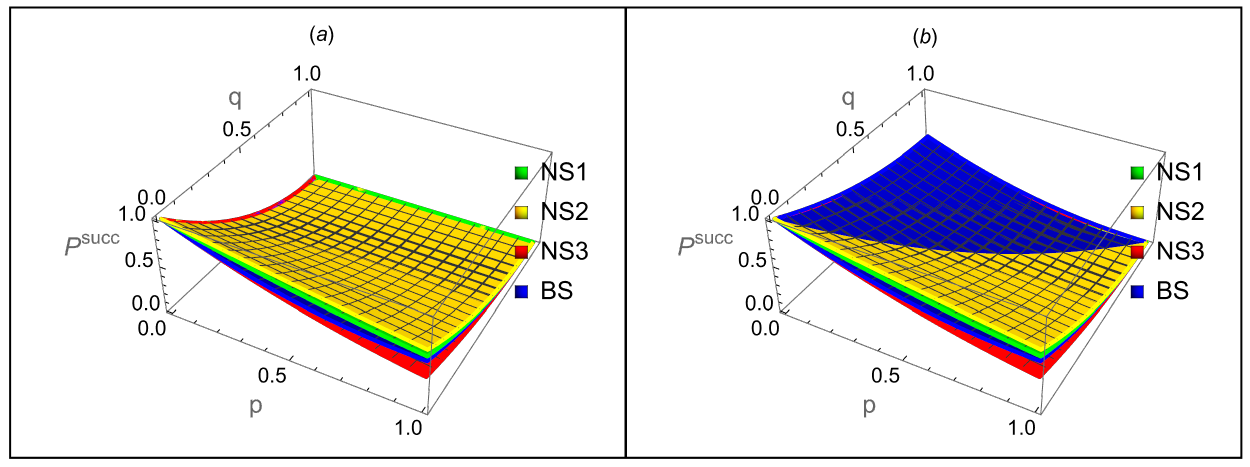}
    \caption{Variation of success probability of $\ket{NS_1}$, $\ket{NS_2}$, $\ket{NS_3'}$, and Bell state ($BS$) under non-Markovian AD and RTN channels in subplots (a) and (b), respectively. Here, the figures are depicted for WM strength ($p$) and QMR strength ($q$) at time $t = 10$. The non-Markovian AD and RTN channel parameters are ($g = 0.01$, $\gamma = 5$) and ($b = 0.05$, $\gamma = 0.001$), respectively.}
    \label{P_success}
\end{figure}

\section{\label{ch4_conclusion}Summary}
In this article, we have investigated the impact of weak measurement (WM) and quantum measurement reversal (QMR) on the quantum correlations and universal quantum teleportation (UQT) of two-qubit $\ket{NS_1}$, $\ket{NS_2}$, $\ket{NS_3'}$, and the Bell states under both non-unital (non-Markovian Amplitude Damping) and unital (non-Markovian Random Telegraph Noise) quantum channels. To this end, we discussed the negative quantum states followed by their quantum correlations, particularly concurrence, discord, and steering, and their maximal fidelity, and fidelity deviation under the influence of noisy quantum channels with(without) WM and QMR. It was shown that WM and QMR brought out better performance of the negative quantum states for the quantum correlations and the UQT. Additionally, the relationship of trade-offs between weak measurement and quantum measurement reversal parameters and success likelihood is examined. Interestingly, it was observed that some surpassed the Bell state's performance over time. It was also observed that the fidelity deviation of negative quantum states was less compared to the Bell state with(without) WM and QMR when states evolved through unital quantum channels. It was found that the $\ket{NS_3'}$ state has a lower fidelity deviation than the Bell state with(without) WM and QMR with high success probability, making it a better choice for UQT.

However, with WM and QMR, the two-qubit $\ket{NS_2}$ state showed zero fidelity deviation under the unital channel. Under the non-unital channel, this state fares better among all the considered states, making it an ideal state for UQT. Also, the $\ket{NS_1}$ state, with the weak measurement and its reversal, outperforms the Bell state under the non-unital channel for UQT. These properties of the two-qubit negative quantum states affected by WM and QMR can be useful in protocols involving quantum correlations and UQT.

\subsection{Limitations and Scope}
The WM and QMR strength parameters $(p, q)$ used in this chapter are optimized at time $t = 0$ for each state independently: $|NS_1\rangle~ (p = 0.17, q = 0.54)$, $|NS_2\rangle~ (p = 0.05, q = 0.74)$, $|NS_3'\rangle~ (p = 0.05, q = 0.05)$, and the Bell state $(p = 0.01, q = 0.01)$. These optimal values are state-specific and channel-specific; they may differ for other noise strengths or qubit parameters. In particular, the success probability of the WM and QMR protocol (Fig.~\ref{P_success}) decreases significantly as p and q increase, meaning that the enhanced performance of $|NS_2>$ with WM and QMR comes at the cost of a lower success probability compared to the Bell state. In any practical implementation, the viability of this trade-off depends on the acceptable post-selection overhead. Furthermore, the analysis in this chapter is restricted to two-qubit systems under two specific non-Markovian channels (AD and RTN). Extensions to multi-qubit systems, other noise models (e.g., depolarizing, dephasing, generalized amplitude damping), and finite-temperature environments would require separate investigation, but we generally expect a similar trend under other dephasing and amplitude-damping channels.


\newpage
\setcounter{chapter}{4} 

\titleformat{\chapter}[display]
{\sffamily\fontsize{27}{27}\bfseries\filleft}{\thechapter}{0pt}{{#1}}  
  
\thispagestyle{empty}

\chapter{Noise-Resilient Negative Quantum States}\label{chap5:Physical_realization}

\section{Introduction}
Entanglement is a unique ingredient in a quantum computing scheme that is pivotal in harnessing the full power of quantum mechanics. In addition to being a resource for speeding up computations, entanglement within quantum systems is also responsible for the robust storage of quantum information in a quantum memory~\citep{Quantumerrorcorrection2013}, secure communication~\citep{Ekert1991Quantumcryptography, Scarani2009The_security}, and quantum enhanced metrology~\citep{Pezz2018Quantummetrology, Braun2018Quantum_enhanced}. Bell states have been considered the typical benchmark for entangled resources. However, their fragility under realistic noise poses a critical barrier. Since the advent of quantum computing, considerable efforts have been directed towards attaining a profound understanding of the entanglement properties inherent in quantum systems. Noise is an imminent threat to quantum information tasks; it is generally addressed using the framework of open quantum systems. Open quantum systems are traditionally modeled using Markovian GKSL (Gorini-Kossakowski-Sudarshan-Lindblad) dynamical evolution~\citep{Breuer2007, Banerjee2018}. An example of such evolution is the depolarizing noise~\citep{nielsen2010quantum}. However, a broad range of real-world noise in physical systems is dominated by non-Markovian noise, such as non-Markovian amplitude damping and random telegraph noise~\citep{Breuer2007, Banerjee2018, SM23, Tiwari2023Impact}. Non-Markovian noise retains memory of past interactions, causing correlations that standard error correction or simplistic noise models cannot fully capture~\citep{lalita2025non_classicality}. Consequently, the Bell states often lose their entanglement before they can be effectively utilized.

This scenario could be addressed using the discrete Wigner function (DWF) formalism. Unlike classical distributions, quantum Wigner functions can take negative values~\citep{W32}, a hallmark of non-classicality~\citep{AZ04}. Although the Wigner function is conventionally applied in the context of continuous variable systems, there exists a less prevalent but potent theory concerning the discrete Wigner function (DWFs)~\citep{wootters2004picturing, gibbons2004discrete}. Analogously to the continuous variable scenario, the states of a system whose discrete Wigner function manifests negative values within the discrete phase space can be represented as eigenstates of the negative eigenvalues of phase-space point operators~\citep{gibbons2004discrete,van2011noise, casaccino2008extrema}. In a recent study, quantum states that exhibit negativity of the discrete Wigner function have been identified to be robust to a wide variety of noise in quantum systems captured by non-Markovian errors~\citep{lalita2023harnessing, Lalita_2024ProtectingQC}. These two-qubit negative quantum states are recognized as optimal candidates for universal quantum teleportation using weak measurements~\citep{Lalita_2024ProtectingQC} within non-Markovian noise environments. Consequently, the physical realization of these two-qubit negative quantum states is imperative, as they hold the potential to augment quantum information protocols by providing resilient resources.

In this work, we present methods to generate these states using operations native to superconducting hardware, specifically single-qubit gates and the $CZ$ gate. Through quantum state tomography, we demonstrate that these states can be prepared with high fidelity on IBM’s quantum hardware under realistic noise conditions. We employ various approaches to assess the noise resilience of the negative quantum states and benchmark them against the standard Bell states. These approaches include fidelity variation analysis, phase sensitivity under $SU(2)$ rotations using quantum Fisher information (QFI), violations of Bell-CHSH inequality, and quantum information measures such as concurrence and teleportation fidelity. Our results highlight the role of negative quantum states as robust entanglement resources for a range of quantum computing and communication applications. Furthermore, we propose a teleportation circuit that utilizes one of the two-qubit negative quantum states as its entanglement resource.

This chapter is organized as follows. In Sec.~\ref{sec:methods}, we present methods for preparing these states on IBM superconducting hardware along with ideal and mitigated state tomography techniques. Section~\ref{sec:results} tests and compares negative quantum states' fidelity, maximal mean QFI, and Bell-CHSH inequality violation under non-Markovian noise with the Bell $\ket{\phi^{+}}$ state. Also, a teleportation circuit using one of the negative quantum states is provided in this section. This is followed by conclusions in Sec.~\ref{sec:conclusion}. \textit{This chapter's content is based on~\citep{lalita2025realizingnegativequantumstates}. \copyright APS. Adapted and reproduced with permission.}

\section{Constructing negative quantum states for IBM hardware}\label{sec:methods}

\subsection{Unitary transformations to realize negative quantum states}
In this work, we particularly implement two-qubit negative quantum states elaborated in~\citep{lalita2023harnessing, Lalita_2024ProtectingQC} on IBM's quantum hardware. To determine the unitary transformations of the two-qubit negative quantum states elaborated earlier in Sec.~\ref{ch3_NQS} of chapter~\ref{chap3:Harnessing}, the Gram-Schmidt procedure is employed~\citep{nielsen2010quantum}. A comprehensive description of this procedure is as follows. \textit{Step 1}: Consider any two-qubit negative quantum states, i.e., either of the $\ket{NS_{1}}$, $\ket{NS_2}$, $\ket{NS_3}$, and $\ket{NS_3^{\prime}}$ as $\ket{V_1}$. Let $\ket{V_1}$ = $\ket{NS_1}$. \textit{Step 2}: To find the other three orthonormal vectors ($\ket{V_2}$, $\ket{V_3}$, $\ket{V_4}$) of $\ket{V_1}$, we take any three linearly independent vectors of $\ket{V_1}$. We pick the standard computational basis vectors $\ket{e_{1}} = \ket{00}, \ket{e_{2}} = \ket{01}, \ket{e} = \ket{10}$ as linearly independent vectors of $\ket{V_1}$. Considering $\ket{O_2} = \ket{e_{1}}$, $\ket{O_3} = \ket{e_{2}}$, $\ket{O_4} = \ket{e_{3}}$, the orthonormal vectors $\ket{V_2}$, $\ket{V_3}$, and $\ket{V_4}$ can be calculated using the Gram-Schmidt decomposition as
\begin{equation}
    \ket{V_{K+1}} = \frac{\ket{O_{K+1}} - \sum_{i=1}^{K}\bra{V_i}\ket{O_{K+1}}\ket{V_i}}{||\ket{O_{K+1}} - \sum_{i=1}^{K}\bra{V_i}\ket{O_{K+1}}\ket{V_i}||}.
\end{equation}
\textit{Step 3}: Now, we have four orthonormal vectors $\ket{V_1}$, $\ket{V_2}$, $\ket{V_3}$, and $\ket{V_4}$, which can span the two-qubit system's Hilbert space.
\textit{Step 4}: Finally, the unitary transformation $U$ from the computational basis set $\{\ket{e_i}\}$ to the orthonormal set $\{\ket{V_i}\}$ is given by
     $U = \sum_{i = 1}^{4} \ket{V_i}\bra{e_i}$, which takes the vector $\ket{00}$ to the state $\ket{NS_1}$.
Similarly, the unitary transformations of the other two-qubit negative quantum states can be calculated. The unitary transformation matrices obtained using the Gram-Schmidt procedure for realizing $\ket{NS_1}$, $\ket{NS_2}$, $\ket{NS_3}$, $\ket{NS_3^{\prime}}$, and $\ket{NS_3^{\prime\prime}}$ states from the $\ket{00}$ state are given as follows:
\small{
    \begin{equation}
    U_{NS_1} =\left(
    \begin{array}{cccc}
     -0.742977+0. i & 0.669317\, +0. i & 0.\, +0. i & 0.\, +0. i \\
     -0.357599+0.357599 i & -0.396953+0.396953 i & 0.655059\, +0. i & 0.\, +0. i \\
     0.101586\, +0.101586 i & 0.112766\, +0.112766 i & 0.\, +0.247581 i & 0.944792\, +0. i \\
     -0.414237+0. i & -0.459824+0. i & -0.504778-0.504778 i & 0.231698\, -0.231698 i \\
    \end{array}
    \right)
    \end{equation}
    
    \begin{equation}
        U_{NS_2} = \left(
    \begin{array}{cccc}
     0.788675\, +0. i & 0.61481\, +0. i & 0.\, +0. i & 0.\, +0. i \\
     -0.288675+0.288675 i & 0.370311\, -0.370311 i & 0.747712\, +0. i & 0.\, +0. i \\
     -0.288675-0.288675 i & 0.370311\, +0.370311 i & 0.\, -0.589702 i & 0.459701\, +0. i \\
     -0.211325+0. i & 0.271086\, +0. i & -0.215846-0.215846 i & -0.627963+0.627963 i \\
    \end{array}
    \right)
    \end{equation}
    
    \begin{equation}
        U_{NS_3} = \left(
    \begin{array}{cccc}
     -0.0508479+0. i & 0.998706\, +0. i & 0.\, +0. i & 0.\, +0. i \\
     0.631483\, -0.228733 i & 0.0321511\, -0.0116456 i & 0.740096\, +0. i & 0.\, +0. i \\
     -0.27958-0.68233 i & -0.0142345-0.03474 i & 0.0277425\, +0.670334 i & 0.0687934\, +0. i \\
     0.0508479\, +0. i & 0.00258886\, +0. i & -0.0434981-0.0157557 i & 0.378252\, -0.923143 i \\
    \end{array}
    \right)
    \end{equation}

    \begin{equation}
        U_{NS_3^{\prime}} = \left(
        \begin{array}{cccc}
         -0.575107+0. i & 0.818078\, +0. i & 0.\, +0. i & 0.\, +0. i \\
         -0.345634+0.310025 i & -0.242979+0.217946 i & 0.823336\, +0. i & 0.\, +0. i \\
         -0.265082-0.229473 i & -0.186352-0.161319 i & -0.0371656-0.293085 i & 0.85384\, +0. i \\
         0.575107\, +0. i & 0.404298\, +0. i & 0.360743\, +0.323577 i & 0.393558\, -0.34069 i \\
        \end{array}
        \right)
    \end{equation}

    \begin{equation}
        U_{NS_3^{\prime\prime}} = \left(
    \begin{array}{cccc}
     0.\, +0. i & 0.\, +0. i & 0.\, +0.707106781i & 0.\, +0.707106781i \\
     0.\, +0.707106781i & 0.\, -0.707106781i & 0.\, +0. i & 0.\, +0. i \\
     +0.707106781+0. i & +0.707106781+0. i & 0.\, +0. i & 0.\, +0. i \\
     0.\, +0. i & 0.\, +0. i & +0.707106781+0. i & -0.707106781+0. i \\
    \end{array}
    \right)
    \end{equation}
    }%
Now, we discuss the quantum circuits corresponding to the unitary transformations of the two-qubit negative quantum states.
\subsection{Optimized circuit design with native gates} \label{subsec:circs}
After obtaining the unitary transformations of negative quantum states, the final optimized circuits for $\ket{NS_1}$, $\ket{NS_2}$, $\ket{NS_3}$, and $\ket{NS_3^{\prime}}$ states using $H$, $R_x$, $R_z$, and $CZ$ gates are constructed and verified using Qiskit~\citep{qiskit2024}. Their corresponding quantum circuits are shown in Fig.~\ref{NS1_NS2_NS3__NS3_prime_circuit}. Additionally, we can observe that the $\ket{NS_3^{\prime\prime}}$ state has a relative phase of $\pi/4$ with respect to the Bell $\ket{\psi^{+}}$ state, see Fig.~\ref{Teleportation_NS3_double_prime}. Moreover, the circuit depth for $\ket{NS_1}$, $\ket{NS_2}$, $\ket{NS_3}$ and $\ket{NS_3^{\prime}}$ is $13$, while for $\ket{NS3^{\prime\prime}}$ it is just $4$. The loss function for all the quantum circuits is also calculated using $\delta = \sqrt{1 - \frac{|{\rm Tr}(U_f^{\dag}U_t)|^2}{d^2}}$~\citep{Younis2020QFASTQS}, where $U_f$ is the operation implemented by the encoded circuit, $U_t$ is the target input, and $d$ is the dimension of the unitary matrix under consideration. On implementing the circuits given in Fig.~\ref{NS1_NS2_NS3__NS3_prime_circuit} on Qiskit's \emph{AerSimulator}, the loss function for all the two-qubit negative quantum states comes out to be $\sim 10^{-5}$. The Schmidt rank of the two-qubit negative states and output states of their corresponding quantum circuits is the same, i.e., $2$. In addition, we perform the tomographic reconstruction of the given quantum circuits on both the simulator and the IBM quantum computer to validate them.

\begin{figure}[H]
    \centering
    \includegraphics[height=95mm,width=0.95\textwidth]{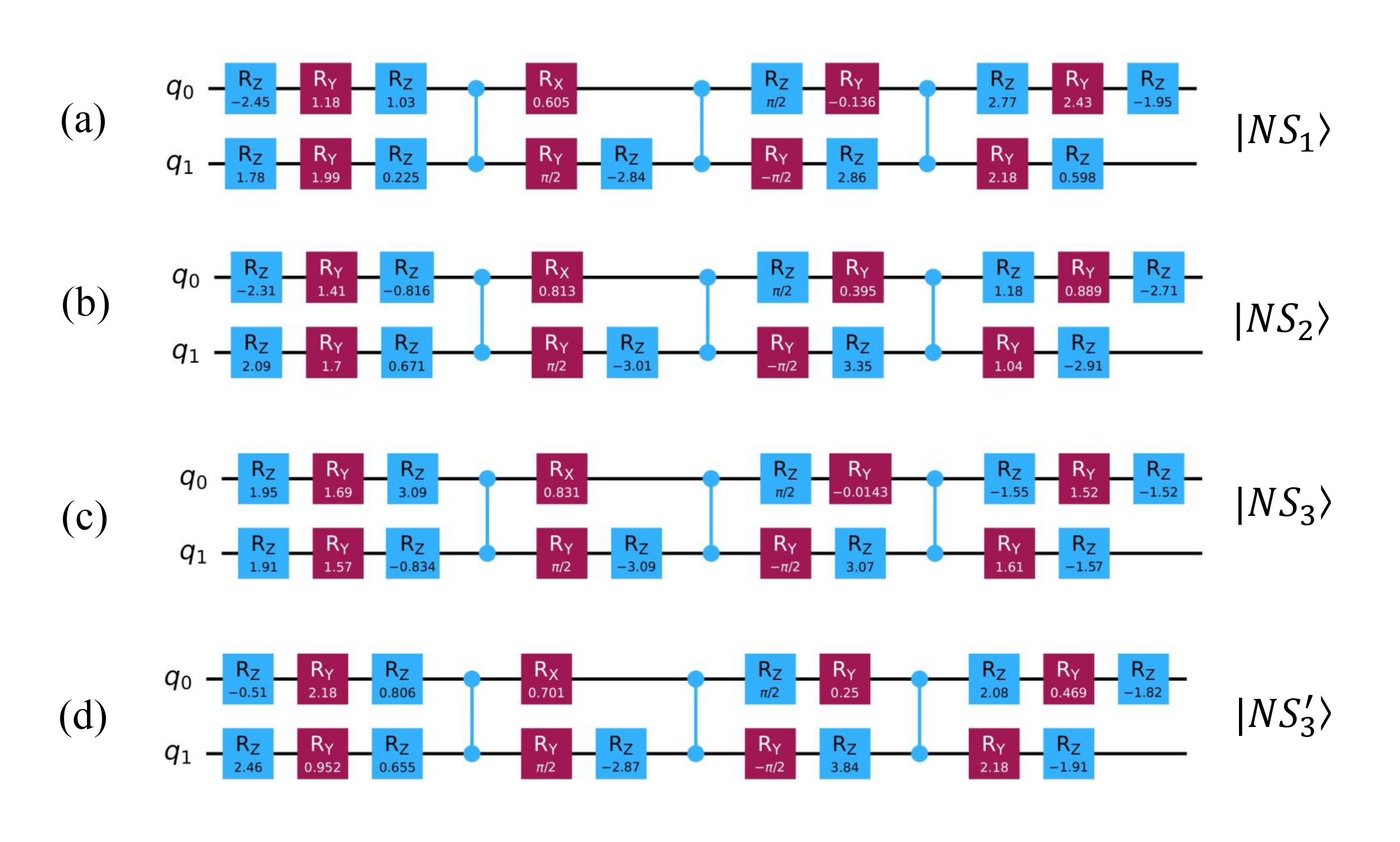}
    \addtocounter{figure}{-1}
    \caption{Quantum circuits to generate the two-qubit $\ket{NS_1}$, $\ket{NS_2}$, $\ket{NS_3}$, and $\ket{NS_3^{\prime}}$ states from the $\ket{00}$ state using $H$, $R_x$, $R_z$, and $CZ$ gates are shown in subfigures (a), (b), (c), and (d) respectively. Here, $q_0$ and $q_1$ represent the qubits in the $\ket{00}$ state.}
    \label{NS1_NS2_NS3__NS3_prime_circuit}
\end{figure}

\subsection{Quantum state tomography and error mitigation on IBM devices}
Tomography is a technique used to construct an image of a hidden object by analyzing several observable projections~\citep{Quantum-State-Tomography1995}. Due to the inherent nature of quantum physics, it is not feasible to directly observe physical objects in their actual state. Instead, we see only the various aspects of the physical objects, like the wave or the particle aspects, which depend on the particular type of measurement. However, it is still possible to determine the quantum state by doing numerous experiments on identically prepared systems and building up good statistics on the outcomes. Suppose that the collection of experiments is completely informative. In that case, it is possible to reconstruct the density matrix of the quantum system, known as quantum state tomography. An infinite number of perfect measurements would be required to determine the state entirely. 

\begin{figure}[!htpb]
    \centering
    \includegraphics[height = 8cm, width=0.50\columnwidth]{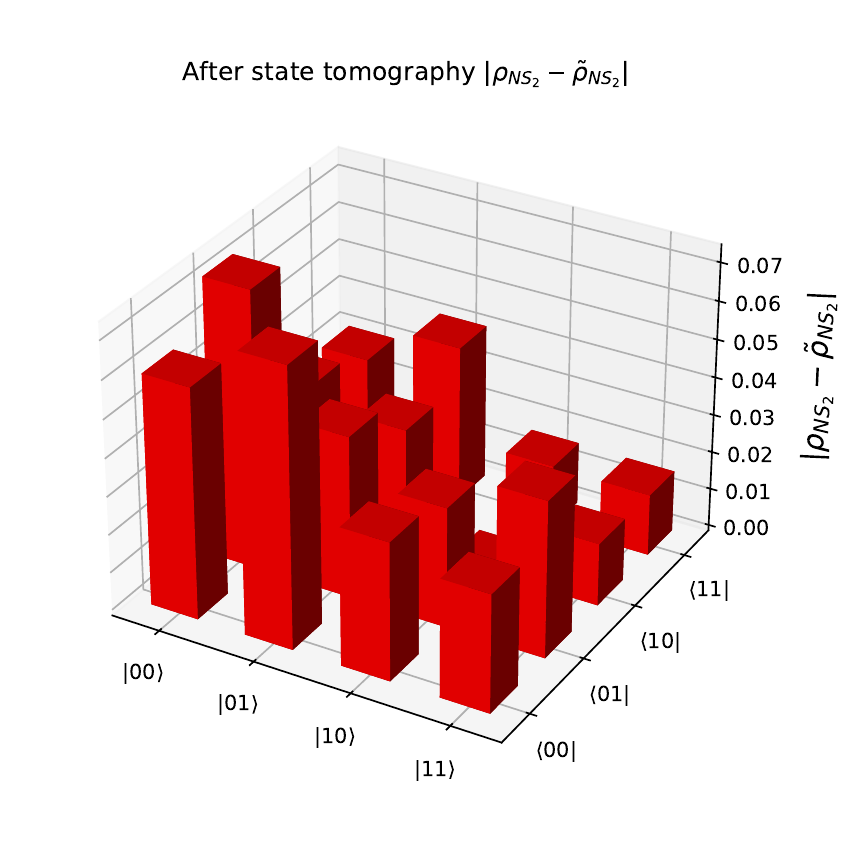}
    \caption{The city plot for the $\ket{NS_2}$ state displays the absolute difference between the components of the original $\rho_{NS_2}$ and the $\tilde\rho_{NS_2}$ obtained after performing a state tomography experiment on the real IBM quantum computer $\it{ibm\_brisbane}$ for $8192$ times.}
    \label{city_plot_NS2_tomo}
\end{figure}

Using Qiskit~\citep{qiskit2024}, we perform the state tomography of two-qubit negative quantum states based on identifying the maximum-likelihood state that aligns with the available data~\citep{Maximum-Likelihood_2012}. The tomography of two-qubit negative quantum states is experimentally implemented by executing their respective quantum circuits for $8192$ times on both the $\textit{ibm\_brisbane}$ quantum computer and the IBM $\textit{AerSimulator}$~\citep{qiskit2024, ibm_brisbane}, resulting in the $\tilde\rho$ state. Figure~\ref{city_plot_NS2_tomo} shows the city plot illustrating the absolute difference between the components of the original state $\rho_{NS_2}$ and the reconstructed state $\tilde\rho_{NS_2}$. As we can observe, the individual elements of the density matrix obtained by state tomography are closer to the actual $\ketbra{NS_2}{NS_2}$ density matrix with a maximum difference of approximately $0.07$. Similarly, we can perform and study the quantum state tomography for other negative quantum states. Further, to quantify the closeness between the state obtained by tomography and the actual negative quantum states, we study their fidelity,~\textit{viz}, given as~\citep{jozsa1994fidelity, nielsen2010quantum} 
\begin{equation}
    F(\rho_{NS_2}, \tilde\rho_{NS_2}) \equiv {\rm Tr}\sqrt{\sqrt{\rho_{NS_2}}\tilde\rho_{NS_2}\sqrt{\rho_{NS_2}}},
    \label{ch5_Fidelity_formula}
\end{equation}
Fidelity estimates the likelihood that one state will successfully undergo a test to be recognized as the other, providing a way to validate the quantum circuits of the negative quantum states. The fidelity values for all the negative quantum states together with the Bell state are provided in TABLE~\ref{New_table}. After tomographic reconstruction, we observe that it is around $0.87-0.91$ on the actual quantum computer and around $0.91-0.93$ on the simulator for the negative quantum states. 
Interestingly, the circuit depth for the negative quantum states is comparatively higher than the Bell state, yet the fidelity values are approximately the same.  
\begin{figure}[!htpb]
    \centering
    \includegraphics[height = 8cm, width=0.50\columnwidth]{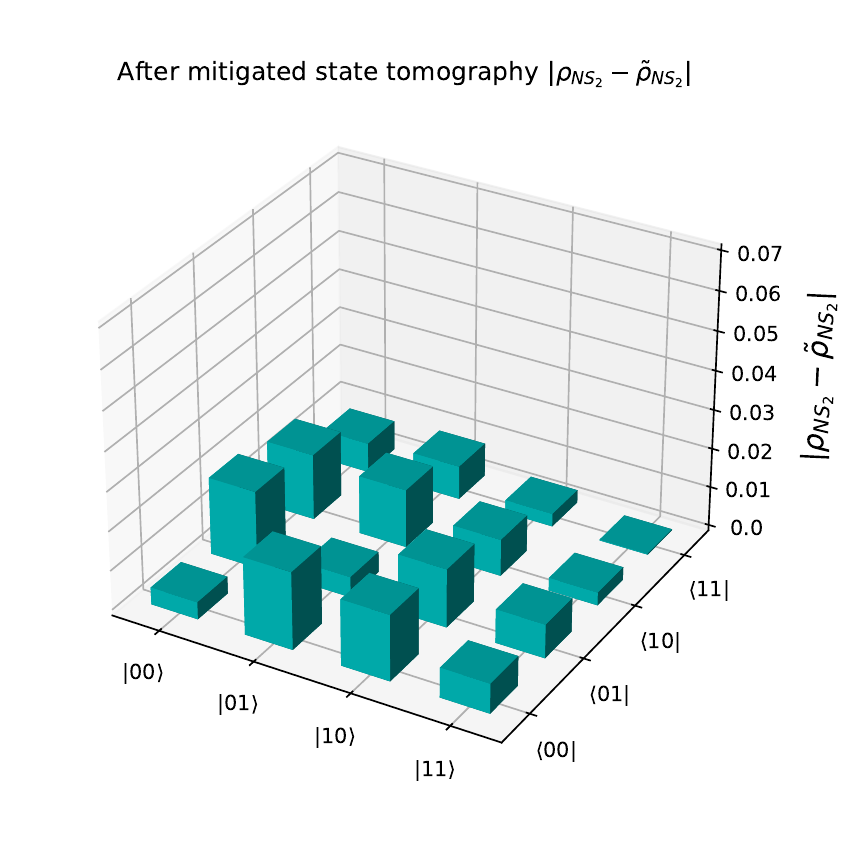}
    \caption{The city plot for the $\ket{NS_2}$ state displays the absolute difference between the components of the original $\rho_{NS_2}$ and the $\tilde\rho_{NS_2}$ obtained after performing a mitigated state tomography experiment on the real IBM quantum computer $\it{ibm\_brisbane}$ for $8192$ times.}
    \label{city_plot_NS2_miti}
\end{figure}

\begin{table}[htbp]
\centering
\resizebox{\textwidth}{!}{%
\begin{tabular}{|c|c|c|c|c|c|c|c|}
\hline
\multirow{3}{*}{\textbf{\begin{tabular}{c}Quantum\\States\end{tabular}}} &
\multirow{3}{*}{\textbf{\begin{tabular}{c}Circuit\\depth\end{tabular}}} &
\multirow{3}{*}{\textbf{\begin{tabular}{c}Circuit\\complexity\end{tabular}}} &
\multirow{3}{*}{\textbf{\begin{tabular}{c}Schmidt\\Rank\end{tabular}}} &
\multicolumn{4}{c|}{\textbf{Fidelity}} \\ \cline{5-8}

 & & & & \multicolumn{2}{c|}{\textbf{\textit{State Tomography}}} & \multicolumn{2}{c|}{\textbf{\textit{Mitigated State Tomography}}} \\ \cline{5-8}

 & & & &
\begin{tabular}{c}On simulator\\(\textit{AerSimulator})\end{tabular} &
\begin{tabular}{c}On IBM quantum\\computer\\(\textit{ibm\_brisbane})\end{tabular} &
\begin{tabular}{c}On simulator\\(\textit{AerSimulator})\end{tabular} &
\begin{tabular}{c}On IBM quantum\\computer\\(\textit{ibm\_brisbane})\end{tabular} \\ \hline

$|NS_1\rangle$ & 13 & 3 & 2 & 0.92 & $0.87 \pm 0.01$ & 0.97 & $0.99 \pm 0.01$ \\ \hline
$|NS_2\rangle$ & 13 & 3 & 2 & 0.92 & $0.88 \pm 0.01$ & 0.97 & $0.98 \pm 0.01$ \\ \hline
$|NS_3\rangle$ & 13 & 3 & 2 & 0.91 & $0.89 \pm 0.01$ & 0.97 & $0.98 \pm 0.01$ \\ \hline
$|NS_3'\rangle$ & 13 & 3 & 2 & 0.91 & $0.89 \pm 0.01$ & 0.96 & $0.98 \pm 0.01$ \\ \hline
$|NS_3''\rangle$ & 4 & 1 & 2 & 0.93 & $0.91 \pm 0.01$ & 0.99 & $0.99 \pm 0.01$ \\ \hline
Bell state ($|\phi^+\rangle$) & 2 & 1 & 2 & 0.93 & $0.90 \pm 0.01$ & 0.99 & $0.99 \pm 0.01$ \\ \hline

\end{tabular}
}
\caption{ Comparison of circuit depth, complexity, Schmidt rank, and fidelity after performing state tomography and mitigated state tomography on IBM $\textit{AerSimulator}$ and real quantum computer $\textit{ibm\_brisbane}$ of the negative quantum states and the Bell state ($\ket{\phi^{+}}$).}
\label{New_table}
\end{table}

All tomography experiments were performed using $8192$ shots per measurement basis. The statistical uncertainty in the reconstructed density matrix elements due to finite shot count is of order $1/\sqrt{8192} ≈ 0.011$.  Device properties (single-qubit gate error ($10^{-3}~–~10^{-4}$), two-qubit CZ gate error ($10^{-2}~–~10^{-3}$), readout error ($10^{-2}$), and $T1/T2$ coherence times $≈ 100–200$ microseconds) are known to fluctuate between calibration cycles (typically every 24 hours), and results may vary if the experiment is reproduced on a different day or different hardware backend. By understanding these errors, we can create a readout error mitigator to improve the accuracy of output distributions and the precision of measurable expectations. We conduct a batched experiment to characterize readout errors and then perform the state tomography. This is known as mitigated state tomography. We carry out the mitigated state tomography for the negative quantum states and the Bell state on the $\textit{ibm\_brisbane}$ quantum computer and $\textit{AerSimulator}$ with 8192 shots. The absolute difference between the components of the original $\rho_{NS_2}$ and the $\tilde\rho_{NS_2}$ obtained after performing mitigated state tomography experiments on the real IBM quantum computer $\it{ibm\_brisbane}$ is depicted in Fig.~\ref{city_plot_NS2_miti}. We observe that the individual matrix elements obtained by mitigated state tomography are closer to the actual values than those obtained by state tomography. The other negative quantum states also show similar behavior. This improvement is quantified using fidelity and shown in TABLE~\ref{New_table}. On the simulator, the fidelity is around $0.96-0.99$, while on the actual quantum computer, it is around $0.98-0.99$. Again, after mitigated state tomography, we observe that the fidelity of the Bell state is similar to that of the negative quantum states, though the circuit depth for the negative quantum states is comparatively higher.

\section{Benchmarking Negative Quantum States Against Bell states}\label{sec:results}

\subsection{Fidelity of Prepared States}
Our experiments reveal that negative states achieve fidelities in the range of $0.87–0.91$ on hardware before mitigation, and $0.98–0.99$ after mitigation. Despite the greater circuit depth required, these values are comparable to those obtained for Bell states. This demonstrates that negative quantum states can be reliably prepared with present-day hardware.

\subsection{Robustness Under (non)-Markovian Noise}
We extend our analysis to assess the robustness of these states under depolarizing noise, as well as under non-Markovian random telegraph noise (RTN) and amplitude damping (AD) noise.

\textit{Depolarizing noise.---} 
Depolarizing noise is a prevalent type of quantum noise in quantum computing, discussed in detail in chapter~\ref{chap2:Preliminaries}. We apply the depolarizing noise after constructing the Bell and two-qubit negative quantum states. To this effect, a random variable $r$ uniformly distributed in the interval $[0,1]$ is generated. Whenever the condition $r < \frac{p}{3}$ is satisfied, a Pauli gate ($X, Y$, or $Z$) is randomly selected and applied to both qubits of the state to simulate the occurrence of an error. The Shor's error correction scheme~\citep{Shor_correction} is subsequently implemented to mitigate these induced errors on both the qubits of the two-qubit state, whereas, during error correction, the depolarizing error is randomly applied on one of the qubits, including the ancillary qubits, restoring the integrity of the quantum information.

To observe the effect of depolarizing noise on the generation of the two-qubit negative quantum states and the Bell state, variation of (1 - $F$) ($F$ is the fidelity between the original state and the state gone through the above process) of the $\ket{NS_1}$, $\ket{NS_2}$, $\ket{NS_3}$, and the $\ket{\phi^{+}}$ Bell state with error probability $p$ is calculated. 
\begin{figure}[H]
    \centering
    \includegraphics[width=1\columnwidth]{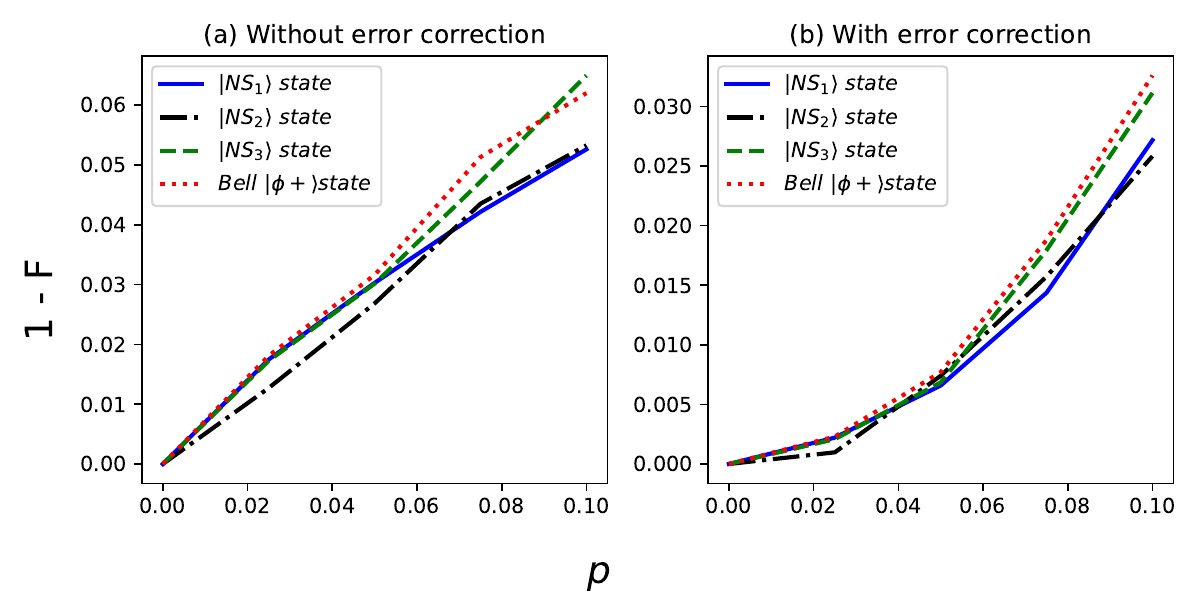}
    \addtocounter{figure}{-1}
    \caption{Variation of ($1 - F$) for $\ket{NS_1}$, $\ket{NS_2}$, $\ket{NS_3}$, and $\ket{\phi^{+}}$ Bell state with depolarizing error probability $p$ (a) without any error correction and (b) after implementing Shor’s error correction.}
    \label{DP_NS1_NS2_NS3}
\end{figure}
Figure~\ref{DP_NS1_NS2_NS3}(a) depicts that the fidelity of all the states decreases sharply without performing any error correction. Conversely, Fig.~\ref{DP_NS1_NS2_NS3}(b) illustrates that after implementing Shor's error correction, all the states remain unaffected by depolarizing noise up to an error probability $0.012$ while for the $\ket{NS_2}$ state this value is $0.025$. It shows that the $\ket{NS_2}$ state is robust among all. Furthermore, the behavior of $\ket{NS_3}$ and the Bell state is analogous, but the $\ket{NS_1}$ state dominates over both with increased $p$.

\textit{Non-Markovian noise.---} 
Non-Markovian noise refers to noise in quantum systems where the system retains the memory of its past states, unlike Markovian noise, which assumes memory-less interactions~\citep{Breuer2007}. In a non-Markovian environment, information lost to the surroundings can flow back into the system, leading to correlations between the current and previous states. Understanding non-Markovian noise is essential for accurately simulating and mitigating errors in quantum computing. Here, we examine the effect of non-Markovian random telegraph noise (RTN) and amplitude damping (AD) noise on the generation of two-qubit negative quantum states and the $\ket{\phi^{+}}$ Bell state. The evolution of a two-qubit system having local interactions with the non-Markovian noisy channels is given by ${\rho}_{AB}(t) = \sum_{i = 0}^{1}\sum_{j = 0}^{1} ({K}_i \otimes {K}_j) {\rho}_{AB}(0) ({K}_i \otimes {K}_j)^{\dag}$, where ${K}_i$'s, and ${K}_j$'s are the Kraus operators of the non-Markovian RTN and AD noise as stated in Sec.~\ref{ch2_noise_models} of chapter~\ref{chap2:Preliminaries}.

\begin{figure}[H]
    \centering
    \includegraphics[width=1\columnwidth]{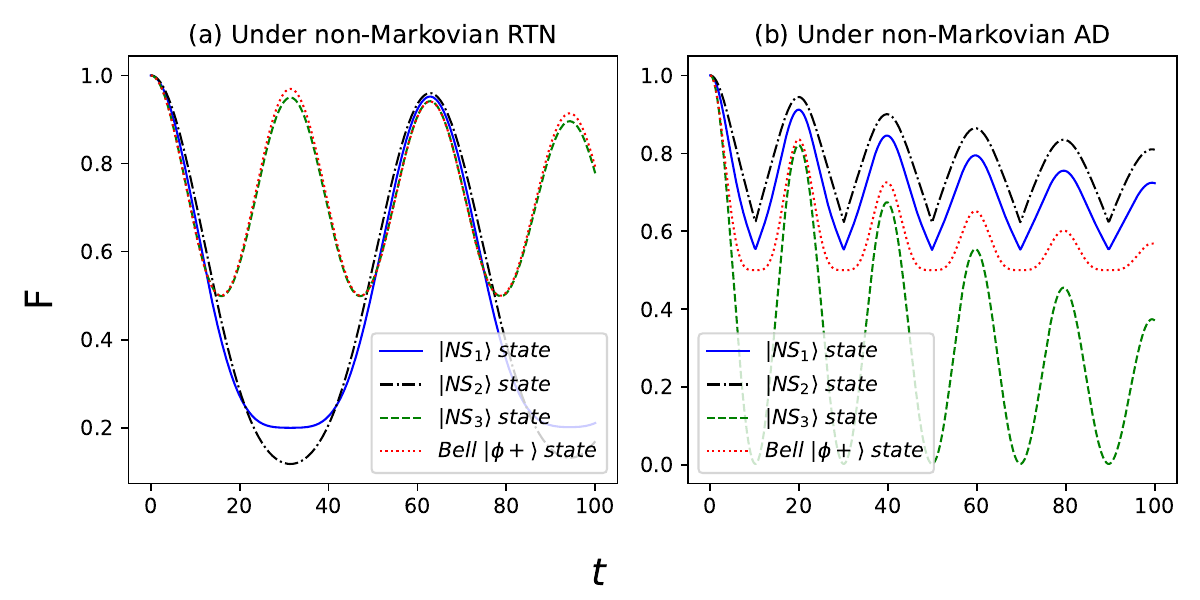}
    \caption{Variation of fidelity of $\ket{NS_1}$, $\ket{NS_2}$, $\ket{NS_3}$ and $\ket{\phi^{+}}$ Bell state with time (a) under non-Markovian RTN for $b = 0.05$ and $\gamma^{RTN} = 0.001$, (b) under non-Markovian AD for $g = 0.01$ and $\gamma^{AD} = 5$.}
    \label{Fidelity_NMRTN_NMAD}
\end{figure}
The variation of fidelity $F$, between the original state and the state after applying the non-Markovian noise, of two-qubit negative quantum states and the Bell state under non-Markovian RTN and AD noise is illustrated in Fig.~\ref{Fidelity_NMRTN_NMAD}. We can observe that, in the presence of non-Markovian RTN noise, the $\ket{NS_3}$ state and the Bell state's fidelity behave equivalently, while under non-Markovian AD noise, the $\ket{NS_2}$ state is the most resilient among all the other considered states.

As part of exploring the applications of two-qubit negative quantum states, we study their role in enhancing parameter estimation sensitivity through quantum Fisher information. Also, we investigate the optimal CHSH inequality violation and present a teleportation circuit based on $\ket{NS_3^{\prime\prime}}$ state, further demonstrating their potential in quantum information processing and communication.

\subsection{Quantum Fisher Information: Preserving Metrological Usefulness \label{Q_F_I}}
Quantum Fisher information (QFI) is the central quantity in quantum metrology~\citep{Bollinger1996Optimalfrequency,peters1999measurement}. It quantifies the ultimate precision limit with which a parameter can be estimated and connected to the lower bound on the variance of an unbiased parameter via the quantum Cram\'er-Rao bound~\citep{helstrom1969quantum}.

For a density matrix $\rho(\phi)$, with a parameter $\phi$ acquired by an $SU(2)$ rotation, i.e., $\rho_{\phi} = U_{\phi} \rho U_{\phi}^{\dag}$, where $U_{\phi} = e^{i \phi J_{\vec{n}}}$ with 
\begin{equation}
    J_{\vec{n}} = \sum_{\alpha = x,y,z}\frac{1}{2}n_{\alpha}\sigma_{\alpha},
\end{equation}
being the angular momentum operator in the $\vec{n}$ direction, and $\sigma_{\alpha}$ are the Pauli matrices. The QFI, for an unbiased estimator $\langle \hat{\phi} \rangle = \phi$, is defined as~\citep{ma2011quantum}
\begin{equation}
    F = Tr[\rho(\phi) L_{\phi}^2],
    \label{QFI}
\end{equation}
where $L_{\phi}$ is the symmetric logarithmic derivative determined by the following equation,
\begin{equation}
    \frac{\partial}{\partial \phi}\rho(\phi) = \frac{1}{2}[\rho(\phi)L_{\phi} + L_{\phi}\rho(\phi)].
    \label{derivative_L_phi}
\end{equation}
Now, using Eq. (\ref{derivative_L_phi}), explicit form of $L_{\phi}$ can be derived and the QFI for a density matrix $\rho$ is thus given as~\citep{ma2011quantum},

\begin{equation}
    F(\rho, J_{\vec{n}}) = \sum_{i \neq j} \frac{2(p_i - p_j)^2}{p_i + p_j}|\langle i|J_{\vec{n}}|j\rangle|^2 = \vec{n}~\mathcal{C}~\vec{n}^T,
\end{equation}
where $\vec{n}$ is a normalized three-dimensional vector. Further, \{$p_i$, $p_j$\} and \{$\ket{i}$, $\ket{j}$\} are the eigenvalues and eigenvectors of $\rho$, respectively. Further, the matrix elements of the symmetric matrix $\mathcal{C}$ can be obtained as 
\begin{equation}
    \mathcal{C}_{kl} = \sum_{i \neq j}\frac{(p_i - p_j)^2}{p_i + p_j}[\langle i|J_k|j\rangle \langle j|J_l|i\rangle + \langle i|J_l|j\rangle \langle j|J_k|i\rangle].
\end{equation}

\begin{figure}[H]
    \centering
    \includegraphics[height=75mm,width=0.75\columnwidth]{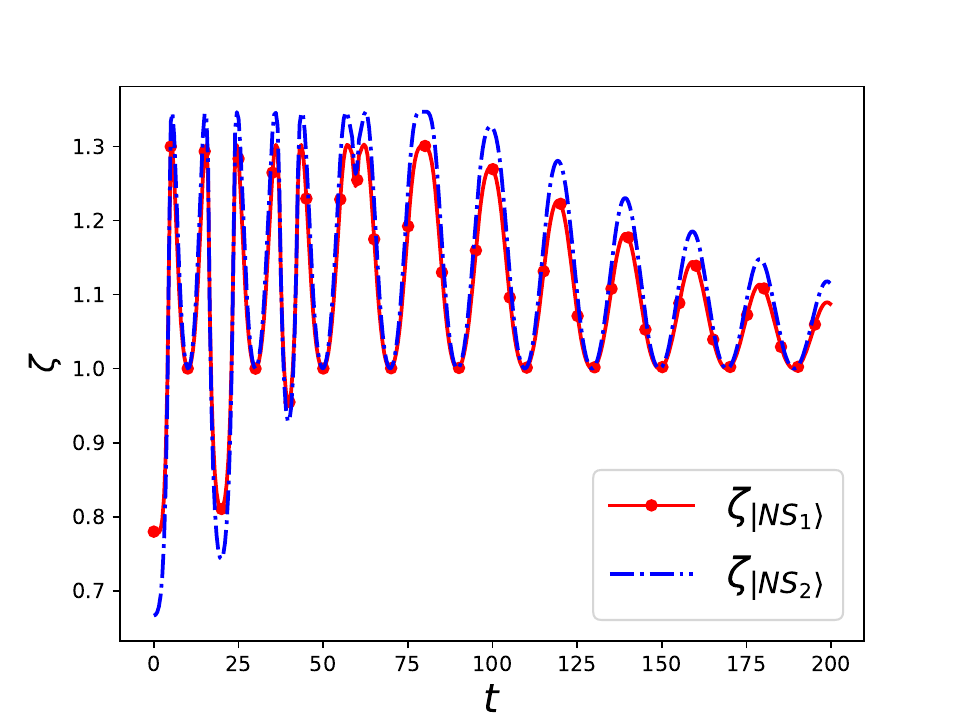}
    \caption{Variation of $\zeta_{\ket{NS_1}}$ and $\zeta_{\ket{NS_2}}$ with respect to time under non-Markovian AD for $g = 0.01$ and $\gamma^{AD} = 5$.}
    \label{NS1_NS2_fisher_info_NMAD}
\end{figure}

The QFI, $F(\rho, J_{\vec{n}})$, measures the sensitivity of the state concerning the rotations along the $\vec{n}$ direction. Additionally, the maximal mean QFI is defined as follows~\citep{hyllus2012fisher},
\begin{equation}
    {\bar{F}}_{max} = \frac{1}{N} \underset{\vec{n}}{\max} F(\rho, J_{\vec{n}}) = \frac{\lambda_{max}}{N},
\end{equation}
where $\lambda_{max}$ is the largest eigenvalue of the symmetric matrix $\mathcal{C}$ and $N$ is the number of two-level particles. The ${\bar{F}}_{max}$ characterizes the phase sensitivity, i.e., the sensitivity of a state with respect to $SU(2)$ rotations, and is independent of the $\vec{n}$. Now, we calculate the ${\bar{F}}_{max}$ of two-qubit negative quantum states and the Bell state in the presence of the non-Markovian AD noise to study the effect of decoherence on their phase sensitivity. Figure \ref{NS1_NS2_fisher_info_NMAD} shows the variation of the ratio of maximal mean QFI for $\ket{NS_i}$ state to that of $\ket{\phi^{+}}$ state, denoted by $\zeta_{\ket{NS_i}}$, under non-Markovian AD noise. In Fig. \ref{NS1_NS2_fisher_info_NMAD}, wherever $\zeta_{\ket{NS_i}} > 1$, we can observe that ${\bar{F}}_{max}$ for $\ket{NS_i}$ state is greater than ${\bar{F}}_{max}$ for $\ket{\phi^{+}}$ state. Figure \ref{NS1_NS2_fisher_info_NMAD} shows that  $\ket{NS_1}$ and $\ket{NS_2}$ states maintain a higher ${\bar{F}}_{max}$ for longer duration in comparison to the Bell $\ket{\phi^{+}}$ state making them better suited for realistic quantum metrology applications under noise. Because the higher maximal mean QFI leads to better phase estimation via the Cram\'er-Rao bound. 
\subsection{CHSH Inequality Violation: Nonlocality Under Noise} \label{sec:Optimal_CHSH_inequality_violation}
\begin{figure}[H]
    \centering
    \includegraphics[height=75mm,width=0.75\columnwidth]{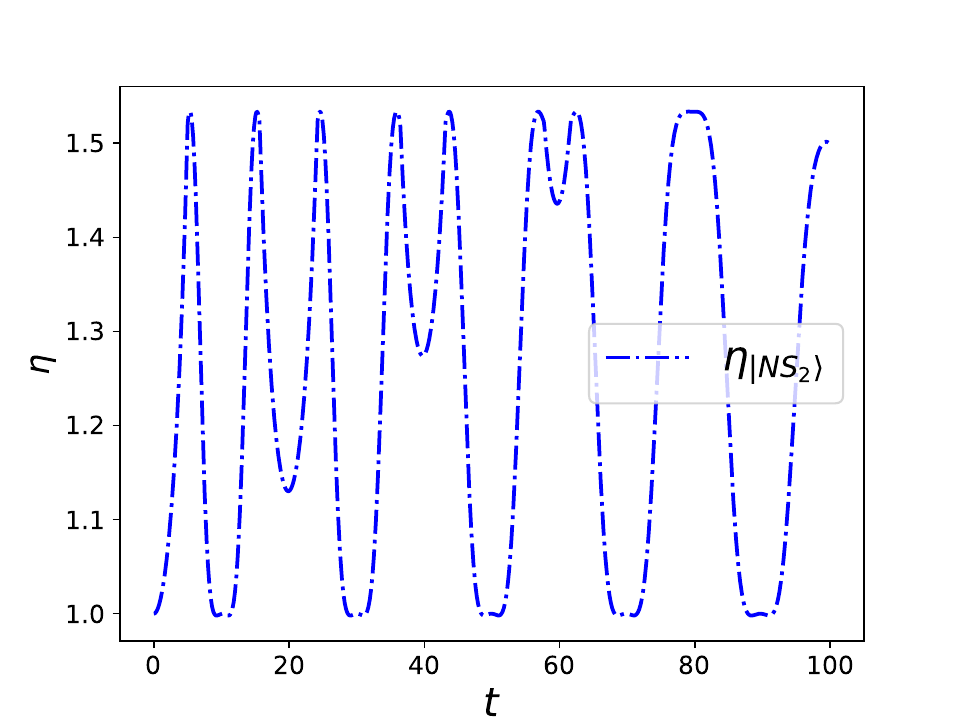}
    \caption{The temporal variation of $\eta_{\ket{NS_2}}$, specifically the ratio $S_{max \ket{NS_2}}/S_{max \ket{\phi^{+}}}$ under non-Markovian AD channel with WM and QMR for $\ket{NS_2}$ ($p = 0.05$, $q = 0.74$), and for Bell $\ket{\phi^{+}}$ state ($p = 0.05$, $q = 0.05$). The non-Markovian AD channel parameters are $g = 0.01$ and $\gamma^{AD} = 5$.}
    \label{NS2_CHSH_maximal_violation_NMAD_with_WM}
\end{figure}

The optimal violation of the Clauser-Horne-Shimony-Holt (CHSH) inequality constitutes a fundamental benchmark for detecting the quantum nonlocality of a general two-qubit state. Given a two-qubit density matrix $\rho$, the maximal CHSH violation is determined through the spectral properties of the matrix $T_{\rho}^{T}T_{\rho}$ as follows~\citep{HORODECKI1995340}, 
\begin{equation}
    S_{max} = 2\sqrt{\lambda_1 + \lambda_2},
    \label{S_max_eq.}
\end{equation}
where $\lambda_1$ and $\lambda_2$ are the largest two eigenvalues of the $T_{\rho}^{T}T_{\rho}$ matrix. The elements of $3 \times 3$ correlation matrix $T_{\rho}$ are defined as $t_{ij} = {\rm Tr}[\rho. (\sigma_i \otimes \sigma_j)]$, where $\sigma_1$, $\sigma_2$, and $\sigma_3$ are the Pauli matrices. We study the optimal CHSH inequality violation of two-qubit negative quantum states and all Bell states using weak measurement (WM) and quantum measurement reversal (QMR), as detailed in Appendix~\ref{WM_QMR}, in the presence of non-Markovian AD noise. Figure~\ref{NS2_CHSH_maximal_violation_NMAD_with_WM} depicts the variation of the ratio of $S_{max}$ for $\ket{NS_i}$ state to the $S_{max}$ for $\ket{\phi^{+}}$ state with WM and QMR, denoted by $\eta_{\ket{NS_i}}$, particularly for the state which perform best among all, i.e., $\ket{NS_2}$ state. We observe that the $\ket{NS_2}$ state significantly outperforms the Bell $\ket{\phi^{+}}$ state in preserving CHSH violation over time. The $\ket{NS_2}$ state shows more frequent revivals above the classical limit $2$, which means it retains non-locality better, while the Bell $\ket{\phi^{+}}$ state often stays below $2$ under non-Markovian AD noise.
Further, a detailed methodology and comprehensive analysis for identifying optimal states, among the Bell states and two-qubit negative quantum states, that facilitate universal quantum teleportation are provided in Appendix~\ref{WM_QMR}.
\subsection{Quantum teleportation using Negative Quantum States} \label{sec:T_circuit}
\begin{figure}[H]
    \centering
    \includegraphics[height=55mm,width=0.95\textwidth]{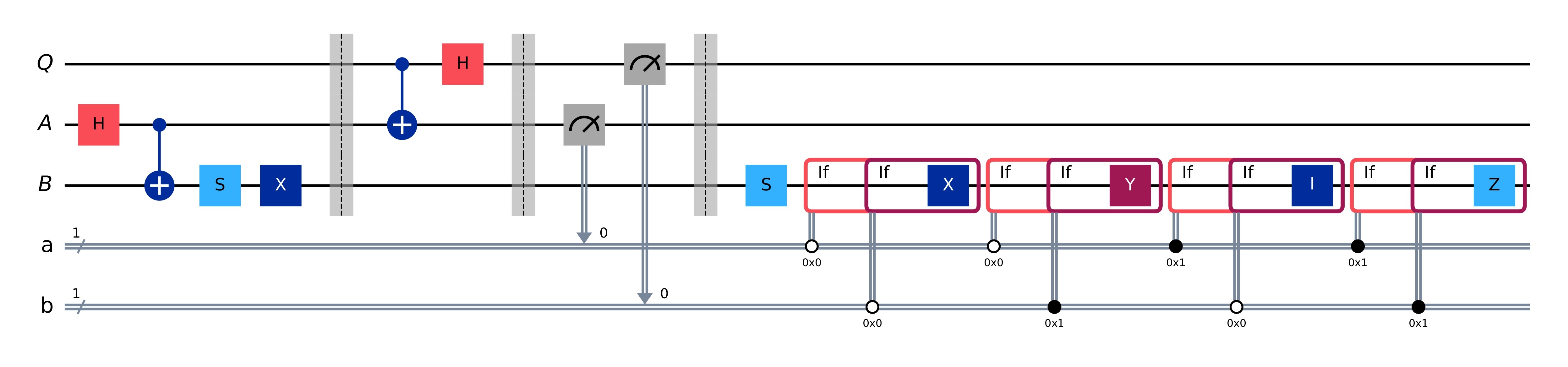}
    \caption{Circuit for implementing quantum teleportation scheme using $\ket{NS_3^{\prime\prime}}$ as an entangled resource. The quantum gates in the blocks annotated with ``if" are applied conditioned on the values of the classical bits corresponding to the measurement outcomes.}
    \label{Teleportation_NS3_double_prime}
\end{figure} 

\begin{table}[h!]
\centering
\begin{tabular}{|c|c|c|}
\hline
ab & Operation on B & Final State of B \\
\hline
00 & $SX$ & $\alpha|0\rangle + \beta|1\rangle$ \\
\hline
01 & $SY$ & $\alpha|0\rangle + \beta|1\rangle$ \\
\hline
10 & $SI$ & $\alpha|0\rangle + \beta|1\rangle$ \\
\hline
11 & $SZ$ & $\alpha|0\rangle + \beta|1\rangle$ \\
\hline
\end{tabular}
\caption{Alice transmits the classical information (ab) to Bob through a classical communication channel. Based on the received classical bits, Bob performs specific unitary operations on his qubit (denoted as operation on $B$). Following the application of these operations, the resulting final state of Bob's qubit (Final state of $B$) corresponds precisely to the unknown quantum information originally intended for teleportation.}
\label{Teleportation_table}
\end{table}

Quantum teleportation is a method used to transfer quantum information from one end to another, even when no direct quantum communications channel connects them~\citep{Teleporting1993}. To accomplish this, the sender, referred to as ``Alice," and the receiver, referred to as ``Bob," must make prior arrangements for a shared pair of Einstein-Podolsky-Rosen (EPR) correlated particles. Alice performs the Bell measurement on her EPR particle and the unknown quantum state. She then transmits the classical outcome of this measurement to Bob. With this knowledge, Bob can transform the state of his EPR particle into an identical copy of the unknown quantum state~\citep{nielsen2010quantum}. The teleportation protocol using the Bell state as an EPR pair has been extensively studied. Since the two-qubit $\ket{NS_3^{\prime\prime}}$ state also has the maximum concurrence, i.e., 1, we now introduce the teleportation protocol using this state as a shared EPR pair. The corresponding quantum teleportation circuit is shown in Fig~\ref{Teleportation_NS3_double_prime}. ``A" (Alice) denotes the sender and ``B" (Bob) the receiver, who share the entangled $\ket{NS3^{\prime\prime}}$ state in Fig.~\ref{Teleportation_NS3_double_prime}. ``Q" represents the unknown qubit to be teleported from Alice to Bob. Following a Bell-state measurement, Alice transmits two classical bits, ``a" and ``b," to Bob. The corrective operations that Bob must apply, determined by the values of ``a" and ``b," are specified by the conditional statements and summarized in Table~II. After applying the appropriate operations, Bob's resulting state for an initial unknown state $\alpha\ket{0} + \beta\ket{1}$ is also listed in Table~II. The teleportation circuit illustrated in Fig.~\ref{Teleportation_NS3_double_prime} is verified on the Qiskit~\citep{qiskit2024} by successfully teleporting random quantum information from Alice to Bob. Moreover, the teleportation protocol can be executed utilizing an alternative $\ket{NS_3}$ and $\ket{NS_3^{\prime}}$ states, as they possess entanglement at the maximal likelihood of the Bell state. Nonetheless, in this instance, identifying the unitary transformations that Bob must implement is challenging.

\section{Summary}\label{sec:conclusion}
In this work, for the first time, we proposed methods for preparing stable entangled states, specifically focusing on two-qubit negative quantum states, bridging theoretical predictions with hardware realization. 
By constructing unitary transformations, designing circuits with native gates, and validating them on IBM hardware via. ideal and error-mitigated quantum state tomography, we established the feasibility of preparing two-qubit negative quantum states. We achieved fidelities equivalent to those of an ideal Bell state. The resilience of these states against (non)-Markovian noise was demonstrated through detailed analyses of fidelity estimation, maximal mean quantum Fisher information, optimal CHSH inequality violation, and performance in universal quantum teleportation, both with and without the application of weak measurement (WM) and quantum measurement reversal (QMR). These findings emphasize the practical relevance of negative quantum states for quantum sensing and metrology, where maintaining precision over extended timescales is critical. Furthermore, the application of WM and QMR techniques effectively extends the coherence time of these states, enhancing their viability in quantum computing, quantum key distribution, quantum teleportation, and superdense coding areas where the preservation of quantum correlations is critical. Future directions include the development of fault-tolerant circuits for the realization of the two-qubit negative states and the generalization of these states to multiqubit architectures to design resilient quantum memories.

\subsection{Scalability Limitations}
The experimental demonstrations in this chapter are limited to two-qubit systems. Extending the negative quantum state framework to multi-qubit systems faces several practical challenges. First, circuit depth scales with system size: the native-gate circuits for two-qubit negative states already require multiple single-qubit rotations and a $CZ$ gate (Fig.~\ref{NS1_NS2_NS3__NS3_prime_circuit}); circuits for three- or four-qubit negative states would have significantly greater depth, approaching or exceeding the coherence times of current NISQ devices. Second, quantum state tomography, used here to verify state preparation, requires $O(4^n)$ measurements for an $n$-qubit system; for $n = 3 or 4$, this becomes experimentally prohibitive without compressed sensing or shadow tomography techniques. Third, error mitigation (zero-noise extrapolation and measurement error mitigation used here) scales in cost with system size and circuit depth. Fourth, the definition of negative quantum states as eigenstates of phase-space point operators for dimensions $d = p^m$ is algebraically well-defined for any prime power. However, the explicit construction of these operators and their eigenstates for $d = 8$ (three qubits) or $d = 16$ (four qubits) is technically involved and has not been worked out in the present thesis. These scalability challenges motivate future work on fault-tolerant circuit decompositions, scalable tomography methods, and the extension of the phase-space point operator construction to higher-dimensional systems.


\newpage
\setcounter{chapter}{5} 

\titleformat{\chapter}[display]
{\sffamily\fontsize{27}{27}\bfseries\filleft}{\thechapter}{0pt}{{#1}}  
  
\thispagestyle{empty}

\chapter{A two-qubit quantum collision model: non-Markovianity and non-classicality}\label{chap6:two_qubit_collision_model}

\section{Introduction}
In this study, we examine such a two-qubit collision model by simulating a sequence of ``collisions" between the two-qubit system and environmental ancillae to describe the system-environment interaction in a controllable manner. Depending on how the environmental degrees of freedom interact, such a ``collision" model can simulate both Markovian and non-Markovian dynamics~\citep{ThermalizingQuantumMachines_2002, ziman2005description, Rybár_2012, McCloskey2014Non-Markovianity, Ciccarello2013Collision-model, Kretschmer2016Collisionmodel, Saha_2024_quantum, Li2024Witnessing}. We also introduce two different schemes to investigate the effects of manipulating the interactions between environmental ancillae and the two-qubit system on the non-Markovianity and non-classicality of the two-qubit system. For that, we study trace distance, Wigner function, non-classical volume, and dynamics of entanglement generation of the two-qubit system using both schemes. The trace distance is diagnostic for non-Markovianity witnessed by information flow, while the Wigner function and non-classical volume reveal phase-space signatures of quantum behavior and deviations from classicality~\citep{Wigner1932Quantum, zavatta2004quantum, Thapliyal2015Quasiprobability}. Moreover, entanglement generation with the number of collisions allows the identification of regimes where quantum correlations are enhanced or degraded by the environment~\citep{Naikoo2019Facets, Tiwari2023QuantumCorrelations}. We also investigate the steady state behavior under varying physical conditions for possible thermodynamic applications~\citep{Hanggi_talkner_review, tiwari2024strong, Thomas2018thermodynamics, Ashutosh2023thermodynamics}.

The chapter organization is as follows. In Sec.~\ref{Collision model}, we introduce the two-qubit collision model and two different interaction schemes, \textit{viz}. scheme A and scheme B. Section~\ref{Non_Markovianity_measure} investigates the non-Markovian nature of the model through a non-Markovianity witness, i.e., trace distance, for both the schemes. It is followed by examining both schemes' non-classical features in Sec.~\ref{Wigner_function} and~\ref{Quantum_correlation}. In Sec.~\ref {Wigner_function}, we analyze the Wigner function and the associated non-classical volume. Section~\ref {Quantum_correlation} explores the emergence of quantum correlations, particularly quantum entanglement, using both schemes for initially separable two-qubit states. Additionally, the steady-state behavior of the system qubits is analyzed in Sec.~\ref{SS_behaviour} for scheme B. Finally, our conclusions are summarized in Sec.~\ref{ch6_conclusion}. \textit{This chapter's information is derived from~\citep{lalita2025non_classicality}. \copyright APS. Adapted and reproduced with permission.}

\section{\label{Collision model} Collision model}
The single-qubit collision model is a powerful framework for studying how a quantum system interacts with its environment. In this model, the system qubit interacts one at a time with a sequence of identical environmental qubits (often called ancillae) through unitary operations~\citep{Ciccarello2022Quantumcollisionmodel, McCloskey2014Non-Markovianity, Ciccarello2013Collision-model, Campbell2018Systemenvironment, Campbell2021Collision, csenyacsa2022entropy}. When the system interacts with individual environmental qubits and then moves on to interact with fresh ones while discarding the previous ones, its behavior follows a Markovian process. However, memory effects can arise if an interaction is introduced between consecutive environmental qubits, especially between the one that just interacted with the system and the next one. This allows some of the system's lost information to be partially retrieved later in the process. These memory effects play a key role in the system's dynamics, which we will explore further. The system's evolution over time is determined by repeatedly following these interaction steps while tracking its state. The flexibility of collision models is particularly valuable, as they allow adjustments to the number of qubits and the nature of interactions, making them adaptable to different problems~\citep{Cattaneo2021Collision, Cattaneo2022ABriefJourney}. 

Here, we consider a collision model for a two-qubit system $S$, composed of qubits $s_1$ and $s_2$, interacting with a stream of identical ancillae one at a time. The system-ancilla interaction is modeled using two schemes,~\textit{viz.} scheme A and scheme B, as depicted in Figs.~\ref{approach_1} and \ref{Scheme_B}, respectively. In scheme A, a stream of ancillae denoted as $a_n^{R}$ interacts with one of the system qubits, say $s_2$, whereas in scheme B, two independent streams of ancillae denoted by $a_n^{L}$ and $a_n^{R}$ interact with the qubits $s_1$, and $s_2$ of the system, respectively. Throughout the paper, the ancillae are considered to be initially in the thermal state 
\begin{eqnarray}\label{eq_ancilla_thermal_state}
    \rho_{0}^{a_n^{e}} = e^{-\beta_{a_n^{e}}H_{a_n^{e}}}/\Tr[e^{-\beta_{a_n^{e}}H_{a_n^{e}}}], ~~~~~\forall ~e \in \{L, R\},
\end{eqnarray}
where $(L,~R)$ denotes the left ($L$) and right ($R$) sides of the two-qubit system, and $\beta_{a_n^{e}} = 1/k_BT_{a_n^{e}}$ are the inverse temperatures of the environment qubits, and we set $\hbar = k_B = 1$. Since ancillae are freshly and independently initialized before interacting with the system, their initial states are well described by single-qubit Gibbs states, and distinct temperatures may be consistently assigned to different ancilla streams~\citep{Landi2021Irreversibleentropy}.

The Hamiltonians that describe the system qubits and environment ancillae are defined as
\begin{eqnarray}
\nonumber
    H_{s_1} = \hbar\omega_{s_1}\sigma_z,
    ~H_{s_2} = \hbar\omega_{s_2}\sigma_z,\\
    H_{a_n^{L}} = \hbar\omega_{a_n^{L}}\sigma_z,
    ~H_{a_n^{R}} = \hbar\omega_{a_n^{R}}\sigma_z.
\end{eqnarray}
Here, $\omega_{s_1}$, $\omega_{s_2}$, $\omega_{a_n^{L}}$, and $\omega_{a_n^{R}}$ are the corresponding transition frequencies of the system and environment qubits, and $\sigma_z$ is the Pauli $Z$ matrix. Throughout the paper, except stated otherwise, we assume that the system qubits and ancillae are in resonance, $\it{i.e.}$, $\omega_{s_1} = \omega_{s_2} = \omega_{a_n^{L}} = \omega_{a_n^{R}} = 1$. The intra-system and system-ancilla interactions of the two-qubit system are governed by the Heisenberg interaction as follows
\begin{eqnarray}
    H_{s_1s_2} &=& g_{s_1s_2}(\sigma_x^{s_1}\sigma_x^{s_2} + \sigma_y^{s_1}\sigma_y^{s_2}),\nonumber \\
    H_{s_ja_n^{e}} &=& g_{s_ja_n^{e}}(\sigma_x^{s_j}\sigma_x^{a_n^{e}} + \sigma_y^{s_j}\sigma_y^{a_n^{e}}), ~~\forall~ j \in \{1, 2\},
    \label{H_int_unitary}
\end{eqnarray}
where $\sigma_{x,y}$ are the Pauli matrices and $g_{s_1s_2}$, $g_{s_ja_n^{e}}$ are the intra-system and system-ancilla coupling constants, respectively, and $e$ can be $L$ or $R$. Unless specified otherwise, we set $g_{s_1s_2} = 0.95$ and $g_{s_{ja_n^{e}}} = 0.85$ (and in one instance $0.5$), which are of the order of the transition frequencies of the system and ancillae qubits, taken here to be one, indicating a strong intra-system and system-ancilla coupling. Additionally, the Hamiltonian for the swap operation that governs the intra-ancilla interactions is
\begin{equation}
    H_{SWAP} = \frac{1}{2}(\Vec{\sigma}{^{a_n^{e}}}.~\Vec{\sigma}{^{a_{n+1}^{e}}} + I_4),
\end{equation}
where $\Vec{\sigma}{^{a_n^{e}}}$ and $\Vec{\sigma}{^{a_{n+1}^{e}}}$ are the Pauli matrices of the two consecutive ancillae that are swapped, and $I_4$ is the $4\times 4$ identity operator. The corresponding unitary operator is
\begin{equation}
    \begin{aligned}
        U_{a_n^{e}, a_{n+1}^{e}} = e^{-i \Theta H_{SWAP}} = \cos(\Theta) I - i\sin(\Theta)H_{SWAP},   
    \end{aligned}
    \label{p_swap_unitary}
\end{equation}
where $e \in \{L,~R\}$, and $\Theta \in [0, \frac{\pi}{2}]$ is the intra-ancilla interaction strength that can be used to alter from Markovian to non-Markovian behavior. In other words, it can introduce and regulate memory effects~\citep{McCloskey2014Non-Markovianity, ThermalizingQuantumMachines_2002}. 

Further, the time evolution of the two-qubit system is governed by the consecutive application of three unitaries, which include the intra-system, system-ancilla, and intra-ancilla interactions, where the latter's presence is necessary for introducing non-Markovianity into the dynamics~\citep{Kretschmer2016Collisionmodel, CM_nm_2017, Campbell2018Systemenvironment}. In both schemes, we keep the evolved joint state of the system and ancilla $\sigma^{Sa^e}_{n + 1}$, see for example Eq.~\eqref{joint_state} below, untouched and use it, as it is, in the next iteration, summarized in detail below.  
\subsection{Scheme A}
\begin{figure}
    \centering
    \includegraphics[height=75mm,width=0.75\linewidth]{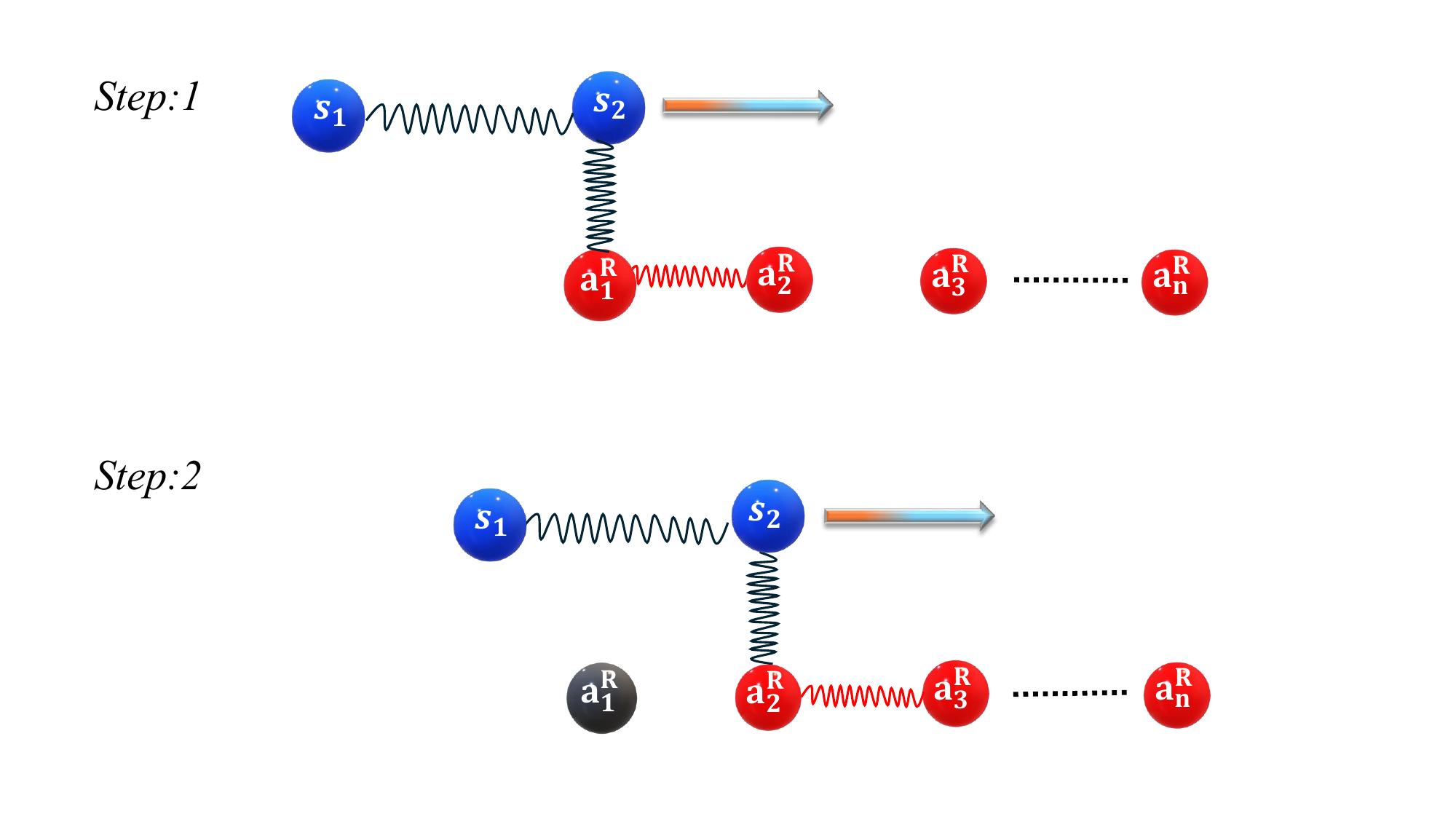}
    \caption{Schematic illustration of Scheme A, where only one system qubit $s_2$ interacts with a sequence of right-ancilla qubits $a_n^R$. In this diagram, the black lines represent Heisenberg-type interactions both between the two qubits of the system and between the system qubit $s_2$ and the ancillae. The red line indicates a partial-swap interaction occurring between successive ancilla qubits.}
    \label{approach_1}
\end{figure}

In this scheme, only the qubit $s_2$ of the two-qubit system $S$ interacts with the ancilla denoted by $a^R_n$ as shown in Fig.~\ref{approach_1}. The corresponding unitary concerning the intra-system and system-ancilla interaction is
\begin{equation}
    U_{s_1s_2, s_2a^R_n} = e^{-i H_A \Delta t},
\end{equation}
where $H_A = H_{s_1} + H_{s_2} + H_{a_n^{R}} + H_{s_1s_2} + H_{s_2a^R_n}$ with $\Delta t$ being the collision time step. The collision time steps used throughout the paper are chosen to be $\Delta t=0.08, 0.1$, or $0.5$. These values are smaller than the characteristic transition frequencies of both the system qubits and the ancillae (set to one), ensuring that the interaction time remains finite and that the discrete-time collisions faithfully approximate the underlying continuous dynamics. Further, the intra-ancilla interaction unitary, $U_{a^R_na^R_{n+1}}$, is given by Eq.~\eqref{p_swap_unitary}. Now, the system's reduced state after the $n^{th}$ step of the scheme A is given by
\begin{equation}
    \rho_{n+1}^S = \Tr_{a^R_na^R_{n+1}}[\sigma^{Sa^R}_{n + 1}],
\end{equation}
where the joint state $\sigma^{Sa^R}_{n + 1}$ of the two-qubit system $S$ and the environmental ancillae $a_n^R$ after evolving through unitaries $U_{s_1s_2, s_2a^R_n}$ and $U_{a^R_na^R_{n+1}}$ is given by
\begin{equation}
    \sigma^{Sa^R}_{n + 1} = U_{a^R_na^R_{n+1}} U_{s_1s_2, s_2a^R_n}(\rho_n^{Sa_n^R} \otimes \rho^{a_{n + 1}^R})U_{s_1s_2, s_2a^R_n}^{\dag} U_{a^R_na^R_{n+1}}^{\dag}.
    \label{joint_state}
\end{equation}
For $n=1$, the initial joint state is $\rho^{Sa^R_1}_1 = \rho^S_0 \otimes \rho^{a^R_1}$. Additionally, $\rho_n^{Sa_n^R} = \Tr_{a^R_{n-1}}[\sigma^{Sa^R}_{n}]$ ensures that correlations are established between $S$ and $a_n^R$ on the $(n - 1)^{th}$ step, i.e., prior to their direct interaction on the $n^\text{th}$ collision. The simulation steps for scheme A are as follows,
\begin{enumerate}
    \item \textit{Initialization:} Prepare the initial system state $\rho^S_0$ and each ancillae qubit in a thermal Gibbs state $\rho_{0}^{a_n^{R}} = e^{-\beta_{a_n^{R}}H_{a_n^{R}}}/\Tr [e^{-\beta_{a_n^{R}}H_{a_n^{R}}}]$.

    \item \textit{State construction:} Construct the initial system–ancilla block used in collisions as
    \[
        \rho^{Sa^R_1}_1 = \rho^S_0 \otimes \rho^{a^R_1}.
    \]

    \item \textit{Unitary evolution per collision:} At every iteration, apply the unitary ($U_{s_1s_2, s_2a^R_n}$) concerning the intra-system and system-ancilla interactions and the ancilla–ancilla partial-swap unitary ($U_{a^R_na^R_{n+1}}$). The updated overall state is given by Eq.~\eqref{joint_state}.

    \item \textit{State update:} Obtain the updated system state $\rho_{n+1}^S$ by tracing out the ancillas and use it for further investigations, and obtain the next step’s system–ancilla block $\rho_n^{Sa_n^R}$ by tracing out the outgoing ancillas.
\end{enumerate}

\subsection{Scheme B}
\begin{figure}
    \centering
    \includegraphics[height=75mm,width=0.75\linewidth]{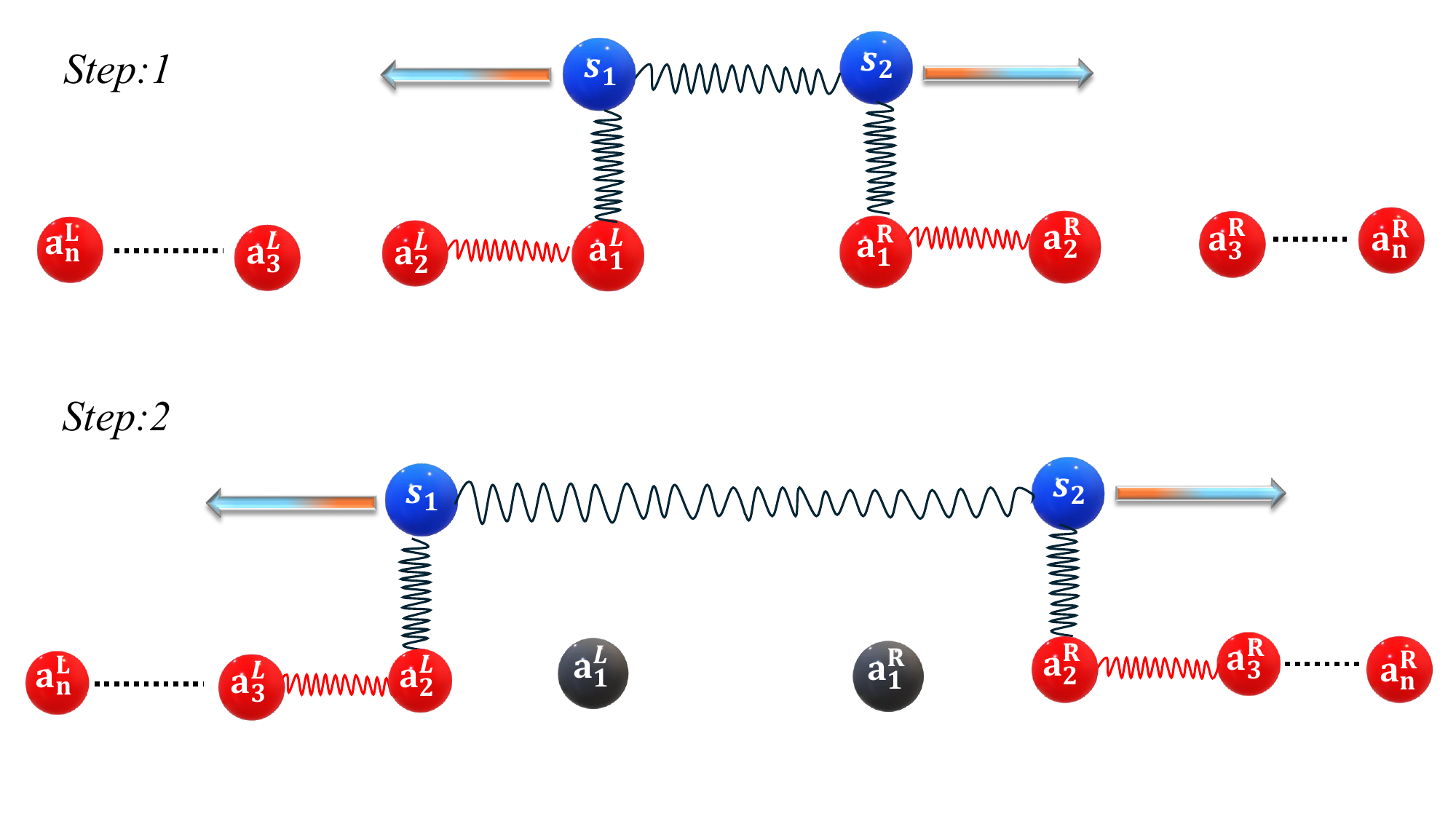}
    \caption{Schematic representation of Scheme B, where system qubits $s_1$ and $s_2$ interact with two independent streams of ancillae $a_n^L$ and $a_n^R$, respectively. In this diagram, black lines indicate Heisenberg-type interactions between the system qubits as well as between each system qubit and its corresponding ancilla. The red line represents a partial swap interaction occurring between successive ancillae in the same stream.}
    \label{Scheme_B}
\end{figure}

In this scheme, qubits $s_1$ and $s_2$ of the two-qubit system $S$ interact with the environment ancillae denoted by $a^L_n$ and $a^R_n$, respectively, as depicted in Fig.~\ref{Scheme_B}. The system's reduced state after the $n^{th}$ step of scheme B is 
\begin{equation}
    \rho_{n+1}^S = \Tr_{a^R_na^R_{n+1}a^L_na^L_{n+1}}[\sigma^{Sa^La^R}_{n + 1}],
\end{equation}
where
\begin{align}
    \sigma^{Sa^La^R}_{n + 1} = U_{a^R_na^R_{n+1}} U_{a^L_na^L_{n+1}} U_{s_1s_2, s_1a^L_n, s_2a^R_n}(\rho_n^{Sa_n^La_n^R} \otimes \rho^{a_{n + 1}^L} \nonumber \\
    \otimes \rho^{a_{n + 1}^R}) U_{s_1s_2, s_1a^L_n, s_2a^R_n}^{\dag} U_{a^L_na^L_{n+1}}^{\dag} U_{a^R_na^R_{n+1}}^{\dag},
    \label{scheme_B_unitary_evolution}
\end{align}
and $\rho_n^{Sa_n^La_n^R} = \Tr_{a^R_{n-1}a^L_{n-1}}[\sigma^{Sa^La^R}_{n}]$.

Here, $\sigma^{sa^La^R}_{n+1}$ is the joint state of the two-qubit system and the environment ancillae after evolving through the unitary matrix corresponding to the intra-system and system environment interactions $U_{s_1s_2, s_1a^L_n, s_2a^R_n} = e^{-iH_B\Delta t}$, where $H_B = H_{s_1} + H_{s_2} + H_{a_n^{L}} + H_{a_n^{R}} + H_{s_1s_2} + H_{s_1a^L_n} + H_{s_2a^R_n}$, followed by the left side intra-ancilla interaction unitary $U_{a^L_na^L_{n+1}}$, and right side intra-ancilla interaction unitary $U_{a^R_na^R_{n+1}}$. The forms of $U_{a^L_na^L_{n+1}}$ and $U_{a^R_na^R_{n+1}}$ are as provided in Eq.~\eqref{p_swap_unitary}. For $n=1$, the initial joint state of the system and ancillae is $\rho^{Sa_1^La_1^R}_1 = \rho^S_0 \otimes \rho^{a^L_1} \otimes \rho^{a^R_1}$. The simulation steps for scheme B are detailed below,
\begin{enumerate}

    \item \textit{Initialization:} Prepare the initial system state $\rho^S_0$ and the left/right ancillae qubits in thermal states $\rho^{a^L_n}$ and $\rho^{a^R_n}$.

    \item \textit{State construction:} Construct the initial system–ancilla block used in collisions as
    \[
        \rho^{Sa_1^La_1^R}_1 = \rho^S_0 \otimes \rho^{a^L_1} \otimes \rho^{a^R_1}.
    \]

    \item \textit{Unitary evolution per collision:} At each iteration, form the six-partite input by tensoring the current block with fresh left and right ancillas, then apply intra-system and system environment interactions unitary, i.e., $U_{s_1s_2, s_1a^L_n, s_2a^R_n}$ followed by the left and right ancilla partial swap unitaries $U_{a^L_na^L_{n+1}}$ and $U_{a^R_na^R_{n+1}}$. Finally, calculate $\sigma^{Sa^La^R}_{n + 1}$ as elaborated in Eq.~\eqref{scheme_B_unitary_evolution}.

    \item \textit{State update:} Obtain the updated system state $\rho_{n+1}^S$ by tracing out the ancillas and use it for further studies, and obtain the next step’s system–ancilla block $\rho_n^{Sa_n^La_n^R}$ by tracing out the outgoing ancillas.
\end{enumerate}

\section{\label{Non_Markovianity_measure}Non-Markovianity witness}
\begin{figure}
    \centering
    \includegraphics[height=49.8mm,width=1.06\columnwidth]{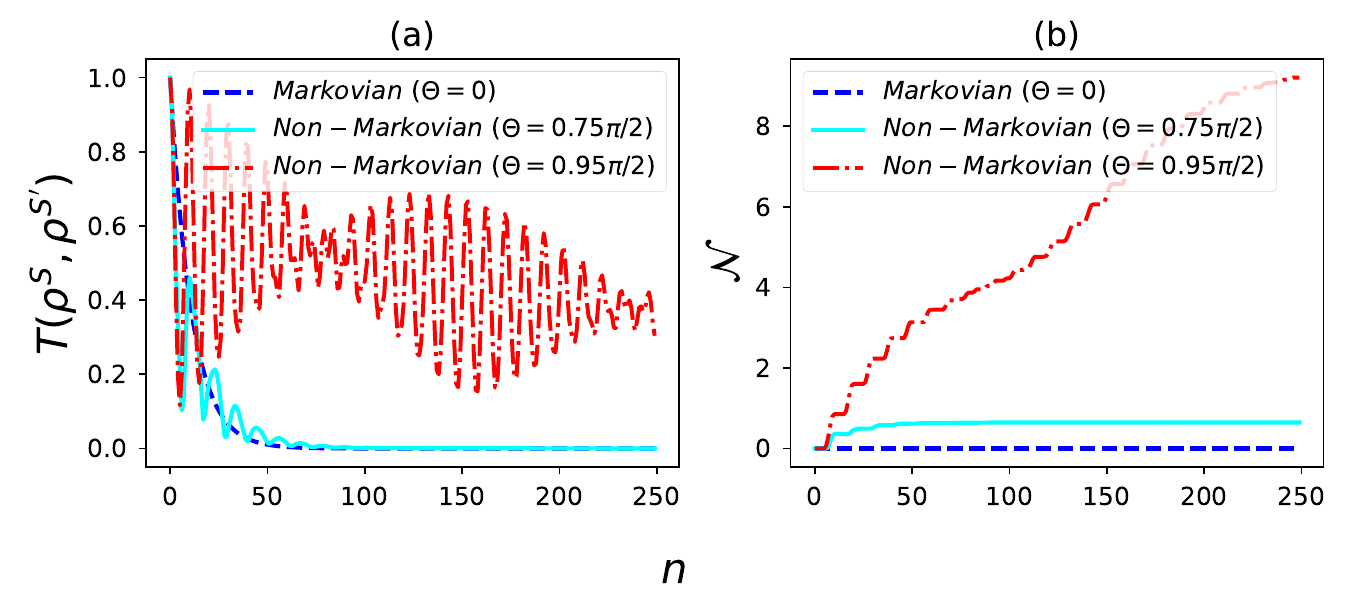}
    \caption{The evolution of the trace distance and non-Markovianity measure ($\mathcal{N}$) between the Bell states $\ket{\phi^{+}}$ and $\ket{\phi^{-}}$ is analyzed using the scheme A as a function of the number of collisions ($n$) in subplots $(a)$ and $(b)$, respectively. In this analysis, the parameters are set as follows: $\omega_{s_1} = \omega_{s_2} = \omega_{a^R_n} = 1$, $g_{s_2a^R_n} = 0.85$, $g_{s_1s_2} = 0.95$, $\Delta t = 0.5$ and $\beta_{a_n^{R}} = 1$. For Markovian dynamics, intra-ancilla interaction strength $\Theta = 0$, and for non-Markovian dynamics $\Theta = 0.75\pi/2, 0.95\pi/2$.}
    \label{Trace_distance_scheme_A}
\end{figure}

\begin{figure}
    \centering
    \includegraphics[height=75mm,width=0.75\columnwidth]{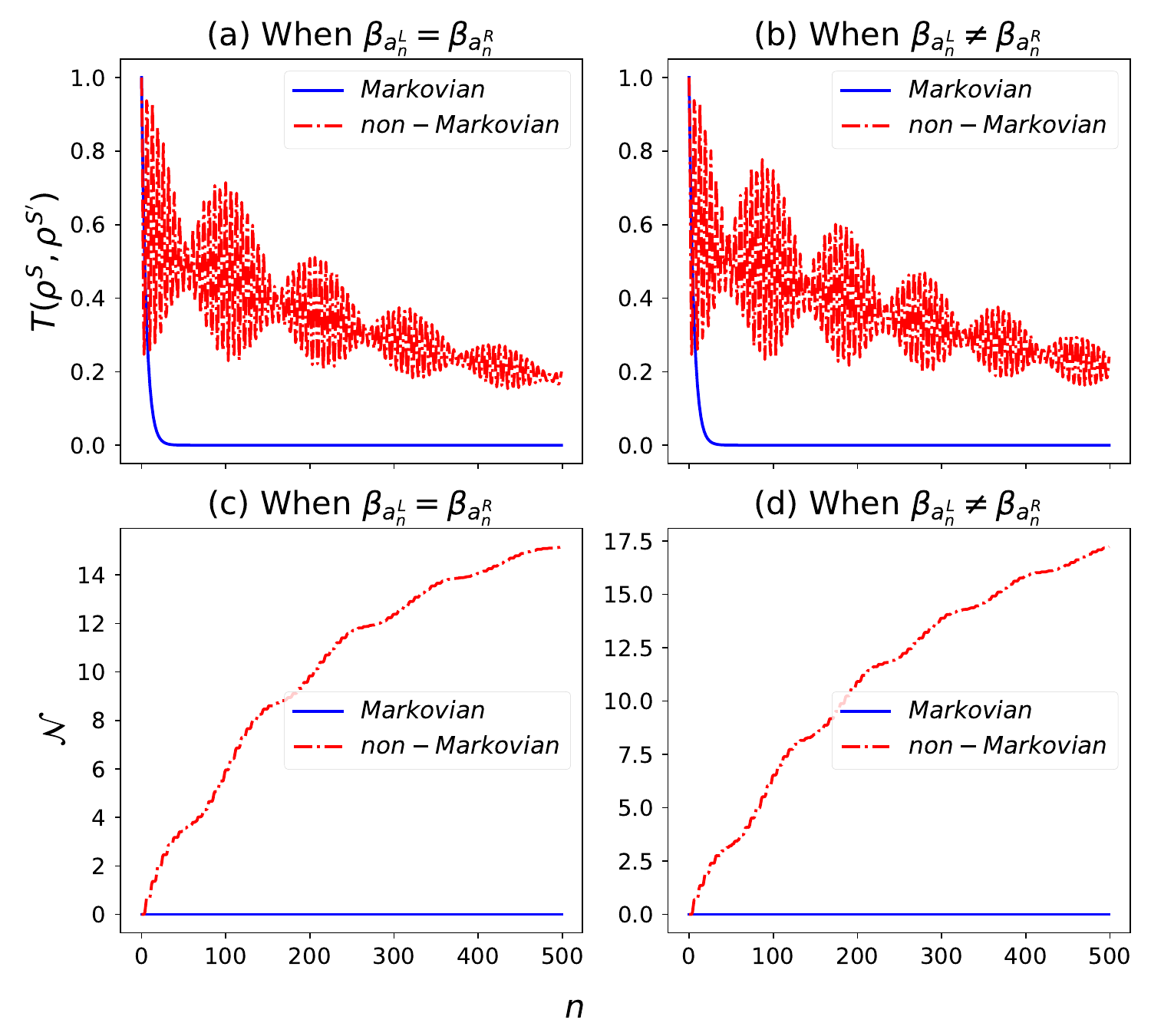}
    \caption{The evolution of the trace distance in the upper panels (a, b) and non-Markovianity measure ($\mathcal{N}$) in the lower panels (c, d) between the Bell states $\ket{\phi^{+}}$ and $\ket{\phi^{-}}$ is analyzed using the scheme B as a function of the number of collisions ($n$). In this analysis, the parameters are set as follows: $\omega_{s_1} = \omega_{s_2} = \omega_{a^L_n} = \omega_{a^R_n} = 1$, $g_{s_2a^R_n} = g_{s_1a^L_n} = 0.85$, $g_{s_1s_2} = 0.95$, $\Delta t = 0.5$. In subplot (a) and (c) $\beta_{a_n^{L}} = \beta_{a_n^{R}} = 1$, and in subplot (b) and (d) $\beta_{a_n^{L}} = 1$, $\beta_{a_n^{R}} = 4$. For Markovian dynamics, intra-ancilla interaction strength $\Theta = 0$, and for non-Markovian dynamics $\Theta = 0.95\pi/2$. }
    \label{Trace_distance_scheme_B}
\end{figure}

The characterization and quantification of non-Markovianity in open quantum systems have garnered significant interest in recent times. Various measures for non-Markovianity have been introduced in literature, which claim to identify memory effects in the open systems' dynamics~\citep{rivas2014quantum, Breuer2016Colloquium, Utagi2020}. We study one of the extensively used non-Markovianity quantifiers, Breuer-Laine-Piilo (BLP) measure~\citep{Breuer2009Measure}. To identify memory effects that come from the non-Markovian nature of the open system dynamics, this method uses the distinguishability of the system's states based on their trace distance from one another. The trace distance $T$ is defined by
\begin{align}
    T(\rho^{S}_{n+1}, \rho^{S'}_{n+1}) &= \frac{1}{2}|| \rho^{S}_{n+1} - \rho^{S'}_{n+1} ||_1 \nonumber \\ 
    &= \frac{1}{2} \Tr\left[\sqrt{(\rho^{S}_{n+1} - \rho^{S'}_{n+1} )^{\dag} (\rho^{S}_{n+1} - \rho^{S'}_{n+1})}\right],
\end{align}
where $||.||_1$ represents the trace norm~\citep{nielsen2010quantum}, and $\rho^{S}_{n+1}$, $\rho^{S'}_{n+1}$ are the states of the two-qubit system at the $n^{th}$ step of scheme A and B, evolved from two different initial states of the system $S$ and $S'$, and depicted in Figs.~\ref{Trace_distance_scheme_A} and \ref{Trace_distance_scheme_B}, respectively. The flow of information between the open quantum system and its surroundings can be attributed to the change in distinguishability between two arbitrary initial states of the open quantum system during the dynamics. If the distinguishability in the form of trace distance $T(\rho^{S}_{n+1}, \rho^{S'}_{n+1})$ experiences brief revivals over time evolution, it indicates an information backflow from the environment to the system, resulting in memory effects. The degree of non-Markovianity of an open quantum system's dynamics can be quantitatively measured by~\citep{Breuer2009Measure}
\begin{equation}
    \mathcal{N} = \max_{(\rho^{S}_{0},\, \rho^{S'}_{0})}\int_{\sigma > 0} \sigma dt,
    \label{N_eq}
\end{equation}

where $\sigma = \frac{dT\left[\rho^S(t), \rho^{S'}(t)\right]}{dt}$, and the optimization in the above equation is carried out over all potential initial state pairs of the open system, $\rho^S_0$ and $\rho^{S'}_0$.  The dynamics of open quantum systems in our work occur in discrete time steps, as we employ a collision model to characterize the system's dynamics.  Consequently, we use a discretized form of Eq.~\eqref{N_eq} as examined in~\citep{Laine2010Measure} to assess the degree of non-Markovianity
\begin{equation}
    \mathcal{N} = \max_{(\rho^{S}_{0},\, \rho^{S'}_{0})}\sum_n [T(\rho^{S}_{n+1},\, \rho^{S'}_{n+1}) - T(\rho^{S}_{n},\, \rho^{S'}_{n})],
\end{equation}

where $n$ is the $n^{th}$ step of the collision model. Figures~\ref{Trace_distance_scheme_A} and~\ref{Trace_distance_scheme_B} show the variation of trace distance $T(\rho^{S}_{n+1}, \rho^{S'}_{n+1})$ and non-Markovianity measure $\mathcal{N}$ with the number of collisions when the initial states of the system $S$ are Bell $\ket{\phi^{+}}$ and $\ket{\phi^{-}}$ states for schemes A and B, respectively. In all the cases, when the intra-ancilla interaction strength ($\Theta$) is zero, the trace distance monotonically decreases with the number of collisions, see Fig.~\ref{Trace_distance_scheme_A}(a), and Figs.~\ref{Trace_distance_scheme_B}(a), (b). The non-Markovianity measure $\mathcal{N}$ also remains constant at zero with the number of collisions, see Fig.~\ref{Trace_distance_scheme_A}(b), and Figs.~\ref{Trace_distance_scheme_B}(c), (d), indicating a Markovian evolution.   

For intermediate and high values of $\Theta$, specifically $\Theta = 0.75 \pi/2$ and $\Theta = 0.95 \pi/2$ in Fig.~\ref{Trace_distance_scheme_A} for scheme A, the trace distance shows an oscillatory behavior. Additionally, the non-Markovianity measure $\mathcal{N}$ is positive for these values of $\Theta$ and increases gradually with the number of collisions, indicating the presence of non-Markovianity.
In scheme B, we consider two cases. In the first case, both the environmental ancillae are at the same temperature, $\beta_{a^L_n} = \beta_{a^R_n}$, Figs.~\ref{Trace_distance_scheme_B}(a) and (c). In the second case, both are at different temperatures, $\beta_{a^L_n} \ne \beta_{a^R_n}$, Figs.~\ref{Trace_distance_scheme_B}(b) and (d). The initial state of the ancillae is according to Eq.~\eqref{eq_ancilla_thermal_state}. Here, for the case of $\Theta = 0.95\pi/2$, we observe an oscillatory evolution of the trace distance. Also, the non-Markovianity measure $\mathcal{N}$ increases with the number of collisions for $\Theta = 0.95\pi/2$, quantifying the non-Markovian behavior of the system. 
Figure~\ref{Trace_distance_scheme_B} illustrates that, in the case when the ancillae are at different temperatures, the non-Markovianity measure attains a higher value in comparison to the case when the ancillae temperatures are equal. It is important to note that non-Markovianity can be quantified using several inequivalent measures, including the Rivas–Huelga–Plenio (RHP) measure~\citep{rivas2014quantum}. In this work, the Breuer–Laine–Piilo (BLP) measure is employed due to its operational robustness and computational tractability. While widely adopted, the BLP measure has certain limitations. First, it depends on an optimization over pairs of initial states; in practice, however, this optimization is typically restricted to a subset of states (e.g., Bell states), which may lead to an underestimation of the actual non-Markovianity. Second, the BLP criterion is sufficient but not necessary for identifying non-Markovian dynamics: processes that are non-Markovian in the sense of complete positive (CP) indivisibility may not exhibit any revival in trace distance. Third, in collision-model frameworks, where the dynamics are inherently discrete, the BLP measure is evaluated via finite differences rather than a continuous-time derivative, introducing an additional approximation. These considerations do not compromise the validity of the present results, but should be taken into account when comparing with alternative characterizations, such as the RHP measure based on CP indivisibility.

Having discussed the non-Markovian evolution of the two-qubit quantum collision model, we now study the non-classicality of the system using the Wigner function, non-classical volume, and entanglement generation. 

\section{\label{Wigner_function} Wigner function}
The Wigner function, $W(\theta, \phi)$, is a phase-space representation of a quantum state that extends classical probability distributions to the quantum domain~\citep{Wigner1932Quantum}. It is a real-valued function that can take negative values, indicating non-classicality~\citep{zavatta2004quantum}. A quasi-probability distribution can be described as a function of the polar $(\theta)$ and azimuthal $(\phi)$ angles using the relationship between spin-like, $SU(2)$ systems and the sphere. When this is extended over the entire basis set, with the spherical harmonics, the $W(\theta, \phi)$ function for a single spin-$j$ state can be written as~\citep{Thapliyal2015Quasiprobability}
\begin{equation}\label{Eq_Wigner1}
    W(\theta, \phi) = \sqrt{\frac{2j + 1}{4\pi}} \sum_{K, Q}\rho_{KQ}Y_{KQ}(\theta, \phi),
\end{equation}
where $K = 0, 1, ..., 2j$, and $Q = -K, -K+1, ....., 0, ....., K-1, K$,~$Y_{KQ}$ are the spherical harmonics and $\rho_{KQ} = \Tr\left[T_{KQ}^{\dag}\rho\right]$, with $\rho$ being the density matrix of the system. Further, multipole operators $T_{KQ}$ are given by
\begin{equation}
    T_{KQ} = \sum_{m, m'} (-1)^{j-m} (2K + 1)^{1/2} \begin{pmatrix}
                                j & K & j\\
                                -m & Q & m'
                               \end{pmatrix} 
                                \ket{j, m}\bra{j, m'},
\end{equation}
where $\begin{pmatrix} 
        j_1 & j_2 & j\\ 
        m_1 & m_2 & m 
        \end{pmatrix} = \frac{(-1)^{j_1 - j_2 - m}}{\sqrt{2j + 1}} \langle j_1m_1j_2m_2|j-m\rangle $ 
is the Wigner $3j$ symbol and $\langle j_1m_1j_2m_2|j-m\rangle$ is the Clebsh-Gordon coefficient. The multipole operators for $j = \frac{1}{2}$ are given in~\citep{Thapliyal2015Quasiprobability}. The Wigner function is also normalized such that $\int W(\theta, \phi) \sin{\theta} d{\theta} d{\phi} = 1$. Likewise, for a two-qubit system each with spin-$j$, the case considered here, the $W(\theta, \phi)$ function is 
\begin{align}
    W({\theta_1}, {\phi_1}, {\theta_2}, {\phi_2}) &= \left(\frac{2j + 1}{4\pi}\right)\sum_{K_1, Q_1}\sum_{K_2, Q_2}\rho^S_{n+1 {K_1}{Q_1}{K_2}{Q_2}}\nonumber \\
    &\times Y_{{K_1}{Q_1}}(\theta_1, \phi_1)Y_{K_2Q_2}{(\theta_2, \phi_2)},
    \label{two_qubit_W_func}
\end{align}
where $\rho^S_{n+1 K_1 Q_1 K_2 Q_2} = \Tr\left[\rho^S_{n+1} T_{K_1 Q_1}^{\dag} T_{K_2 Q_2}^{\dag}\right]$ and it is also normalized for $\theta_1, \theta_2 \in [0, \pi]$, and $\phi_1, \phi_2 \in [0, 2\pi]$.

\begin{figure}
    \centering
    \includegraphics[height=45mm,width=1\columnwidth]{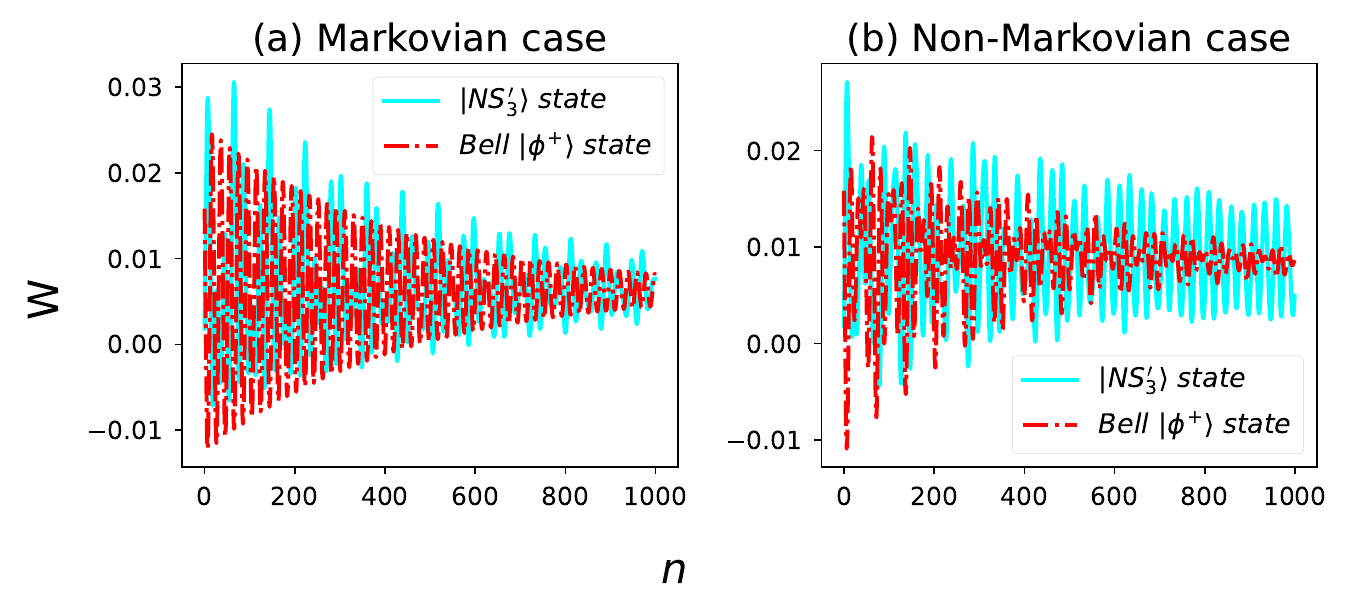}
    \caption{Variation of the Wigner function of the $\ket{NS_3^{\prime}}$, and Bell $\ket{\phi^{+}}$ states using the scheme A with the number of collisions. Here $\theta_1 = \theta_2 = \pi/2$, $\phi_1 = \phi_2 = \pi/6$, $\omega_{s_1} = \omega_{s_2} = \omega_{a^R_n} = 1$, $g_{s_2a^R_n} = 0.85$, $g_{s_1s_2} = 0.95$, $\Delta t = 0.08$ and $\beta_{a_n^{R}} = 1$. Subplot (a) is for Markovian dynamics when $\Theta = 0$, and subplot (b) is for non-Markovian dynamics when $\Theta = 0.95 \times \pi/2$.}
    \label{W_func_H_int_two_qubit_aR_all_states_scheme_A}
\end{figure}

\begin{figure}
    \centering
    \includegraphics[height=75mm,width=0.85\columnwidth]{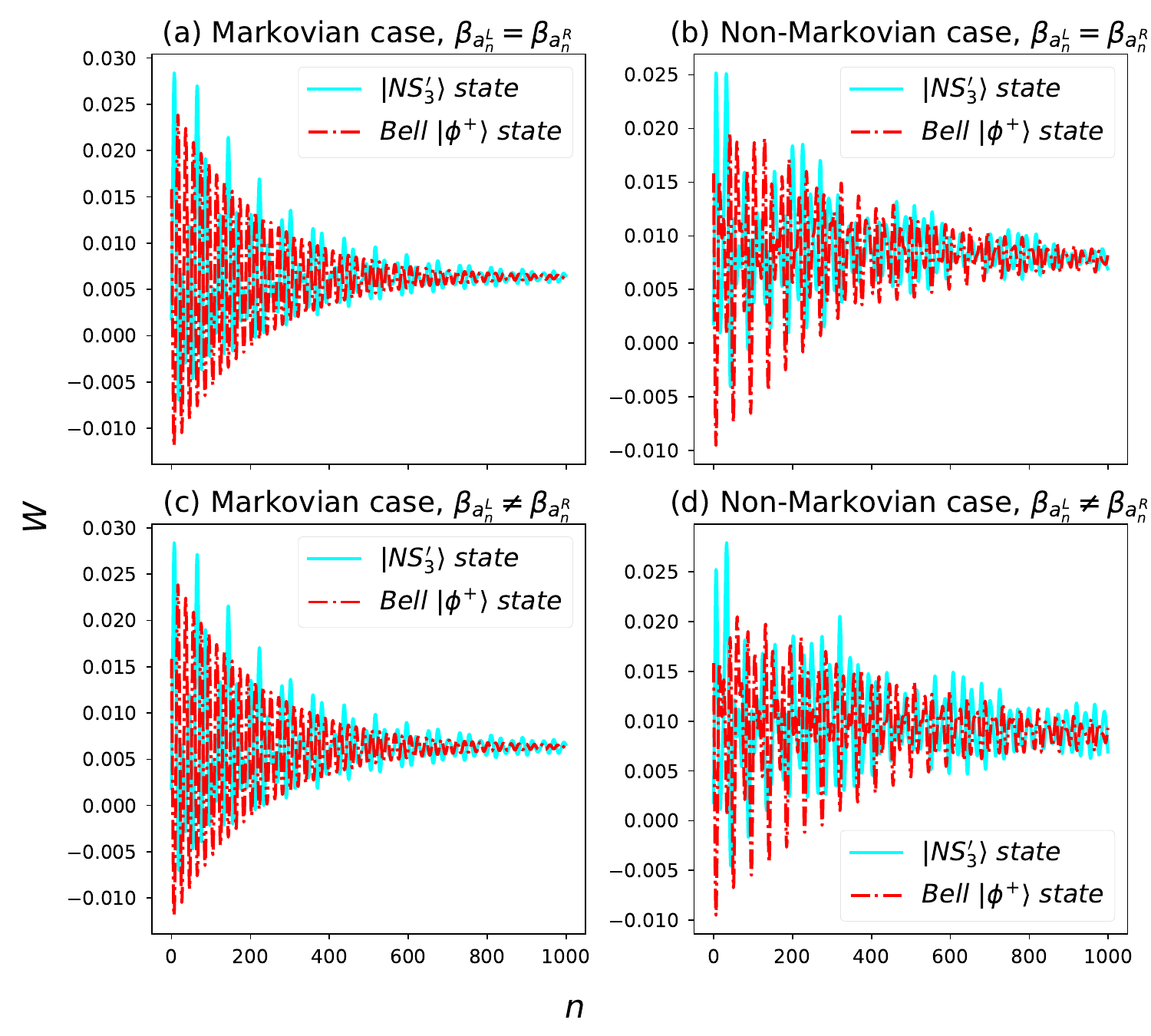}
    \caption{Variation of the Wigner function of the $\ket{NS_3^{\prime}}$, and Bell $\ket{\phi^{+}}$ states using the scheme B with the number of collisions. In this analysis, the parameters are set as follows: $\theta_1 = \theta_2 = \pi/2$, $\phi_1 = \phi_2 = \pi/6$, $\omega_{s_1} = \omega_{s_2} = \omega_{a^L_n} = \omega_{a^R_n} = 1$, $g_{s_2a^R_n} = g_{s_1a^L_n} = 0.85$, $g_{s_1s_2} = 0.95$, $\Delta t = 0.08$. In subplot (a) and (b) $\beta_{a_n^{L}} = \beta_{a_n^{R}} = 1$, and in subplot (c) and (d) $\beta_{a_n^{L}} = 1$, $\beta_{a_n^{R}} = 4$. For Markovian dynamics, intra-ancilla interaction strength $\Theta = 0$, and for non-Markovian dynamics $\Theta = 0.95\pi/2$.}
    \label{W_func_H_int_two_qubit_aR_all_states_scheme_B}
\end{figure}

Figures~\ref{W_func_H_int_two_qubit_aR_all_states_scheme_A},~\ref{W_func_H_int_two_qubit_aR_all_states_scheme_B} illustrate the variation of Wigner function of the two-qubit $\ket{NS_3^{\prime}}$ state~\citep{lalita2023harnessing, Lalita_2024ProtectingQC, lalita2025realizingnegativequantumstates} (detailed in Sec.~\ref{ch3_NQS} of chapter~\ref{chap3:Harnessing}) and Bell $\ket{\phi^{+}}$ state with the number of collisions using the schemes A and B, respectively. The $\ket{NS_3^{\prime}}$ is one of the two-qubit negative quantum states~\citep{lalita2023harnessing, Lalita_2024ProtectingQC, lalita2025realizingnegativequantumstates} known to offer enhanced robustness against non-Markovian noise for reliable quantum information processing tasks such as teleportation in realistic, noisy environments. The values of the polar angles $(\theta_1, \theta_2)$ and azimuthal angles $(\phi_1, \phi_2)$ are chosen to clearly demonstrate the oscillations and negativity of the Wigner function.

In scheme A, Fig.~\ref{W_func_H_int_two_qubit_aR_all_states_scheme_A}(b), for the non-Markovian evolution, the amplitude of oscillation of the Wigner function persists longer for the $\ket{NS_3^{\prime}}$ state in comparison to the Bell $\ket{\phi^{+}}$ state. Also, decay in amplitude of the Wigner function for the $\ket{NS_3^{\prime}}$ and $\ket{\phi^{+}}$ states in the non-Markovian case is comparatively lesser than the Markovian case as depicted by Fig. \ref{W_func_H_int_two_qubit_aR_all_states_scheme_A} (a) and (b). In Scheme B, Fig.~\ref{W_func_H_int_two_qubit_aR_all_states_scheme_B}, the Wigner function exhibits similar qualitative behavior for both identical and different ancilla temperatures. The decay of the amplitude of oscillation for the Wigner function is gradual and relatively ordered in the Markovian case than in the non-Markovian scenario, see Fig. \ref{W_func_H_int_two_qubit_aR_all_states_scheme_B} (a), (b), (c), and (d).  

In all the cases, we observe that the variation of the Wigner function with the number of collisions for both the two-qubit states shows negative values, indicating quantumness. Now, to quantify the quantumness, we study the non-classical volume.  
\subsection{Non-classical Volume}
The Wigner function's negative values serve as a non-classicality signature. Nevertheless, the negative values do not offer a quantitative measure of non-classicality. The non-classical volume, introduced in~\citep{Anatole2004Negativity}, is used as a quantitative measure of quantumness in a given quantum system. For a two-qubit quantum system, it is given by
\begin{equation}
    \delta = \int |W({\theta_1}, {\phi_1}, {\theta_2}, {\phi_2})| \sin({\theta_1}) \sin({\theta_2}) d{\theta_1} d{\theta_2} d{\phi_1} d{\phi_2} - 1,
\end{equation}
where $W({\theta_1}, {\phi_1}, {\theta_2}, {\phi_2})$ is the Wigner function for a two-qubit quantum system defined above in Eq.~\eqref{two_qubit_W_func}. A non-zero value of $\delta$ indicates that the system is non-classical. 

\begin{figure}
    \centering
    \includegraphics[height=45mm,width=1\columnwidth]{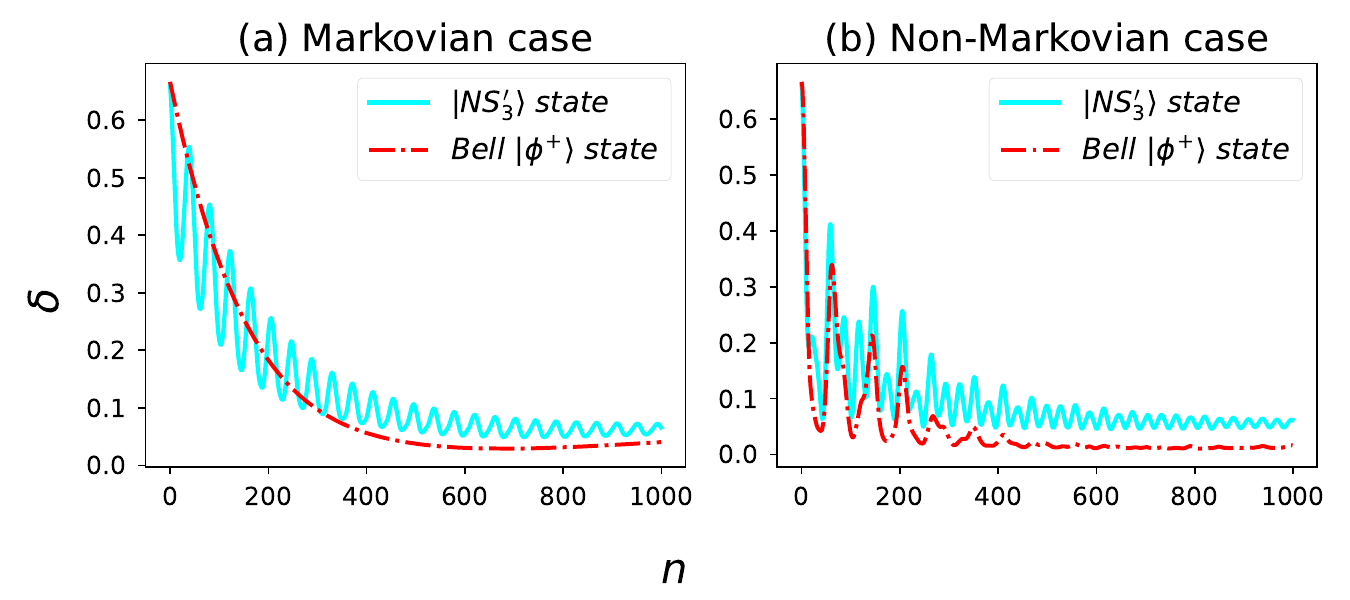}
    \caption{Variation of the non-classical volume of the $\ket{NS_3^{\prime}}$, and Bell $\ket{\phi^{+}}$ states with the number of collisions using scheme A. Here $\omega_{s_1} = \omega_{s_2} = \omega_{a^R_n} = 1$, $g_{s_2a^R_n} = 0.85$, $g_{s_1s_2} = 0.95$, $\Delta t = 0.08$ and $\beta_{a_n^{R}} = 1$. Subplot (a) is for Markovian dynamics when $\Theta = 0$, and subplot (b) is for non-Markovian dynamics when $\Theta = 0.95 \times \pi/2$.}
    \label{non_classical_vol_H_int_two_qubit_aR_all_states_scheme_A}
\end{figure}

\begin{figure}
    \centering
    \includegraphics[height=75mm,width=0.75\columnwidth]{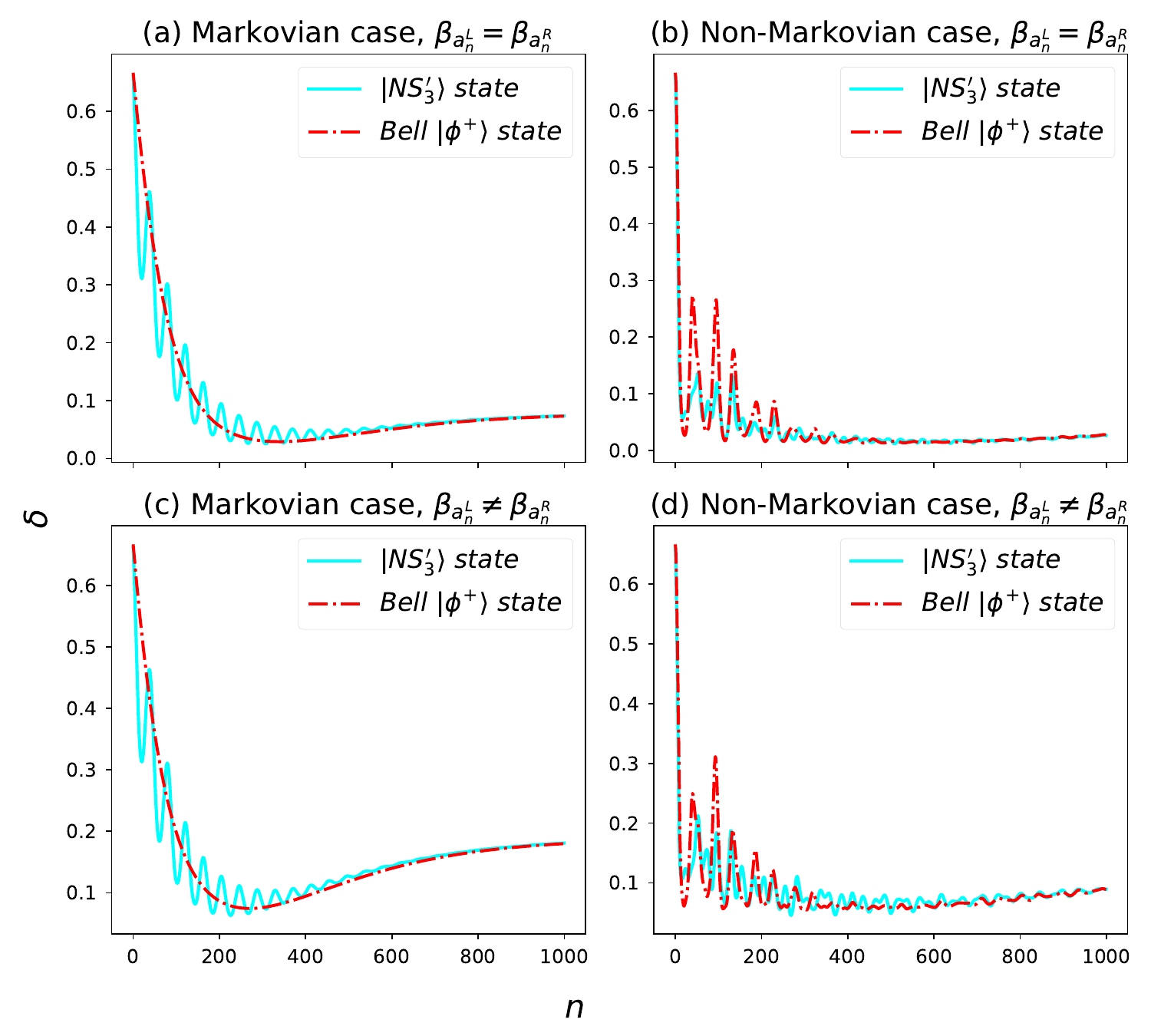}
    \caption{Variation of the non-classical volume of the $\ket{NS_3^{\prime}}$, and Bell $\ket{\phi^{+}}$ states with the number of collisions using scheme B. In this analysis, the parameters are set as follows: $\omega_{s_1} = \omega_{s_2} = \omega_{a^L_n} = \omega_{a^R_n} = 1$, $g_{s_2a^R_n} = g_{s_1a^L_n} = 0.85$, $g_{s_1s_2} = 0.95$, $\Delta t = 0.08$. In subplot (a) and (b) $\beta_{a_n^{L}} = \beta_{a_n^{R}} = 1$, and in subplot (c) and (d) $\beta_{a_n^{L}} = 1$, $\beta_{a_n^{R}} = 4$. For Markovian dynamics, intra-ancilla interaction strength $\Theta = 0$, and for non-Markovian dynamics $\Theta = 0.95\pi/2$.}
    \label{non_classical_vol_H_int_two_qubit_aR_all_states_scheme_B}
\end{figure}

The variation of non-classical volume of the $\ket{NS_3^{\prime}}$ and Bell $\ket{\phi^{+}}$ states, using the scheme A and B, is depicted in Figs.~\ref{non_classical_vol_H_int_two_qubit_aR_all_states_scheme_A},~\ref{non_classical_vol_H_int_two_qubit_aR_all_states_scheme_B}, respectively, for both (non-)Markovian evolution. 
We observe that for all the cases, both states have the same initial value of non-classical volume. The variation of non-classical volume for the Bell state is oscillatory in the non-Markovian scenario and non-oscillatory in the Markovian scenario of both schemes A and B.
Interestingly, the $\ket{NS_3^{\prime}}$ state's non-classical volume shows oscillatory behavior with the number of collisions, even in the Markovian case, see Figs.~\ref{non_classical_vol_H_int_two_qubit_aR_all_states_scheme_A}(a),~\ref{non_classical_vol_H_int_two_qubit_aR_all_states_scheme_B}(a) and (c). This behavior could be attributed to the presence of complex numbers in this state (see Sec.~\ref{ch3_NQS} of chapter~\ref{chap3:Harnessing}), which causes phase-dependent interference effects in the Wigner function negativity~\citep{li2021controllable}. In scheme A of the collision model, the $\ket{NS_3^{\prime}}$ state sustains its non-classical behavior longer in both the Markovian and non-Markovian cases, as can be seen from Fig.~\ref{non_classical_vol_H_int_two_qubit_aR_all_states_scheme_A}. On the other hand, in scheme B, at longer duration, both states show equivalent variation of non-classical volume, Fig.~\ref{non_classical_vol_H_int_two_qubit_aR_all_states_scheme_B}.
We now study how schemes A and B and (non-)Markovian evolution affect quantum correlations in the system qubits.  
\section{\label{Quantum_correlation} Quantum correlations}
Quantum non-local correlations are one of the most profound and intrinsically non-classical features of a quantum system, exhibiting phenomena that lack any counterpart in classical physics. Central to these correlations is entanglement, which serves as a foundational element in describing quantum correlations within composite quantum systems. Specifically, in the case of a bipartite system composed of two qubits, entanglement can be quantitatively characterized by concurrence \citep{Wootters1998Entanglement}
\begin{equation}
   \begin{aligned}
     C({\rho}^S_{n+1}) = \max \{0, \lambda_{1} - \lambda_{2} - \lambda_{3} - \lambda_{4}\},
      \end{aligned}
      \label{ch6_concur_eq.}
\end{equation}
where $\lambda_{i}$'s are the eigenvalues of $\sqrt{\sqrt{{\rho}^S_{n+1}} \tilde{{\rho}}^S_{n+1} \sqrt{{\rho}^S_{n+1}}}$, such that $\lambda_{1} \geq \lambda_{2} \geq \lambda_{3} \geq \lambda_{4}$, and $\Tilde{{\rho}}^S_{n+1} = (\sigma_{y} \otimes \sigma_{y}) {\rho}_{n+1}^{S*} (\sigma_{y} \otimes \sigma_{y})$, where ${\rho}_{n+1}^{S*}$ is the complex conjugate of ${\rho}^S_{n+1}$. Further, to investigate whether entanglement can emerge between initially separable qubits of the system as a result of sufficiently strong intra-system and system-ancilla interaction strengths, we analyze how the concurrence evolves with the number of collisions of the two-qubit collision model. 

\begin{figure}
    \centering
    \includegraphics[height=45mm,width=1\columnwidth]{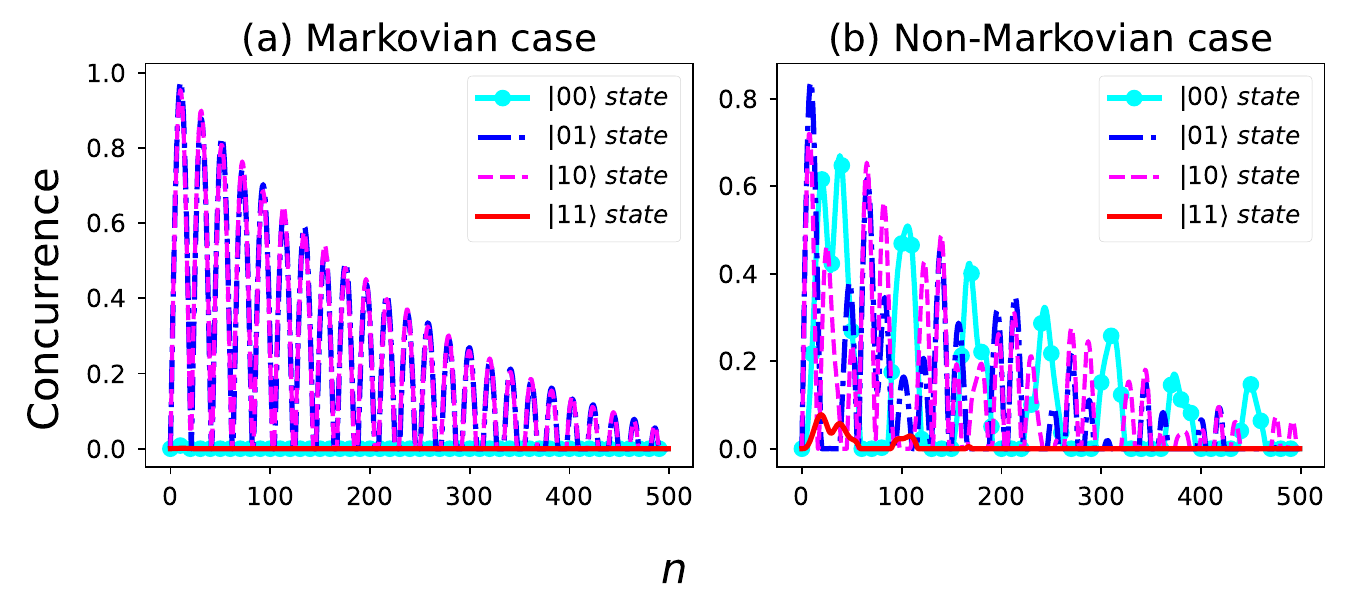}
    \caption{Variation of the concurrence of the $\ket{00}$, $\ket{01}$, $\ket{10}$ and $\ket{11}$ states using the scheme A with the number of collisions. Here $\omega_{s_1} = \omega_{s_2} = \omega_{a^R_n} = 1$, $g_{s_2a^R_n} = 0.85$, $g_{s_1s_2} = 0.95$, $\Delta t = 0.08$ and $\beta_{a_n^{R}} = 1$. Subplot (a) is for Markovian dynamics when $\Theta = 0$, and subplot (b) is for non-Markovian dynamics when $\Theta = 0.95 \times \pi/2$.}
    \label{concurrence_H_int_two_qubit_aR_all_states_scheme_A}
\end{figure}

\begin{figure}
    \centering
    \includegraphics[height=85mm,width=0.85\columnwidth]{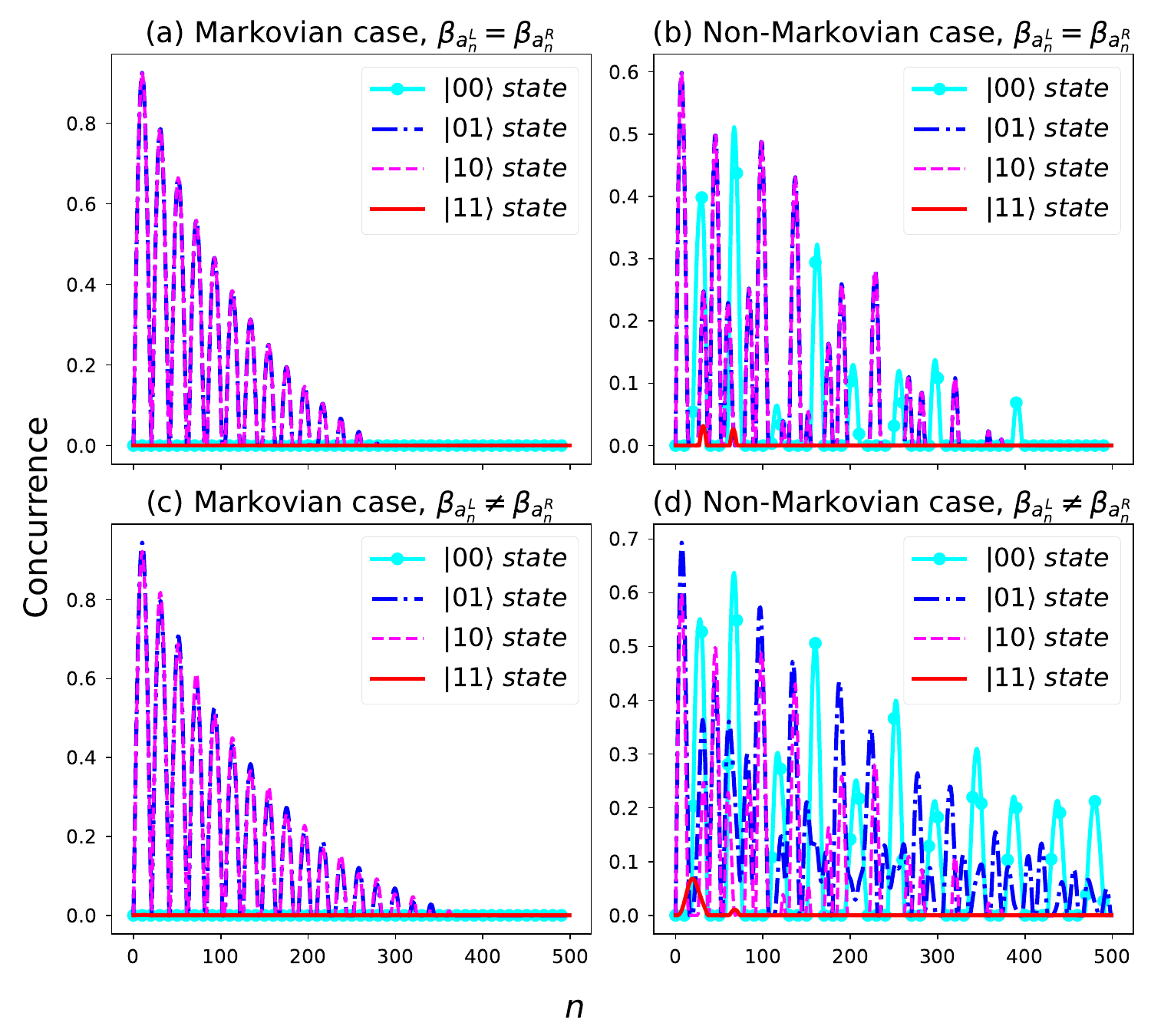}
    \caption{Variation of the concurrence of the $\ket{00}$, $\ket{01}$, $\ket{10}$ and $\ket{11}$ states with the number of collisions using scheme B. In this analysis, the parameters are set as follows: $\omega_{s_1} = \omega_{s_2} = \omega_{a^L_n} = \omega_{a^R_n} = 1$, $g_{s_2a^R_n} = g_{s_1a^L_n} = 0.85$, $g_{s_1s_2} = 0.95$, $\Delta t = 0.08$. In subplot (a) and (b) $\beta_{a_n^{L}} = \beta_{a_n^{R}} = 1$, and in subplot (c) and (d) $\beta_{a_n^{L}} = 1$, $\beta_{a_n^{R}} = 4$. For Markovian dynamics, intra-ancilla interaction strength $\Theta = 0$, and for non-Markovian dynamics $\Theta = 0.95\pi/2$.}
    \label{concurrence_H_int_two_qubit_aR_aL_all_states_scheme_B}
\end{figure}

Figures~\ref{concurrence_H_int_two_qubit_aR_all_states_scheme_A} and~\ref{concurrence_H_int_two_qubit_aR_aL_all_states_scheme_B} illustrate the variation of concurrence with the number of collisions using scheme A and B, respectively, for the $\ket{00}$, $\ket{01}$, $\ket{10}$ and $\ket{11}$ states as the initial separable states of the two-qubit system. In the Markovian regime, for both schemes A and B, no appreciable entanglement is generated for the $\ket{00}$ and $\ket{11}$ states under the specified intra-system and system-ancilla interaction strengths, see Figs.~\ref{concurrence_H_int_two_qubit_aR_all_states_scheme_A}(a), ~\ref{concurrence_H_int_two_qubit_aR_aL_all_states_scheme_B}(a), and~\ref{concurrence_H_int_two_qubit_aR_aL_all_states_scheme_B}(c). Interestingly, in the non-Markovian regime, for both schemes A and B, there is significant entanglement generation for the $\ket{00}$ state, as can be seen from Figs.~\ref{concurrence_H_int_two_qubit_aR_all_states_scheme_A}(b),~\ref{concurrence_H_int_two_qubit_aR_aL_all_states_scheme_B}(b), and  ~\ref{concurrence_H_int_two_qubit_aR_aL_all_states_scheme_B}(d). Also, in the non-Markovian regime, for $\ket{11}$ initial state, entanglement is seen to be generated initially, but it is not substantial.
The asymmetry between the initial excited and ground states in entanglement generation could be attributed to the non-Markovian nature of the evolution~\citep{entanglement_generation_paper}. The excited state emits to the ancialle, which in turn re-emits. This exchange creates oscillations in the population and the coherence, leading to entanglement revival. The absence of excitations in the initial ground state leads to little entanglement generation during the dynamics.

Substantial entanglement growth is observed for the $\ket{01}$ and $\ket{10}$ states in both (non-)Markovian regimes of schemes A and B. This sustains for a few hundred collisions before it becomes zero, because of open system effects.  Although the $\ket{01}$ and $\ket{10}$ states initially display the highest concurrence, the $\ket{00}$ state shows more sustained entanglement over time in both schemes within the non-Markovian framework, see Figs.~\ref{concurrence_H_int_two_qubit_aR_all_states_scheme_A}(b), ~\ref{concurrence_H_int_two_qubit_aR_aL_all_states_scheme_B}(b), and~\ref{concurrence_H_int_two_qubit_aR_aL_all_states_scheme_B}(d). Furthermore, for the specified intra-ancilla and system-ancilla coupling strengths, scheme B yields a more robust and sustained entanglement over a larger number of collisions when the ancillae are maintained at different temperatures in both the Markovian and non-Markovian regimes, as illustrated in Figs.~\ref{concurrence_H_int_two_qubit_aR_aL_all_states_scheme_B}(c) and~\ref{concurrence_H_int_two_qubit_aR_aL_all_states_scheme_B}(d).

In the subsequent section, we study the steady state behavior of the system qubits and the impact of ancilla temperature on both qubits of the two-qubit collision model. 
\section{\label{SS_behaviour} Steady state behaviour}
The steady state of an open quantum system refers to the asymptotic state in which the system’s density matrix does not change due to repeated interactions with its environment~\citep{Breuer2007, Strasberg2017Quantum}. This state, often independent of the initial conditions, may correspond to thermal equilibrium (Gibbs state) or a non-equilibrium stationary state, depending on the system-environment dynamics. Observing whether the system relaxes to a Gibbs state of the form $\rho\propto e^{-\beta H_s}$ helps to evaluate the applicability of thermodynamic principles to small quantum systems. Here, we determine the steady state of a two-qubit system by evolving it under scheme B. The steady state is where the difference between successive reduced system states, $\rho^S_{n+1} - \rho^S_n$, vanishes; here $n$ denotes the $n^{th}$ collision. Here, we investigate the thermalization of the single and two-qubit systems. For the two-qubit system in scheme B, the steady state is obtained by identifying the point at which the reduced system state converges, that is, when the difference $\rho^S_{n+1} - \rho^S_n$ vanishes. We maintain the temperature of both ancillae at the same value, denoted as $\beta_{a^L_n} = \beta_{a^R_n} = \beta$. Then, to determine whether the steady state of the two-qubit system corresponds to its thermal state, we calculate the fidelity $F$ between the two-qubit system's steady state and the Gibbs states associated with the two-qubit system Hamiltonian ($H_S = H_{s_1} + H_{s_2} + H_{s_1s_2}$) by using~\citep{jozsa1994fidelity, nielsen2010quantum} 
\begin{equation}
    F(\rho^{ss}, \sigma_{GS}) \equiv {\rm Tr}\sqrt{\sqrt{\rho^{ss}}\sigma_{GS}\sqrt{\rho^{ss}}},
    \label{Fidelity_formula}
\end{equation}
where $\rho^{ss}$ represents the steady state of the two-qubit system and $\sigma_{GS} = e^{-\beta H_S}/\Tr\left[e^{-\beta H_S}\right]$ represents the Gibbs states associated with the two-qubit system Hamiltonian.
We benchmark this with the fidelity between the system's steady state and Gibbs state corresponding to the Hamiltonian of the mean force~\citep{Talkner2020colloquium, pathania2024}
\begin{align}
H^*_S = \frac{1}{\beta} \ln{\frac{\Tr_b\left[e^{-\beta H_T}\right]}{\Tr_b\left[e^{-\beta H_b}\right]}},
\end{align}
where $H_T = H_{s_1} + H_{s_2} + H_{a_n^{L}} + H_{a_n^{R}} + H_{s_1s_2} + H_{s_1a^L_n} + H_{s_2a^R_n}$, and $H_b = H_{a_n^{L}} + H_{a_n^{R}}$ in scheme B, as elaborated in Sec.~\ref{Collision model}. The corresponding fidelity $\tilde F$ is given by 
\begin{align}
    \tilde F(\rho^{ss}, \sigma^*_{GS}) \equiv {\rm Tr}\sqrt{\sqrt{\rho^{ss}}\sigma^*_{GS}\sqrt{\rho^{ss}}},
    \label{HMF_Fidelity_formula}
\end{align}
where $\sigma^*_{GS} = e^{-\beta H^*_S}/\Tr\left[e^{-\beta H^*_S}\right]$ represents the Gibbs states obtained using the Hamiltonian of Mean Force.
\begin{figure}
    \centering
    \includegraphics[height=65mm,width=0.85\columnwidth]{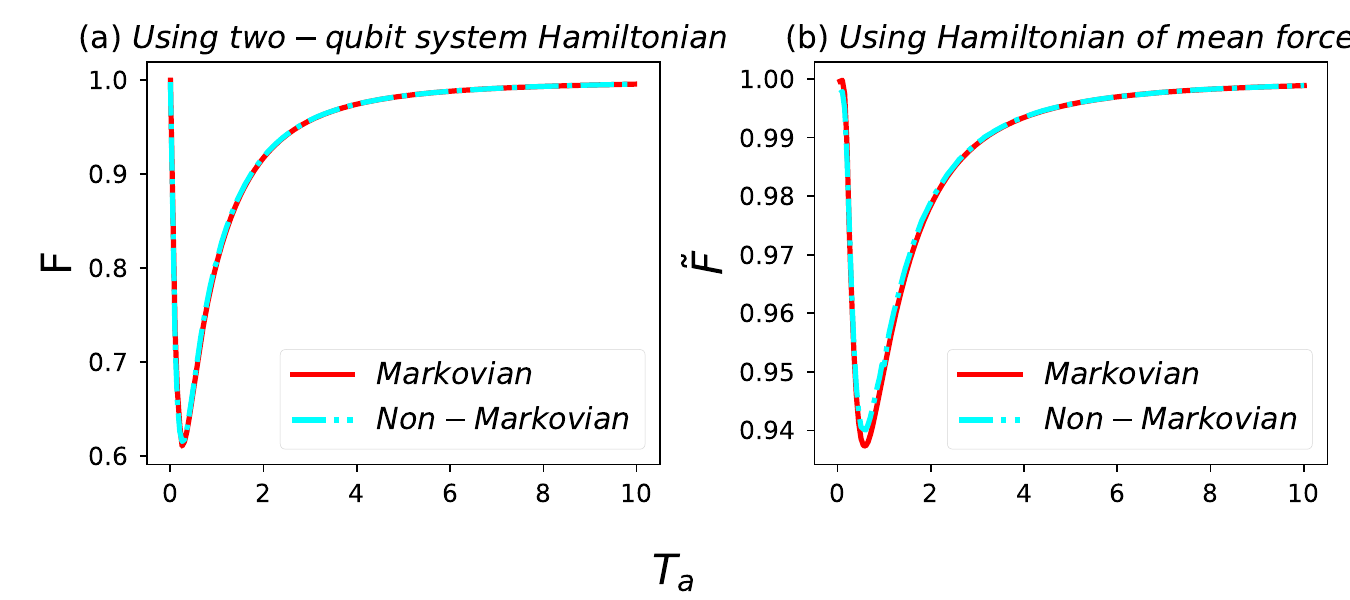}
    \caption{Evolution of fidelity between the two-qubit system's steady state and the Gibbs state corresponding to the two-qubit system Hamiltonian and Hamiltonian of mean force in subplots (a) and (b), respectively, under scheme B within the context of Markovian ($\Theta = 0$) and non-Markovian dynamics ($\Theta = 0.95\pi/2$) with the temperature of ancillae. Here, the temperature of ancillae, i.e., $T_{a^{L}_n} = 1/\beta_{a^{L}_n}$, $T_{a^{R}_n} = 1/\beta_{a^{R}_n}$, and $T_{a^{L}_n} = T_{a^{R}_n} = T_a$. The parameters in both the subplots are set as follows: $\omega_{s_1} = \omega_{s_2} = \omega_{a^L} = \omega_{a^R} = 1$,  $g_{s_1s_2} = 0.95$, $g_{s_1a^L} = g_{s_2a^R} = 0.5$, and $\Delta t = 0.1$.}
    \label{ss_fidelity_all_cases_two_qubit_together}
\end{figure}
Figure~\ref{ss_fidelity_all_cases_two_qubit_together} examines how the fidelities $F$ and $\tilde F$, defined in Eqs.~\eqref{Fidelity_formula} and~\eqref{HMF_Fidelity_formula}, respectively, vary with the temperature of the ancillae, under (non-)Markovian dynamics. In all scenarios, the system qubits
are strongly coupled, with coupling strength comparable to their transition frequencies, i.e, around 0.95, and the initial state is taken as $\ket{00}$. In the low temperature regime, the fidelity $F$ between the steady state and the Gibbs state corresponding to the two-qubit system Hamiltonian exhibits a sharp dip, see Fig.~\ref{ss_fidelity_all_cases_two_qubit_together}(a). As we increase the temperature, $F$ rises and saturates to unity, indicating that at high temperatures, the system's steady state is equivalent to the Gibbs state corresponding to the system Hamiltonian. Further, in Fig.~\ref{ss_fidelity_all_cases_two_qubit_together}(b), we plot the variation of the fidelity $\tilde F$ between the steady state and the Gibbs state obtained using the Hamiltonian of Mean Force. The pattern of variation is similar to the previous case; however, the local minima in the variation of $\tilde F$ are significantly higher than the local minima of $F$ (in the previous case). This behavior indicates that the fidelity between the steady state and the system's thermal equilibrium state significantly improves when the Hamiltonian of Mean Force is used. At high temperatures, both $F$ and $\tilde F$ saturate to unity, indicating that in this regime, the steady state, Gibbs state corresponding to the system Hamiltonian, and Gibbs state obtained using the Hamiltonian of Mean Force are equivalent. This could be attributed to the fact that at high temperatures, the system's Hamiltonian of Mean Force converges to the bare system Hamiltonian~\citep{Hanggi_talkner_review, Miller2018, tiwari2024strong, pathania2024}.  

Additionally, we examine whether the steady state of each qubit individually matches a thermal state of its bare system Hamiltonian. This is achieved by calculating the fidelity, using Eq.~\eqref{Fidelity_formula}, between each individual qubit's steady state and the corresponding Gibbs states associated with the bare system Hamiltonian of that qubit.
\begin{figure}
    \centering
    \includegraphics[height=75mm,width=0.75\columnwidth]{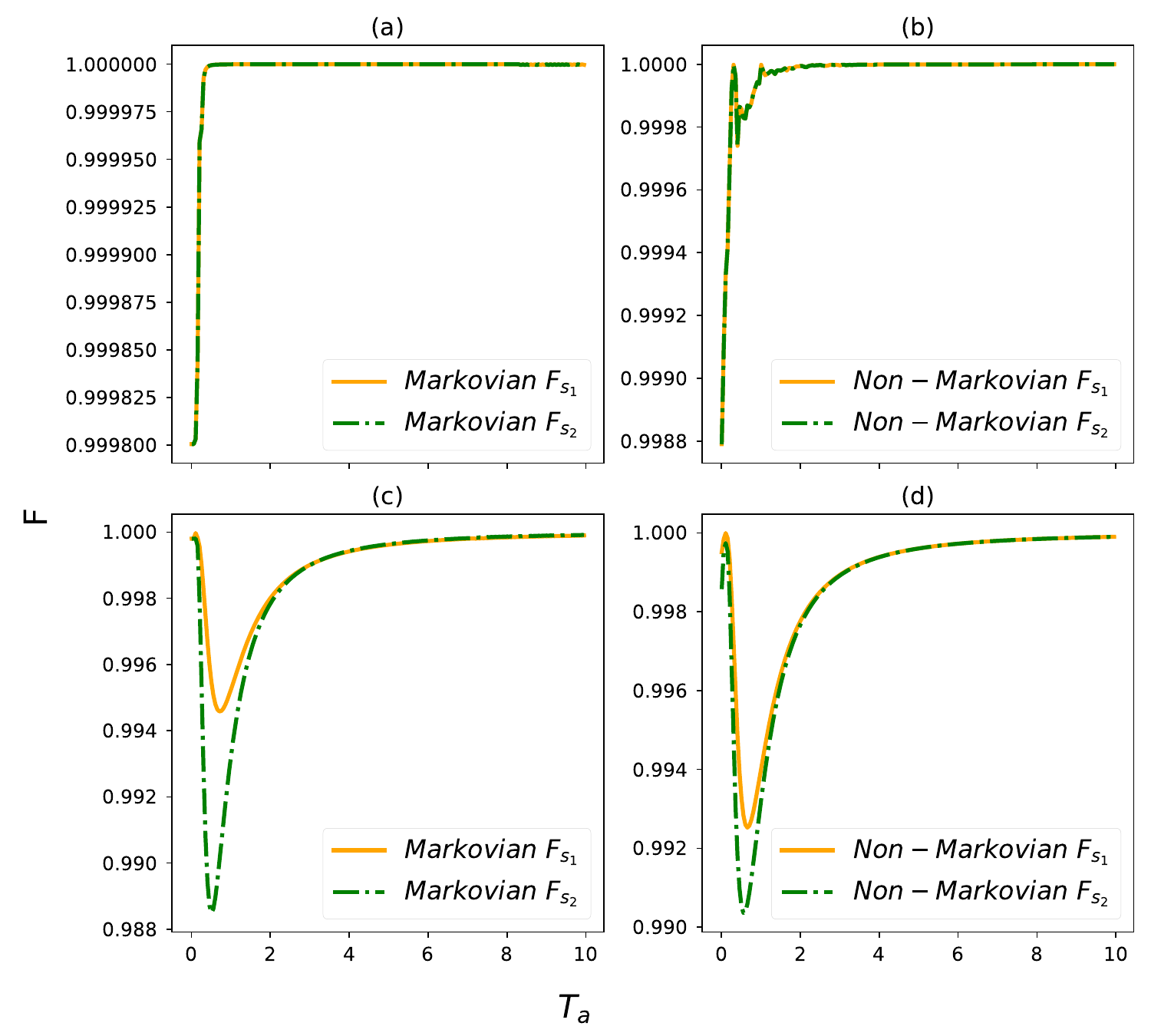}
    \caption{Evolution of fidelity between the each system's qubit individual steady state and the Gibbs state corresponding to its bare Hamiltonian with the temperature of ancillae, i.e., $T_{a^{L}_n} = 1/\beta_{a^{L}_n}$, and $T_{a^{R}_n} = 1/\beta_{a^{R}_n}$, under scheme B within the context of Markovian ($\Theta = 0$) and non-Markovian dynamics ($\Theta = 0.95\pi/2$). Here $T_{a^{L}_n} = T_{a^{R}_n} = T_a$. The parameters are set as follows: in subplots (a) and (b) $\omega_{s_1} = \omega_{s_2} = \omega_{a^L} = \omega_{a^R} = 1$,  $g_{s_1s_2} = 0.95$, in subplots (c) and (d) $\omega_{s_1} = \omega_{a^L} = 0.5$, $\omega_{s_2} = \omega_{a^R} = 1$, $g_{s_1s_2} = 0.95$ and for all plots $g_{s_1a^L} = g_{s_2a^R} = 0.5$, and $\Delta t = 0.1$.}
    \label{ss_fidelity_all_cases}
\end{figure}
In Fig.~\ref{ss_fidelity_all_cases}, we plot the variation of fidelity with the temperature of the ancillae when the initial state of the two-qubit system is $\ket{00}$. Figures~\ref{ss_fidelity_all_cases}(a) and~\ref{ss_fidelity_all_cases}(b) represent the case where all the ancillae and system qubits have the same transition frequencies. In contrast, Figs.~\ref{ss_fidelity_all_cases}(c) and (d) depict a scenario where the transition frequencies of the qubits differ, although the transition frequencies are related such that $\omega_{a^L_n} = \omega_{s_1}$ and $\omega_{a^R_n} = \omega_{s_2}$. In all the subplots, the coupling between the system qubits is taken to be of the order of their transition frequencies, approximately 0.95, denoting a strong coupling between them. Here, in the low-temperature regime, we observe that the fidelity between the steady state and the Gibbs state corresponding to the bare system Hamiltonian is lower than one and exhibits fluctuations. From Figs.~\ref{ss_fidelity_all_cases}(a) and (b), we observe that when all the ancillae and system qubits have the same transition frequencies, the difference in fidelity between the Markovian and non-Markovian cases is not very significant. On the other hand, when all the ancillae and system qubits' transition frequencies are different, the deviation in fidelity for the Markovian and non-Markovian cases becomes more pronounced. As the temperature increases, the fidelity stabilizes and eventually reaches unity in all cases. This denotes that the steady state becomes equal to the Gibbs state corresponding to the bare system Hamiltonian at higher temperatures. Typically, the thermal steady state of the system is characterized by the Gibbs state corresponding to an effective Hamiltonian, which differs from the bare system Hamiltonian at low temperatures and in the strong coupling regime. This becomes equal to the bare system Hamiltonian as the coupling decreases and with higher temperature, as observed in the previous case of two-qubits.

\section{\label{ch6_conclusion}Summary}
The quantum collision model has emerged as a robust and conceptually simple framework for simulating open quantum dynamics of single-qubit systems. However, its extension to multipartite systems remains comparatively underexplored. In this work, we have explored a two-qubit quantum collision model. Two schemes were introduced to model the two-qubit collision model. In scheme A, a stream of ancillae interacted with only one of the qubits of the two-qubit system. In scheme B, both the qubits interacted with independent streams of ancillae. Here, an ancilla interacts with the system and subsequently interacts with the next ancilla destined to interact with the system. This interaction between ancillae was shown to introduce memory effects into the system. This non-Markovian feature was explored using the BLP measure, where the trace distance between two different initial states showed oscillatory evolution. Additionally, the non-Markovianity measure displayed a positive, gradually increasing value with the number of collisions, effectively quantifying the degree of non-Markovianity in both schemes. The non-classical features of the model were studied first using the Wigner function and non-classical volume. For this, the two-qubit negative quantum state $\ket{NS_3^\prime}$ and the Bell states were used as the initial states of the system. It was found that, in scheme A, the $\ket{NS_3^\prime}$ state was advantageous, as it was robust against the decay of non-classicality for a longer duration. Next, the behavior of quantum correlations, particularly entanglement, was investigated using the concept of concurrence. In the Markovian case, entanglement was not significantly generated for the most excited and the ground states of the system. However, in the non-Markovian case, entanglement was generated for the most excited state, and it was robust for a larger number of collisions compared to the other states.

\subsection{Limitations and Scope}
The results of this chapter are obtained for specific parameter choices: intra-system coupling $g_{s_1 s_2} = 0.95$, system-ancilla coupling $g_{s_j a_n^e} = 0.85$ (and $0.5$ in steady-state analysis), collision time step $\Delta t = 0.08$ or $0.1$, and inverse temperatures $\beta_{a_n^e} = 1$ or $4$. These parameters are chosen to represent the strong-coupling, finite-temperature regime. The qualitative conclusions, that intra-ancilla interactions ($\Theta > 0$) generate non-Markovian dynamics, that the $|NS_3'\rangle$ state is more robust to non-classicality decay than the Bell state in Scheme A, and that entanglement is generated from separable initial states in the non-Markovian case, may change quantitatively for different coupling regimes. In particular, in the weak-coupling limit ($g_{s_j a_n^e} << \omega_s$), the Markovian and non-Markovian dynamics become increasingly similar, and the advantage of the $|NS_3'\rangle$ state may diminish. Additionally, the non-Markovianity is quantified solely using the BLP trace-distance measure for the specific initial pair $|\phi^+\rangle$ and $|\phi^{-}\rangle$; as discussed in Sec .~\ref {Non_Markovianity_measure}, this measure may underestimate the true non-Markovianity for other initial pairs. Future work should include parameter sweeps to delineate the boundary between regimes where the memory-induced non-classicality enhancement persists.


\newpage
\setcounter{chapter}{6} 

\titleformat{\chapter}[display]
{\sffamily\fontsize{27}{27}\bfseries\filleft}{\thechapter}{0pt}{{#1}}  
  
\thispagestyle{empty}

\chapter{Interrelation of Non-Classicality, Entropy, Irreversibility and Work extraction in Open Quantum Systems}\label{chap7:Interrelation}

\section{Introduction}
In this work, we investigate two broad classes of open quantum system models, namely, spin–spin and spin–boson interaction models, to encompass a wide range of physically relevant and experimentally realizable scenarios. The spin–spin interaction category includes the quantum collision model~\citep{ThermalizingQuantumMachines_2002, ziman2005description, Rybár_2012, Ciccarello2013Collision-model, McCloskey2014Non-Markovianity, Campbell2018Systemenvironment, csenyacsa2022entropy, lalita2025non_classicality} and the central spin model~\citep{NV2000Theory, Breuer2004Non_Markovian, He2019Exact, Mukhopadhyay2017Dynamics, Tiwari2022Dynamics, tiwari2024strong}, both of which are instrumental in describing system–environment coupling in finite-dimensional Hilbert spaces. The spin–boson interaction category, on the other hand, includes models such as the non-Markovian amplitude damping channel~\citep{Garraway1997Decay, nielsen2010quantum, Breuer_2012_foundations}, the Markovian generalized amplitude damping channel~\citep{Srikanth2008Squeezed, omkar2013dissipative}, and the Jaynes–Cummings model~\citep{Larson2021TheJaynes–Cummings, Jaynes1963Comparison, Garraway1997Decay}, which collectively capture memory effects and energy exchange between discrete quantum systems and bosonic reservoirs. Here, the collision models hold particular significance among spin–spin interaction frameworks due to their ability to simulate open-system dynamics through repeated and controllable system-environment interactions~\citep {Ciccarello2013Collision-model, McCloskey2014Non-Markovianity, lalita2025non_classicality}. Additionally, they can be demonstrated as numerically exact techniques~\citep{lacroix2025making}. Moreover, the collision models possess an intrinsic thermalization capability, enabling the system to asymptotically reach a thermal steady state determined by the statistical properties of the ancillas, thereby providing a microscopic and operational perspective on quantum thermalization~\citep{arisoy2019thermalization, lalita2025non_classicality, BANERJEE2023Thermalization}.

In the collision models, the system qubit interacts sequentially with a stream of ancillary qubits representing the environment. This discrete interaction sequence provides a natural physical interpretation of the dynamical map formalism and allows the transition between Markovian and non-Markovian behavior by tuning inter-ancilla correlations. Moreover, the collision model possesses an intrinsic thermalization capability, enabling the system to asymptotically reach a thermal steady state determined by the statistical properties of the ancillas, thereby providing a microscopic and operational perspective on quantum thermalization. Because of its operational simplicity, versatility, and thermalizing nature, the collision model has become a valuable theoretical and experimental tool for investigating decoherence mechanisms, quantum thermodynamic cycles, and information flow in open quantum systems.

An essential concept underlying the analysis of the above-specified models is non-classicality \citep{Wigner1932Quantum, Sudarshan1963Equivalence, Thapliyal2015Quasiprobability}, which embodies purely quantum features such as superposition, coherence, and entanglement that have no classical counterpart. Non-classicality serves as a fundamental resource in quantum technologies \citep{dowling2003quantum}, enabling quantum advantage in quantum computation~\citep{Horodecki2009QuantumEntanglement, Chandrashekar2007Symmetries}, secure communication~\citep{Ekert1991Quantumcryptography, Scarani2009The_security}, and precision measurement~\citep{Vittorio2004Quantum_Enhanced, Pezz2018Quantummetrology}. In the context of open quantum systems, it plays a dual role: while environmental interactions typically degrade non-classical correlations through decoherence, the persistence or regeneration of non-classical features, especially in non-Markovian regimes, offers deep insights into reversibility, memory effects, and energy exchange processes. Quantitative witness such as the non-classical volume~\citep{Anatole2004Negativity} are therefore indispensable for connecting the microscopic quantum behavior of systems to their macroscopic thermodynamic consequences.

A comprehensive characterization of open-system dynamics aims to bridge the informational and thermodynamic viewpoints. Four quantities, non-classical volume, von Neumann entropy, entropy production, and ergotropy, jointly provide complementary insights into this connection. The non-classical volume quantifies coherence and superposition~\citep{Anatole2004Negativity}; the von Neumann entropy characterizes statistical uncertainty and information loss~\citep{nielsen2010quantum}; entropy production captures the degree of irreversibility~\citep{Esposito2010Threefaces}; and ergotropy represents the extractable work through cyclic unitary operations~\citep{AEAllahverdyan_2004Maximalwork}. Together, these quantities elucidate how quantum resources transform into thermodynamic quantities. Recent investigations in quantum information, quantum thermodynamics, and quantum batteries have further highlighted the intricate connections between them ~\citep{Manfredi2000Entropy, Perarnau2015Extractable, Francica2020QuantumCoherence, Medina2025Anomalous, pathania2025quantum}. 

Building upon these insights, the present study systematically explores the interplay among non-classicality, von Neumann entropy, entropy production, and ergotropy across both spin–spin and spin–boson interaction models. By comparing the collision and central spin models, the non-Markovian amplitude damping and Markovian generalized amplitude damping channels, and the Jaynes–Cummings model, this work aims to uncover universal correspondences linking non-classicality, irreversibility, and extractable work. Although the inverse behavior of non-classical volume and von Neumann entropy might appear intuitively expected, there is no a priori guarantee that this trend persists across non-Markovian memory effects, thermal vs. non-thermal reservoirs, or strong-coupling regimes. We aim to demonstrate its robustness across five fundamentally different models and establish it as a structural feature of open-system evolution. Further, the comparison between ergotropy (a state function) and entropy production (a process-dependent quantity) is not intended as a quantitative relation. Rather, our motivation is to test whether a qualitative thermodynamic correspondence exists, namely, whether states that experience larger irreversible information loss during evolution tend to exhibit systematically lower extractable work, irrespective of the microscopic mechanism generating irreversibility. Such an investigation contributes toward a deeper understanding of how environmental structure, memory effects, and system–bath coupling jointly determine open quantum systems' informational and thermodynamic evolution.

The chapter is organized as follows. Section~\ref{Preliminaries} introduces the four key quantities, i.e., non-classical volume, von Neumann entropy, entropy production, and ergotropy. Section~\ref{spin-spin} analyzes spin–spin interaction models, namely the quantum collision and central spin models. Section~\ref{spin-boson} extends the investigation to spin–boson interactions, including the non-Markovian and generalized amplitude-damping channels and the Jaynes–Cummings model. Section~\ref{ch7_conclusion} concludes with a discussion on the conceptual implications of these findings for the design of resource-efficient quantum thermodynamic devices. \textit{This chapter's contents comes from~\citep{lalita2025interrelation}.}

\section{\label{Preliminaries} Preliminaries}
\subsection{Non-classical Volume}
The non-classical volume is a quantitative witness of non-classicality, initially introduced in the context of phase space quasi-probability distributions~\citep{Anatole2004Negativity}. It is defined using the Wigner function, $W(\theta, \phi)$, representing the quantum state in continuous phase space~\citep{Wigner1932Quantum, Thapliyal2015Quasiprobability}. Quantum states with non-classical characteristics, such as superposition and entanglement, can create regions of negative values of the Wigner functions, whereas classical states have a non-negative Wigner function. 

The non-classical volume, denoted by ${\delta}$, is computed as follows~\citep{Thapliyal2015Quasiprobability, lalita2025non_classicality},
\begin{equation}
    {\delta} = \int |W({\theta}, {\phi})| \sin{\theta} d{\theta} d{\phi} - 1.
    \label{NV_formula}
\end{equation}
This witness vanishes for states that can be described by classical probability distributions (e.g., coherent states) and increases with the degree of non-classicality. Importantly, ${\delta}$ is basis-independent, making it suitable for comparing different systems or studying the dynamical degradation of non-classicality under noise. It has found applications in quantifying decoherence in optical fields and qubits evolving under various open-system models~\citep{Thapliyal2016tomograms}.

\subsection{\label{Entropy}Von Neumann Entropy}
A fundamental concept in quantum information, the von Neumann entropy extends the classical Shannon entropy to quantum systems. For a quantum state represented by a density operator $\rho$, the von Neumann entropy is defined as~\citep{nielsen2010quantum} 
\begin{equation}
    S(\rho) = - \mathrm{Tr}(\rho \log \rho).
    \label{von_neumann_entropy}
\end{equation}
It captures the degree of uncertainty or mixedness associated with the state. For this reason, pure states have zero entropy, while maximally mixed states attain the highest entropy allowed by the system's Hilbert space dimension. 

Unlike classical entropy, the von Neumann entropy accounts for quantum superposition and entanglement, as a key tool for characterizing correlations between subsystems. In bipartite systems, it plays a central role in quantifying entanglement through the entropy of reduced density matrices~\citep{Horodecki2009quantum_ent}. Moreover, its concavity and invariance under unitary transformations underline its consistency as an information-theoretic measure. The von Neumann entropy not only bridges quantum physics and information theory but also appears in the thermodynamics of open quantum systems~\citep{Breuer2007}, and black hole physics~\citep{Bekenstein1973Blackholes, Jha2025probing}, making it indispensable for understanding both the informational and physical aspects of quantum mechanics.

\subsection{\label{Entropy production} Entropy production}
The quantum formulation of entropy production begins with the joint unitary evolution of a system $S$ and its environment $E$, initially in states $\rho_S$ and $\rho_E$. Their composite state after interaction through a global unitary operator $U$ is
\begin{equation}
    \rho'_{SE} = U(\rho_S \otimes \rho_E)U^\dagger.
\end{equation}
The system's reduced state follows from tracing out the environment, $\rho'_S = \mathcal{E}(\rho_S) = \mathrm{tr}_E\{\rho'_{SE}\}$, and it is precisely this partial trace that introduces irreversibility, since discarding inaccessible environmental degrees of freedom and system-environment correlations prevents exact reversibility. Entropy production in this framework is quantified as~\citep{Esposito2010Threefaces, Landi2021Irreversibleentropy}
\begin{equation}
    \Sigma = I^{\rho'_{SE}}(S: E) + S(\rho'_E \| \rho_E),
\end{equation}
where the first term is the mutual information created between the system and its environment,
$I^{\rho'_{SE}}(S: E) = S(\rho'_S) + S(\rho'_E) - S(\rho'_{SE})$, and the second term is the relative entropy between the final and initial environment states, $S(\rho \| \sigma) = \Tr[\rho \ln \rho - \rho \ln \sigma]$. These contributions together can be written more compactly as
\begin{equation}
    \Sigma = S(\rho'_{SE} \| \rho'_S \otimes \rho_E),
    \label{entropy_production_final}
\end{equation}
where $\rho'_{SE}$ and $\rho'_S$ represent the evolved states of the composite system-environment and the system, respectively. Furthermore, $\rho_E$ is the initial state of the environment fed to the system during evolution at each step. Altogether, these relations establish entropy production as a non-negative measure of lost information, arising from correlations with the environment and irretrievable changes in it.

\subsection{\label{Ergotropy} Ergotropy}
Ergotropy quantifies the maximum extractable work from a quantum state under unitary operations, without changing its entropy~\citep{AEAllahverdyan_2004Maximalwork}. For open quantum systems, ergotropy is determined by using the evolved state ($\rho'_s$) of the system as the input for the computation. This enables us to ascertain the limitations on work extraction through unitary transformations, including environmental influences~\citep{cakmak2020Ergotropy}. The ergotropy of a $d$-dimensional quantum system is given by~\citep{AEAllahverdyan_2004Maximalwork, Tiwari2023Impact}
\begin{equation}
    \mathcal{W}(\rho'_s) = \text{Tr}[\rho'_s H_s] - \min_{U} \text{Tr}[U \rho'_s U^\dagger Hs],
\end{equation} 
where the minimization identifies the passive state of $\rho'_s$. The passive state is obtained by rearranging the eigenvalues of $\rho'_s$ in decreasing order and aligning them with the eigenvectors corresponding to the increasing eigenvalues of $H_s$. This ensures no further work can be extracted through unitary operations.

For a single-qubit system, the Hamiltonian is generally expressed as $H_s = \frac{\omega_0}{2} \sigma_z$, where $\omega_0$ is the qubit frequency and $\sigma_z$ the Pauli operator. The mean energy of the system at time $t$ is $E(t) = \text{Tr}[\rho'_s H_s]$, while the corresponding passive energy is defined as
$E_{\text{pas}}(t) = \sum_i r_i \epsilon_i$, with $r_i$ being the ordered eigenvalues of $\rho'_s$, largest to smallest, \textit{viz.} $r_1 \geq r_2 \geq r_3 ...\geq r_d$ and $\epsilon_i$ the energy eigenvalues of $H_s$ smallest to largest, \textit{i.e.}, $\epsilon_1 \leq \epsilon_2 \leq \epsilon_3 ...\leq \epsilon_d$.
Finally, the ergotropy for a single-qubit state takes the explicit form~\citep{tiwari2024strong, Andolina_OQS_battery}
\begin{equation}
    \mathcal{W}(t) = \frac{\omega_0}{2} \left(r'_{z} + \sqrt{(r_{x}^{'2} + r_{y}^{'2} + r_{z}^{'2})} \right),
    \label{ergotropy_formula}
\end{equation}
where $r'_x, r'_y, r'_z$ are the Bloch vector components of $\rho'_s$.

\section{\label{spin-spin}Spin-Spin interaction Models}
This section explores the above-specified quantities under the spin-spin interaction models, particularly the quantum collision and central spin models. 
\subsection{\label{ch7_Collision model} Collision model}
\begin{figure}
    \centering
    \includegraphics[height=75mm,width=0.75\linewidth]{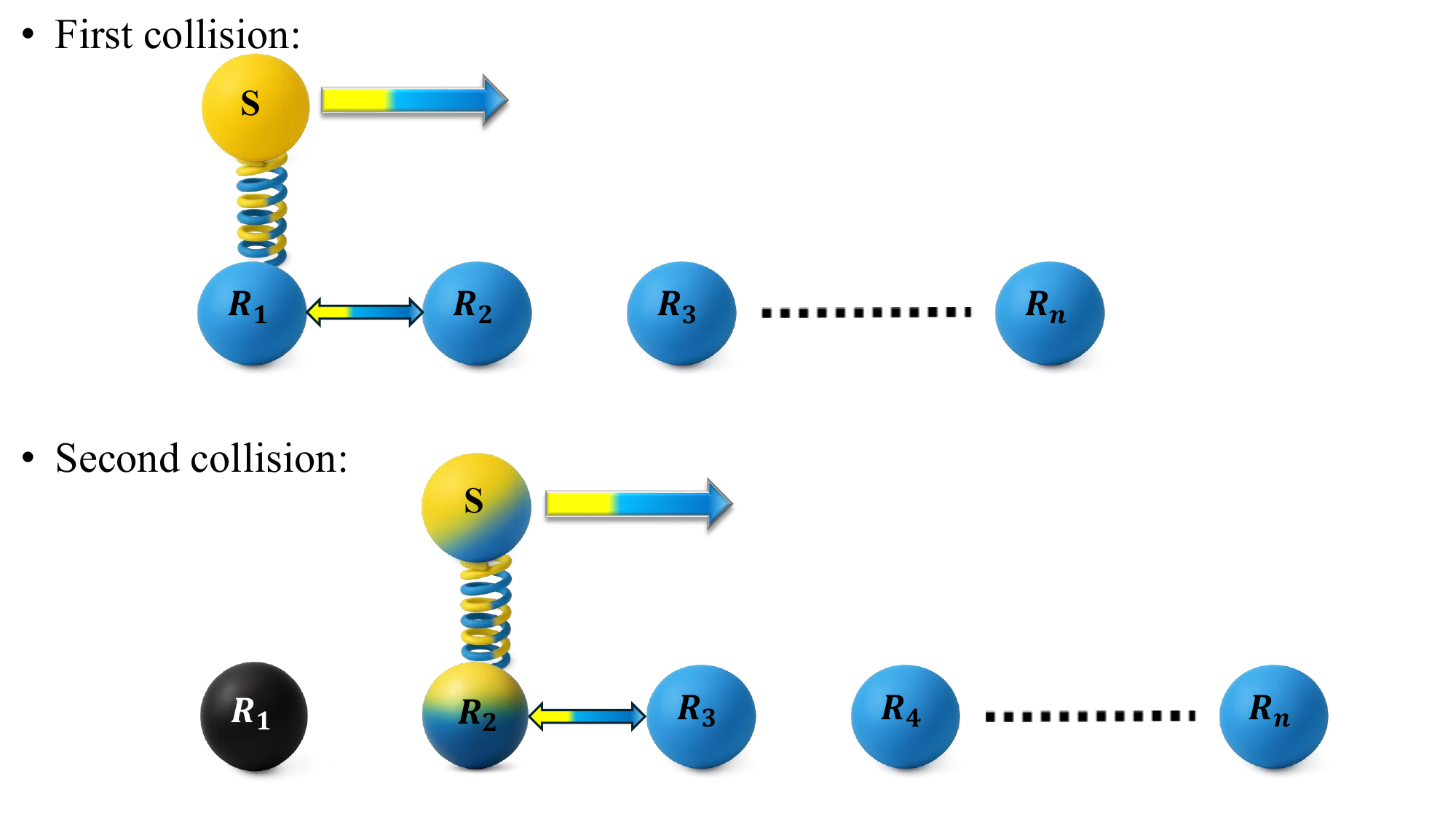}
    \caption{This diagram illustrates a single-qubit collision model in which the system qubit, denoted as $S$, interacts with a sequence of ancilla qubits $R_n$. The spiral lines represent a Heisenberg-type interaction between the system qubit $S$ and the ancillae. In contrast, the double-sided arrowed straight lines indicate a partial-swap interaction between successive ancilla qubits.}
    \label{Single_qubit_collision_model}
\end{figure}
The collision model provides a transparent and versatile way to describe the reduced dynamics of an open quantum system. The idea is to model the reservoir not as a single large environment but as a collection of small, identical subunits (called ancillas) interacting with the system $S$ one at a time. In this framework, the reservoir is represented by a sequence of ancillas $\{R_1, R_2, R_3, \dots \}$, each interacting with the system $S$ for a finite time interval $\tau$ ~\citep{Ciccarello2013Collision-model, McCloskey2014Non-Markovianity, Campbell2018Systemenvironment,  Ciccarello2022Quantumcollisionmodel}, as shown in Fig.~\ref{Single_qubit_collision_model}. Significantly, every ancilla interacts only once with the system before being discarded, and the next ancilla in the sequence takes its place as illustrated by the black ball ($R_1$) in the second collision in Fig.~\ref{Single_qubit_collision_model}. The total Hamiltonian of the combined setup is written as
\begin{equation}
    H_T^{CM} = H_S + H_{R_n} + H_{SR_n},
\end{equation}
where $H_S = \frac{{\hbar\omega}_{s}}{2}\sigma_z$ is the Hamiltonian of the system of interest, $H_{R_n} = \frac{\hbar\omega_{R_n}}{2}\sigma_z$ represents each reservoir's subunit Hamiltonian, and the Heisenberg interaction $H_{SR_n} = g_{SR}(\sigma_x^{S}\sigma_x^{R_n} + \sigma_y^{S}\sigma_y^{R_n})$,  accounts for the system–reservoir interaction. Further, $\omega_s$, $\omega_{R_n}$, and $g_{SR}$ denote the system frequency, ancilla frequency, and system-reservoir coupling strength, respectively. Now, to study the evolution of system $S$, let us consider the $n^{th}$ collision between the system and the ancilla $R_n$. The joint unitary evolution operator for this process is given by
\begin{equation}
    U_{SR_n} = \exp\left[-i~\left(H_T^{CM}\right)\tau \right],
\end{equation}
where $\tau$ is the collision (interaction) time. If $\rho_S^n$ and $\rho_{R_n}$, respectively, denote the states of the system and the ancilla $R_n$ before their interaction, then the global state of the pair after the collision is $\rho_{SR_n}' = U_{SR_n} \, \big(\rho_S^n \otimes \rho_{R_n}\big) \, U_{SR_n}^\dagger$. The state of the system after this step is obtained by tracing out the ancilla as
\begin{equation}
    \rho_S^{n+1} = \mathrm{Tr}_{R_n}\!\left[\rho_{SR_n}'\right].
\end{equation}
This construction makes the system's dynamics iterative; each new step depends on the system's state from the previous step and the fresh ancilla it encounters.
In our case, we assume that all ancillas are prepared identically in the same thermal state, i.e., 
$\rho_{R_n}(0) = e^{-\beta_{R_n}H_{R_n}}/\Tr[e^{-\beta_{R_n}H_{R_n}}]$, and $\beta_{R_n} = 1/k_BT_{R_n}$ is the inverse temperatures of the environment qubit. Further, we set $\hbar = k_B = 1$ throughout the paper. In its simplest version, once an ancilla has interacted with the system, it is discarded and plays no further role. The environment, therefore, has no memory, and the system dynamics are purely Markovian. Each collision is independent, and the reduced evolution is described by a sequence of completely positive trace-preserving (CPTP) maps. However, one of the strengths of the collision model is that it can also be extended to describe non-Markovian dynamics by introducing correlations into the environment~\citep{McCloskey2014Non-Markovianity, ThermalizingQuantumMachines_2002}. This is achieved by allowing inter-ancilla interactions between successive system–ancilla collisions.
Here, we consider the scheme where the system still interacts with the ancillas in the same sequential manner described above. However, this ancilla is not immediately discarded after the system has interacted with ancilla $R_n$. Instead, it can undergo an additional unitary interaction with the next ancilla $R_{n+1}$ before $R_{n+1}$ collides with the system, as illustrated in Fig.~\ref{Single_qubit_collision_model}. Let the unitary interaction between the successive reservoir subunits be governed by a partial swap operation described as~\citep{nielsen2010quantum, loss1998quantum}
\begin{equation}
    \begin{aligned}
        U_{R_n, R_{n+1}} = \cos(\Theta) I - i\sin(\Theta)H_{swap},   
    \end{aligned}
    \label{ch7_p_swap_unitary}
\end{equation}
where $\Theta \in [0, \frac{\pi}{2}]$, and $H_{swap} = \frac{1}{2}(\Vec{\sigma}{^{R_n}}.~\Vec{\sigma}{^{R_{n+1}}} + I_4)$. This modification introduces correlations between successive ancillas. The key effect is that some information about the system, which has been imprinted on $R_n$ during its collision, can be passed on to $R_{n+1}$. When $R_{n+1}$ subsequently interacts with the system, this stored information can flow back, leading to non-Markovian dynamics.

Now, we investigate the dynamics of non-classical volume $\delta$, Eq.~\eqref{NV_formula}, von-Neumann entropy $S$, Eq.~\eqref{von_neumann_entropy}, entropy production $\Sigma$, Eq.~\eqref{entropy_production_final}, and ergotropy $\mathcal{W}$, Eq.~\eqref{ergotropy_formula}, for this model. The initial state of the system evolving under the non-Markovian collision model is taken to be the maximally negative quantum state, $\ket{NS_1}$ state. 
The single-qubit $\ket{NS_1}$ state is the eigenstate corresponding to the most negative eigenvalue of the phase space point operator according to the single-qubit discrete Wigner function formalism, detailed in~\citep{lalita2023harnessing, Lalita_2024ProtectingQC, lalita2025realizingnegativequantumstates}. The explicit form of the single-qubit $\ket{NS_1}$ state, used in this work, in the Bloch vector form, is provided in chapter~\ref {chap3:Harnessing}, Sec.~\ref{ch3_NQS}.

\begin{figure}
    \centering
    \includegraphics[width=1\linewidth]{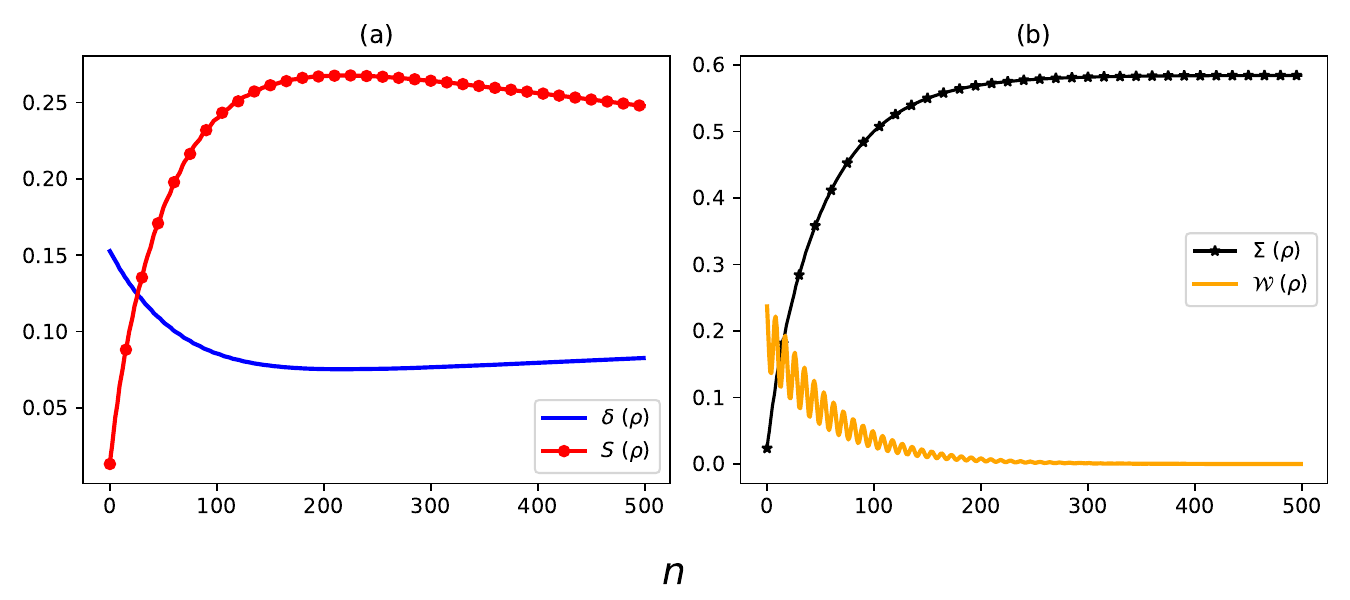}
    \caption{Variation of non-classical volume ($\delta$), von-Neumann entropy ($S$) in subplot $(a)$, and entropy production ($\Sigma$), ergotropy ($\mathcal{W}$) in subplot $(b)$ with the number of collisions of the single-qubit maximally negative quantum state using a non-Markovian collision model. The parameters are: $\omega_s = 1.5$, $\omega_R = 1$, $\beta = 50$, $g_{SR} = 0.5$, $\Theta = 0.98\frac{\pi}{2}$ and $\tau = 0.5$.}
    \label{non_Markovian_collision_model}
\end{figure}

Figure~\ref{non_Markovian_collision_model} illustrates the variation of non-classical volume, von-Neumann entropy, entropy production, and ergotropy with the number of collisions $n$ of the single-qubit non-Markovian collision model. An intriguing contrasting behavior is observed between the non-classical volume $\delta$ and von Neumann entropy $S$, as shown in Fig. \ref{non_Markovian_collision_model}(a). Specifically, a decrease in non-classicality corresponds to an increase in the system’s randomness. Figure \ref{non_Markovian_collision_model}(b) highlights a similar opposite relation between entropy production and ergotropy, demonstrating that reduced irreversibility facilitates greater work extraction. Moreover, a closer examination of Figs. \ref{non_Markovian_collision_model}($a$) and ($b$) reveals that higher von Neumann entropy is accompanied by increased entropy production, while a decline in non-classical volume is associated with a corresponding reduction in ergotropy.  

\subsection{\label{Central spin model} Central spin model}
The central spin model is a fundamental framework in quantum many-body physics. It is often described as a single two-level system (the central spin) interacting with a bath of surrounding spins, arranged symmetrically around it~\citep{NV2000Theory, Tiwari2022Dynamics, tiwari2024strong, Breuer2004Non_Markovian}. For this model, the Hamiltonian of the composite system is given by
\begin{equation}
    H = H_S + H_B + V,
\end{equation}
with individual contributions (for $\hbar = 1$)
\begin{equation}
    H = \frac{\omega_0}{2}\sigma_z^0 + \frac{\omega}{N}J_z + \frac{{\epsilon'}}{\sqrt{N}}\left(\sigma_x^0 J_x + \sigma_y^0 J_y\right).
\end{equation}
Here, $\omega_0$ is the transition frequency of the central spin, $\omega/N$ is the scaled frequency of the collective bath and $N$ is the number of bath spins, ${\epsilon'}$ is the system–bath coupling strength, $\sigma_\alpha^0$ $(\alpha = x,y,z)$ are Pauli operators of the central spin, and $J_\alpha = \frac{1}{2}\sum_{i=1}^N \sigma_\alpha^{(i)}$ are collective angular momentum operators of the bath spins. The global unitary evolution of the joint system–bath state for an initial separable condition $\rho_{SB}(0) = \rho_S(0) \otimes \rho_B(0)$ is given by
\begin{equation}
    \rho_{SB}(t) = e^{-iHt}\,\rho_{SB}(0)\,e^{iHt}.
\end{equation}
The reduced dynamics of the central spin is then obtained by tracing out the bath degrees of freedom as
\begin{equation}
    \rho'_S = \text{Tr}_B\left[e^{-iHt}\,\rho_{SB}(0)\,e^{iHt}\right].
\end{equation}
The bath spins are initially considered to be in a thermal state (Gibbs state), obtained using the spectral decomposition of the bath Hamiltonian
$H_B = \frac{\omega}{N} J_z = \frac{\omega}{2}\sum_{n=0}^N \left(1 - \frac{2n}{N}\right)|n\rangle\langle n|$ as
\begin{equation}
    \rho_B(0) = \frac{e^{-\beta H_B}}{Z} = \frac{1}{Z}\sum_{n=0}^N e^{-\frac{\beta \omega}{2}\left(1 - \frac{2n}{N}\right)}|n\rangle\langle n|,
\end{equation}
where $Z = \sum_{n=0}^N e^{-\frac{\beta \omega}{2}\left(1 - \frac{2n}{N}\right)}$ with inverse temperature $\beta = 1/k_BT$ and $\ket{n}$ is the standard computational basis.
This formulation sets the stage for deriving the exact dynamics of the central spin. The time-dependent reduced state $\rho'_S$ can be explicitly calculated by diagonalizing the total Hamiltonian $H$ numerically. This exact solvability makes the central spin model a powerful tool for probing strong-coupling and non-Markovian quantum thermodynamics. 

\begin{figure}
    \centering
    \includegraphics[width=1\linewidth]{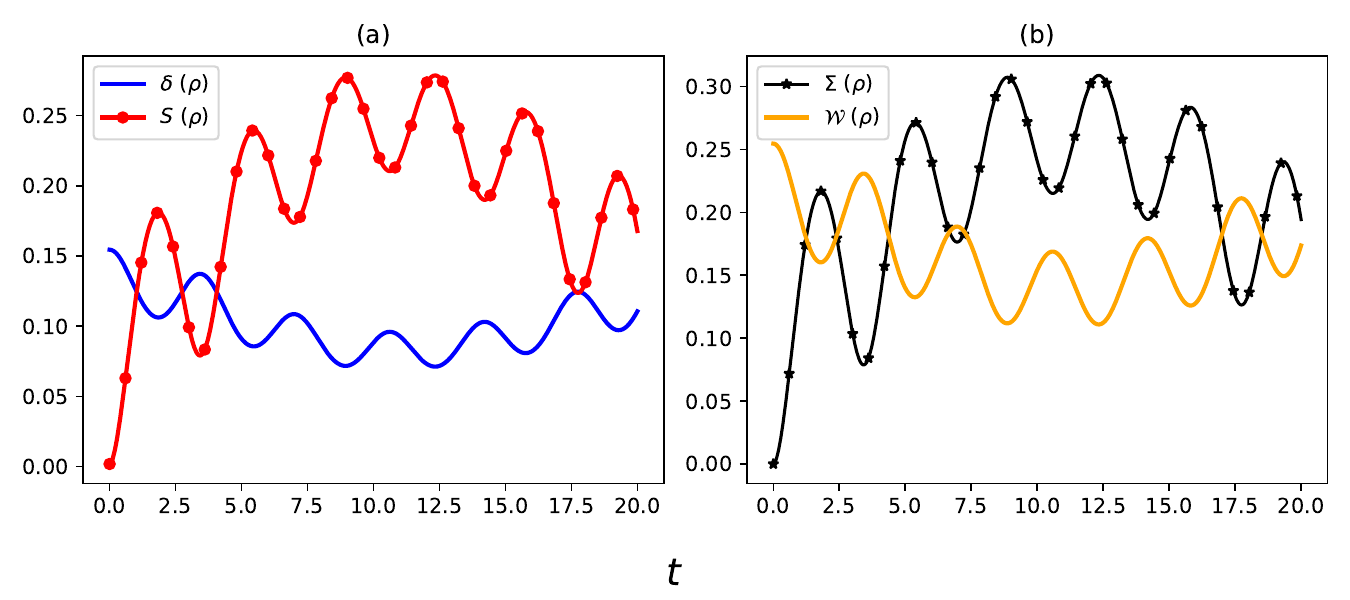}
    \caption{Variation of non-classical volume ($\delta$), von-Neumann entropy ($S$) in subplot $(a)$, and entropy production ($\Sigma$), ergotropy ($\mathcal{W}$) in subplot $(b)$ with time using the central spin model of the single-qubit maximally negative quantum state. The parameters are: $\omega_0 = 1.5$, $\omega = 1$, $\beta = 100$, ${\epsilon'} = 0.5$ and $N = 50$.}
    \label{Central_spin_model}
\end{figure}

The dynamics of the non-classical volume and above-specified thermodynamic quantities for the single-qubit $\ket{NS_1}$ state using the central spin model are studied using Eqs. \eqref{NV_formula}, \eqref{von_neumann_entropy}, \eqref{entropy_production_final}, and \eqref{ergotropy_formula}. Within this spin–spin interaction model, we also observe a contrasting relationship between $\delta$ and $S$, as well as between $\Sigma$ and $\mathcal{W}$, as shown in Fig.~\ref{Central_spin_model}, which mirrors the correspondence found in the collision model. Moreover, the mutual interdependence among $S$ and $\Sigma$, $\delta$, and $\mathcal{W}$ is preserved, highlighting the consistent thermodynamic structure underlying the system’s dynamics in a thermal spin environment.

\section{\label{spin-boson}Spin-boson interaction Models}
Next, we study the interrelation between non-classical volume and the above-specified quantum thermodynamic quantities for a spin interacting with the bosonic environment. In particular, we consider the non-Markovian amplitude damping (NMAD) channel, the generalized amplitude damping (GAD) channel, and the Jaynes-Cummings Model (JCM).

\subsection{\label{Garraway Model} Non-Markovian amplitude damping model}
This model studies the decay of a two-level atom interacting with a bosonic reservoir~\citep{Garraway1997Decay, Breuer_2012_foundations}. The total Hamiltonian governing the system bath setup is given by
\begin{equation}
    H = H_s \otimes I_B + I_s \otimes H_B + H_I,
\end{equation}
where $H_s$ is the system Hamiltonian, $H_B$ is the bath Hamiltonian, and $H_I$ describes their interaction. For the qubit system,
\begin{equation}
    H_s = \frac{\omega_0}{2} \sigma_z,
\end{equation}
where $\omega_0$ is the system qubit frequency. The environment is modeled as a collection of harmonic oscillators,
\begin{equation}
    H_B = \sum_k \omega_k a_k^\dagger a_k,
\end{equation}
where $a_k^\dagger$ and $a_k$ are bosonic creation and annihilation operators satisfying $[a_k,a_{k'}^\dagger] = \delta_{k,k'}$. The interaction Hamiltonian is given by
\begin{equation}
    H_I = \sum_k \left( g_k \sigma^+ \otimes a_k + g_k^* \sigma^- \otimes a_k^\dagger \right),
\end{equation}
where $g_k$ is the coupling constant and $\sigma^+ = |e\rangle\langle g|$ and $\sigma^- = |g\rangle\langle e|$ denote the atomic raising and lowering operators, with $\ket{g}$ and $\ket{e}$ being the ground and excited states, respectively. Here, the initial state of the bath is taken to be the vacuum state. The total number of excitations, for this model, is conserved, which in this case is one. The reduced dynamics of the system follow the following master equation~\citep{Breuer1999Stochastic}
\begin{equation}
    \begin{aligned}
       \frac{d}{dt}\rho_s(t) &= -\frac{i}{2} S(t)[\sigma^+\sigma^-,\rho_s(t)]\\
       &+ \gamma(t)\Big(\sigma^- \rho_s(t) \sigma^+ - \tfrac{1}{2}\{\sigma^+\sigma^-,\rho_s(t)\}\Big), 
    \end{aligned}
    \label{dynamics_eq}
\end{equation}
where $\gamma(t) = -2 \,Re\left(\frac{\dot{G}(t)}{G(t)}\right)$ and $S(t) = -2 \,Im\left(\frac{\dot{G}(t)}{G(t)}\right)$ are the decay rate and the time-dependent frequency shift, respectively. For a Lorentzian spectral density of the bath in resonance with the qubit frequency, the function $G(t)$ becomes
\begin{equation}
    G(t) = e^{-\lambda t/2}\!\left[ \cosh\!\left(\tfrac{lt}{2}\right) + \tfrac{\lambda}{l}\sinh\!\left(\tfrac{lt}{2}\right) \right],
\end{equation}
where $l = \sqrt{\lambda^2 - 2\gamma_0 \lambda}$, with $\gamma_0$ being the system-bath coupling strength and $\lambda$ being the spectral width of the bath. Further, $-2\frac{\dot{G}(t)}{G(t)} = 2\left(\frac{\gamma_0}{ \sqrt{1 - \tfrac{2\gamma_0}{\lambda}} \, \coth\!\left(\tfrac{1}{2}\lambda t\sqrt{1-\frac{2\gamma_0}{\lambda}}\right) + 1 }\right)$.
For $\lambda < 2\gamma_0$, the decay rate becomes negative in certain intervals, leading to non-Markovian amplitude damping (NMAD) evolution, while $\lambda > 2\gamma_0$ gives time-dependent Markovian dynamics, and $\lambda \gg \gamma_0$ reduces to the standard time-independent amplitude damping channel.
Moreover, the qubit state at time $t$, $\rho_s(t)$ using Eq.~\eqref{dynamics_eq} can be expressed in Bloch vector form as
\begin{equation}
    \rho_s(t) = \tfrac{1}{2} \begin{pmatrix} 1+z(t) & x(t)-iy(t) \\ x(t)+iy(t) & 1-z(t) \end{pmatrix},
    \label{rho_bloch_vector}
\end{equation}
with $x(t) = \text{Tr}[\sigma_x \rho(t)]$, $y(t) = \text{Tr}[\sigma_y \rho(t)]$, and $z(t) = \text{Tr}[\sigma_z \rho(t)]$. Solving the dynamics using Eq. \eqref{rho_bloch_vector} gives
\begin{equation}
   \begin{aligned}
       \rho_{00}(t) &= \big(1-|G(t)|^2\big)\rho_{11}(0) + \rho_{00}(0), \\
    \rho_{01}(t) &= \rho_{01}(0)\, G^*(t), \\
    \rho_{10}(t) &= \rho_{10}(0)\, G(t), \\
    \rho_{11}(t) &= \rho_{11}(0)|G(t)|^2,
    \label{rho_elements}
   \end{aligned}
\end{equation}
where $\rho_{ij}$'s are the elements of the density matrix $\rho(t)$. On comparing Eqs. \eqref{rho_bloch_vector} and \eqref{rho_elements}, the Bloch vector components are given as
\begin{equation}
   \begin{aligned}
    x(t) = 2\,Re[\rho_{10}(0) G(t)], \\
    y(t) = -2\,Im[\rho_{10}(0) G(t)], \\
    z(t) = 2\rho_{11}(0)|G(t)|^2 - 1.
   \end{aligned}
\end{equation}

\begin{figure}
    \centering
    \includegraphics[width=1\linewidth]{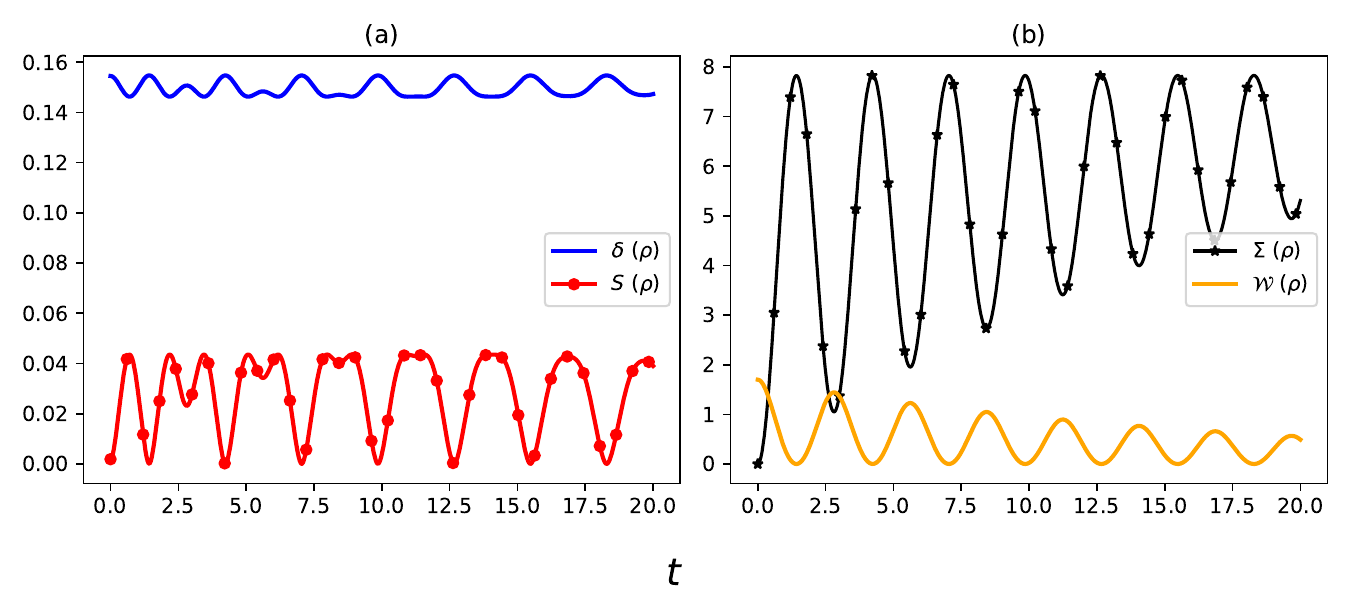}
    \caption{Variation of non-classical volume ($\delta$), von-Neumann entropy ($S$) in subplot $(a)$, and entropy production ($\Sigma$), ergotropy ($\mathcal{W}$) in subplot $(b)$ with time of the single-qubit maximally negative quantum state under the NMAD channel. The parameters are: $\omega_0 = 10$, $\lambda = 0.05$ and $\gamma_0 = 50$.}
    \label{NMAD_channel}
\end{figure}
Figure~\ref{NMAD_channel} shows the variation of the non-classical volume, von-Neumann entropy, entropy production, and ergotropy with time under the NMAD channel. The system's initial state is taken to be the $\ket{NS_1}$ state. The NMAD channel satisfies the global fixed point condition, viz., $U[\rho_{s}^{*} \otimes \rho_E]U^{\dag} = \rho_{s}^{*} \otimes \rho_E$, where $\rho_{s}^{*}$ is the steady state of the system and $\rho_E$ is the initial state of the bath. For this condition, the formula for the entropy production becomes~\citep{Landi2021Irreversibleentropy},
\begin{equation}
    \Sigma = S(\rho_s \| \rho_{s}^{*}) - S(\rho'_s \| \rho_{s}^{*}),
    \label{entropy_production_rho_th}
\end{equation}
where $\rho_s$ and $\rho_s'$ are the initial and evolved states of the system.

From Fig.~\ref{NMAD_channel}(a), it can be observed that the variations of $\delta$ and $S$ exhibit a contradictory relationship. Similarly, the dynamics of $\Sigma$ and $\mathcal{W}$ also display contradictory behavior in the spin–boson interaction model, as shown in Fig.~\ref{NMAD_channel}(b), in close analogy with the spin–spin interaction model. However, a notable difference arises in their mutual dependence. In this case, the fluctuations of $S$ and $\Sigma$ no longer follow a similar trend, and the oscillations of $\delta$ and $\mathcal{W}$ are also not synchronized, as illustrated in Fig.~\ref{NMAD_channel}(a) and (b). A reason for this deviation could be the initial state of the bosonic bath, which is taken to be the ground state here. 

To take into account the impact of the thermal state as the initial state of the bosonic bath in the spin-boson interaction, we study the dynamics of the non-classical volume and the above-considered thermodynamic quantities under the generalized amplitude channel and the Jaynes-Cummings model.

\subsection{\label{GAD channel} Generalized amplitude damping channel}
The generalized amplitude damping (GAD) channel models a Markovian evolution due to a finite-temperature environment. It captures decay from the excited to the ground and thermal excitation from the ground to the excited states. The corresponding master equation is give by~\citep{Breuer2007, Srikanth2008Squeezed, omkar2013dissipative},
\begin{equation}
    \begin{aligned}
        \frac{d}{dt}\rho_s(t) &= -i[H_s,\rho_s(t)] \\
                            &+ \gamma(N^{th} + 1)\Big(\sigma^- \rho_s(t) \sigma^+ - \tfrac{1}{2}\{\sigma^+\sigma^-,\rho_s(t)\}\Big) \\
                            &+ \gamma N^{th} \Big(\sigma^+ \rho_s(t) \sigma^- - \tfrac{1}{2}\{\sigma^-\sigma^+,\rho_s(t)\}\Big),
    \label{GAD_dynamics_eq}
    \end{aligned}
\end{equation}
where $H_s = \frac{\omega_0}{2}\sigma_z$ is the system Hamiltonian, $\gamma$ is the dissipative constant, and $N^{th} = \frac{1}{e^{\beta \omega_0}- 1}$ is the mean thermal photon number of the bath at a finite temperature $T$. Also, the bosonic bath in this channel is taken to be the thermal state, i.e., $\rho_{B}(0) = e^{-\beta H_b}/\Tr[e^{-\beta H_b}]$, and $\beta = 1/k_BT$. The GAD channel also satisfies the global fixed point condition, and its entropy production is calculated using Eq.~\eqref{entropy_production_rho_th}.

\begin{figure}
    \centering
    \includegraphics[width=1\linewidth]{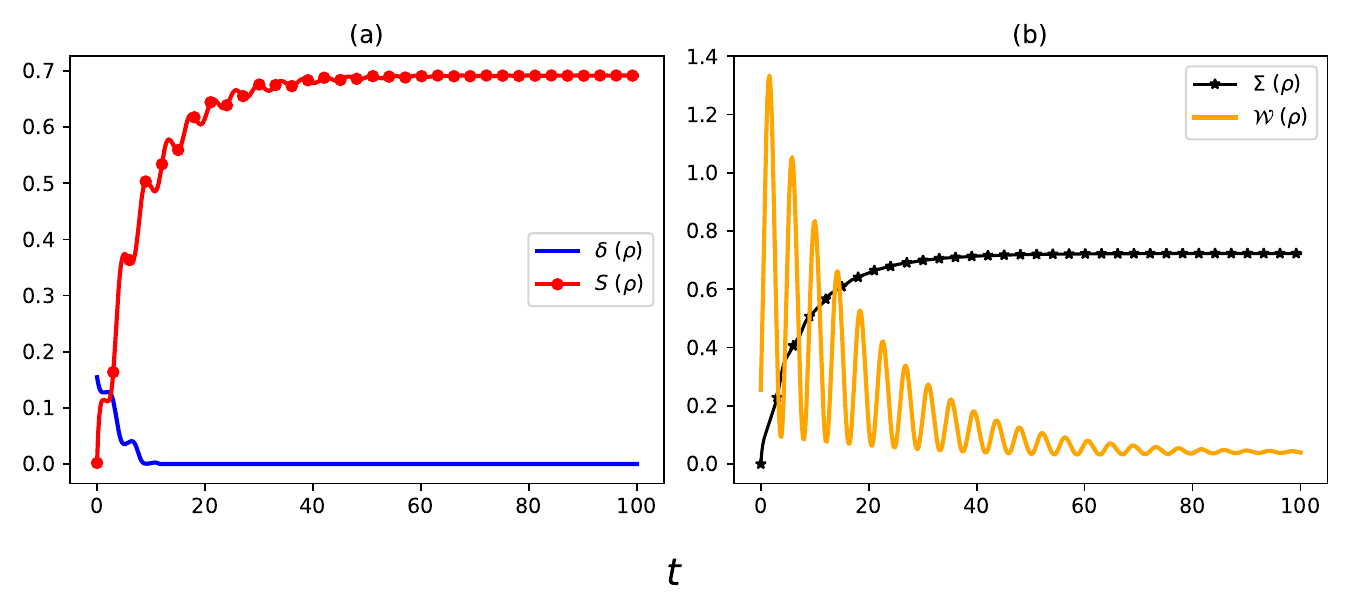}
    \caption{Variation of non-classical volume ($\delta$), von-Neumann entropy ($S$) in subplot $(a)$, and entropy production ($\Sigma$), ergotropy ($\mathcal{W}$) in subplot $(b)$ with time of the one-qubit maximally negative quantum state. The evolution is governed by the Markovian generalized amplitude damping master equation. The parameters are: $\omega_0 = 1.5$, $\beta = 1$, and $g = 0.05$.}
    \label{Markovian_GAD_channel}
\end{figure}

Figure~\ref{Markovian_GAD_channel} illustrates the variation of $\delta$, $S$, $\Sigma$, and $\mathcal{W}$ of the single-qubit $\ket{NS_1}$ state with time under the Markovian GAD channel. We can observe from Fig.~\ref{Markovian_GAD_channel}(a) that $\delta$ and $S$ behave in an opposite manner. The $\Sigma$ and $\mathcal{W}$ also exhibit contradictory behavior as illustrated by Fig.~\ref{Markovian_GAD_channel}(b). Moreover, the rise and fall patterns of $\delta$ and $\mathcal{W}$ and those of $S$ and $\Sigma$ remain in agreement with the trends observed in the collision and central spin models. This consistency arises because the bosonic reservoir is initialized in a thermal state in the GAD channel, which permits energy exchange between the system and the bath. 
 
\subsection{\label{Jaynes-Cummings model}Jaynes-Cummings model}
The Jaynes-Cummings model is a single-mode Garraway model, i.e., a qubit inside a single bosonic mode of frequency $\omega_c$~\citep{Jaynes1963Comparison, Garraway1997Decay, Larson2021TheJaynes–Cummings}. The Hamiltonian of the Jaynes-Cummings model is
\begin{equation}
    H_{\mathrm{JC}}=\tfrac{\omega_0}{2}\sigma_z + \omega_c a^\dagger a + g(\sigma^+ a + \sigma^- a^\dagger),
\end{equation}
where $\tfrac{\omega_0}{2}\sigma_z$, $\omega_c a^\dagger a$, and $g(\sigma^+ a + \sigma^{-} a^\dagger)$ are the system $H_s$, bath $H_b$, and interaction $H_{sb}$ Hamiltonians, respectively. Here, the initial state of the bosonic bath is taken to be the thermal state, i.e., $\rho_{B}(0) = e^{-\beta H_b}/\Tr[e^{-\beta H_b}]$, and $\beta = 1/k_BT$, where $T$ is the temperature of the bosonic bath.

\begin{figure}
    \centering
    \includegraphics[width=1\linewidth]{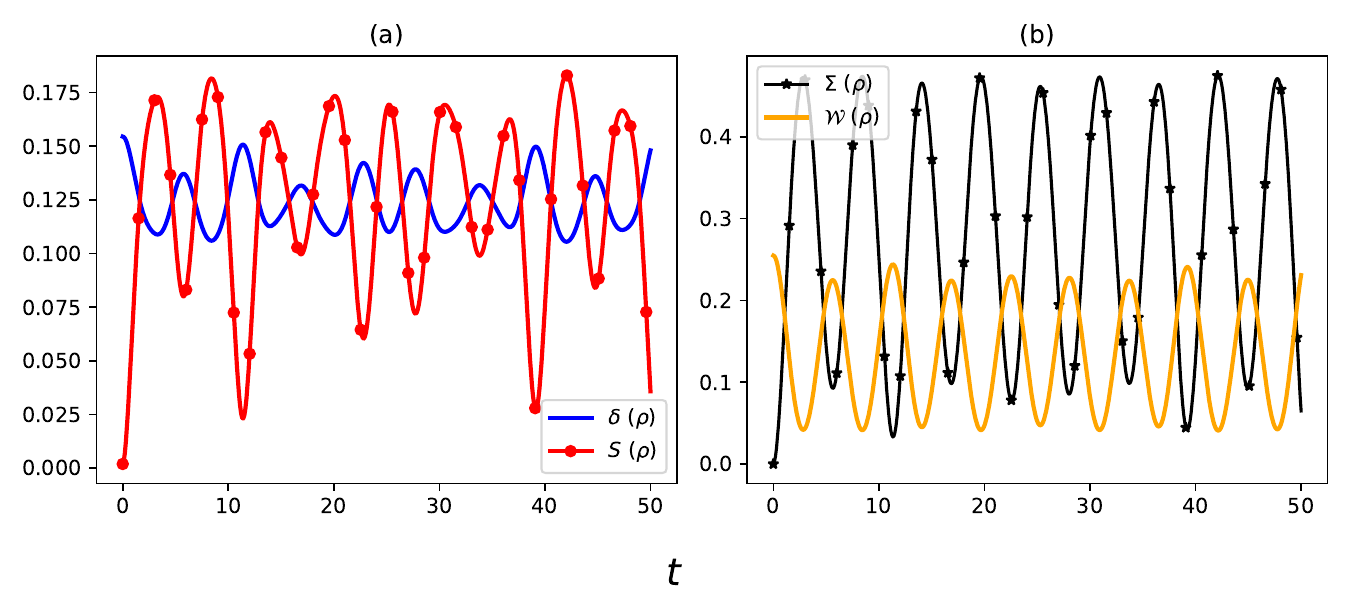}
     \caption{Variation of non-classical volume ($\delta$), von-Neumann entropy ($S$) in subplot $(a)$, and entropy production ($\Sigma$), ergotropy ($\mathcal{W}$) in subplot $(b)$ with time of the one-qubit maximally negative quantum state using the Jaynes-Cummings model. The parameters are: $\omega_0 = 1.5$, $\omega_c = 1$, $\beta = 3$, $g = 0.5$, and $\tau = 0.5$.}
    \label{Jaynes_cummings_model}
\end{figure}

The dynamics of $\delta$, $S$, $\Sigma$, and $\mathcal{W}$ for the single-qubit $\ket{NS_1}$ state evolved using the Jaynes–Cummings Hamiltonian are analyzed using Eqs. \eqref{NV_formula}, \eqref{von_neumann_entropy}, \eqref{entropy_production_final}, and \eqref{ergotropy_formula}. As shown in Fig.~\ref{Jaynes_cummings_model}(a) and (b), this model also exhibits a consistent contradictory relationship, viz., $\delta$ evolves in opposition to $S$, while $\Sigma$ displays contradictory behavior with respect to $\mathcal{W}$. Moreover, the rise and fall patterns of $\delta$ and $\mathcal{W}$ and those of $S$ and $\Sigma$ remain in agreement with the trends observed in the collision model, central spin model, and the GAD channel. This consistency arises because the bosonic reservoir is initialized in a thermal state in the Jaynes-Cummings model as well.

\section{\label{ch7_conclusion}Summary}
A comparative analysis of different open quantum system models, namely the collision model, the central spin model, the spin–boson interaction model under the non-Markovian amplitude damping (NMAD) channel, the Markovian generalized amplitude damping (GAD) channel, and the Jaynes–Cummings model revealed a remarkable consistency in the thermodynamic interplay among non-classical volume $\delta$, von Neumann entropy $S$, entropy production $\Sigma$, and ergotropy $\mathcal{W}$. 
In all the cases, $\delta$ evolved in opposition to $S$. At the same time, $\Sigma$ displayed a contradictory relation with $\mathcal{W}$, emphasizing that enhanced randomness corresponds to reduced non-classicality and loss in accessible information and vice-versa. In contrast, reduced irreversibility allows for more work to be extracted.

Despite this universal structure, notable distinctions emerged depending on the nature of the reservoir. In the NMAD channel, where the bosonic bath was restricted to its ground state, correlated fluctuations between $\delta$ and $\mathcal{W}$, and $S$ and $\Sigma$ lose synchronization. In contrast, the collision model, central spin model, GAD channel, and Jaynes–Cummings model involving thermal reservoirs exhibited sustained, correlated fluctuations, preserving the interdependence among the four quantities.

These results highlight a universal interrelation among fundamental quantum thermodynamic and information quantifiers across different system environment interaction models. The non-classical volume $\delta$ evolves in opposition to von Neumann entropy $S$, while entropy production $\Sigma$ contrasts with ergotropy $\mathcal{W}$. The degree of correlation between entropy and entropy production, as well as between non-classical volume and ergotropy, also depends on the bath state, highlighting the crucial role of reservoir preparation in determining these quantifiers.

\subsection{Limitations and Scope of the Thermodynamic Correspondences}
The correspondences between non-classical volume, entropy, entropy production, and ergotropy established in this chapter are qualitative: they demonstrate a consistent directional relationship (non-classical volume decreases as entropy increases; entropy production increases as ergotropy decreases) across five distinct open-system models for various initial states. This consistency is the key finding, and it suggests these are structural features of open-system evolution rather than model-specific accidents. However, several important qualifications apply. The correspondences do not represent quantitative equalities or bounds; the rates of change of these quantities differ across models and are not compared here. Providing quantitative bounds on these correspondences, for example, a lower bound on ergotropy given the entropy production, or an upper bound on non-classical volume given the von Neumann entropy, would require model-specific analytical calculations. This constitutes a meaningful direction for future theoretical work, potentially connecting to recent results in quantum thermodynamics on the relationship between coherence and work extraction~\citep{Quantum_Coherence2020Francica}.


\newpage
\setcounter{chapter}{7} 

\titleformat{\chapter}[display]
{\sffamily\fontsize{27}{27}\bfseries\filleft}{\thechapter}{0pt}{{#1}}  
  
\thispagestyle{empty}

\chapter{Conclusion}\label{chap8:conclusion}
The investigations presented in this thesis aim to understand how genuinely quantum features behave when quantum systems interact with noisy environments, and how such features may be preserved, recovered, or meaningfully utilized. Across five interconnected studies, each focusing on a different facet of open-system dynamics, the thesis gradually developed a unified picture of non-classicality as a dynamic, controllable, and operationally significant resource. The narrative that unfolded began with the discrete phase-space representation of quantum states, moved through measurement-based protection schemes and real-device implementation, expanded into microscopic dynamical modeling, and finally connected these insights to the thermodynamic structure underlying irreversibility and work extraction.

The first part of the thesis, devoted to \textit{Harnessing quantumness of states using discrete Wigner functions under (non)-Markovian quantum channels}~\citep{lalita2023harnessing}, laid the conceptual foundation by employing the discrete Wigner representation as a highly expressive and operationally meaningful tool. In this framework, the appearance of negative values serves as a direct signature of non-classicality, enabling one to visualize and quantify quantum interference in a manner not easily accessible through density matrices alone. By evolving qubit, qutrit, and two-qubit states under both Markovian and non-Markovian noise models, the study revealed a vibrant picture of how negativity decays, persists, and in some cases revives. The distinction between environments with and without memory played a central role: non-Markovian channels, endowed with information backflow, slowed the disappearance of negativity and enabled temporary recovery of non-classical features. Within this landscape, the negative quantum states emerged as remarkably resilient, preserving quantum characteristics more effectively than commonly used Bell states. We demonstrated that certain negative states, derived from phase-space point operators, exhibit greater resilience than Bell states in measures of entanglement (concurrence, Figs.~\ref{concurMAD},\ref{concurNMAD}; coherence, Fig.~\ref{coherence_NMAD}) and teleportation fidelity under non-Markovian amplitude damping noise at the longer duration.(Figs.~\ref{mad_fid_final}, \ref{fidelityNMAD}; Chapter~\ref{chap3:Harnessing}). 

The second study, \textit{Protecting quantum correlations of negative quantum states using weak measurement under non-Markovian noise}~\citep{Lalita_2024ProtectingQC}, sought to build upon this inherent resilience by testing whether quantum correlations could be actively safeguarded during noisy evolution. Weak measurement and its reversal offered a subtle but powerful strategy. By gently perturbing the system before environmental interaction and subsequently reversing that perturbation, it became possible to counteract some of the decohering effects of noise. When this strategy was applied to the negative quantum states identified in the earlier chapter, the results were striking. Not only were entanglement and quantum correlations preserved for longer durations, but in several cases the combination of non-Markovian memory and the WM-QMR protocol produced performance levels in teleportation fidelity and correlation preservation that exceeded those of standard Bell states (Figs.~\ref{concur_NMAD}-\ref{concur_NMAD_same_p_and_q}). This revealed an important synergy between environmental memory and measurement-induced protection, illuminating a pathway for enhancing quantum information processing without the overhead of full error-correction schemes.

These insights were applied in practice in the third study, \textit{Noise-Resilient Negative Quantum States}~\citep{lalita2025realizingnegativequantumstates}, where the robustness of negative quantum states was tested using IBM's superconducting quantum processors. Constructing these states on real hardware required translating them into experimentally feasible circuits by calculating their corresponding unitary transformations and subsequently verifying their quantum circuits via quantum state tomography. Despite the inherent device noise, particularly the depolarizing noise and the non-Markovian AD, which are very prominent, the negative states persisted with high fidelity (Figs.~\ref{DP_NS1_NS2_NS3}, \ref{Fidelity_NMRTN_NMAD}). Even more striking was their performance in operational measures: they exhibited more substantial CHSH inequality violations, higher quantum Fisher information, and more reliable teleportation fidelity than Bell states in the presence of non-Markovian AD noise (Secs.~\ref{sec:Optimal_CHSH_inequality_violation}, \ref{Q_F_I}). This hardware-level verification demonstrated that these states are not merely theoretical constructs but viable candidates for noise-resilient protocols on near-term quantum devices (TABLE~\ref{New_table}).

The fourth study, \textit{Non-classicality of two-qubit quantum collision model: non-Markovian effects}~\citep{lalita2025non_classicality}, approached the problem of non-classicality from a more dynamical and microscopic perspective by introducing two-qubit collision models for the first time (Figs.~\ref{approach_1}, ~\ref{Scheme_B}). Collision models offer a uniquely transparent view of open-system evolution by depicting the environment as a stream of ancillary subsystems that interact sequentially with the system. By introducing tunable interactions among the ancillae, it becomes possible to embed controllable memory directly into the environment. This non-Markovian feature was explored using the BLP measure, where the trace distance between two different initial states showed oscillatory evolution (Figs.~\ref{Trace_distance_scheme_A}, \ref{Trace_distance_scheme_B}). Within this framework, the evolution of non-classicality could be traced with satisfactory resolution. Memory-induced effects generated revivals of entanglement that were absent in Markovian regimes (Figs.~\ref{concurrence_H_int_two_qubit_aR_all_states_scheme_A}, \ref{concurrence_H_int_two_qubit_aR_aL_all_states_scheme_B}). The collision model also revealed how environmental temperature, system-ancilla coupling strengths, and inter-ancilla correlations collaborate to shape long-term steady states and short-term oscillatory behavior (Figs.~\ref{ss_fidelity_all_cases_two_qubit_together}, \ref{ss_fidelity_all_cases}). This mechanistic insight sheds light on the deeper reasons behind the resilience observed in earlier chapters, positioning non-Markovianity as a physically meaningful resource rather than merely a mathematical abstraction.

All these ideas converged naturally in the final study, \textit{Interrelation of Non-Classicality, Entropy, Irreversibility and Work Extraction in Open Quantum Systems}~\citep{lalita2025interrelation}. Here, the thesis broadened its perspective to examine how quantum features relate to fundamental thermodynamic quantities. By probing non-classical volume, von Neumann entropy, entropy production, and ergotropy across a diverse selection of spin-spin and spin-boson models, including collision models, generalized and non-Markovian amplitude damping channels, the Jaynes-Cummings model, and the central spin model, the study uncovered remarkably qualitative relationships. As systems interacted with their environments, non-classical volume typically decreased as entropy increased, revealing a fundamental tension between quantum structure and statistical disorder  (Figs.~\ref{non_Markovian_collision_model}a-\ref{Jaynes_cummings_model}a). Meanwhile, entropy production, representing irreversibility, tended to rise as the system's ability to yield practical work diminished (Figs.~\ref{non_Markovian_collision_model}b-\ref{Jaynes_cummings_model}b). These parallel trends highlighted a profound structural unity: the degradation of non-classicality, the growth of disorder, the onset of irreversibility, and the decline of extractable work are intertwined manifestations of open-system dynamics. This realization broadened the thesis's contribution from quantum information science into quantum thermodynamics, suggesting that quantum advantages in information processing and energy manipulation may stem from common underlying principles.

Throughout the entire thesis, a coherent narrative emerges. Non-classicality, whether expressed through discrete Wigner negativity, entanglement, coherence, or non-classical volume, is not an inherently fragile trait destined to vanish under environmental influence. Instead, its evolution depends intricately on the structure of the environment, the nature of system-environment interactions, and the interventions applied to the system. Negative quantum states emerged as robust and operationally meaningful resources in the presence of (non)-Markovian AD noise. Weak measurement provided a practical means of protection against non-Markovian AD and RTN noise at the cost of reduced success probability. Moreover, crucially, experimental validation confirmed that these theoretical insights hold in real quantum hardware. Collision models revealed the microscopic origins of memory effects. Thermodynamic analysis connected these behaviors to the fundamental limits imposed by entropy and work. 

In summary, this thesis demonstrates that quantumness can be guided, preserved, and meaningfully exploited through deliberate design of interactions and informed strategies grounded in both information-theoretic and thermodynamic principles. As quantum technologies move toward practical deployment, understanding and engineering the evolution of non-classicality will become increasingly essential. The results presented here contribute not only to the scientific foundation but also to the broader vision of building quantum devices that operate reliably in the presence of unavoidable noise.

\section{Practical Criteria for Choosing Negative Quantum States over Bell States}
Based on the analyses presented in Chapters~\ref{chap3:Harnessing}, \ref{chap4:Protecting}, and \ref{chap5:Physical_realization}, we delineate the regimes in which two-qubit negative quantum states provide a preferable entanglement resource compared to Bell states.

\textbf{Conditions favoring negative quantum states:}

\begin{itemize}
\item \textit{Noise characteristics:} Under both Markovian and non-Markovian amplitude damping (non-unital) channels, the states $|NS_1\rangle$ and $|NS_2\rangle$ exhibit prolonged preservation of concurrence, quantum discord, and coherence relative to Bell states, with and without the application of weak measurement (WM) and quantum measurement reversal (QMR). Moreover, under a non-Markovian random telegraph noise (RTN) channel (unital), the superiority of $|NS_2\rangle$ is achieved for WM strength $p=0.05$ and QMR strength $q=0.74$. This enhancement is accompanied by a finite, albeit reduced, success probability (Fig.~\ref{P_success}), which remains acceptable in scenarios where post-selection is viable (Figs.~\ref{coherence_NMRTN}-\ref{fidelityNMAD}, \ref{concur_NMAD}-\ref{concur_NMAD_same_p_and_q}, TABLES~\ref{table1},~\ref{table2}). Further, the states $|NS_1\rangle$ and $|NS_2\rangle$ can be generated robustly in the presence of depolarizing noise (with increasing error probability) as well as under time-evolving (non-)Markovian amplitude damping channels (Figs.~\ref{DP_NS1_NS2_NS3}, \ref{Fidelity_NMRTN_NMAD}).

\item \textit{Universal quantum teleportation:} The state $|NS_2\rangle$ achieves near-vanishing fidelity deviation under non-Markovian RTN in the presence of WM and QMR, thereby constituting an optimal candidate for universal quantum teleportation in such environments (Fig.~\ref{FD_NMRTN}, Table~\ref{table2}).

\item \textit{Operational metrics (CHSH violation and quantum Fisher information):} The state $|NS_2\rangle$ consistently yields higher CHSH inequality violation ratios compared to the Bell state $|\phi^+\rangle$ under non-Markovian amplitude damping, both with and without WM and QMR (Sec.~\ref{sec:Optimal_CHSH_inequality_violation}). This renders it advantageous for entanglement-enhanced metrological and sensing applications (Sec.~\ref{Q_F_I}).

\end{itemize}
 
These criteria should be regarded as indicative guidelines within the parameter regimes explored. A comprehensive characterization of the full noise-parameter landscape remains an open problem for future investigation.

\section*{Future Directions}

The investigations presented in this thesis open several promising avenues for future research:

\begin{enumerate}
    \item The qutrit negative quantum states identified in this work belong to the class of magic states; therefore, a systematic analysis of their contextuality is a natural extension. In particular, it would be worthwhile to examine whether weak measurement and quantum measurement reversal protocols can preserve contextuality and other non-classical features under (non-)Markovian noise.

    \item The explicit quantum circuits developed for two-qubit negative quantum state preparation facilitate their implementation across diverse experimental platforms beyond superconducting systems. In this context, nuclear magnetic resonance (NMR) techniques provide a viable route, offering fine control over coherence and system–environment interactions.

    \item A rigorous resource-theoretic framework for negative quantum states remains to be established. This includes identifying suitable free operations and formulating meaningful resource monotones.

    \item The observed robustness of two-qubit negative quantum states against noise, relative to standard Bell states, motivates their investigation in quantum communication tasks. In particular, their potential advantage in increasing the success probability of nonlocality-assisted, error-free communication over noisy classical channels merits detailed study.

    \item Quantum collision models may be extended to analyze quantum correlations in collective neutrino oscillations, where many-body coherence and environmental effects are significant. Such an approach could provide new insights into non-classical features of neutrino dynamics.

    \item The proposed two-qubit collision model framework is well-suited for studying information scrambling and correlation spreading. Tracking the evolution of initially localized information through successive collisions may yield insights into irreversibility, correlation growth, and emergent thermalization.

    \item The equilibrium and steady-state behavior of two-qubit collision models requires further investigation. In particular, the emergence and explicit construction of the Hamiltonian of mean force (HMF) within this framework would strengthen connections to strong-coupling quantum thermodynamics.

    \item Extensions of collision models to multipartite (non-)Markovian settings would enable controlled exploration of memory effects, entanglement propagation, and correlation structures in larger quantum systems.

    \item The role of quasi-probability distributions in quantum thermodynamics presents a rich direction for future work. Notably, the Kirkwood–Dirac (KD) distribution may provide a unified phase-space framework for quantifying work, heat, and entropy production in collisional quantum heat engines.

    \item Finally, establishing quantitative bounds linking non-classical volume, von Neumann entropy, entropy production, and ergotropy remains an important open problem.
\end{enumerate}


\appendix 


\newpage
\setcounter{chapter}{0} 

\titleformat{\chapter}[display]
{\fontsize{27}{27}\bfseries\filleft}{ \thechapter}{0pt}{{#1}}  

\thispagestyle{empty}

\chapter{Appendix A}\label{appendices:AppendixA}
\section{\label{WM_QMR} Optimal CHSH inequality violation, Concurrence, and Teleportation fidelity under non-Markovian AD noise with(without) weak measurement}
\begin{figure}[h]
    \centering
    \includegraphics[width=1\columnwidth]{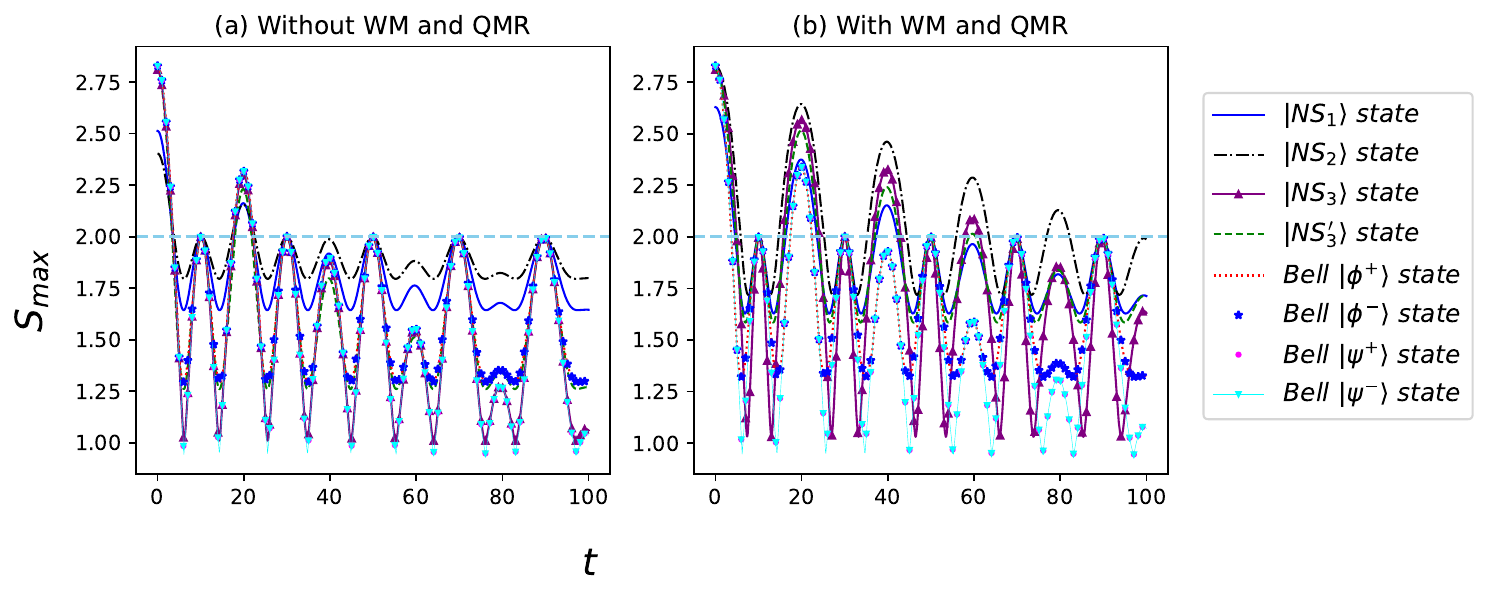}
    \caption{Variation of optimal CHSH inequality violation of $\ket{NS_1}$, $\ket{NS_2}$, $\ket{NS_3}$, $\ket{NS_3^{\prime}}$ and all the Bell states under non-Markovian AD channel without WM and QMR in subplot (a) and with WM and QMR in subplot (b) with time. Here, for $\ket{NS_1}$ ($p = 0.17$, $q = 0.54$), for $\ket{NS_2}$ ($p = 0.05$, $q = 0.74$), for $\ket{NS_3}$ ($p = 0.54$, $q = 0.54$), for $\ket{NS_3^{\prime}}$ ($p = 0.58$, $q = 0.58$), for Bell $\ket{\phi^{+}}$, $\ket{\phi^{-}}$ states ($p = 0.05$, $q = 0.05$) and for Bell $\ket{\psi^{+}}$, $\ket{\psi^{-}}$ states ($p = 0.01$, $q = 0.05$). The non-Markovian AD channel parameters are $g = 0.01$ and $\gamma^{AD} = 5$.}
    \label{smax_NMAD}
\end{figure}
The physical framework for preserving quantum correlations of two-qubit entangled states using weak measurement (WM) and quantum measurement reversal (QMR) is detailed in chapter~\ref{chap4:Protecting} Sec.~\ref{ch4_Model}.  In~\citep{Lalita_2024ProtectingQC}, $\ket{\phi^{+}}$ Bell state and the two-qubit negative quantum states elaborated in Sec.~\ref{ch3_NQS} of chapter~\ref{chap3:Harnessing}, are taken into account to find the optimal state for universal quantum teleportation in the presence of non-Markovian RTN and AD noise with(without) weak measurement (WM)~\citep{oreshkov2005weak, katz2008reversal, breuer2002theory}. Here, we extend this study to find the most suitable candidate for optimal CHSH inequality violation and universal quantum teleportation among all four Bell states (given below) and two-qubit negative quantum states in the presence of non-Markovian AD noise.
\begin{eqnarray}
    \begin{aligned}
        \ket{\phi^{+}} &= {1/\sqrt{2}}\left(~1~~0~~0~~1~\right)^T;
        \ket{\phi^{-}} = {1/\sqrt{2}}\left(~1~~0~~0~~{-1}~\right)^T; \nonumber \\
        \ket{\psi^{+}} &= {1/\sqrt{2}}\left(~0~~1~~1~~0~\right)^T; \nonumber
        \ket{\psi^{-}} = {1/\sqrt{2}}\left(~0~~1~~{-1}~~0~\right)^T. 
    \end{aligned}
    \label{Bell_states}
\end{eqnarray}
\begin{figure}
    \centering
    \includegraphics[width=1\columnwidth]{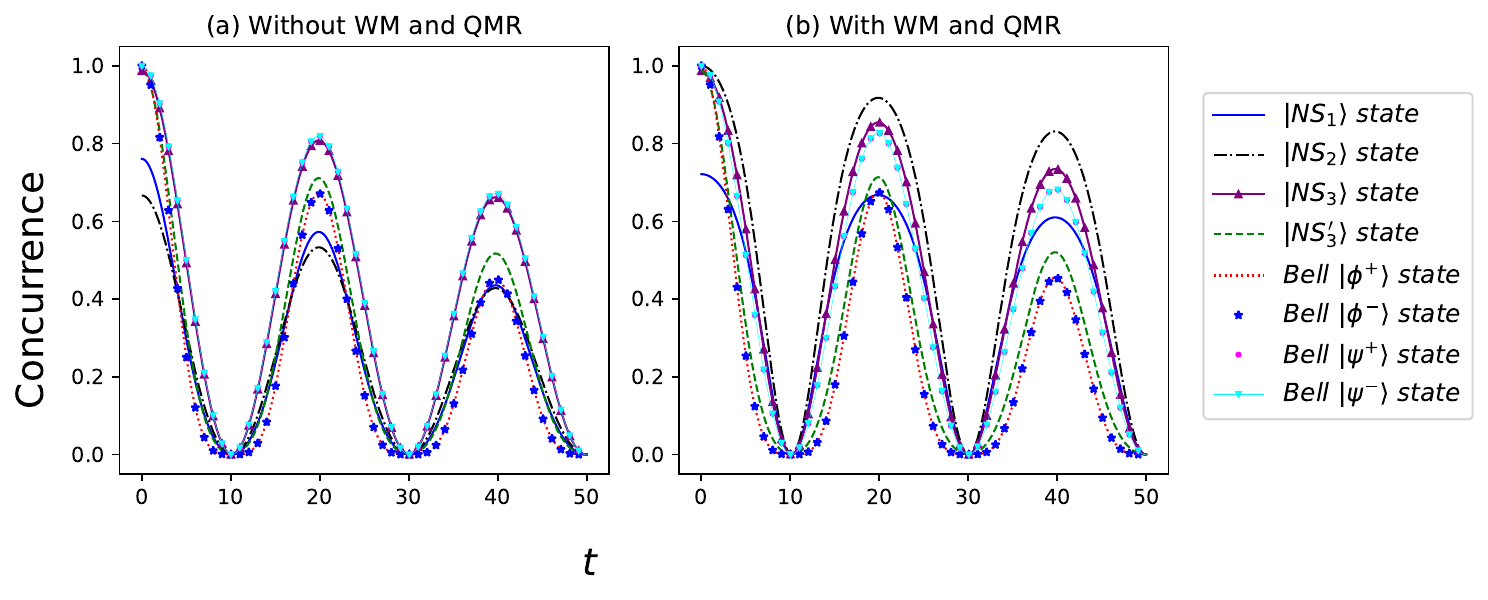}
    \caption{Variation of concurrence of $\ket{NS_1}$, $\ket{NS_2}$, $\ket{NS_3}$, $\ket{NS_3^{\prime}}$ and all the Bell states under non-Markovian AD channel without WM and QMR in subplot (a) and with WM and QMR in subplot (b) with time. Here, for $\ket{NS_1}$ ($p = 0.17$, $q = 0.54$), for $\ket{NS_2}$ ($p = 0.05$, $q = 0.74$), for $\ket{NS_3}$ ($p = 0.05$, $q = 0.05$), for $\ket{NS_3^{\prime}}$ ($p = 0.3$, $q = 0.3$), for Bell $\ket{\phi^{+}}$, $\ket{\phi^{-}}$ states ($p = 0.01$, $q = 0.01$) and for Bell $\ket{\psi^{+}}$, $\ket{\psi^{-}}$ states ($p = 0.01$, $q = 0.05$). The non-Markovian AD channel parameters are $g = 0.01$ and $\gamma^{AD} = 5$.}
    \label{concur_NMAD_all_states}
\end{figure}%
\begin{figure}
    \centering
    \includegraphics[width=1\columnwidth]{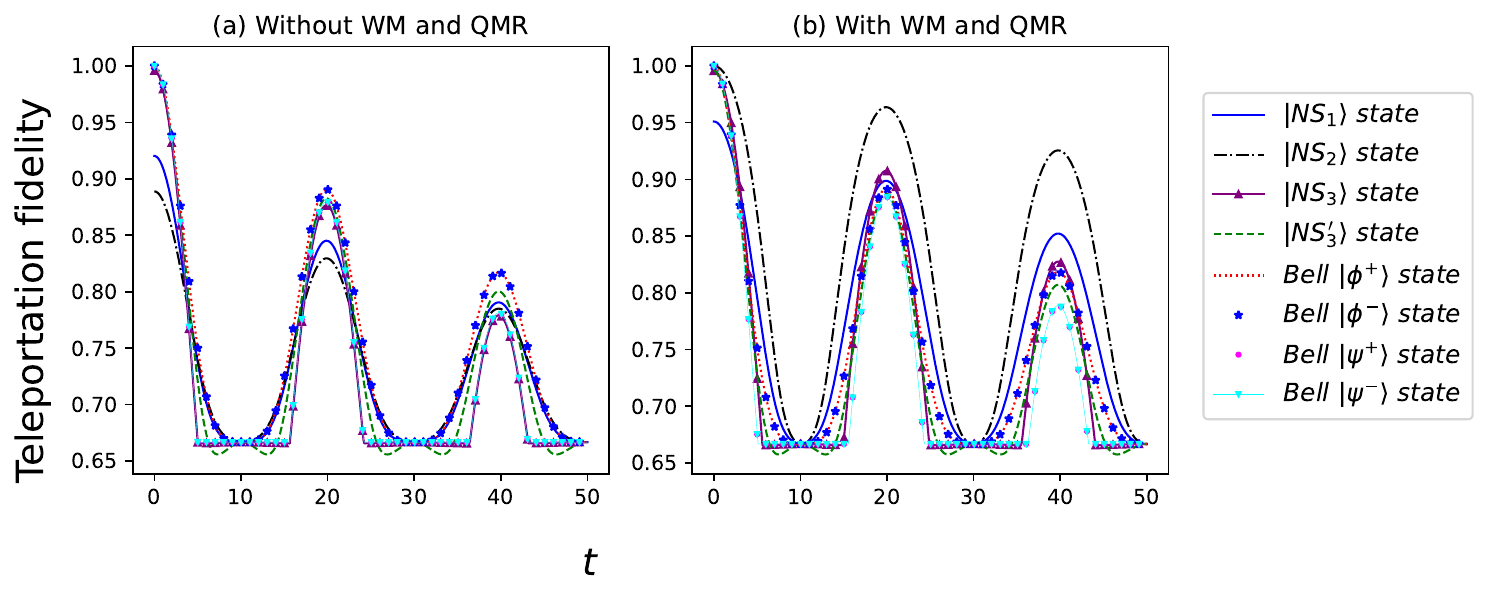}
    \caption{Variation of teleportation fidelity of $\ket{NS_1}$, $\ket{NS_2}$, $\ket{NS_3}$, and Bell state under non-Markovian AD channel without WM and QMR in subplot (a), and with WM and QMR in subplot (b) with time. Here, for $\ket{NS_1}$ ($p = 0.17$, $q = 0.54$), for $\ket{NS_2}$ ($p = 0.05$, $q = 0.74$), for $\ket{NS_3}$ ($p = 0.05$, $q = 0.05$), for $\ket{NS_3^{\prime}}$ ($p = 0.3$, $q = 0.3$), for Bell $\ket{\phi^{+}}$, $\ket{\phi^{-}}$ states ($p = 0.01$, $q = 0.01$) and for Bell $\ket{\psi^{+}}$, $\ket{\psi^{-}}$ states ($p = 0.01$, $q = 0.05$). The non-Markovian AD channel parameters are $g = 0.01$ and $\gamma^{AD} = 5$.}
    \label{TFid_NMAD}
\end{figure}
\begin{figure}
    \centering
    \includegraphics[width=1\columnwidth]{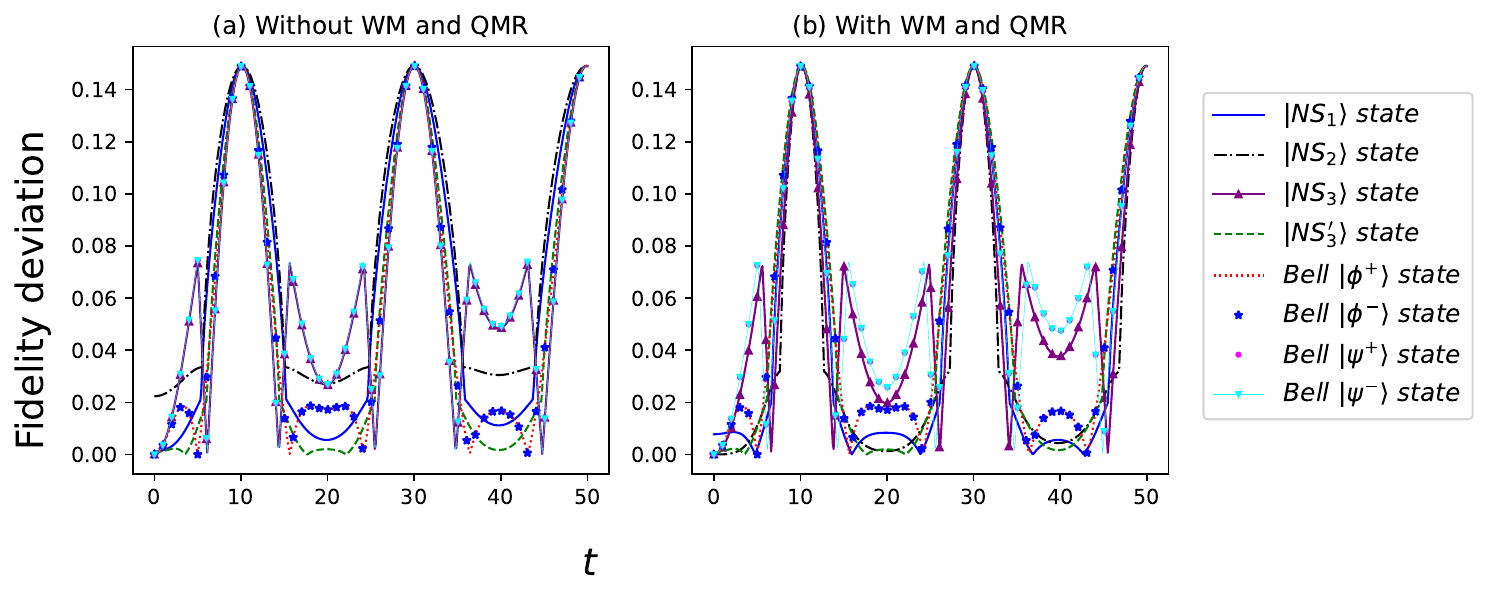}
    \caption{Variation of fidelity deviation of $\ket{NS_1}$, $\ket{NS_2}$, $\ket{NS_3}$, and Bell state under non-Markovian AD channel without WM and QMR in subplot (a), and with WM and QMR in subplot (b) with time. Here, for $\ket{NS_1}$ ($p = 0.17$, $q = 0.54$), for $\ket{NS_2}$ ($p = 0.05$, $q = 0.74$), for $\ket{NS_3}$ ($p = 0.05$, $q = 0.05$), for optimal $NS3$ ($p = 0.3$, $q = 0.3$), for Bell $\ket{\phi^{+}}$, $\ket{\phi^{-}}$ states ($p = 0.01$, $q = 0.01$) and for Bell $\ket{\psi^{+}}$, $\ket{\psi^{-}}$ states ($p = 0.01$, $q = 0.05$). The non-Markovian AD channel parameters are $g = 0.01$ and $\gamma^{AD} = 5$.}
    \label{FD_NMAD_all_states}
\end{figure}

The optimal CHSH inequality violation, denoted by $S_{max}$, Eq.~(\ref{S_max_eq.}) in the main text, is computed using the final state ${\rho}_f(t)$ obtained after sequential application of WM, non-Markovian amplitude damping (AD) noise, and QMR for all Bell states and selected two-qubit negative quantum states. The temporal evolution of $S_{max}$ under the influence of WM and QMR in the presence of non-Markovian AD noise is depicted in Fig.~\ref{smax_NMAD}(b). In contrast, Fig.~\ref{smax_NMAD}(a) depicts the temporal behavior of $S_{max}$ for the same set of states when only non-Markovian AD noise is present, without WM and QMR interventions.
From Fig.~\ref{smax_NMAD}, it is evident that in the absence of WM and QMR, the states $|NS_1\rangle$ and $|NS_2\rangle$ do not achieve maximal CHSH inequality violation (i.e., $2\sqrt{2}$), yet they exhibit slower decoherence compared to the states that initially achieve maximal violation. However, when WM and QMR are applied, the $|NS_2\rangle$ state attains maximal CHSH inequality violation and displays more frequent and pronounced random fluctuations above the classical bound of $2$ than any other considered state. Furthermore, the $|NS_3\rangle$ and $|NS_3'\rangle$ states also consistently exhibit stronger nonlocal correlations over extended durations relative to the Bell states.

Additionally, building upon our earlier work~\citep{Lalita_2024ProtectingQC}, we have expanded our analysis to identify the optimal state for achieving universal quantum teleportation by considering all Bell states and the aforementioned two-qubit negative quantum states. Quantifying the entanglement content is critical for universal quantum teleportation. Thus, we calculate concurrence~\citep{Wootters1998Entanglement}, a prominent measure of entanglement, for all the states considered. The variation of concurrence for the above Bell states and the negative quantum states under non-Markovian AD noise with(without) weak measurement is shown in Fig.~\ref{concur_NMAD_all_states}. We observe that the $\ket{NS_3}$ state's concurrence is equivalent to the Bell $\ket{\psi^{+}}$ and $\ket{\psi^{-}}$ states under non-Markovian AD noise without WM, whereas the $\ket{NS_2}$ state is the most robust among all the considered states. Similarly, variations of the teleportation fidelity~\citep{horodecki1996teleportation} and fidelity deviation~\citep{ghosal2020optimal} for all the considered states under the non-Markovian AD channel are provided in Fig.~\ref{TFid_NMAD} and Fig.~\ref{FD_NMAD_all_states}, respectively. Out of all the Bell states and two-qubit negative quantum states, the $\ket{NS_2}$ state is the most suitable state for universal quantum teleportation with WM under both non-Markovian AD and RTN channels.



\newpage

\phantomsection
\addcontentsline{toc}{chapter}{List of Publications}
\chapter*{Publications}


This thesis is based on the following publications:
\begin{enumerate}
    \item \textbf{J. Lalita}, P. S. Iyer, and S. Banerjee, 
    \textit{Noise-resilient negative quantum states}, 
    \href{https://link.aps.org/doi/10.1103/g3f6-p18d}
    {\textbf{Physical Review A} 113, 022427 (2026)}.
    \item \textbf{J. Lalita} and S. Banerjee, 
    \textit{Two-qubit quantum collision model: non-Markovianity and non-classicality}, 
    \href{https://link.aps.org/doi/10.1103/8rtv-ftrr}{\textbf{Physical Review A} 113, 012225 (2026)}.
    \item \textbf{J. Lalita} and S. Banerjee, 
    \textit{Interrelation of Non-Classicality, Entropy, Irreversibility and Work Extraction in Open Quantum Systems}, 
    \href{https://doi.org/10.48550/arXiv.2510.15140}{arXiv:2510.1514 (2025)}. (Under review in \textbf{Physical Review E})
    \item \textbf{J. Lalita} and S. Banerjee, 
    \textit{Protecting quantum correlations of negative quantum states using weak measurement under non-Markovian noise}, 
    \href{https://doi.org/10.1088/1402-4896/ad273e}{\textbf{Physica Scripta} 99, 035116 (2024)}.
    \item \textbf{J. Lalita}, K. G. Paulson, and S. Banerjee, 
    \textit{Harnessing quantumness using discrete Wigner functions under (non)-Markovian channels}, 
    \href{https://doi.org/10.1002/andp.202300139}{\textbf{Annalen der Physik} 535, 2300139 (2023)}.
\end{enumerate}

\bibliographystyle{myphd}
\bibliography{chapters/refs}

@article{Sudarshan1961Stochastic,
  title = {Stochastic Dynamics of Quantum-Mechanical Systems},
  author = {Sudarshan, E. C. G. and Mathews, P. M. and Rau, Jayaseetha},
  journal = {Phys. Rev.},
  volume = {121},
  issue = {3},
  pages = {920--924},
  numpages = {0},
  year = {1961},
  month = {Feb},
  publisher = {American Physical Society},
  doi = {10.1103/PhysRev.121.920},
  url = {https://link.aps.org/doi/10.1103/PhysRev.121.920}
}

@article{Lindblad1976Onthegenerators,
       author = {Lindblad, G.},
        title = "{On the generators of quantum dynamical semigroups}",
      journal = {Communications in Mathematical Physics},
         year = 1976,
        month = jun,
       volume = {48},
       number = {2},
        pages = {119-130},
          doi = {10.1007/BF01608499},
       url = {https://ui.adsabs.harvard.edu/abs/1976CMaPh..48..119L}
}

@article{gorini1976completely,
  title={Completely positive dynamical semigroups of N-level systems},
  author={Gorini, Vittorio and Kossakowski, Andrzej and Sudarshan, Ennackal Chandy George},
  journal={Journal of Mathematical Physics},
  volume={17},
  number={5},
  pages={821--825},
  year={1976},
  publisher={American Institute of Physics},
  url={https://doi.org/10.1063/1.522979}
}

@article{Wolf2008Assissing,
  title = {Assessing Non-Markovian Quantum Dynamics},
  author = {Wolf, M. M. and Eisert, J. and Cubitt, T. S. and Cirac, J. I.},
  journal = {Phys. Rev. Lett.},
  volume = {101},
  issue = {15},
  pages = {150402},
  numpages = {4},
  year = {2008},
  month = {Oct},
  publisher = {American Physical Society},
  doi = {10.1103/PhysRevLett.101.150402},
  url = {https://link.aps.org/doi/10.1103/PhysRevLett.101.150402}
}

@book{alicki2007quantum,
  title={Quantum dynamical semigroups and applications},
  author={Alicki, Robert and Lendi, Karl},
  volume={717},
  year={2007},
  publisher={Springer}
}

@article{Hall2014Canonical,
  title = {Canonical form of master equations and characterization of non-Markovianity},
  author = {Hall, Michael J. W. and Cresser, James D. and Li, Li and Andersson, Erika},
  journal = {Phys. Rev. A},
  volume = {89},
  issue = {4},
  pages = {042120},
  numpages = {11},
  year = {2014},
  month = {Apr},
  publisher = {American Physical Society},
  doi = {10.1103/PhysRevA.89.042120},
  url = {https://link.aps.org/doi/10.1103/PhysRevA.89.042120}
}

@book{kraus1983states,
  title={States, effects, and operations fundamental notions of quantum theory: Lectures in mathematical physics at the university of Texas at Austin},
  author={Kraus, Karl and B{\"o}hm, Arno and Dollard, John D and Wootters, WH},
  year={1983},
  publisher={Springer}
}

@book{shankar2012principles,
  title={Principles of quantum mechanics},
  author={Shankar, Ramamurti},
  year={2012},
  publisher={Springer Science \& Business Media}
}

@article{schwinger1960special,
  title={The special canonical group},
  author={Schwinger, Julian},
  journal={Proceedings of the National Academy of Sciences},
  volume={46},
  number={10},
  pages={1401--1415},
  year={1960}
}

@book{Schleich2001,
  title        = {Quantum Optics in Phase Space},
  author       = {Schleich, Wolfgang P.},
  year         = {2001},
  publisher    = {Wiley-VCH},
  address      = {Berlin},
}

@article{case2008wigner,
  title={Wigner functions and Weyl transforms for pedestrians},
  author={Case, William B},
  journal={American Journal of Physics},
  volume={76},
  number={10},
  pages={937--946},
  year={2008},
  publisher={AIP Publishing},
  url={https://doi.org/10.1119/1.2957889}
}

@article{howard2014contextuality,
  title={Contextuality supplies the ‘magic’for quantum computation},
  author={Howard, Mark and Wallman, Joel and Veitch, Victor and Emerson, Joseph},
  journal={Nature},
  volume={510},
  number={7505},
  pages={351--355},
  year={2014},
  publisher={Nature Publishing Group UK London},
  url={https://doi.org/10.1038/nature13460},
  doi={10.1038/nature13460}
}

@article{schmid2022uniqueness,
  title = {Uniqueness of Noncontextual Models for Stabilizer Subtheories},
  author = {Schmid, David and Du, Haoxing and Selby, John H. and Pusey, Matthew F.},
  journal = {Phys. Rev. Lett.},
  volume = {129},
  issue = {12},
  pages = {120403},
  numpages = {6},
  year = {2022},
  month = {Sep},
  publisher = {American Physical Society},
  doi = {10.1103/PhysRevLett.129.120403},
  url = {https://link.aps.org/doi/10.1103/PhysRevLett.129.120403}
}

@article{paulson2022quantum,
  title={Quantum speed limit time: role of coherence},
  author={Paulson, K G and Banerjee, Subhashish},
  journal={Journal of Physics A: Mathematical and Theoretical},
  volume={55},
  number={50},
  pages={505302},
  year={2022},
  publisher={IOP Publishing},
  doi = {10.1088/1751-8121/acaadb},
  url = {https://doi.org/10.1088/1751-8121/acaadb}
}

@article{malpani2020impact,
  title={Impact of photon addition and subtraction on nonclassical and phase properties of a displaced Fock state},
  author={Malpani, Priya and Thapliyal, Kishore and Alam, Nasir and Pathak, Anirban and Narayanan, V and Banerjee, Subhashish},
  journal={Optics Communications},
  volume={459},
  pages={124964},
  year={2020},
  publisher={Elsevier},
  url={https://doi.org/10.1016/j.optcom.2019.124964}
}

@article{jain2020qutritmagic,
  title = {Qutrit and ququint magic states},
  author = {Jain, Akalank and Prakash, Shiroman},
  journal = {Phys. Rev. A},
  volume = {102},
  issue = {4},
  pages = {042409},
  numpages = {19},
  year = {2020},
  month = {Oct},
  publisher = {American Physical Society},
  doi = {10.1103/PhysRevA.102.042409},
  url = {https://link.aps.org/doi/10.1103/PhysRevA.102.042409}
}

@article{xi2015quantum,
  title={Quantum coherence and correlations in quantum system},
  author={Xi, Zhengjun and Li, Yongming and Fan, Heng},
  journal={Scientific reports},
  volume={5},
  number={1},
  pages={1--9},
  year={2015},
  publisher={Springer},
  url={https://doi.org/10.1038/srep10922},
  doi={10.1038/srep10922}
}

@article{streltsov2017colloquium,
  title = {Colloquium: Quantum coherence as a resource},
  author = {Streltsov, Alexander and Adesso, Gerardo and Plenio, Martin B.},
  journal = {Rev. Mod. Phys.},
  volume = {89},
  issue = {4},
  pages = {041003},
  numpages = {34},
  year = {2017},
  month = {Oct},
  publisher = {American Physical Society},
  doi = {10.1103/RevModPhys.89.041003},
  url = {https://link.aps.org/doi/10.1103/RevModPhys.89.041003}
}

@article{zhao20191,
  title = {${l}_{1}$-norm coherence of assistance},
  author = {Zhao, Ming-Jing and Ma, Teng and Quan, Quan and Fan, Heng and Pereira, Rajesh},
  journal = {Phys. Rev. A},
  volume = {100},
  issue = {1},
  pages = {012315},
  numpages = {6},
  year = {2019},
  month = {Jul},
  publisher = {American Physical Society},
  doi = {10.1103/PhysRevA.100.012315},
  url = {https://link.aps.org/doi/10.1103/PhysRevA.100.012315}
}

@article{hu2018quantum,
  title={Quantum coherence and geometric quantum discord},
  author={Hu, Ming-Liang and Hu, Xueyuan and Wang, Jieci and Peng, Yi and Zhang, Yu-Ran and Fan, Heng},
  journal={Physics Reports},
  volume={762},
  pages={1--100},
  year={2018},
  publisher={Elsevier},
  url={https://doi.org/10.1016/j.physrep.2018.07.004}
}

@article{hudson1974wigner,
  title={When is the Wigner quasi-probability density non-negative?},
  author={Hudson, Robin L},
  journal={Reports on Mathematical Physics},
  volume={6},
  number={2},
  pages={249--252},
  year={1974},
  publisher={Elsevier}
}

@article{hillery1984distribution,
  title={Distribution functions in physics: Fundamentals},
  author={Hillery, MOSM and O'Connell, Robert F and Scully, Marlan O and Wigner, Eugene P},
  journal={Physics reports},
  volume={106},
  number={3},
  pages={121--167},
  year={1984},
  publisher={Elsevier},
  url={https://doi.org/10.1016/0370-1573(84)90160-1}
}

@article{paulson2021hierarchy,
  title={Hierarchy of quantum correlations under non-Markovian dynamics},
  author={Paulson, K G and Panwar, Ekta and Banerjee, Subhashish and Srikanth, R},
  journal={Quantum Information Processing},
  volume={20},
  number={4},
  pages={1--26},
  year={2021},
  publisher={Springer},
  url={https://doi.org/10.1007/s11128-021-03061-9},
  doi={10.1007/s11128-021-03061-9}
}

@article{vacchini2012classical,
  title={A classical appraisal of quantum definitions of non-Markovian dynamics},
  author={Vacchini, Bassano},
  journal={Journal of Physics B: Atomic, Molecular and Optical Physics},
  volume={45},
  number={15},
  pages={154007},
  year={2012},
  publisher={IOP Publishing},
  doi = {10.1088/0953-4075/45/15/154007},
  url = {https://doi.org/10.1088/0953-4075/45/15/154007}
}

@article{girolami2014observable,
  title = {Observable Measure of Quantum Coherence in Finite Dimensional Systems},
  author = {Girolami, Davide},
  journal = {Phys. Rev. Lett.},
  volume = {113},
  issue = {17},
  pages = {170401},
  numpages = {5},
  year = {2014},
  month = {Oct},
  publisher = {American Physical Society},
  doi = {10.1103/PhysRevLett.113.170401},
  url = {https://link.aps.org/doi/10.1103/PhysRevLett.113.170401}
}

@article{baumgratz2014quantifying,
  title = {Quantifying Coherence},
  author = {Baumgratz, T. and Cramer, M. and Plenio, M. B.},
  journal = {Phys. Rev. Lett.},
  volume = {113},
  issue = {14},
  pages = {140401},
  numpages = {5},
  year = {2014},
  month = {Sep},
  publisher = {American Physical Society},
  doi = {10.1103/PhysRevLett.113.140401},
  url = {https://link.aps.org/doi/10.1103/PhysRevLett.113.140401}
}

@article{li2023optimal,
  title={Optimal diagonal qutrit gates for creating Wigner negativity},
  author={Li, Xiaohui and Luo, Shunlong},
  journal={Physics Letters A},
  pages={128620},
  year={2023},
  publisher={Elsevier},
  url={https://doi.org/10.1016/j.physleta.2022.128620}
}

@article{goyal2016geometry,
  title={Geometry of the generalized Bloch sphere for qutrits},
  author={Goyal, Sandeep K and Simon, B Neethi and Singh, Rajeev and Simon, Sudhavathani},
  journal={Journal of Physics A: Mathematical and Theoretical},
  volume={49},
  number={16},
  pages={165203},
  year={2016},
  publisher={IOP Publishing},
  doi = {10.1088/1751-8113/49/16/165203},
  url = {https://doi.org/10.1088/1751-8113/49/16/165203}
}

@article{brierley2009all,
  title={All mutually unbiased bases in dimensions two to five},
  author={Brierley, Stephen and Weigert, Stefan and Bengtsson, Ingemar},
  journal={arXiv preprint arXiv:0907.4097},
  year={2009},
  url={https://dl.acm.org/doi/10.5555/2011464.2011470}
}

@article{utagi2020ping,
  title={Ping-pong quantum key distribution with trusted noise: non-Markovian advantage},
  author={Utagi, Shrikant and Srikanth, R and Banerjee, Subhashish},
  journal={Quantum Information Processing},
  volume={19},
  number={10},
  pages={1--12},
  year={2020},
  publisher={Springer},
  url={https://doi.org/10.1007/s11128-020-02874-4},
  doi={10.1007/s11128-020-02874-4}
}

@article{dutta2016entanglement,
  title={Entanglement criteria for noise resistance of two-qudit states},
  author={Dutta, Arijit and Ryu, Junghee and Laskowski, Wies{\l}aw and {\.Z}ukowski, Marek},
  journal={Physics Letters A},
  volume={380},
  number={27-28},
  pages={2191--2199},
  year={2016},
  publisher={Elsevier},
  url={https://doi.org/10.1016/j.physleta.2016.04.043}
}

@article{galvao2005discrete,
  title = {Discrete Wigner functions and quantum computational speedup},
  author = {Galv\~ao, Ernesto F.},
  journal = {Phys. Rev. A},
  volume = {71},
  issue = {4},
  pages = {042302},
  numpages = {6},
  year = {2005},
  month = {Apr},
  publisher = {American Physical Society},
  doi = {10.1103/PhysRevA.71.042302},
  url = {https://link.aps.org/doi/10.1103/PhysRevA.71.042302}
}

@article{veitch2014resource,
doi = {10.1088/1367-2630/16/1/013009},
url = {https://doi.org/10.1088/1367-2630/16/1/013009},
year = {2014},
month = {jan},
publisher = {IOP Publishing},
volume = {16},
number = {1},
pages = {013009},
author = {Veitch, Victor and Hamed Mousavian, S A and Gottesman, Daniel and Emerson, Joseph},
title = {The resource theory of stabilizer quantum computation},
journal = {New Journal of Physics}
}

@article{kenfack2004negativity,
doi = {10.1088/1464-4266/6/10/003},
url = {https://doi.org/10.1088/1464-4266/6/10/003},
year = {2004},
month = {aug},
publisher = {},
volume = {6},
number = {10},
pages = {396},
author = {Anatole Kenfack and Karol Życzkowski},
title = {Negativity of the Wigner function as an indicator of non-classicality},
journal = {Journal of Optics B: Quantum and Semiclassical Optics}
}

@article{teklu2015nonlinearity,
  title={Nonlinearity and nonclassicality in a nanomechanical resonator},
  author={Teklu, Berihu and Ferraro, Alessandro and Paternostro, Mauro and Paris, Matteo GA},
  journal={EPJ Quantum Technology},
  volume={2},
  pages={1--10},
  year={2015},
  publisher={Springer},
  url={https://doi.org/10.1140/epjqt/s40507-015-0029-x}
}

@article{gibbons2004discrete,
  title = {Discrete phase space based on finite fields},
  author = {Gibbons, Kathleen S. and Hoffman, Matthew J. and Wootters, William K.},
  journal = {Phys. Rev. A},
  volume = {70},
  issue = {6},
  pages = {062101},
  numpages = {23},
  year = {2004},
  month = {Dec},
  publisher = {American Physical Society},
  doi = {10.1103/PhysRevA.70.062101},
  url = {https://link.aps.org/doi/10.1103/PhysRevA.70.062101}
}

@article{lalita2023harnessing,
  title={Harnessing Quantumness of States using Discrete Wigner Functions under (non)-Markovian Quantum Channels},
  author={Lalita, Jai and Paulson, K G and Banerjee, Subhashish},
  journal={Annalen der Physik},
  pages={2300139},
  year={2023},
  publisher={Wiley Online Library},
  doi={https://doi.org/10.1002/andp.202300139},
  url={https://onlinelibrary.wiley.com/doi/abs/10.1002/andp.202300139}
}

@article{daffer2004depolarizing,
  title = {Depolarizing channel as a completely positive map with memory},
  author = {Daffer, Sonja and W\'odkiewicz, Krzysztof and Cresser, James D. and McIver, John K.},
  journal = {Phys. Rev. A},
  volume = {70},
  issue = {1},
  pages = {010304},
  numpages = {4},
  year = {2004},
  month = {Jul},
  publisher = {American Physical Society},
  doi = {10.1103/PhysRevA.70.010304},
  url = {https://link.aps.org/doi/10.1103/PhysRevA.70.010304}
}

@inproceedings{schrodinger1935discussion,
  title={Discussion of probability relations between separated systems},
  author={Schr{\"o}dinger, Erwin},
  booktitle={Mathematical Proceedings of the Cambridge Philosophical Society},
  volume={31},
  pages={555--563},
  year={1935},
  organization={Cambridge University Press},
  doi ={https://doi.org/10.1017/S0305004100013554}
}

@inproceedings{schrodinger1936probability,
  title={Probability relations between separated systems},
  author={Schr{\"o}dinger, Erwin},
  booktitle={Mathematical Proceedings of the Cambridge Philosophical Society},
  volume={32},
  pages={446--452},
  year={1936},
  organization={Cambridge University Press},
  doi ={https://doi.org/10.1017/S0305004100019137}
}

@article{brunner2014bell,
  title = {Bell nonlocality},
  author = {Brunner, Nicolas and Cavalcanti, Daniel and Pironio, Stefano and Scarani, Valerio and Wehner, Stephanie},
  journal = {Rev. Mod. Phys.},
  volume = {86},
  issue = {2},
  pages = {419--478},
  numpages = {60},
  year = {2014},
  month = {Apr},
  publisher = {American Physical Society},
  doi = {10.1103/RevModPhys.86.419},
  url = {https://link.aps.org/doi/10.1103/RevModPhys.86.419}
}

@article{horodecki2009quantum,
  title = {Quantum entanglement},
  author = {Horodecki, Ryszard and Horodecki, Pawe\l{} and Horodecki, Micha\l{} and Horodecki, Karol},
  journal = {Rev. Mod. Phys.},
  volume = {81},
  issue = {2},
  pages = {865--942},
  numpages = {0},
  year = {2009},
  month = {Jun},
  publisher = {American Physical Society},
  doi = {10.1103/RevModPhys.81.865},
  url = {https://link.aps.org/doi/10.1103/RevModPhys.81.865}
}

@article{fan2022quantum,
  title = {Quantum steering as resource of quantum teleportation},
  author = {Fan, Yi and Jia, Chuanlei and Qiu, Liang},
  journal = {Phys. Rev. A},
  volume = {106},
  issue = {1},
  pages = {012433},
  numpages = {6},
  year = {2022},
  month = {Jul},
  publisher = {American Physical Society},
  doi = {10.1103/PhysRevA.106.012433},
  url = {https://link.aps.org/doi/10.1103/PhysRevA.106.012433}
}

@article{costa2016quantification,
  title = {Quantification of Einstein-Podolsky-Rosen steering for two-qubit states},
  author = {Costa, A. C. S. and Angelo, R. M.},
  journal = {Phys. Rev. A},
  volume = {93},
  issue = {2},
  pages = {020103},
  numpages = {5},
  year = {2016},
  month = {Feb},
  publisher = {American Physical Society},
  doi = {10.1103/PhysRevA.93.020103},
  url = {https://link.aps.org/doi/10.1103/PhysRevA.93.020103}
}

@article{bennett1993teleporting,
  title = {Teleporting an unknown quantum state via dual classical and Einstein-Podolsky-Rosen channels},
  author = {Bennett, Charles H. and Brassard, Gilles and Cr\'epeau, Claude and Jozsa, Richard and Peres, Asher and Wootters, William K.},
  journal = {Phys. Rev. Lett.},
  volume = {70},
  issue = {13},
  pages = {1895--1899},
  numpages = {0},
  year = {1993},
  month = {Mar},
  publisher = {American Physical Society},
  doi = {10.1103/PhysRevLett.70.1895},
  url = {https://link.aps.org/doi/10.1103/PhysRevLett.70.1895}
}

@article{horodecki1996teleportation,
title = {Teleportation, Bell's inequalities and inseparability},
journal = {Physics Letters A},
volume = {222},
number = {1},
pages = {21-25},
year = {1996},
issn = {0375-9601},
doi = {https://doi.org/10.1016/0375-9601(96)00639-1},
url = {https://www.sciencedirect.com/science/article/pii/0375960196006391},
author = {Ryszard Horodecki and Michał Horodecki and Paweł Horodecki}
}

@article{badziag2000local,
  title = {Local environment can enhance fidelity of quantum teleportation},
  author={Badziag, Piotr and Horodecki, Michał and Horodecki, Paweł and Horodecki, Ryszard},
  journal = {Phys. Rev. A},
  volume = {62},
  issue = {1},
  pages = {012311},
  numpages = {7},
  year = {2000},
  month = {Jun},
  publisher = {American Physical Society},
  doi = {10.1103/PhysRevA.62.012311},
  url = {https://link.aps.org/doi/10.1103/PhysRevA.62.012311}
}

@article{bang2018fidelity,
doi = {10.1088/1751-8121/aaac35},
url = {https://dx.doi.org/10.1088/1751-8121/aaac35},
year = {2018},
month = {feb},
publisher = {IOP Publishing},
volume = {51},
number = {13},
pages = {135302},
author = {Jeongho Bang and Junghee Ryu and Dagomir Kaszlikowski},
title = {Fidelity deviation in quantum teleportation},
journal = {Journal of Physics A: Mathematical and Theoretical}
}

@article{horodecki1999general,
  title = {General teleportation channel, singlet fraction, and quasidistillation},
  author = {Horodecki, Micha\l{} and Horodecki, Pawe\l{} and Horodecki, Ryszard},
  journal = {Phys. Rev. A},
  volume = {60},
  issue = {3},
  pages = {1888--1898},
  numpages = {0},
  year = {1999},
  month = {Sep},
  publisher = {American Physical Society},
  doi = {10.1103/PhysRevA.60.1888},
  url = {https://link.aps.org/doi/10.1103/PhysRevA.60.1888}
}

@article{kim2012protecting,
  title={Protecting entanglement from decoherence using weak measurement and quantum measurement reversal},
  author={Kim, Yong-Su and Lee, Jong-Chan and Kwon, Osung and Kim, Yoon-Ho},
  journal={Nature Physics},
  volume={8},
  number={2},
  pages={117--120},
  year={2012},
  publisher={Nature Publishing Group UK London},
  url = {https://doi.org/10.1038/nphys2178}
}

@article{korotkov2006undoing,
  title = {Undoing a Weak Quantum Measurement of a Solid-State Qubit},
  author = {Korotkov, Alexander N. and Jordan, Andrew N.},
  journal = {Phys. Rev. Lett.},
  volume = {97},
  issue = {16},
  pages = {166805},
  numpages = {4},
  year = {2006},
  month = {Oct},
  publisher = {American Physical Society},
  doi = {10.1103/PhysRevLett.97.166805},
  url = {https://link.aps.org/doi/10.1103/PhysRevLett.97.166805}
}

@article{ollivier2001quantum,
  title = {Quantum Discord: A Measure of the Quantumness of Correlations},
  author = {Ollivier, Harold and Zurek, Wojciech H.},
  journal = {Phys. Rev. Lett.},
  volume = {88},
  issue = {1},
  pages = {017901},
  numpages = {4},
  year = {2001},
  month = {Dec},
  publisher = {American Physical Society},
  doi = {10.1103/PhysRevLett.88.017901},
  url = {https://link.aps.org/doi/10.1103/PhysRevLett.88.017901}
}

@article{bennett1999quantum,
  title = {Quantum nonlocality without entanglement},
  author = {Bennett, Charles H. and DiVincenzo, David P. and Fuchs, Christopher A. and Mor, Tal and Rains, Eric and Shor, Peter W. and Smolin, John A. and Wootters, William K.},
  journal = {Phys. Rev. A},
  volume = {59},
  issue = {2},
  pages = {1070--1091},
  numpages = {0},
  year = {1999},
  month = {Feb},
  publisher = {American Physical Society},
  doi = {10.1103/PhysRevA.59.1070},
  url = {https://link.aps.org/doi/10.1103/PhysRevA.59.1070}
}

@article{ramkarthik2020quantum,
  title={Quantum discord and logarithmic negativity in the generalized N-qubit Werner state},
  author={Ramkarthik, MS and Tiwari, Devvrat and Barkataki, Pranay},
  journal={International Journal of Theoretical Physics},
  volume={59},
  pages={4040--4057},
  year={2020},
  publisher={Springer},
  doi = {10.1007/s10773-020-04663-2},
  url =  {https://doi.org/10.1007/s10773-020-04663-2}
}

@article{leonhardt1997measuring,
title = {Measuring the quantum state of light},
journal = {Progress in Quantum Electronics},
volume = {19},
number = {2},
pages = {89-130},
year = {1995},
issn = {0079-6727},
doi = {https://doi.org/10.1016/0079-6727(94)00007-L},
url = {https://www.sciencedirect.com/science/article/pii/007967279400007L},
author = {U. Leonhardt and H. Paul}
}

@article{agarwal1981relation,
  title = {Relation between atomic coherent-state representation, state multipoles, and generalized phase-space distributions},
  author = {Agarwal, G. S.},
  journal = {Phys. Rev. A},
  volume = {24},
  issue = {6},
  pages = {2889--2896},
  numpages = {0},
  year = {1981},
  month = {Dec},
  publisher = {American Physical Society},
  doi = {10.1103/PhysRevA.24.2889},
  url = {https://link.aps.org/doi/10.1103/PhysRevA.24.2889}
}

@article{agarwal1998state,
  title = {State reconstruction for a collection of two-level systems},
  author = {Agarwal, G. S.},
  journal = {Phys. Rev. A},
  volume = {57},
  issue = {1},
  pages = {671--673},
  year = {1998},
  month = {Jan},
  publisher = {American Physical Society},
  doi = {10.1103/PhysRevA.57.671},
  url = {https://link.aps.org/doi/10.1103/PhysRevA.57.671}
}

@article{cohen1986joint,
  title={Joint Wigner distribution for spin-1/2 particles},
  author={Cohen, Leon and Scully, Marlan O},
  journal={Foundations of physics},
  volume={16},
  number={4},
  pages={295--310},
  year={1986},
  publisher={Springer},
  url={https://doi.org/10.1007/BF01882690},
  doi={10.1007/BF01882690}
}

@article{galetti1988extended,
title = {An extended Weyl-Wigner transformation for special finite spaces},
journal = {Physica A: Statistical Mechanics and its Applications},
volume = {149},
number = {1},
pages = {267-282},
year = {1988},
issn = {0378-4371},
doi = {https://doi.org/10.1016/0378-4371(88)90219-1},
url = {https://www.sciencedirect.com/science/article/pii/0378437188902191},
author = {D. Galetti and A.F.R. {de Toledo Piza}}
}

@article{leonhardt1996discrete,
  title = {Discrete Wigner function and quantum-state tomography},
  author = {Leonhardt, Ulf},
  journal = {Phys. Rev. A},
  volume = {53},
  issue = {5},
  pages = {2998--3013},
  numpages = {0},
  year = {1996},
  month = {May},
  publisher = {American Physical Society},
  doi = {10.1103/PhysRevA.53.2998},
  url = {https://link.aps.org/doi/10.1103/PhysRevA.53.2998}
}

@article{wootters2004picturing,
  author={Wootters, W. K.},
  journal={IBM Journal of Research and Development}, 
  title={Picturing qubits in phase space}, 
  year={2004},
  volume={48},
  number={1},
  pages={99-110},
  doi={10.1147/rd.481.0099}
}

@article{chaturvedi2005wigner,
  title={Wigner distributions for finite-dimensional quantum systems: An algebraic approach},
  author={Chaturvedi, S and Ercolessi, Elisa and Marmo, Giuseppe and Morandi, Giuseppe and Mukunda, N and Simon, R},
  journal={Pramana},
  volume={65},
  pages={981--993},
  year={2005},
  publisher={Springer},
  url={https://doi.org/10.1007/BF02705275},
  doi={10.1007/BF02705275}
}

@article{paz2005qubits,
  title = {Qubits in phase space: Wigner-function approach to quantum-error correction and the mean-king problem},
  author = {Paz, Juan Pablo and Roncaglia, Augusto Jos\'e and Saraceno, Marcos},
  journal = {Phys. Rev. A},
  volume = {72},
  issue = {1},
  pages = {012309},
  numpages = {19},
  year = {2005},
  month = {Jul},
  publisher = {American Physical Society},
  doi = {10.1103/PhysRevA.72.012309},
  url = {https://link.aps.org/doi/10.1103/PhysRevA.72.012309}
}

@article{lopez2003phase,
  title = {Phase-space approach to the study of decoherence in quantum walks},
  author = {L\'opez, Cecilia C. and Paz, Juan Pablo},
  journal = {Phys. Rev. A},
  volume = {68},
  issue = {5},
  pages = {052305},
  numpages = {9},
  year = {2003},
  month = {Nov},
  publisher = {American Physical Society},
  doi = {10.1103/PhysRevA.68.052305},
  url = {https://link.aps.org/doi/10.1103/PhysRevA.68.052305}
}

@article{paz2004quantum,
  title = {Quantum algorithms for phase-space tomography},
  author = {Paz, Juan Pablo and Roncaglia, Augusto Jos\'e and Saraceno, Marcos},
  journal = {Phys. Rev. A},
  volume = {69},
  issue = {3},
  pages = {032312},
  numpages = {9},
  year = {2004},
  month = {Mar},
  publisher = {American Physical Society},
  doi = {10.1103/PhysRevA.69.032312},
  url = {https://link.aps.org/doi/10.1103/PhysRevA.69.032312}
}

@article{koniorczyk2001wigner,
  title = {Wigner-function description of quantum teleportation in arbitrary dimensions and a continuous limit},
  author = {Koniorczyk, M. and Bu\ifmmode \check{z}\else \v{z}\fi{}ek, V. and Janszky, J.},
  journal = {Phys. Rev. A},
  volume = {64},
  issue = {3},
  pages = {034301},
  numpages = {4},
  year = {2001},
  month = {Aug},
  publisher = {American Physical Society},
  doi = {10.1103/PhysRevA.64.034301},
  url = {https://link.aps.org/doi/10.1103/PhysRevA.64.034301}
}

@article{paz2002discrete,
  title = {Discrete Wigner functions and the phase-space representation of quantum teleportation},
  author = {Paz, Juan Pablo},
  journal = {Phys. Rev. A},
  volume = {65},
  issue = {6},
  pages = {062311},
  numpages = {8},
  year = {2002},
  month = {Jun},
  publisher = {American Physical Society},
  doi = {10.1103/PhysRevA.65.062311},
  url = {https://link.aps.org/doi/10.1103/PhysRevA.65.062311}
}

@article{pittenger2005wigner,
doi = {10.1088/0305-4470/38/26/012},
url = {https://dx.doi.org/10.1088/0305-4470/38/26/012},
year = {2005},
month = {jun},
publisher = {},
volume = {38},
number = {26},
pages = {6005},
author = {Arthur O Pittenger and Morton H Rubin},
title = {Wigner functions and separability for finite systems},
journal = {Journal of Physics A: Mathematical and General}
}

@article{korotkov2010decoherence,
  title = {Decoherence suppression by quantum measurement reversal},
  author = {Korotkov, Alexander N. and Keane, Kyle},
  journal = {Phys. Rev. A},
  volume = {81},
  issue = {4},
  pages = {040103},
  numpages = {4},
  year = {2010},
  month = {Apr},
  publisher = {American Physical Society},
  doi = {10.1103/PhysRevA.81.040103},
  url = {https://link.aps.org/doi/10.1103/PhysRevA.81.040103}
}

@article{sabale2023towards,
  title={Towards realization of universal quantum teleportation using weak measurements},
  author={Sabale, Vivek Balasaheb and Kumar, Atul and Banerjee, Subhasish},
  journal={arXiv preprint arXiv:2307.09231},
  year={2023}
}

@book{breuer2002theory,
  title={The Theory of Open Quantum Systems},
  author={Breuer, H.P. and Petruccione, F.},
  isbn={9780198520634},
  lccn={2002075713},
  url={https://books.google.co.in/books?id=0Yx5VzaMYm8C},
  year={2002},
  publisher={Oxford University Press}
}

@book{weiss2012quantum,
  title={Quantum Dissipative Systems},
  author={Weiss, U.},
  isbn={9789814374910},
  lccn={2012418311},
  series={G - Reference,Information and Interdisciplinary Subjects Series},
  url={https://books.google.co.in/books?id=qgfuFZxvGKQC},
  year={2012},
  publisher={World Scientific}
}

@article{de2017dynamics,
  title = {Dynamics of non-Markovian open quantum systems},
  author = {de Vega, In\'es and Alonso, Daniel},
  journal = {Rev. Mod. Phys.},
  volume = {89},
  issue = {1},
  pages = {015001},
  numpages = {58},
  year = {2017},
  month = {Jan},
  publisher = {American Physical Society},
  doi = {10.1103/RevModPhys.89.015001},
  url = {https://link.aps.org/doi/10.1103/RevModPhys.89.015001}
}

@article{li2018concepts,
  title={Concepts of quantum non-Markovianity: A hierarchy},
  author={Li Li and Michael J. W. Hall and Howard M. Wiseman},
  journal={Physics Reports},
  year={2017},
  url={https://api.semanticscholar.org/CorpusID:20209678}
}

@article{bouwmeester1997experimental,
  title={Experimental quantum teleportation},
  author={Bouwmeester, Dik and Pan, Jian-Wei and Mattle, Klaus and Eibl, Manfred and Weinfurter, Harald and Zeilinger, Anton},
  journal={Nature},
  volume={390},
  number={6660},
  pages={575--579},
  year={1997},
  publisher={Nature Publishing Group UK London}, 
  urk={https://doi.org/10.1038/37539},
  doi={10.1038/37539}
}

@article{masanes2011secure,
  title={Secure device-independent quantum key distribution with causally independent measurement devices},
  author={Masanes, Lluis and Pironio, Stefano and Ac{\'\i}n, Antonio},
  journal={Nature communications},
  volume={2},
  number={1},
  pages={238},
  year={2011},
  publisher={Nature Publishing Group UK London},
  url={https://doi.org/10.1038/ncomms1244},
  doi={10.1038/ncomms1244}
}

@article{giovannetti2011advances,
  title={Advances in quantum metrology},
  author={Giovannetti, Vittorio and Lloyd, Seth and Maccone, Lorenzo},
  journal={Nature photonics},
  volume={5},
  number={4},
  pages={222--229},
  year={2011},
  publisher={Nature Publishing Group UK London},
  url={https://doi.org/10.1038/nphoton.2011.35},
  doi={10.1038/nphoton.2011.35}
}

@article{cormick2006classicality,
  title = {Classicality in discrete Wigner functions},
  author = {Cormick, Cecilia and Galv\~ao, Ernesto F. and Gottesman, Daniel and Paz, Juan Pablo and Pittenger, Arthur O.},
  journal = {Phys. Rev. A},
  volume = {73},
  issue = {1},
  pages = {012301},
  numpages = {9},
  year = {2006},
  month = {Jan},
  publisher = {American Physical Society},
  doi = {10.1103/PhysRevA.73.012301},
  url = {https://link.aps.org/doi/10.1103/PhysRevA.73.012301}
}

@article{lalita2025interrelation,
  title={Interrelation of Non-Classicality, Entropy, Irreversibility and Work extraction in Open Quantum Systems},
  author={Lalita, Jai and Banerjee, Subhashish},
  journal={arXiv preprint arXiv:2510.15140},
  year={2025}
}

@article{henderson2001classical,
doi = {10.1088/0305-4470/34/35/315},
url = {https://dx.doi.org/10.1088/0305-4470/34/35/315},
year = {2001},
month = {aug},
publisher = {},
volume = {34},
number = {35},
pages = {6899},
author = {L Henderson and  V Vedral},
title = {Classical, quantum and total correlations},
journal = {Journal of Physics A: Mathematical and General}
}

@article{adhikari2012operational,
  title={Operational meaning of discord in terms of teleportation fidelity},
  author={Satyabrata Adhikari and Subhashis Banerjee},
  journal={Physical Review A},
  year={2012},
  volume={86},
  pages={062313},
  url={https://api.semanticscholar.org/CorpusID:118416180}
}

@article{thapliyal2017quantum,
  title={Quantum cryptography over non-Markovian channels},
  author={Thapliyal, Kishore and Pathak, Anirban and Banerjee, Subhashish},
  journal={Quantum Information Processing},
  volume={16},
  pages={1--21},
  year={2017},
  publisher={Springer},
  url={https://doi.org/10.1007/s11128-017-1567-1},
  doi={10.1007/s11128-017-1567-1}
}

@article{Chakrabarty2010study,
  title={A study of quantum correlations in open quantum systems},
  author={Indranil Chakrabarty and Subhashis Banerjee and Nana Siddharth},
  journal={Quantum Inf. Comput.},
  year={2010},
  volume={11},
  pages={541-562},
  url={https://api.semanticscholar.org/CorpusID:15930121}
}

@article{monroe2021weak,
  title = {Weak Measurement of a Superconducting Qubit Reconciles Incompatible Operators},
  author = {Monroe, Jonathan T. and Yunger Halpern, Nicole and Lee, Taeho and Murch, Kater W.},
  journal = {Phys. Rev. Lett.},
  volume = {126},
  issue = {10},
  pages = {100403},
  numpages = {7},
  year = {2021},
  month = {Mar},
  publisher = {American Physical Society},
  doi = {10.1103/PhysRevLett.126.100403},
  url = {https://link.aps.org/doi/10.1103/PhysRevLett.126.100403}
}

@article{xiao2013protecting,
  title={Protecting qutrit-qutrit entanglement by weak measurement and reversal},
  author={Xiao, Xing and Li, Yan-Ling},
  journal={The European Physical Journal D},
  volume={67},
  pages={1--7},
  year={2013},
  publisher={Springer},
  url={https://doi.org/10.1140/epjd/e2013-40036-3},
  doi={10.1140/epjd/e2013-40036-3}
}

@article{sun2017recovering,
doi = {10.1088/1612-202X/aa8e86},
url = {https://dx.doi.org/10.1088/1612-202X/aa8e86},
year = {2017},
month = {nov},
publisher = {IOP Publishing},
volume = {14},
number = {12},
pages = {125204},
author = {Wen-Yang Sun and Dong Wang and Zhi-Yong Ding and Liu Ye},
title = {Recovering the lost steerability of quantum states within non-Markovian environments by utilizing quantum partially collapsing measurements},
journal = {Laser Physics Letters}
}

@article{aharonov1988result,
  title = {How the result of a measurement of a component of the spin of a spin-1/2 particle can turn out to be 100},
  author = {Aharonov, Yakir and Albert, David Z. and Vaidman, Lev},
  journal = {Phys. Rev. Lett.},
  volume = {60},
  issue = {14},
  pages = {1351--1354},
  numpages = {0},
  year = {1988},
  month = {Apr},
  publisher = {American Physical Society},
  doi = {10.1103/PhysRevLett.60.1351},
  url = {https://link.aps.org/doi/10.1103/PhysRevLett.60.1351}
}

@article{dressel2014colloquium,
  title = {Colloquium: Understanding quantum weak values: Basics and applications},
  author = {Dressel, Justin and Malik, Mehul and Miatto, Filippo M. and Jordan, Andrew N. and Boyd, Robert W.},
  journal = {Rev. Mod. Phys.},
  volume = {86},
  issue = {1},
  pages = {307--316},
  numpages = {10},
  year = {2014},
  month = {Mar},
  publisher = {American Physical Society},
  doi = {10.1103/RevModPhys.86.307},
  url = {https://link.aps.org/doi/10.1103/RevModPhys.86.307}
}

@article{lahiri2021exploring,
  title={Exploring the extent of validity of quantum work fluctuation theorems in the presence of weak measurements},
  author={Lahiri, Sourabh and Banerjee, Subhashish and Jayannavar, AM},
  journal={Quantum Information Processing},
  volume={20},
  number={11},
  pages={372},
  year={2021},
  publisher={Springer},
  url= {https://ui.adsabs.harvard.edu/abs/2021QuIP...20..372L},
  doi={10.1007/s11128-021-03260-4}
}

@article{luo2008quantum,
  title = {Quantum discord for two-qubit systems},
  author = {Luo, Shunlong},
  journal = {Phys. Rev. A},
  volume = {77},
  issue = {4},
  pages = {042303},
  numpages = {6},
  year = {2008},
  month = {Apr},
  publisher = {American Physical Society},
  doi = {10.1103/PhysRevA.77.042303},
  url = {https://link.aps.org/doi/10.1103/PhysRevA.77.042303}
}

@article{boschi1998experimental,
  title = {Experimental Realization of Teleporting an Unknown Pure Quantum State via Dual Classical and Einstein-Podolsky-Rosen Channels},
  author = {Boschi, D. and Branca, S. and De Martini, F. and Hardy, L. and Popescu, S.},
  journal = {Phys. Rev. Lett.},
  volume = {80},
  issue = {6},
  pages = {1121--1125},
  numpages = {0},
  year = {1998},
  month = {Feb},
  publisher = {American Physical Society},
  doi = {10.1103/PhysRevLett.80.1121},
  url = {https://link.aps.org/doi/10.1103/PhysRevLett.80.1121}
}

@article{gisin2007quantum,
  title={Quantum communication},
  author={Gisin, Nicolas and Thew, Rob},
  journal={Nature photonics},
  volume={1},
  number={3},
  pages={165--171},
  year={2007},
  publisher={Nature Publishing Group UK London},
  url = {https://doi.org/10.1038/nphoton.2007.22}
}

@article{jin2010experimental,
  title={Experimental free-space quantum teleportation},
  author={Jin, Xian-Min and Ren, Ji-Gang and Yang, Bin and Yi, Zhen-Huan and Zhou, Fei and Xu, Xiao-Fan and Wang, Shao-Kai and Yang, Dong and Hu, Yuan-Feng and Jiang, Shuo and others},
  journal={Nature photonics},
  volume={4},
  number={6},
  pages={376--381},
  year={2010},
  publisher={Nature Publishing Group UK London},
  url={http://dx.doi.org/10.1038/nphoton.2010.87}
}

@article{zeilinger2018quantum,
  title={Quantum teleportation, onwards and upwards},
  author={Zeilinger, Anton},
  journal={Nature Physics},
  volume={14},
  number={1},
  pages={3--4},
  year={2018},
  publisher={Nature Publishing Group UK London},
  url={https://doi.org/10.1038/nphys4339}
}

@article{jozsa1993teleporting,
  title = {Teleporting an unknown quantum state via dual classical and Einstein-Podolsky-Rosen channels},
  author = {Bennett, Charles H. and Brassard, Gilles and Cr\'epeau, Claude and Jozsa, Richard and Peres, Asher and Wootters, William K.},
  journal = {Phys. Rev. Lett.},
  volume = {70},
  issue = {13},
  pages = {1895--1899},
  numpages = {0},
  year = {1993},
  month = {Mar},
  publisher = {American Physical Society},
  doi = {10.1103/PhysRevLett.70.1895},
  url = {https://link.aps.org/doi/10.1103/PhysRevLett.70.1895}
}

@article{ghosal2021characterizing,
  title = {Characterizing qubit channels in the context of quantum teleportation},
  author = {Ghosal, Arkaprabha and Das, Debarshi and Banerjee, Subhashish},
  journal = {Phys. Rev. A},
  volume = {103},
  issue = {5},
  pages = {052422},
  numpages = {18},
  year = {2021},
  month = {May},
  publisher = {American Physical Society},
  doi = {10.1103/PhysRevA.103.052422},
  url = {https://link.aps.org/doi/10.1103/PhysRevA.103.052422}
}

@article{pramanik2013improving,
  title={Improving the fidelity of teleportation through noisy channels using weak measurement},
  author={Pramanik, Tanumoy and Majumdar, AS},
  journal={Physics Letters A},
  volume={377},
  number={44},
  pages={3209--3215},
  year={2013},
  publisher={Elsevier}, 
  url={https://www.sciencedirect.com/science/article/pii/S0375960113009444}
}

@article{SM23,
  title={Quantum non-Markovianity: Overview and recent developments},
  author={Shrikant, U and Mandayam, Prabha},
  journal={Frontiers in Quantum Science and Technology},
  volume={2},
  pages={1134583},
  year={2023},
  publisher={Frontiers Media SA},
  url={https://doi.org/10.3389/frqst.2023.1134583}
}

@article{AZ04,
doi = {10.1088/1464-4266/6/10/003},
url = {https://dx.doi.org/10.1088/1464-4266/6/10/003},
year = {2004},
month = {aug},
publisher = {},
volume = {6},
number = {10},
pages = {396},
author = {Anatole Kenfack and  Karol Życzkowski},
title = {Negativity of the Wigner function as an indicator of non-classicality},
journal = {Journal of Optics B: Quantum and Semiclassical Optics}
}

@article{W32,
  title = {On the Quantum Correction For Thermodynamic Equilibrium},
  author = {Wigner, E.},
  journal = {Phys. Rev.},
  volume = {40},
  issue = {5},
  pages = {749--759},
  numpages = {0},
  year = {1932},
  month = {Jun},
  publisher = {American Physical Society},
  doi = {10.1103/PhysRev.40.749},
  url = {https://link.aps.org/doi/10.1103/PhysRev.40.749}
}

@article{Lawrence2002MUB,
  title = {Mutually unbiased binary observable sets on N qubits},
  author = {Lawrence, Jay and Brukner, \ifmmode \check{C}\else \v{C}\fi{}aslav and Zeilinger, Anton},
  journal = {Phys. Rev. A},
  volume = {65},
  issue = {3},
  pages = {032320},
  numpages = {5},
  year = {2002},
  month = {Feb},
  publisher = {American Physical Society},
  doi = {10.1103/PhysRevA.65.032320},
  url = {https://link.aps.org/doi/10.1103/PhysRevA.65.032320}
}

@article{bandyopadhyay2002MUB,
  title={A new proof for the existence of mutually unbiased bases},
  author={Bandyopadhyay and Boykin and Roychowdhury and Vatan},
  journal={Algorithmica},
  volume={34},
  pages={512--528},
  year={2002},
  publisher={Springer},
  doi = {10.1007/s00453-002-0980-7},
  url = {https://rdcu.be/dde4t}
}

@article{pittenger2004mutually,
  title={Mutually unbiased bases, generalized spin matrices and separability},
  author={Pittenger, Arthur O and Rubin, Morton H},
  journal={Linear algebra and its applications},
  volume={390},
  pages={255--278},
  year={2004},
  publisher={Elsevier},
  url={https://doi.org/10.1016/j.laa.2004.04.025}
}

@article{wootters1987wigner,
title = {A Wigner-function formulation of finite-state quantum mechanics},
journal = {Annals of Physics},
volume = {176},
number = {1},
pages = {1-21},
year = {1987},
issn = {0003-4916},
doi = {https://doi.org/10.1016/0003-4916(87)90176-X},
url = {https://www.sciencedirect.com/science/article/pii/000349168790176X},
author = {William K Wootters}
}

@article{van2011noise,
  title = {Noise thresholds for higher-dimensional systems using the discrete Wigner function},
  author = {van Dam, Wim and Howard, Mark},
  journal = {Phys. Rev. A},
  volume = {83},
  issue = {3},
  pages = {032310},
  numpages = {13},
  year = {2011},
  month = {Mar},
  publisher = {American Physical Society},
  doi = {10.1103/PhysRevA.83.032310},
  url = {https://link.aps.org/doi/10.1103/PhysRevA.83.032310}
}

@article{casaccino2008extrema,
  title = {Extrema of discrete Wigner functions and applications},
  author = {Casaccino, Andrea and Galv\~ao, Ernesto F. and Severini, Simone},
  journal = {Phys. Rev. A},
  volume = {78},
  issue = {2},
  pages = {022310},
  numpages = {6},
  year = {2008},
  month = {Aug},
  publisher = {American Physical Society},
  doi = {10.1103/PhysRevA.78.022310},
  url = {https://link.aps.org/doi/10.1103/PhysRevA.78.022310}
}

@book{lidl1994introduction,
  title={Introduction to finite fields and their applications},
  author={Lidl, Rudolf and Niederreiter, Harald},
  year={1994},
  publisher={Cambridge university press},
  doi = {https://doi.org/10.1017/CBO9781139172769}
}

@book{nielsen2010quantum,
  title={Quantum Computation and Quantum Information: 10th Anniversary Edition},
  author={Nielsen, M.A. and Chuang, I.L.},
  isbn={9781139495486},
  url={https://books.google.co.in/books?id=-s4DEy7o-a0C},
  year={2010},
  publisher={Cambridge University Press}
}

@article{Younis2020QFASTQS,
  title={QFAST: Quantum Synthesis Using a Hierarchical Continuous Circuit Space},
  author={Ed Younis and Koushik Sen and Katherine A. Yelick and Costin Iancu},
  journal={arXiv: Quantum Physics},
  year={2020},
  url={https://api.semanticscholar.org/CorpusID:212644706}
}

@article{jozsa1994fidelity,
  title={Fidelity for mixed quantum states},
  author={Jozsa, Richard},
  journal={Journal of modern optics},
  volume={41},
  number={12},
  pages={2315--2323},
  year={1994},
  publisher={Taylor \& Francis},
  url={https://doi.org/10.1080/09500349414552171}
}

@misc{qiskit2024,
      title={Quantum computing with {Q}iskit},
      author={Javadi-Abhari, Ali and Treinish, Matthew and Krsulich, Kevin and Wood, Christopher J. and Lishman, Jake and Gacon, Julien and Martiel, Simon and Nation, Paul D. and Bishop, Lev S. and Cross, Andrew W. and Johnson, Blake R. and Gambetta, Jay M.},
      year={2024},
      doi={10.48550/arXiv.2405.08810},
      eprint={2405.08810},
      archivePrefix={arXiv},
      primaryClass={quant-ph}
}

@article{Maximum-Likelihood_2012,
  title = {Efficient Method for Computing the Maximum-Likelihood Quantum State from Measurements with Additive Gaussian Noise},
  author = {Smolin, John A. and Gambetta, Jay M. and Smith, Graeme},
  journal = {Phys. Rev. Lett.},
  volume = {108},
  issue = {7},
  pages = {070502},
  numpages = {4},
  year = {2012},
  month = {Feb},
  publisher = {American Physical Society},
  doi = {10.1103/PhysRevLett.108.070502},
  url = {https://link.aps.org/doi/10.1103/PhysRevLett.108.070502}
}

@article{Teleporting1993,
  title = {Teleporting an unknown quantum state via dual classical and Einstein-Podolsky-Rosen channels},
  author = {Bennett, Charles H. and Brassard, Gilles and Cr\'epeau, Claude and Jozsa, Richard and Peres, Asher and Wootters, William K.},
  journal = {Phys. Rev. Lett.},
  volume = {70},
  issue = {13},
  pages = {1895--1899},
  numpages = {0},
  year = {1993},
  month = {Mar},
  publisher = {American Physical Society},
  doi = {10.1103/PhysRevLett.70.1895},
  url = {https://link.aps.org/doi/10.1103/PhysRevLett.70.1895}
}

@article{Quantumerrorcorrection2013,
doi = {10.1088/0034-4885/76/7/076001},
url = {https://dx.doi.org/10.1088/0034-4885/76/7/076001},
year = {2013},
month = {jun},
publisher = {IOP Publishing},
volume = {76},
number = {7},
pages = {076001},
author = {Simon J Devitt and William J Munro and Kae Nemoto},
title = {Quantum error correction for beginners},
journal = {Reports on Progress in Physics}
}

@article{Quantum-State-Tomography1995,
  title = {Quantum-State Tomography and Discrete Wigner Function},
  author = {Leonhardt, Ulf},
  journal = {Phys. Rev. Lett.},
  volume = {74},
  issue = {21},
  pages = {4101--4105},
  numpages = {0},
  year = {1995},
  month = {May},
  publisher = {American Physical Society},
  doi = {10.1103/PhysRevLett.74.4101},
  url = {https://link.aps.org/doi/10.1103/PhysRevLett.74.4101}
}

@misc{ibm_brisbane,
  author       = {Quantum, IBM},
  title        = {IBM Quantum {B}risbane backend},
  year         = 2024,
  note         = {Accessed via IBM Quantum Experience},
  howpublished = {\url{https://quantum-computing.ibm.com/}},
}

@article{oreshkov2005weak,
  title = {Weak Measurements Are Universal},
  author = {Oreshkov, Ognyan and Brun, Todd A.},
  journal = {Phys. Rev. Lett.},
  volume = {95},
  issue = {11},
  pages = {110409},
  numpages = {4},
  year = {2005},
  month = {Sep},
  publisher = {American Physical Society},
  doi = {10.1103/PhysRevLett.95.110409},
  url = {https://link.aps.org/doi/10.1103/PhysRevLett.95.110409}
}

@article{katz2008reversal,
  title = {Reversal of the Weak Measurement of a Quantum State in a Superconducting Phase Qubit},
  author = {Katz, Nadav and Neeley, Matthew and Ansmann, M. and Bialczak, Radoslaw C. and Hofheinz, M. and Lucero, Erik and O'Connell, A. and Wang, H. and Cleland, A. N. and Martinis, John M. and Korotkov, Alexander N.},
  journal = {Phys. Rev. Lett.},
  volume = {101},
  issue = {20},
  pages = {200401},
  numpages = {4},
  year = {2008},
  month = {Nov},
  publisher = {American Physical Society},
  doi = {10.1103/PhysRevLett.101.200401},
  url = {https://link.aps.org/doi/10.1103/PhysRevLett.101.200401}
}

@article{kim2009reversing,
  title={Reversing the weak quantum measurement for a photonic qubit},
  author={Kim, Yong-Su and Cho, Young-Wook and Ra, Young-Sik and Kim, Yoon-Ho},
  journal={Optics express},
  volume={17},
  number={14},
  pages={11978--11985},
  year={2009},
  publisher={Optica Publishing Group},
  url = {https://doi.org/10.1364/OE.17.011978}
}

@article{kumar2018non,
  title={Non-Markovian evolution: a quantum walk perspective},
  author={Kumar, N Pradeep and Banerjee, Subhashish and Srikanth, R and Jagadish, Vinayak and Petruccione, Francesco},
  journal={Open Systems \& Information Dynamics},
  volume={25},
  number={03},
  pages={1850014},
  year={2018},
  publisher={World Scientific},
  url = {https://doi.org/10.1142/S1230161218500142}
}

@article{Bellomo2007NMAD,
  title = {Non-Markovian Effects on the Dynamics of Entanglement},
  author = {Bellomo, B. and Lo Franco, R. and Compagno, G.},
  journal = {Phys. Rev. Lett.},
  volume = {99},
  issue = {16},
  pages = {160502},
  numpages = {4},
  year = {2007},
  month = {Oct},
  publisher = {American Physical Society},
  doi = {10.1103/PhysRevLett.99.160502},
  url = {https://link.aps.org/doi/10.1103/PhysRevLett.99.160502}
}

@book{Banerjee2018,
  title = {Open Quantum Systems: Dynamics of Nonclassical Evolution},
  ISBN = {9789811331824},
  ISSN = {2366-8857},
  url = {http://dx.doi.org/10.1007/978-981-13-3182-4},
  DOI = {10.1007/978-981-13-3182-4},
  journal = {Texts and Readings in Physical Sciences},
  publisher = {Springer Singapore},
  author = {Banerjee,  Subhashish},
  year = {2018}
}

@article{HORODECKI1995340,
title = {Violating Bell inequality by mixed spin-12 states: necessary and sufficient condition},
journal = {Physics Letters A},
volume = {200},
number = {5},
pages = {340-344},
year = {1995},
issn = {0375-9601},
doi = {https://doi.org/10.1016/0375-9601(95)00214-N},
url = {https://www.sciencedirect.com/science/article/pii/037596019500214N},
author = {R. Horodecki and P. Horodecki and M. Horodecki}
}

@article{helstrom1969quantum,
  title={Quantum detection and estimation theory},
  author={Helstrom, Carl W},
  journal={Journal of Statistical Physics},
  volume={1},
  pages={231--252},
  year={1969},
  publisher={Springer}
}

@article{hyllus2012fisher,
  title = {Fisher information and multiparticle entanglement},
  author = {Hyllus, Philipp and Laskowski, Wies\l{}aw and Krischek, Roland and Schwemmer, Christian and Wieczorek, Witlef and Weinfurter, Harald and Pezz\'e, Luca and Smerzi, Augusto},
  journal = {Phys. Rev. A},
  volume = {85},
  issue = {2},
  pages = {022321},
  numpages = {10},
  year = {2012},
  month = {Feb},
  publisher = {American Physical Society},
  doi = {10.1103/PhysRevA.85.022321},
  url = {https://link.aps.org/doi/10.1103/PhysRevA.85.022321}
}

@article{ma2011quantum,
  title = {Quantum Fisher information of the Greenberger-Horne-Zeilinger state in decoherence channels},
  author = {Ma, Jian and Huang, Yi-xiao and Wang, Xiaoguang and Sun, C. P.},
  journal = {Phys. Rev. A},
  volume = {84},
  issue = {2},
  pages = {022302},
  numpages = {7},
  year = {2011},
  month = {Aug},
  publisher = {American Physical Society},
  doi = {10.1103/PhysRevA.84.022302},
  url = {https://link.aps.org/doi/10.1103/PhysRevA.84.022302}
}

@INPROCEEDINGS{Shor_correction,
  author={Shor, P.W.},
  booktitle={Proceedings of 37th Conference on Foundations of Computer Science}, 
  title={Fault-tolerant quantum computation}, 
  year={1996},
  volume={},
  number={},
  pages={56-65},
  doi={10.1109/SFCS.1996.548464}}

@article{Bollinger1996Optimalfrequency,
  title = {Optimal frequency measurements with maximally correlated states},
  author = {Bollinger, J. J . and Itano, Wayne M. and Wineland, D. J. and Heinzen, D. J.},
  journal = {Phys. Rev. A},
  volume = {54},
  issue = {6},
  pages = {R4649--R4652},
  numpages = {0},
  year = {1996},
  month = {Dec},
  publisher = {American Physical Society},
  doi = {10.1103/PhysRevA.54.R4649},
  url = {https://link.aps.org/doi/10.1103/PhysRevA.54.R4649}
}

@article{peters1999measurement,
  title={Measurement of gravitational acceleration by dropping atoms},
  author={Peters, Achim and Chung, Keng Yeow and Chu, Steven},
  journal={Nature},
  volume={400},
  number={6747},
  pages={849--852},
  year={1999},
  publisher={Nature Publishing Group UK London}, 
  url={https://doi.org/10.1038/23655}
}

@article{he2020enhancing,
  title={Enhancing entanglement of assistance using weak measurement and quantum measurement reversal in correlated amplitude damping channel},
  author={He, Zhi and Zeng, Hao-Sheng},
  journal={Quantum Information Processing},
  volume={19},
  pages={1--13},
  year={2020},
  publisher={Springer},
  url={https://doi.org/10.1007/s11128-020-02791-6}
}

@article{ghosal2020optimal,
  title = {Optimal two-qubit states for quantum teleportation vis-\`a-vis state properties},
  author = {Ghosal, Arkaprabha and Das, Debarshi and Roy, Saptarshi and Bandyopadhyay, Somshubhro},
  journal = {Phys. Rev. A},
  volume = {101},
  issue = {1},
  pages = {012304},
  numpages = {10},
  year = {2020},
  month = {Jan},
  publisher = {American Physical Society},
  doi = {10.1103/PhysRevA.101.012304},
  url = {https://link.aps.org/doi/10.1103/PhysRevA.101.012304}
}

@article{Ekert1991Quantumcryptography,
  title = {Quantum cryptography based on Bell's theorem},
  author = {Ekert, Artur K.},
  journal = {Phys. Rev. Lett.},
  volume = {67},
  issue = {6},
  pages = {661--663},
  numpages = {0},
  year = {1991},
  month = {Aug},
  publisher = {American Physical Society},
  doi = {10.1103/PhysRevLett.67.661},
  url = {https://link.aps.org/doi/10.1103/PhysRevLett.67.661}
}

@article{Pezz2018Quantummetrology,
  title = {Quantum metrology with nonclassical states of atomic ensembles},
  author = {Pezz\`e, Luca and Smerzi, Augusto and Oberthaler, Markus K. and Schmied, Roman and Treutlein, Philipp},
  journal = {Rev. Mod. Phys.},
  volume = {90},
  issue = {3},
  pages = {035005},
  numpages = {70},
  year = {2018},
  month = {Sep},
  publisher = {American Physical Society},
  doi = {10.1103/RevModPhys.90.035005},
  url = {https://link.aps.org/doi/10.1103/RevModPhys.90.035005}
}

@ARTICLE{Braun2018Quantum_enhanced,
  title     = "Quantum-enhanced measurements without entanglement",
  author    = "Braun, Daniel and Adesso, Gerardo and Benatti, Fabio and
               Floreanini, Roberto and Marzolino, Ugo and Mitchell, Morgan W
               and Pirandola, Stefano",
  journal   = "Rev. Mod. Phys.",
  publisher = "American Physical Society (APS)",
  volume    =  90,
  number    =  3,
  month     =  sep,
  year      =  2018,
  url = "https://link.aps.org/licenses/aps-default-license",
  language  = "en"
}

@article{Scarani2009The_security,
  title     = "The security of practical quantum key distribution",
  author    = "Scarani, Valerio and Bechmann-Pasquinucci, Helle and Cerf,
               Nicolas J and Du{\v s}ek, Miloslav and L{\"u}tkenhaus, Norbert
               and Peev, Momtchil",
  journal   = "Rev. Mod. Phys.",
  publisher = "American Physical Society (APS)",
  volume    =  81,
  number    =  3,
  pages     = "1301--1350",
  month     =  sep,
  year      =  2009,
  url = "http://link.aps.org/licenses/aps-default-license",
  language  = "en"
}

@book{Breuer2007,
  title = {The Theory of Open Quantum Systems},
  ISBN = {9780191706349},
  url = {http://dx.doi.org/10.1093/acprof:oso/9780199213900.001.0001},
  DOI = {10.1093/acprof:oso/9780199213900.001.0001},
  publisher = {Oxford University PressOxford},
  author = {Breuer,  Heinz-Peter and Petruccione,  Francesco},
  year = {2007},
  month = jan 
}

@article{tiwari2024strong,
    author = {Tiwari, Devvrat and Bose, Baibhab and Banerjee, Subhashish},
    title = {Strong coupling non-Markovian quantum thermodynamics of a finite-bath system},
    journal = {The Journal of Chemical Physics},
    volume = {162},
    number = {11},
    pages = {114104},
    year = {2025},
    month = {03},
    issn = {0021-9606},
    doi = {10.1063/5.0254029},
    url = {https://doi.org/10.1063/5.0254029}
}

@Article{Utagi2020,
author={Utagi, Shrikant
and Srikanth, R.
and Banerjee, Subhashish},
title={Temporal self-similarity of quantum dynamical maps as a concept of memorylessness},
journal={Scientific Reports},
year={2020},
month={Sep},
day={14},
volume={10},
number={1},
pages={15049},
issn={2045-2322},
doi={10.1038/s41598-020-72211-3},
url={https://doi.org/10.1038/s41598-020-72211-3}
}

@article{CM_1963,
  title = {Relaxation Phenomena in Spin and Harmonic Oscillator Systems},
  author = {Rau, Jayaseetha},
  journal = {Phys. Rev.},
  volume = {129},
  issue = {4},
  pages = {1880--1888},
  numpages = {0},
  year = {1963},
  month = {Feb},
  publisher = {American Physical Society},
  doi = {10.1103/PhysRev.129.1880},
  url = {https://link.aps.org/doi/10.1103/PhysRev.129.1880}
}

@article{CM_nm_2017,
  title = {Non-Markovianity, coherence, and system-environment correlations in a long-range collision model},
  author={Çakmak, B. and Pezzutto, M. and Paternostro, M. and Müstecaplıoğlu, O. E.},
  journal = {Phys. Rev. A},
  volume = {96},
  issue = {2},
  pages = {022109},
  numpages = {8},
  year = {2017},
  month = {Aug},
  publisher = {American Physical Society},
  doi = {10.1103/PhysRevA.96.022109},
  url = {https://link.aps.org/doi/10.1103/PhysRevA.96.022109}
}

@article{ziman2005description,
  title={Description of quantum dynamics of open systems based on collision-like models},
  author={Ziman, M{\"a}rio and {\v{S}}telmachovi{\v{c}}, Peter and Bu{\v{z}}ek, Vladim{\'\i}r},
  journal={Open systems \& information dynamics},
  volume={12},
  pages={81--91},
  year={2005},
  publisher={Springer},
  url={https://doi.org/10.1007/s11080-005-0488-0}
}

@article{Rybár_2012,
doi = {10.1088/0953-4075/45/15/154006},
url = {https://dx.doi.org/10.1088/0953-4075/45/15/154006},
year = {2012},
month = {jul},
publisher = {IOP Publishing},
volume = {45},
number = {15},
pages = {154006},
author = {Rybár, Tomáš and Filippov, Sergey N and Ziman, Mário and Bužek, Vladimír},
title = {Simulation of indivisible qubit channels in collision models},
journal = {Journal of Physics B: Atomic, Molecular and Optical Physics}
}

@article{McCloskey2014Non-Markovianity,
  title = {Non-Markovianity and system-environment correlations in a microscopic collision model},
  author = {McCloskey, Ruari and Paternostro, Mauro},
  journal = {Phys. Rev. A},
  volume = {89},
  issue = {5},
  pages = {052120},
  numpages = {6},
  year = {2014},
  month = {May},
  publisher = {American Physical Society},
  doi = {10.1103/PhysRevA.89.052120},
  url = {https://link.aps.org/doi/10.1103/PhysRevA.89.052120}
}

@article{Ciccarello2013Collision-model,
  title = {Collision-model-based approach to non-Markovian quantum dynamics},
  author = {Ciccarello, F. and Palma, G. M. and Giovannetti, V.},
  journal = {Phys. Rev. A},
  volume = {87},
  issue = {4},
  pages = {040103},
  numpages = {5},
  year = {2013},
  month = {Apr},
  publisher = {American Physical Society},
  doi = {10.1103/PhysRevA.87.040103},
  url = {https://link.aps.org/doi/10.1103/PhysRevA.87.040103}
}

@article{Kretschmer2016Collisionmodel,
  title = {Collision model for non-Markovian quantum dynamics},
  author = {Kretschmer, Silvan and Luoma, Kimmo and Strunz, Walter T.},
  journal = {Phys. Rev. A},
  volume = {94},
  issue = {1},
  pages = {012106},
  numpages = {9},
  year = {2016},
  month = {Jul},
  publisher = {American Physical Society},
  doi = {10.1103/PhysRevA.94.012106},
  url = {https://link.aps.org/doi/10.1103/PhysRevA.94.012106}
}

@article{rivas2014quantum,
  title={Quantum non-Markovianity: characterization, quantification and detection},
  author={Rivas, {\'A}ngel and Huelga, Susana F and Plenio, Martin B},
  journal={Reports on Progress in Physics},
  volume={77},
  number={9},
  pages={094001},
  year={2014},
  publisher={IOP Publishing},
  url={https://iopscience.iop.org/article/10.1088/0034-4885/77/9/094001}
}

@article{Breuer2016Colloquium,
  title = {Colloquium: Non-Markovian dynamics in open quantum systems},
  author = {Breuer, Heinz-Peter and Laine, Elsi-Mari and Piilo, Jyrki and Vacchini, Bassano},
  journal = {Rev. Mod. Phys.},
  volume = {88},
  issue = {2},
  pages = {021002},
  numpages = {24},
  year = {2016},
  month = {Apr},
  publisher = {American Physical Society},
  doi = {10.1103/RevModPhys.88.021002},
  url = {https://link.aps.org/doi/10.1103/RevModPhys.88.021002}
}

@article{Breuer2009Measure,
  title = {Measure for the Degree of Non-Markovian Behavior of Quantum Processes in Open Systems},
  author = {Breuer, Heinz-Peter and Laine, Elsi-Mari and Piilo, Jyrki},
  journal = {Phys. Rev. Lett.},
  volume = {103},
  issue = {21},
  pages = {210401},
  numpages = {4},
  year = {2009},
  month = {Nov},
  publisher = {American Physical Society},
  doi = {10.1103/PhysRevLett.103.210401},
  url = {https://link.aps.org/doi/10.1103/PhysRevLett.103.210401}
}

@article{Wigner1932Quantum,
  title = {On the Quantum Correction For Thermodynamic Equilibrium},
  author = {Wigner, E.},
  journal = {Phys. Rev.},
  volume = {40},
  issue = {5},
  pages = {749--759},
  numpages = {0},
  year = {1932},
  month = {Jun},
  publisher = {American Physical Society},
  doi = {10.1103/PhysRev.40.749},
  url = {https://link.aps.org/doi/10.1103/PhysRev.40.749}
}

@article{zavatta2004quantum,
  title={Quantum-to-classical transition with single-photon-added coherent states of light},
  author={Zavatta, Alessandro and Viciani, Silvia and Bellini, Marco},
  journal={science},
  volume={306},
  number={5696},
  pages={660--662},
  year={2004},
  publisher={American Association for the Advancement of Science},
  url={https://doi.org/10.1126/science.1103190}
}

@article{Thapliyal2015Quasiprobability,
    title = {Quasiprobability distributions in open quantum systems: Spin-qubit systems},
    journal = {Annals of Physics},
    volume = {362},
    pages = {261-286},
    year = {2015},
    issn = {0003-4916},
    doi = {https://doi.org/10.1016/j.aop.2015.07.029},
    url = {https://www.sciencedirect.com/science/article/pii/S0003491615002948},
    author = {Kishore Thapliyal and Subhashish Banerjee and Anirban Pathak and S. Omkar and V. Ravishankar}
}

@article{Anatole2004Negativity,
doi = {10.1088/1464-4266/6/10/003},
url = {https://dx.doi.org/10.1088/1464-4266/6/10/003},
year = {2004},
month = {aug},
publisher = {},
volume = {6},
number = {10},
pages = {396},
author = {Anatole Kenfack and Karol Życzkowski},
title = {Negativity of the Wigner function as an indicator of non-classicality},
journal = {Journal of Optics B: Quantum and Semiclassical Optics}
}

@article{Wootters1998Entanglement,
  title = {Entanglement of Formation of an Arbitrary State of Two Qubits},
  author = {Wootters, William K.},
  journal = {Phys. Rev. Lett.},
  volume = {80},
  issue = {10},
  pages = {2245--2248},
  numpages = {0},
  year = {1998},
  month = {Mar},
  publisher = {American Physical Society},
  doi = {10.1103/PhysRevLett.80.2245},
  url = {https://link.aps.org/doi/10.1103/PhysRevLett.80.2245}
}

@article{Strasberg2017Quantum,
  title = {Quantum and Information Thermodynamics: A Unifying Framework Based on Repeated Interactions},
  author = {Strasberg, Philipp and Schaller, Gernot and Brandes, Tobias and Esposito, Massimiliano},
  journal = {Phys. Rev. X},
  volume = {7},
  issue = {2},
  pages = {021003},
  numpages = {33},
  year = {2017},
  month = {Apr},
  publisher = {American Physical Society},
  doi = {10.1103/PhysRevX.7.021003},
  url = {https://link.aps.org/doi/10.1103/PhysRevX.7.021003}
}

@article{Pezzutto2016Implications,
doi = {10.1088/1367-2630/18/12/123018},
url = {https://dx.doi.org/10.1088/1367-2630/18/12/123018},
year = {2016},
month = {dec},
publisher = {IOP Publishing},
volume = {18},
number = {12},
pages = {123018},
author = {Pezzutto, Marco and Paternostro, Mauro and Omar, Yasser},
title = {Implications of non-Markovian quantum dynamics for the Landauer bound},
journal = {New Journal of Physics}
}

@article{Campbell2021Collision,
doi = {10.1209/0295-5075/133/60001},
url = {https://dx.doi.org/10.1209/0295-5075/133/60001},
year = {2021},
month = {may},
publisher = {EDP Sciences, IOP Publishing and Società Italiana di Fisica},
volume = {133},
number = {6},
pages = {60001},
author = {Campbell, Steve and Vacchini, Bassano},
title = {Collision models in open system dynamics: A versatile tool for deeper insights?},
journal = {Europhysics Letters}
}

@ARTICLE{wootters1989MUB,
       author = {{Wootters}, William K. and {Fields}, Brian D.},
        title = "{Optimal state-determination by mutually unbiased measurements}",
      journal = {Annals of Physics},
         year = 1989,
        month = may,
       volume = {191},
       number = {2},
        pages = {363-381},
          doi = {10.1016/0003-4916(89)90322-9},
       adsurl = {https://ui.adsabs.harvard.edu/abs/1989AnPhy.191..363W}
}

@article{durt2010mutually,
  title={On mutually unbiased bases},
  author={Durt, Thomas and Englert, Berthold-Georg and Bengtsson, Ingemar and {\.Z}yczkowski, Karol},
  journal={International journal of quantum information},
  volume={8},
  number={04},
  pages={535--640},
  year={2010},
  publisher={World Scientific}
}

@article{Campbell2018Systemenvironment, 
  title = {System-environment correlations and Markovian embedding of quantum non-Markovian dynamics},
  author = {Campbell, Steve and Ciccarello, Francesco and Palma, G. Massimo and Vacchini, Bassano},
  journal = {Phys. Rev. A},
  volume = {98},
  issue = {1},
  pages = {012142},
  numpages = {11},
  year = {2018},
  month = {Jul},
  publisher = {American Physical Society},
  doi = {10.1103/PhysRevA.98.012142},
  url = {https://link.aps.org/doi/10.1103/PhysRevA.98.012142}
}

@article{csenyacsa2022entropy,
  title={Entropy Production in Non-Markovian Collision Models: Information Backflow vs. System-Environment Correlations},
  author={{\c{S}}enya{\c{s}}a, H{\"u}seyin T and Kesgin, {\c{S}}ahinde and Karpat, G{\"o}ktu{\u{g}} and {\c{C}}akmak, Bar{\i}{\c{s}}},
  journal={Entropy},
  volume={24},
  number={6},
  pages={824},
  year={2022},
  publisher={Mdpi},
  URL = {https://www.mdpi.com/1099-4300/24/6/824}
}

@misc{pathania2024,
      title={Quantum Thermodynamics of Open Quantum Systems: Nature of Thermal Fluctuations}, 
      author={Neha Pathania and Devvrat Tiwari and Subhashish Banerjee},
      year={2024},
      eprint={2407.21584},
      archivePrefix={arXiv},
      primaryClass={quant-ph},
      url={https://arxiv.org/abs/2407.21584}, 
}

@article{Hanggi_talkner_review,
  title = {Colloquium: Statistical mechanics and thermodynamics at strong coupling: Quantum and classical},
  author = {Talkner, Peter and H\"anggi, Peter},
  journal = {Rev. Mod. Phys.},
  volume = {92},
  issue = {4},
  pages = {041002},
  numpages = {26},
  year = {2020},
  month = {Oct},
  publisher = {American Physical Society},
  doi = {10.1103/RevModPhys.92.041002},
  url = {https://link.aps.org/doi/10.1103/RevModPhys.92.041002}
}

@Article{Miller2018,
author={Miller, H. J. D.
and Anders, J.},
title={Energy-temperature uncertainty relation in quantum thermodynamics},
journal={Nature Communications},
year={2018},
month={Jun},
day={06},
volume={9},
number={1},
pages={2203},
issn={2041-1723},
doi={10.1038/s41467-018-04536-7}
}

@article{Ciccarello2022Quantumcollisionmodel,
title = {Quantum collision models: Open system dynamics from repeated interactions},
journal = {Physics Reports},
volume = {954},
pages = {1-70},
year = {2022},
note = {Quantum collision models: Open system dynamics from repeated interactions},
issn = {0370-1573},
doi = {https://doi.org/10.1016/j.physrep.2022.01.001},
url = {https://www.sciencedirect.com/science/article/pii/S0370157322000035},
author = {Francesco Ciccarello and Salvatore Lorenzo and Vittorio Giovannetti and G. Massimo Palma}
}

@article{Rodrigues2019thermodynamics,
  title = {Thermodynamics of Weakly Coherent Collisional Models},
  author = {Rodrigues, Franklin L. S. and De Chiara, Gabriele and Paternostro, Mauro and Landi, Gabriel T.},
  journal = {Phys. Rev. Lett.},
  volume = {123},
  issue = {14},
  pages = {140601},
  numpages = {6},
  year = {2019},
  month = {Oct},
  publisher = {American Physical Society},
  doi = {10.1103/PhysRevLett.123.140601},
  url = {https://link.aps.org/doi/10.1103/PhysRevLett.123.140601}
}

@article{Thomas2018thermodynamics,
  title = {Thermodynamics of non-Markovian reservoirs and heat engines},
  author = {Thomas, George and Siddharth, Nana and Banerjee, Subhashish and Ghosh, Sibasish},
  journal = {Phys. Rev. E},
  volume = {97},
  issue = {6},
  pages = {062108},
  numpages = {8},
  year = {2018},
  month = {Jun},
  publisher = {American Physical Society},
  doi = {10.1103/PhysRevE.97.062108},
  url = {https://link.aps.org/doi/10.1103/PhysRevE.97.062108}
}

@article{Ashutosh2023thermodynamics,
title = {Thermodynamics of one and two-qubit nonequilibrium heat engines running between squeezed thermal reservoirs},
journal = {Physica A: Statistical Mechanics and its Applications},
volume = {623},
pages = {128832},
year = {2023},
issn = {0378-4371},
doi = {https://doi.org/10.1016/j.physa.2023.128832},
url = {https://www.sciencedirect.com/science/article/pii/S0378437123003874},
author = {Ashutosh Kumar and Sourabh Lahiri and Trilochan Bagarti and Subhashish Banerjee}
}

@article{Naikoo2019Facets,
  title = {Facets of quantum information under non-Markovian evolution},
  author = {Naikoo, Javid and Dutta, Supriyo and Banerjee, Subhashish},
  journal = {Phys. Rev. A},
  volume = {99},
  issue = {4},
  pages = {042128},
  numpages = {7},
  year = {2019},
  month = {Apr},
  publisher = {American Physical Society},
  doi = {10.1103/PhysRevA.99.042128},
  url = {https://link.aps.org/doi/10.1103/PhysRevA.99.042128}
}

@article{Tiwari2023QuantumCorrelations,
author = {Tiwari, Devvrat and Paulson, Kavalambramalil G. and Banerjee, Subhashish},
title = {Quantum Correlations and Speed Limit of Central Spin Systems},
journal = {Annalen der Physik},
volume = {535},
number = {2},
pages = {2200452},
doi = {https://doi.org/10.1002/andp.202200452},
url = {https://onlinelibrary.wiley.com/doi/abs/10.1002/andp.202200452},
year = {2023}
}

@article{Landi2021Irreversibleentropy,
  title = {Irreversible entropy production: From classical to quantum},
  author = {Landi, Gabriel T. and Paternostro, Mauro},
  journal = {Rev. Mod. Phys.},
  volume = {93},
  issue = {3},
  pages = {035008},
  numpages = {58},
  year = {2021},
  month = {Sep},
  publisher = {American Physical Society},
  doi = {10.1103/RevModPhys.93.035008},
  url = {https://link.aps.org/doi/10.1103/RevModPhys.93.035008}
}

@article{DeChiara_2018Reconciliation,
    doi = {10.1088/1367-2630/aaecee},
    url = {https://dx.doi.org/10.1088/1367-2630/aaecee},
    year = {2018},
    month = {nov},
    publisher = {IOP Publishing},
    volume = {20},
    number = {11},
    pages = {113024},
    author = {De Chiara, Gabriele and Landi, Gabriel and Hewgill, Adam and Reid, Brendan and Ferraro, Alessandro and Roncaglia, Augusto J and Antezza, Mauro},
    title = {Reconciliation of quantum local master equations with thermodynamics},
    journal = {New Journal of Physics}
}

@article{Cattaneo2021Collision,
  title = {Collision Models Can Efficiently Simulate Any Multipartite Markovian Quantum Dynamics},
  author = {Cattaneo, Marco and De Chiara, Gabriele and Maniscalco, Sabrina and Zambrini, Roberta and Giorgi, Gian Luca},
  journal = {Phys. Rev. Lett.},
  volume = {126},
  issue = {13},
  pages = {130403},
  numpages = {8},
  year = {2021},
  month = {Apr},
  publisher = {American Physical Society},
  doi = {10.1103/PhysRevLett.126.130403},
  url = {https://link.aps.org/doi/10.1103/PhysRevLett.126.130403}
}

@article{Cattaneo2022ABriefJourney,
author = {Cattaneo, Marco and Giorgi, Gian Luca and Zambrini, Roberta and Maniscalco, Sabrina},
title = {A Brief Journey through Collision Models for Multipartite Open Quantum Dynamics},
journal = {Open Systems \& Information Dynamics},
volume = {29},
number = {03},
pages = {2250015},
year = {2022},
doi = {10.1142/S1230161222500159},
URL = { https://doi.org/10.1142/S1230161222500159}
}

@article{Saha_2024_quantum,
doi = {10.1088/1367-2630/ad212f},
url = {https://dx.doi.org/10.1088/1367-2630/ad212f},
year = {2024},
month = {feb},
publisher = {IOP Publishing},
volume = {26},
number = {2},
pages = {023011},
author = {Saha, Tanmay and Das, Arpan and Ghosh, Sibasish},
title = {Quantum homogenization in non-Markovian collisional model},
journal = {New Journal of Physics}
}

@article{Li2024Witnessing,
  title = {Witnessing non-Markovianity with Gaussian quantum steering in a collision model},
  author = {Li, Yan and Li, Xingli and Jin, Jiasen},
  journal = {Phys. Rev. A},
  volume = {109},
  issue = {5},
  pages = {052201},
  numpages = {13},
  year = {2024},
  month = {May},
  publisher = {American Physical Society},
  doi = {10.1103/PhysRevA.109.052201},
  url = {https://link.aps.org/doi/10.1103/PhysRevA.109.052201}
}

@article{Laine2010Measure,
  title = {Measure for the non-Markovianity of quantum processes},
  author = {Laine, Elsi-Mari and Piilo, Jyrki and Breuer, Heinz-Peter},
  journal = {Phys. Rev. A},
  volume = {81},
  issue = {6},
  pages = {062115},
  numpages = {8},
  year = {2010},
  month = {Jun},
  publisher = {American Physical Society},
  doi = {10.1103/PhysRevA.81.062115},
  url = {https://link.aps.org/doi/10.1103/PhysRevA.81.062115}
}

@article{li2021controllable,
  title={Controllable phase-dependent Wigner-function negativity at steady state via parametric driving and feedback},
  author={Li, Jiahua and Ding, Chunling and Wu, Ying},
  journal={Journal of Applied Physics},
  volume={129},
  number={12},
  year={2021},
  publisher={AIP Publishing},
  doi={https://doi.org/10.1063/5.0041406}
}

@article{Talkner2020colloquium,
  title = {Colloquium: Statistical mechanics and thermodynamics at strong coupling: Quantum and classical},
  author = {Talkner, Peter and H\"anggi, Peter},
  journal = {Rev. Mod. Phys.},
  volume = {92},
  issue = {4},
  pages = {041002},
  numpages = {26},
  year = {2020},
  month = {Oct},
  publisher = {American Physical Society},
  doi = {10.1103/RevModPhys.92.041002},
  url = {https://link.aps.org/doi/10.1103/RevModPhys.92.041002}
}

@article{entanglement_generation_paper,
  title = {Non-Markovian Effects on the Dynamics of Entanglement},
  author = {Bellomo, B. and Lo Franco, R. and Compagno, G.},
  journal = {Phys. Rev. Lett.},
  volume = {99},
  issue = {16},
  pages = {160502},
  numpages = {4},
  year = {2007},
  month = {Oct},
  publisher = {American Physical Society},
  doi = {10.1103/PhysRevLett.99.160502},
  url = {https://link.aps.org/doi/10.1103/PhysRevLett.99.160502}
}

@article{lalita2025non_classicality,
  title = {Two-qubit quantum collision model: non-Markovianity and nonclassicality},
  author = {Lalita, Jai and Banerjee, Subhashish},
  journal = {Phys. Rev. A},
  volume = {113},
  issue = {1},
  pages = {012225},
  numpages = {11},
  year = {2026},
  month = {Jan},
  publisher = {American Physical Society},
  doi = {10.1103/8rtv-ftrr},
  url = {https://link.aps.org/doi/10.1103/8rtv-ftrr}
}

@article{Kochen_Specker_contextuality2022Budroni,
  title = {Kochen-Specker contextuality},
  author = {Budroni, Costantino and Cabello, Ad\'an and G\"uhne, Otfried and Kleinmann, Matthias and Larsson, Jan-\AA{}ke},
  journal = {Rev. Mod. Phys.},
  volume = {94},
  issue = {4},
  pages = {045007},
  numpages = {62},
  year = {2022},
  month = {Dec},
  publisher = {American Physical Society},
  doi = {10.1103/RevModPhys.94.045007},
  url = {https://link.aps.org/doi/10.1103/RevModPhys.94.045007}
}

@article{Quantum_Coherence2020Francica,
  title = {Quantum Coherence and Ergotropy},
  author = {Francica, G. and Binder, F. C. and Guarnieri, G. and Mitchison, M. T. and Goold, J. and Plastina, F.},
  journal = {Phys. Rev. Lett.},
  volume = {125},
  issue = {18},
  pages = {180603},
  numpages = {8},
  year = {2020},
  month = {Oct},
  publisher = {American Physical Society},
  doi = {10.1103/PhysRevLett.125.180603},
  url = {https://link.aps.org/doi/10.1103/PhysRevLett.125.180603}
}

@article{Horodecki2009quantum_ent,
  title = {Quantum entanglement},
  author = {Horodecki, Ryszard and Horodecki, Pawe\l{} and Horodecki, Micha\l{} and Horodecki, Karol},
  journal = {Rev. Mod. Phys.},
  volume = {81},
  issue = {2},
  pages = {865--942},
  numpages = {0},
  year = {2009},
  month = {Jun},
  publisher = {American Physical Society},
  doi = {10.1103/RevModPhys.81.865},
  url = {https://link.aps.org/doi/10.1103/RevModPhys.81.865}
}

@article{Esposito2010Threefaces,
  title = {Three faces of the second law. I. Master equation formulation},
  author = {Esposito, Massimiliano and Van den Broeck, Christian},
  journal = {Phys. Rev. E},
  volume = {82},
  issue = {1},
  pages = {011143},
  numpages = {10},
  year = {2010},
  month = {Jul},
  publisher = {American Physical Society},
  doi = {10.1103/PhysRevE.82.011143},
  url = {https://link.aps.org/doi/10.1103/PhysRevE.82.011143}
}

@article{AEAllahverdyan_2004Maximalwork,
doi = {10.1209/epl/i2004-10101-2},
url = {https://dx.doi.org/10.1209/epl/i2004-10101-2},
year = {2004},
month = {aug},
publisher = {},
volume = {67},
number = {4},
pages = {565},
author = {A. E. Allahverdyan and R. Balian and Th. M. Nieuwenhuizen},
title = {Maximal work extraction from finite quantum systems},
journal = {Europhysics Letters}
}

@article{cakmak2020Ergotropy,
  title = {Ergotropy from coherences in an open quantum system},
  author = {Cakmak Baris},
  journal = {Phys. Rev. E},
  volume = {102},
  issue = {4},
  pages = {042111},
  numpages = {11},
  year = {2020},
  month = {Oct},
  publisher = {American Physical Society},
  doi = {10.1103/PhysRevE.102.042111},
  url = {https://link.aps.org/doi/10.1103/PhysRevE.102.042111}
}

@article{ThermalizingQuantumMachines_2002,
  title = {Thermalizing Quantum Machines: Dissipation and Entanglement},
  author = {Scarani, Valerio and Ziman, M\'ario and \ifmmode \check{S}\else \v{S}\fi{}telmachovi\ifmmode \check{c}\else \v{c}\fi{}, Peter and Gisin, Nicolas and Bu\ifmmode \check{z}\else \v{z}\fi{}ek, Vladim\'{\i}r},
  journal = {Phys. Rev. Lett.},
  volume = {88},
  issue = {9},
  pages = {097905},
  numpages = {4},
  year = {2002},
  month = {Feb},
  publisher = {American Physical Society},
  doi = {10.1103/PhysRevLett.88.097905},
  url = {https://link.aps.org/doi/10.1103/PhysRevLett.88.097905}
}

@article{Breuer_2012_foundations,
doi = {10.1088/0953-4075/45/15/154001},
url = {https://dx.doi.org/10.1088/0953-4075/45/15/154001},
year = {2012},
month = {jul},
publisher = {IOP Publishing},
volume = {45},
number = {15},
pages = {154001},
author = {Breuer, Heinz-Peter},
title = {Foundations and measures of quantum non-Markovianity},
journal = {Journal of Physics B: Atomic, Molecular and Optical Physics}
}

@article{Garraway1997Decay,
  title = {Decay of an atom coupled strongly to a reservoir},
  author = {Garraway, B. M.},
  journal = {Phys. Rev. A},
  volume = {55},
  issue = {6},
  pages = {4636--4639},
  numpages = {0},
  year = {1997},
  month = {Jun},
  publisher = {American Physical Society},
  doi = {10.1103/PhysRevA.55.4636},
  url = {https://link.aps.org/doi/10.1103/PhysRevA.55.4636}
}

@article{Breuer1999Stochastic,
  title = {Stochastic wave-function method for non-Markovian quantum master equations},
  author = {Breuer, Heinz-Peter and Kappler, Bernd and Petruccione, Francesco},
  journal = {Phys. Rev. A},
  volume = {59},
  issue = {2},
  pages = {1633--1643},
  numpages = {0},
  year = {1999},
  month = {Feb},
  publisher = {American Physical Society},
  doi = {10.1103/PhysRevA.59.1633},
  url = {https://link.aps.org/doi/10.1103/PhysRevA.59.1633}
}

@article{Jaynes1963Comparison,
  author={Jaynes, E.T. and Cummings, F.W.},
  journal={Proceedings of the IEEE}, 
  title={Comparison of quantum and semiclassical radiation theories with application to the beam maser}, 
  year={1963},
  volume={51},
  number={1},
  pages={89-109},
  doi={10.1109/PROC.1963.1664}
}

@article{Thapliyal2016tomograms,
title = {Tomograms for open quantum systems: In(finite) dimensional optical and spin systems},
journal = {Annals of Physics},
volume = {366},
pages = {148-167},
year = {2016},
issn = {0003-4916},
doi = {https://doi.org/10.1016/j.aop.2016.01.010},
url = {https://www.sciencedirect.com/science/article/pii/S0003491616000129},
author = {Kishore Thapliyal and Subhashish Banerjee and Anirban Pathak}
}

@article{Jha2025probing,
  title = {Probing the quantum speed limit and entanglement in flavor oscillations of neutrino-antineutrino system in curved spacetime},
  author = {Jha, Abhishek Kumar and Dutta, Mriganka and Pathak, Mayank and Banerjee, Subhashish and Mukhopadhyay, Banibrata},
  journal = {Phys. Rev. D},
  volume = {112},
  issue = {4},
  pages = {043045},
  numpages = {21},
  year = {2025},
  month = {Aug},
  publisher = {American Physical Society},
  doi = {10.1103/w9gw-j8zv},
  url = {https://link.aps.org/doi/10.1103/w9gw-j8zv}
}

@article{Bekenstein1973Blackholes,
  title = {Black Holes and Entropy},
  author = {Bekenstein, Jacob D.},
  journal = {Phys. Rev. D},
  volume = {7},
  issue = {8},
  pages = {2333--2346},
  numpages = {0},
  year = {1973},
  month = {Apr},
  publisher = {American Physical Society},
  doi = {10.1103/PhysRevD.7.2333},
  url = {https://link.aps.org/doi/10.1103/PhysRevD.7.2333}
}

@article{Tiwari2023Impact,
AUTHOR={Tiwari, Devvrat  and Banerjee, Subhashish },
TITLE={Impact of non-Markovian evolution on characterizations of quantum thermodynamics},
JOURNAL={Frontiers in Quantum Science and Technology},
VOLUME={2},
YEAR={2023},
pages={1207552},
URL={https://www.frontiersin.org/journals/quantum-science-and-technology/articles/10.3389/frqst.2023.1207552},
DOI={10.3389/frqst.2023.1207552},
ISSN={2813-2181}
}

@article{pathania2025quantum,
  title={Quantum thermodynamics of open quantum systems: Nature of thermal fluctuations},
  author={Pathania, Neha and Tiwari, Devvrat and Banerjee, Subhashish},
  journal={Quantum Information Processing},
  volume={24},
  number={9},
  pages={290},
  year={2025},
  publisher={Springer}, 
  doi={10.1007/s11128-025-04903-6},
  url={https://doi.org/10.1007/s11128-025-04903-6}
}

@article{loss1998quantum,
  title={Quantum computation with quantum dots},
  author={Loss, Daniel and DiVincenzo, David P},
  journal={Physical Review A},
  volume={57},
  number={1},
  pages={120},
  year={1998},
  publisher={APS}, 
  doi={10.1103/PhysRevA.57.120},
  url={https://doi.org/10.1103/PhysRevA.57.120}
}

@article{Lalita_2024ProtectingQC,
doi = {10.1088/1402-4896/ad273e},
url = {https://dx.doi.org/10.1088/1402-4896/ad273e},
year = {2024},
month = {feb},
publisher = {IOP Publishing},
volume = {99},
number = {3},
pages = {035116},
author = {Jai Lalita and Subhashish Banerjee},
title = {Protecting quantum correlations of negative quantum states using weak measurement under non-Markovian noise},
journal = {Physica Scripta},
}

@article{lalita2025realizingnegativequantumstates,
  title = {Noise-resilient negative quantum states},
  author = {Lalita, Jai and Iyer, Pavithran and Banerjee, Subhashish},
  journal = {Phys. Rev. A},
  volume = {113},
  issue = {2},
  pages = {022427},
  numpages = {12},
  year = {2026},
  month = {Feb},
  publisher = {American Physical Society},
  doi = {10.1103/g3f6-p18d},
  url = {https://link.aps.org/doi/10.1103/g3f6-p18d}
}

@article{Srikanth2008Squeezed,
  title = {Squeezed generalized amplitude damping channel},
  author = {Srikanth, R. and Banerjee, Subhashish},
  journal = {Phys. Rev. A},
  volume = {77},
  issue = {1},
  pages = {012318},
  numpages = {9},
  year = {2008},
  month = {Jan},
  publisher = {American Physical Society},
  doi = {10.1103/PhysRevA.77.012318},
  url = {https://link.aps.org/doi/10.1103/PhysRevA.77.012318}
}

@article{czerwinski2022dynamics,
  title={Dynamics of open quantum systems—Markovian semigroups and beyond},
  author={Czerwinski, Artur},
  journal={Symmetry},
  volume={14},
  number={8},
  pages={1752},
  year={2022},
  publisher={MDPI}, 
  url={https://doi.org/10.3390/sym14081752}
}

@article{utagi2020temporal,
  title={Temporal self-similarity of quantum dynamical maps as a concept of memorylessness},
  author={Utagi, Shrikant and Srikanth, R and Banerjee, Subhashish},
  journal={Scientific Reports},
  volume={10},
  number={1},
  pages={1--10},
  year={2020},
  publisher={Nature Publishing Group},
  doi={10.1038/s41598-020-72211-3},
  url={https://doi.org/10.1038/s41598-020-72211-3}
}

@article{Breuer2004Non_Markovian,
  title = {Non-Markovian dynamics in a spin star system: Exact solution and approximation techniques},
  author = {Breuer, Heinz-Peter and Burgarth, Daniel and Petruccione, Francesco},
  journal = {Phys. Rev. B},
  volume = {70},
  issue = {4},
  pages = {045323},
  numpages = {10},
  year = {2004},
  month = {Jul},
  publisher = {American Physical Society},
  doi = {10.1103/PhysRevB.70.045323},
  url = {https://link.aps.org/doi/10.1103/PhysRevB.70.045323}
}

@article{He2019Exact,
  title = {Exact quantum dynamics of XXZ central spin problems},
  author = {He, Wen-Bin and Chesi, Stefano and Lin, Hai-Qing and Guan, Xi-Wen},
  journal = {Phys. Rev. B},
  volume = {99},
  issue = {17},
  pages = {174308},
  numpages = {6},
  year = {2019},
  month = {May},
  publisher = {American Physical Society},
  doi = {10.1103/PhysRevB.99.174308},
  url = {https://link.aps.org/doi/10.1103/PhysRevB.99.174308}
}

@article{Mukhopadhyay2017Dynamics,
  title = {Dynamics and thermodynamics of a central spin immersed in a spin bath},
  author = {Mukhopadhyay, Chiranjib and Bhattacharya, Samyadeb and Misra, Avijit and Pati, Arun Kumar},
  journal = {Phys. Rev. A},
  volume = {96},
  issue = {5},
  pages = {052125},
  numpages = {13},
  year = {2017},
  month = {Nov},
  publisher = {American Physical Society},
  doi = {10.1103/PhysRevA.96.052125},
  url = {https://link.aps.org/doi/10.1103/PhysRevA.96.052125}
}

@book{Larson2021TheJaynes–Cummings,
author = {Larson, Jonas and Mavrogordatos, Themistoklis},
title = {The Jaynes–Cummings Model and Its Descendants},
publisher = {IOP Publishing},
year = {2021},
series = {2053-2563},
isbn = {978-0-7503-3447-1},
url = {https://doi.org/10.1088/978-0-7503-3447-1},
doi = {10.1088/978-0-7503-3447-1}
}

@article{lacroix2025making,
  title={Making quantum collision models exact},
  author={Lacroix, Thibaut and Cilluffo, Dario and Huelga, Susana F and Plenio, Martin B},
  journal={Communications Physics},
  volume={8},
  number={1},
  pages={268},
  year={2025},
  publisher={Nature Publishing Group UK London},
  url={https://doi.org/10.1038/s42005-025-02201-2},
  doi={10.1038/s42005-025-02201-2}
}

@article{Sudarshan1963Equivalence,
  title = {Equivalence of Semiclassical and Quantum Mechanical Descriptions of Statistical Light Beams},
  author = {Sudarshan, E. C. G.},
  journal = {Phys. Rev. Lett.},
  volume = {10},
  issue = {7},
  pages = {277--279},
  numpages = {0},
  year = {1963},
  month = {Apr},
  publisher = {American Physical Society},
  doi = {10.1103/PhysRevLett.10.277},
  url = {https://link.aps.org/doi/10.1103/PhysRevLett.10.277}
}

@article{Vittorio2004Quantum_Enhanced,
author = {Vittorio Giovannetti  and Seth Lloyd  and Lorenzo Maccone },
title = {Quantum-Enhanced Measurements: Beating the Standard Quantum Limit},
journal = {Science},
volume = {306},
number = {5700},
pages = {1330-1336},
year = {2004},
doi = {10.1126/science.1104149},
URL = {https://www.science.org/doi/abs/10.1126/science.1104149}
}

@article{dowling2003quantum,
  title={Quantum technology: the second quantum revolution},
  author={Dowling, Jonathan P and Milburn, Gerard J},
  journal={Philosophical Transactions of the Royal Society of London. Series A: Mathematical, Physical and Engineering Sciences},
  volume={361},
  number={1809},
  pages={1655--1674},
  year={2003},
  publisher={The Royal Society},
  url={https://doi.org/10.1098/rsta.2003.1227}
}

@article{Horodecki2009QuantumEntanglement,
  title = {Quantum entanglement},
  author = {Horodecki, Ryszard and Horodecki, Pawe\l{} and Horodecki, Micha\l{} and Horodecki, Karol},
  journal = {Rev. Mod. Phys.},
  volume = {81},
  issue = {2},
  pages = {865--942},
  numpages = {0},
  year = {2009},
  month = {Jun},
  publisher = {American Physical Society},
  doi = {10.1103/RevModPhys.81.865},
  url = {https://link.aps.org/doi/10.1103/RevModPhys.81.865}
}

@article{BANERJEE2023Thermalization,
title = {Thermalization in quenched open quantum cosmology},
journal = {Nuclear Physics B},
volume = {996},
pages = {116368},
year = {2023},
issn = {0550-3213},
doi = {https://doi.org/10.1016/j.nuclphysb.2023.116368},
url = {https://www.sciencedirect.com/science/article/pii/S0550321323002973},
author = {Subhashish Banerjee and Sayantan Choudhury and Satyaki Chowdhury and Johannes Knaute and Sudhakar Panda and K. Shirish}
}

@article{arisoy2019thermalization,
  title={Thermalization of finite many-body systems by a collision model},
  author={Ar{\i}soy, Onat and Campbell, Steve and M{\"u}stecapl{\i}o{\u{g}}lu, {\"O}zg{\"u}r E},
  journal={Entropy},
  volume={21},
  number={12},
  pages={1182},
  year={2019},
  publisher={MDPI}, 
  url={https://doi.org/10.3390/e21121182}
}

@article{Medina2025Anomalous,
  title = {Anomalous Discharging of Quantum Batteries: The Ergotropic Mpemba Effect},
  author = {Medina, Ivan and Culhane, Ois\'{\i}n and Binder, Felix C. and Landi, Gabriel T. and Goold, John},
  journal = {Phys. Rev. Lett.},
  volume = {134},
  issue = {22},
  pages = {220402},
  numpages = {6},
  year = {2025},
  month = {Jun},
  publisher = {American Physical Society},
  doi = {10.1103/PhysRevLett.134.220402},
  url = {https://link.aps.org/doi/10.1103/PhysRevLett.134.220402}
}

@article{Manfredi2000Entropy,
  title = {Entropy and Wigner functions},
  author = {Manfredi, G. and Feix, M. R.},
  journal = {Phys. Rev. E},
  volume = {62},
  issue = {4},
  pages = {4665--4674},
  numpages = {0},
  year = {2000},
  month = {Oct},
  publisher = {American Physical Society},
  doi = {10.1103/PhysRevE.62.4665},
  url = {https://link.aps.org/doi/10.1103/PhysRevE.62.4665}
}

@article{omkar2013dissipative,
  title={Dissipative and non-dissipative single-qubit channels: dynamics and geometry},
  author={Omkar, S and Srikanth, R and Banerjee, Subhashish},
  journal={Quantum information processing},
  volume={12},
  number={12},
  pages={3725--3744},
  year={2013},
  publisher={Springer}, 
  doi={10.1007/s11128-013-0628-3},
  url={https://doi.org/10.1007/s11128-013-0628-3}
}

@article{Chandrashekar2007Symmetries,
  title = {Symmetries and noise in quantum walk},
  author = {Chandrashekar, C. M. and Srikanth, R. and Banerjee, Subhashish},
  journal = {Phys. Rev. A},
  volume = {76},
  issue = {2},
  pages = {022316},
  numpages = {15},
  year = {2007},
  month = {Aug},
  publisher = {American Physical Society},
  doi = {10.1103/PhysRevA.76.022316},
  url = {https://link.aps.org/doi/10.1103/PhysRevA.76.022316}
}

@article{Tiwari2022Dynamics,
  title = {Dynamics of two central spins immersed in spin baths},
  author = {Tiwari, Devvrat and Datta, Shounak and Bhattacharya, Samyadeb and Banerjee, Subhashish},
  journal = {Phys. Rev. A},
  volume = {106},
  issue = {3},
  pages = {032435},
  numpages = {15},
  year = {2022},
  month = {Sep},
  publisher = {American Physical Society},
  doi = {10.1103/PhysRevA.106.032435},
  url = {https://link.aps.org/doi/10.1103/PhysRevA.106.032435}
}

@article{Perarnau2015Extractable,
  title = {Extractable Work from Correlations},
  author = {Perarnau-Llobet, Mart\'{\i} and Hovhannisyan, Karen V. and Huber, Marcus and Skrzypczyk, Paul and Brunner, Nicolas and Ac\'{\i}n, Antonio},
  journal = {Phys. Rev. X},
  volume = {5},
  issue = {4},
  pages = {041011},
  numpages = {14},
  year = {2015},
  month = {Oct},
  publisher = {American Physical Society},
  doi = {10.1103/PhysRevX.5.041011},
  url = {https://link.aps.org/doi/10.1103/PhysRevX.5.041011}
}

@article{NV2000Theory,
doi = {10.1088/0034-4885/63/4/204},
url = {https://doi.org/10.1088/0034-4885/63/4/204},
year = {2000},
month = {apr},
publisher = {},
volume = {63},
number = {4},
pages = {669},
author = {N V Prokof'ev and P C E Stamp},
title = {Theory 
of the spin bath},
journal = {Reports on Progress in Physics}
}

@article{Francica2020QuantumCoherence,
  title = {Quantum Coherence and Ergotropy},
  author = {Francica, G. and Binder, F. C. and Guarnieri, G. and Mitchison, M. T. and Goold, J. and Plastina, F.},
  journal = {Phys. Rev. Lett.},
  volume = {125},
  issue = {18},
  pages = {180603},
  numpages = {8},
  year = {2020},
  month = {Oct},
  publisher = {American Physical Society},
  doi = {10.1103/PhysRevLett.125.180603},
  url = {https://link.aps.org/doi/10.1103/PhysRevLett.125.180603}
}

@article{Andolina_OQS_battery,
  title = {Charger-mediated energy transfer for quantum batteries: An open-system approach},
  author = {Farina, Donato and Andolina, Gian Marcello and Mari, Andrea and Polini, Marco and Giovannetti, Vittorio},
  journal = {Phys. Rev. B},
  volume = {99},
  issue = {3},
  pages = {035421},
  numpages = {15},
  year = {2019},
  month = {Jan},
  publisher = {American Physical Society},
  doi = {10.1103/PhysRevB.99.035421},
  url = {https://link.aps.org/doi/10.1103/PhysRevB.99.035421}
}


\end{document}